\documentclass[12pt,a4paper, oneside]{book}
\usepackage{subfigure}
\usepackage{graphicx}  
\usepackage{float}

\usepackage{filehook}

\usepackage[toc,page]{appendix}

\usepackage{titlesec}
\titleformat{\chapter}[display]
{\normalfont\bfseries\filcenter}
{\LARGE\thechapter}
{1ex}
{\titlerule[2pt]
\vspace{2ex}
\LARGE}
[\vspace{1ex}
{\titlerule[2pt]}]
\usepackage{amsmath,amssymb,amsthm}
\usepackage{mathtools}
\usepackage[mathscr]{euscript}
\usepackage{epsf}
\usepackage{graphicx}
\usepackage{framed}
\usepackage{wrapfig, framed, caption}
\usepackage{cite}
\usepackage{xcolor}
\usepackage{caption}
\usepackage{footmisc}
\usepackage{url}
\usepackage{fancyheadings}
\usepackage{xparse}

\let\oldquote\quote
\let\endoldquote\endquote

\RenewDocumentEnvironment{quote}{om}
  {\oldquote}
  {\par\nobreak\smallskip
   \hfill(#2\IfValueT{#1}{~---~#1})\endoldquote 
   \addvspace{\bigskipamount}}

\usepackage[T1]{fontenc}
\usepackage[ Sonny ]{fncychap}

\usepackage[utf8]{inputenc}
\usepackage[colorlinks=true,linktocpage=true,linkcolor=blue,citecolor=blue]{hyperref}

\ChNameVar{\Large\fontfamily{put}\selectfont\color{black}}
\ChNumVar{\fontsize{60}{62}\fontfamily{pzc}\selectfont\textcolor{black}}
\ChTitleVar{\Huge\fontfamily{phv}\selectfont\scshape\color{black}}
\def\bwt{\begin{widetext}}
\def\ewt{\end{widetext}}
\def\be{\begin{equation}}
\def\ee{\end{equation}}
\def\bea{\begin{eqnarray}}
\def\eea{\end{eqnarray}}
\def\bean{\begin{eqnarray*}}
\def\eean{\end{eqnarray*}}
\def\bary{\begin{array}}
\def\eary{\end{array}}
\def\bit{\begin{itemize}}
\def\eit{\end{itemize}}
\def\GeV{\,{\rm GeV}}

\def\su5u1{SU(5) \times U(1)}
\def\fsu5u1{SU(5) \times U(1)'}
\def\so10{SO(10)}
\def\sq20{SO(10) \times SO(10)}

\def\bwt{\begin{widetext}}
\def\ewt{\end{widetext}}
\def\be{\begin{equation}}
\def\ee{\end{equation}}
\def\bea{\begin{eqnarray}}
\def\eea{\end{eqnarray}}
\def\bean{\begin{eqnarray*}}
\def\eean{\end{eqnarray*}}
\def\bary{\begin{array}}
\def\eary{\end{array}}
\def\bit{\begin{itemize}}
\def\eit{\end{itemize}}
\def\GeV{\rm GeV}

\def\su5u1{SU(5) \times U(1)}
\def\fsu5u1{SU(5) \times U(1)'}
\def\so10{SO(10)}
\def\sq20{SO(10) \times SO(10)}

\begin{document}

\pagestyle{fancy}
\pagenumbering{roman}
\begin{center}
{\LARGE \bf {\color{black} {\Huge S}TUDY {\huge O}F {\huge P}OTENTIAL  \\ [0.5cm]
{\huge S}ELF-ANNIHILATION {\huge S}IGNAL {\huge F}ROM \\ [0.5cm]
{\huge D}ARK  {\huge M}ATTER {\huge P}ARTICLES {\huge I}N \\ [0.5cm]
{\huge S}OME {\huge P}ROSPECTIVE \\ [0.5cm]
{\huge A}STROPHYSICAL {\huge D}ARK {\huge M}ATTER \\ [0.5cm]
{\huge S}OURCES}}\\ [4.0cm]
{\large\bf Thesis Submitted For The Degree Of \\[0.3cm]
{\bf DOCTOR OF PHILOSOPHY (SCIENCE)} \\[0.3cm]
in \\[0.3cm]
Physics (Theoretical)\\[0.3cm]
by}\\[0.3cm]
{\Large\bf {\huge P}OOJA {\huge B}HATTACHARJEE}\\

\vskip 4.0cm

{\large {\bf DEPARTMENT OF PHYSICS\\[0.3cm]
UNIVERSITY OF CALCUTTA\\[0.3cm]}}
{\Large\bf2020} \\
\end{center}

\thispagestyle{empty} 
\chapter*{Dedication}
\addcontentsline{toc}{chapter}{Dedication} 



\begin{center}
I would like to dedicate this thesis to my loving parents\\
{\it \textbf{Bina Bhattacharjee} $\&$\\
\textbf{Sudipta Bhattacharjee}}\\
for their endless love, support and encouragement
\end{center}



\chapter*{Acknowledgement}
\addcontentsline{toc}{chapter}{Acknowledgement} 




\noindent Throughout the journey of my PhD and even in the writing of this thesis, I have fortunately received a great deal of support and assistance by so many people, directly or indirectly, that it is a very difficult task for me to acknowledge them all. So, before proceeding further, I must say that the following is not at all a comprehensive endeavour.\\

\noindent First of all, I want to thank my PhD supervisor Dr. Parthasarathi Joarder for providing me with the opportunity to work in the field of astrophysics. Without his support, attention and guidance, none of this would have been possible. His insightful feedback, motivation, continuous support and above all constant questioning pushed me to sharpen my thinking and brought my work to a higher level. In this process, I have gradually learnt the subject with more accuracy by recognising the pre-possessed contradictions or misconceptions in me.\\

\noindent I also acknowledge the fruitful interactions with my joint supervisor Prof. Dhruba Gupta. Several discussions with him have enriched my comprehension of the subject.\\

\noindent I would also like to pay my special regards to Prof. Pratik Majumdar, without whom my PhD career would not have been taken-off. His perceptive feedback, consistent patience, guidance and all insightful questioning shape my PhD work. Specifically, I deeply appreciate his approach of rigorously participating in thoughtful argumentative dialogue in order to elicit the ideas and underlying presuppositions within me. This intense disciplined questioning, yet insouciant approach deeply help me in examining any thoughts logically and determining the validity of those ideas. Without his persistent help, the goal of this thesis would not have been realised. I am also grateful to Prof. Mousumi Das for her valuable insights into my research. I gratefully acknowledge the fruitful interactions with her.\\

\noindent I would like to acknowledge my senior colleague, Dr. Sayan Biswas, for his insightful academic discussion in various stages of my PhD. \\

\noindent I would also like to thank my other collaborators, Prof. Dilip Kumar Ghosh, Prof. Debajyoti Choudhury, Prof. Subinoy Das, Dr. Tulun Ergin, Dr. Kasinath Das, Dr. Lab Saha, for their valuable inputs in my research career. \\

\noindent I will forever be indebted towards my two colleagues and friends Sananda Raychaudhuri and Kaushik Naskar. Without them, this journey would not have been completed. I will sincerely miss the friendly, supportive and relaxed environment we maintained over these years. Without their constant support and intense academic or non-academic discussions would not have made this bumpy road to a smooth one.\\

\noindent I am grateful to my friends in Bose Institute: Pracheta Singha, Sumana Bhattacharjee, Deeptak Biswas, Souradeep Sasmal, Debarshi Das and Som Kanjilal.\\

\noindent In addition, I want to thank the Bose Institute for the infrastructure provided to pursue my PhD. I am also grateful to DST-INSPIRE for financial support. Without their support and funding, this thesis could not have reached its goal.\\

\noindent I gratefully acknowledge my teachers from school and college- Dr. Arup Roy, Dr. Satadal Bhattacharyya, Dr. Upendranath Nandi, Mr. Rabindranath Sasmal, Mr. Swapan Samanta, Ms. Jasmine Sinha for their guidance and inspiration. \\

\noindent I take this opportunity to mention the support and enormous encouragement I have received from my loving husband, Dr. Chowdhury Aminul Islam. He was always there to pamper my idiotic questions, sort out my philosophical confusions and clarify any difficulties I faced. Due to his continuous supports and suggestions, I never had a deficiency of interesting and stimulating problems to work on. I would like to whole-heartedly acknowledge that he was always able to put things in perspective and provide inspiration when the rigours of research and thesis writing seemed too much to handle. \\

\noindent I would like to mention the love and care I have received from my family members- my Dida, Dadu, Jethu, Barama, Mama, Mami. At this moment I remember my Jethu, Late Mr. Sanjay Bhattacharjee, who was very supportive of my academic and personal life. My mama, Shyamal Ghosh, was my first teacher. They are the ones who have constantly encouraged me to opt for a research career. My brother, Argha, was always there to provide me with support and affection. I do not have words to express my gratitude to my Maa (Ms. Bina Bhattacharjee) and Baba (Mr. Sudipta Bhattacharjee). Without their constant inspiration, motivation and moral support, I could not continue my PhD work. Thank you both for giving me strength to reach for the stars and chase my dreams.\\

\noindent At this moment I cannot name them all who were always there to provide me with support and affection. You know who you are, and I thank you all wholeheartedly for being there.

\chapter*{Abstract}
\addcontentsline{toc}{chapter}{Abstract} 
\noindent With the growing interest in indirect detection for dark matter signature, the thesis aims to investigate the signal originating from the self-annihilation of dark matter candidates. The methods for targeting the dark matter signal is two-fold, on one hand, we explore the gamma rays resulting from dark matter particles. On the other hand, we focus on complementary radio properties.\\

\noindent To begin with, for the basic understanding of dark matter, it is first important to characterise their nature and all possible scenarios to identify them. While such characterisation has
been briefly discussed in Chapter 1, Chapter 2 provides the methodology and its related mathematical formalism to study the dark matter signal in the multiwavelength spectrum.\\

\noindent Next, in Chapters 3 and 4, by considering the appropriate scenario, the working principle of the Fermi Large Area Telescope and its detailed analysis mechanism have been discussed.\\

\noindent Finally, from Chapters 5 to 8, various aspects of dark matter properties obtained from our detailed analysis have been discussed. In Chapter 5, we investigate the gamma-ray signal from the location of dark matter rich dwarf galaxy, Triangulum II and try to examine whether it can provide strong limits on theoretical dark matter models and the thermal relic annihilation rate. In Chapter 6, we report the faint emission from the location of Tucana-II and in that sense try to investigate whether dark matter is related to such an excess.\\

\noindent Next in Chapters 7 and 8, the electromagnetic radiation over a wide range, from gamma-ray down to radio frequencies appearing from the Low surface brightness galaxies and ultra faint dwarf spheroidal galaxies have been discussed. Moreover, we explore the projected sensitivity of the upcoming Square Kilometer Array and Cherenkov Telescope Array in probing the synchrotron and gamma-ray emission from them.


\chapter*{List of Publications}
\addcontentsline{toc}{chapter}{List of Publications} 




\vspace{0.5cm}
\noindent{\large{\bf Publications relevant to the Thesis:}}

\begin{itemize}
			{\item[1.] \emph{Constraints on dark matter models from 
the observation of Triangulum-II with the Fermi Large Area Telescope}, Sayan 
Biswas, {\bf Pooja Bhattacharjee}, Pratik Majumdar, Mousumi Das, Subinoy Das, 
and Partha Sarathi Joarder, {\color{black} Journal of Cosmology and 
Astroparticle Physics {\bf 11}, 003 (2017); arXiv:1705.00426 [astro-ph.HE] (2017)}.}	 
			\vspace{2mm}
			{\item[2.] \emph{Analysis of Fermi-LAT data from 
Tucana-II: possible constraints on the 
Dark Matter models with an intriguing hint of a signal},  {\bf Pooja 
Bhattacharjee}, Sayan Biswas, Pratik Majumdar, and Partha Sarathi Joarder,  
{\color{black} Journal of Cosmology and Astroparticle Physics {\bf 08}, 028 
(2019); arXiv:1804.07542 [astro-ph.HE] (2018)}.}
			\vspace{2mm}
			{\item[3.] \emph{Multiwavelength analysis of low surface 
brightness galaxies to study
 possible dark matter signature}, {\bf Pooja Bhattacharjee}, Pratik Majumdar, 
Mousumi Das, Subinoy Das, Partha Sarathi Joarder, and Sayan Biswas, 
{\color{black} Monthly Notices of the Royal 
Astronomical Society {\bf 501}, 4238 
(2021); arXiv:1911.00369 [astro-ph.HE] (2019)}.}\vspace{2mm}
			{\item[4.] \emph{Gamma-ray and Synchrotron Radiation 
from Dark Matter annihilations in Ultra-faint Dwarf Galaxies}, {\bf Pooja 
Bhattacharjee}, Debajyoti Choudhury, Kasinath Das, Dilip Kumar Ghosh, and Pratik 
Majumdar, {\color{black} Journal of Cosmology and Astroparticle Physics {\bf 06}, 041 
(2021); arXiv:2011.08917 [hep-ph] (2020)}.}
			\end{itemize}
		
\vspace{0.5cm}
\noindent{\large{\bf Additional publications during the Ph.D. thesis but not forming part of it:}}

\begin{itemize}
			{\item[1.] \emph{Investigating the region of 3C 397 in 
High Energy Gamma rays}, {\bf Pooja Bhattacharjee}, Pratik Majumdar, Tulun 
Ergin, Lab Saha, Partha Sarathi Joarder, {\color{black} Proceedings of the 
International Astronomical Union {\bf 12}, 316 (2017)}; arXiv:1801.05961 [astro-ph.HE] (2018). }	
			\vspace{2mm}
			{\item[2.] \emph{Probing the star formation origin of 
gamma rays from 3FHL J1907.0+0713}, Tulun Ergin, Lab Saha, {\bf Pooja 
Bhattacharjee}, Hidetoshi Sano, Shuta Tanaka, Pratik Majumdar, Ryo Yamazaki, 
Yasuo Fukui, {\color{black} Monthly Notices of the Royal 
Astronomical Society {\bf 501}, 4226 
(2021)}; arXiv:2012.07357 [astro-ph.HE] (2020).}
			\end{itemize}


\tableofcontents
{\countdef\interlinepenalty1000
\listoffigures
}
{\countdef\interlinepenalty1000
\listoftables
}
\newpage

\pagenumbering{arabic} 

\chapter{Introduction}
\section{Introduction to Dark Matter}

\begin{quote}{Harley White}
\noindent Dark matter seems to be \\
What isn't there to be seen \\
In between\\
What we see.\\

\noindent They dub it dark since you cannot detect it\\
Nor can they inspect it\\
With telescopy.\\

\noindent Yet, while it can't be described\\
It cannot be denied\\
For equations that irk\\
To work.\\
\end{quote}

\noindent \textit{In our Universe, all the visible things i.e. Planets, 
stars, asteroids, galaxies constitute less than $5\%$ of the total universe. So 
what are the remaining parts? What does constitute the rest of our Universe? 
This 
is the mystery and beauty of our Universe. Several astrophysical and 
experimental research suggest that a large part of the universe is composed of 
a strange substance known as `dark matter'.} \\

\noindent In human history, one of the most extraordinary intellectual 
achievements is to build the standard model (SM) of Particle Physics. Most of the 
particles 
were being discovered during the second half of the 20th century. 
Experimentally and theoretically, we found that the SM is an 
answer to a 
question as old as civilization itself.\\

\noindent Now, the question is what are the fundamental elements of matter? The 
SM gives us a very explicit representation of the fundamental elements of all 
the matters that are detected in our terrestrial laboratories. We also have an 
exact theoretical argument in a detailed mathematical form which explains how the 
fundamental particles will act. In terms of the understanding of our Universe, 
one of the most revealing discoveries is the baryonic matter, mostly in form of 
protons and neutrons. But unfortunately, they are not the dominant form of material in our Universe. 
Rather, a new mysterious form of ``invisible matter'' or we can say ``dark matter (DM)'' fills our 
Universe 
and from observational evidence, it has been found that they are 
roughly five times more abundant than ordinary matter.  \\

\noindent Unfortunately, the particle content of the SM - the quarks, the leptons, the mediators of the
interactions and the Higgs particle can not fill in the role of DM. This is evident from the
cosmological observations.\\

\noindent Accumulated observational data over the past century has established that visible matter
(baryonic matter) constitutes only 4.6$\%$ of the total substance in the Universe, while DM is theorized
to account for 24$\%$, dark energy accounting for the remaining 71.4$\%$. In 
Fig.~1.1, we have shown the content of baryonic matter, DM and dark energy. 
The invisible matter is termed as DM because it neither emits nor 
absorbs any detectable electromagnetic radiation and hence it is very difficult 
to study or identify it. It is not possible to directly detect the DM
by any traditional telescopes, but there are enough pieces of evidence for the 
existence of DM \cite{Tegmark:2003uf}. Interestingly, the existence of 
missing mass is robustly supported by macroscopic evidences, but the microscopic 
nature and composition of DM are still in much debate. There are many ongoing experiments 
which are dedicated to directly detect and study
the nature of DM candidates, but none have yet 
succeeded. To have a complete understanding of DM, we need to study 
several branches of physics and astronomy.\\

\begin{figure}
\centering
\includegraphics[width=0.5\linewidth]{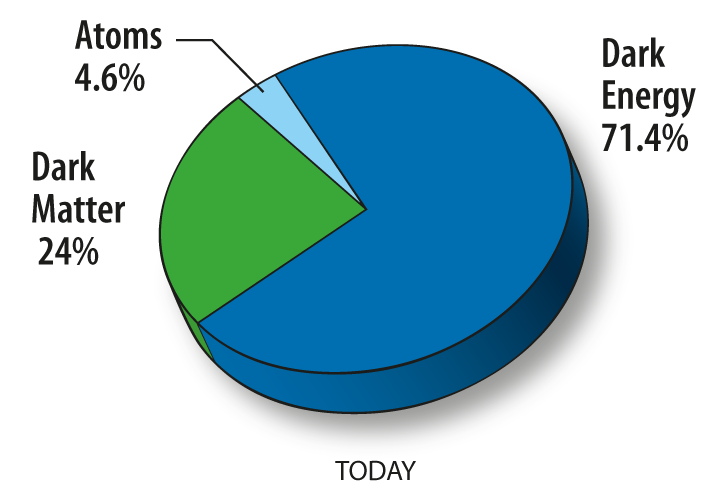}
\caption{\small{The multiple components that compose our universe.
Dark energy comprises 71.4$\%$ of the mass energy density of the universe, DM comprises 24$\%$, and
atomic matter makes up 4.6$\%$.}}
\end{figure}

\section{Brief Overview of Thesis}
 
\noindent The thesis is organized as follows:\\ 
 
\noindent To start with, in the following section~1.3 of this chapter, I present a brief introduction to the indirect evidence of DM. Next, in 
section~1.4, I discuss the possible DM candidates. Then in section~1.5, a brief review on 
DM annihilation process has been discussed. For my thesis, I study the DM 
signal resulting from the pair-annihilation. In this section, I would 
introduce the reader to the theoretically favoured annihilation final states 
and 
how we can obtain the emission (for example $\gamma$-ray and radio emission) as an end product of 
annihilation channels. In section~1.6, I briefly give a summary of the DM detection methodology. Direct detection, indirect detection and collider 
searches are three popular methods to search for the signature of the DM candidates. For my thesis, I solely focused on the indirect detection 
method.\\ 
 
\noindent Next in chapter 2, a brief introduction to the methods of 
multiwavelength searches for DM is given. The mathematical formalism, 
the notations and the other necessary concepts that I will explain in this 
chapter would be used later on my thesis. First, in section 2.1, I discuss the 
possible DM dense regions. Later in sections 2.3 and 2.4, I explain how 
we can study the electromagnetic radiation over a wide range, from gamma-ray down to 
radio frequencies appearing from DM annihilation. \\ 
 
\noindent For my thesis, I concentrate on the DM signature from some 
particular DM sources through indirect detection. For this 
purpose, we need dedicated and sensitive instruments. In Chapter 3, I describe the working 
principle of the Fermi Large Area Telescope (Fermi-LAT) in detail. Fermi-LAT is a gamma-ray 
space telescope that covers the entire celestial sphere. In view of indirect 
detection of DM signal, Fermi-LAT is one of the most sensitive
gamma-ray 
telescopes. For most of my works, I have analysed the gamma-ray data observed by the 
Fermi-LAT. The detector and its working principle are described in Chapter 3 along with a review of its
performance.\\ 
 
\noindent Next, in Chapter 4, I give an overview of the Likelihood function for 
Fermi-LAT data. The details of the mathematical formulation for Likelihood 
function and its methodology are explained in this chapter. Here, I will also 
explain how to estimate the upper limits if we could not detect any signal from 
the source. I use this formulation for my work to estimate the possible 
signature of DM annihilation. \\ 
 
\noindent In Chapter 5, we study Triangulum-II, a newly discovered 
ultra-faint dwarf galaxy, which is assumed to be rich in DM. We 
examine 
the gamma-ray signal from the location of Triangulum-II and from that data try 
to check whether this galaxy can provide strong annihilation rate than other 
well studied sources. We show that Triangulum-II would provide very stringent 
limits on the theoretical DM models and thermal annihilation rate, 
even better than some well studied dwarf spheroidal galaxies (dSphs).\\ 
 
\noindent In Chapter 6, we study the Tucana-II. Like Triangulum-II, Tucana-II is a DM-dominated satellite galaxy of our Milky Way. We examine the gamma-ray 
data from its location and unlike most of the dwarf galaxies, we 
observe a faint emission. Then we first study the maximum significance of this 
emission and check how this excess would vary with the DM mass, 
annihilation channels and periods of exposure. Furthermore, we 
investigate the origin of such emission and our study shows that such excess is mostly coming from the Tucana-II location and most
likely related to DM annihilation.\\ 
 
\noindent Next, in Chapter 7, we study four Low surface brightness galaxies (LSB). Unlike the 
earlier two chapters., for this work, we use the multiwavelength approach to 
investigate the DM signature. LSB galaxies have very diffuse and low 
surface density stellar disks and their extended HI (neutral hydrogen) rotation curves indicate the 
presence of very massive DM halos. We analyze the Fermi-LAT data 
for high energy gamma rays and radio flux upper limits from Very Large Array (VLA) at a frequency of
1.4 GHz to obtain upper limits on annihilation cross-section $\langle\sigma 
v\rangle$ at $95\%$ confidence level (CL) in a model-independent way. From this study, we show that 
for LSB galaxies radio cross-section rate would be competitive with the limits 
predicted from Fermi-LAT. We further discuss the projected sensitivity of the 
upcoming ground-based telescope, the Cherenkov Telescope Array (CTA) and radio 
telescope, the Square Kilometer Array (SKA) and investigate whether they can probe the radiation from LSB galaxies.\\ 
 
\noindent In Chapter 8, 
we 
consider 
14 recently discovered ultra faint dwarf galaxies and study the electromagnetic radiation over a 
wide 
range, from gamma-ray down to radio frequencies appearing from them. We also, 
check the $\langle\sigma v\rangle$ at $95\%$ 
CL for the gamma-ray and radio flux upper limits observed by Fermi-LAT, 
Giant Metrewave Radio (GMRT) 
and VLA. We study the uncertainty in the synchrotron and gamma-ray fluxes 
arising from various astrophysical parameters. Furthermore, we discuss the 
projected sensitivity of the SKA radio telescope in probing 
the synchrotron radiation from the aforementioned dSphs.

\section{Observational Evidence of Dark Matter}

\begin{quote}{Harley White}
\noindent Dark matter exerts gravitational pull.\\
It glues stars together, makes galaxies full.\\
Unlike normal matter it plays hide and seek\\
And so much of it's interactively weak...\\
\end{quote}

\noindent The very first observational hint of the DM, or ``missing 
mass'' began in early 1930. In 1932, Jan Oort observed a bizarre motion of the 
stars of our Milky Way and that hinted the presence of some form of non-luminous 
matter which is far more massive than anyone had ever predicted 
\cite{Oort:1932gat}. By studying the Doppler shifts of each moving star in 
galactic place, Oort calculated the velocities of these stars. The calculation showed 
that stars were moving so quickly to escape from the gravitational pull of 
Milky 
Way. That made Oort to suspect the pressence of massive pull in the galactic plane which can hold the stars to their orbits 
\cite{Oort:1932gat}. \\

\noindent Just one year after the Oort's finding, in 1933, Swiss astronomer F. 
Zwicky, examined a much larger system, Coma Cluster. From Doppler 
effect, he measured the velocity dispersion for member galaxies of the Coma 
galaxy cluster and noticed that the member galaxies were 
moving much faster than we could expect from their luminous components 
\cite{Zwicky:1933gu, Zwicky:1937zza}. Zwicky measured the velocity dispersions 
of each member galaxy (i.e. kinetic energy) and then by employing the virial 
theorem, he estimated the total mass of the Coma cluster. With Virial theorem, 
he established the relation between the total mass of the galaxy cluster and the averaged square 
of the velocities of each galaxy. He then observed that in order to maintain the equilibrium in Coma cluster, a large amount of ``Dunkle 
Materie'' or DM must be present to theoretically explain the large 
velocity dispersion of the system. \\

\noindent The virial theorem denotes the following relation between 
gravitational energy and kinetic energy. The expression is:
\begin{equation}
<T>= - \frac{1}{2} <U>
\end{equation}
\noindent The virial theorem (equation 1.1) states that for a spherically 
symmetric system, the total kinetic energy ($T$) is equal to minus 
$\frac{1}{2}$ 
times the total gravitational potential energy ($U$) \cite{Cari_ena_2012}. 
Hence, if we know the kinetic energy of the system, we can calculate the 
gravitational potential energy, and then the total mass of the system can be 
easily estimated. If the obtained mass of the system is greater than the mass 
of 
the total luminous matter, then there must be some invisible i.e non-luminous 
matter present in the system. The invisible matter can only interact 
gravitationally. Hence, from virial theorem, Zwicky observed that the total 
mass 
of the cluster was about 400 times greater than the luminous mass. This result 
led him to propose that there must be some source of invisible matter that 
created such a difference with the observational estimation. Study of the Virgo cluster soon produced very similar results \cite{Smith:1936mlg}.\\

\noindent Next, roughly around 40 years later the discoveries of Oort and 
Zwicky, beginning in the 1970s, Vera Rubin, Alberto Bosma, and others studied the orbital 
velocities of stars in spiral galaxies (\cite{Rubin:1970zza, Freeman:1970mx, Einasto:1974njh, Ostriker:1974lna, roberts:1975bhh, roberts:1976bhf, bosma:1978ghf, Rubin:1978kmz, Rubin:1980zd}). Rubin and her collaborators separately performed an extended 
study of rotational curves for around 60 individual galaxies 
\cite{rubin:1983hgu}. They performed detailed 
measurements of the Doppler shift for their targets and determined their 
orbital velocities. Their studies also showed an extreme deviation from the theoretical prediction based on Newtonian gravity and other baryonic matter interactions \cite{rubin:1983hgu}. They found that the spiral galaxies have flat 
rotation curves extending out to radii of tens of kpc and their orbital velocities 
did not decrease as expected. From the flat rotational curve, Rubin estimated that the 
galaxies 
have contained almost 10 times more matter than the visible one. This 
remarkable finding just confirmed the earlier claims by Zwicky. Rubin also predicted 
that there might be an unobserved huge spherical halo of DM which 
surrounds the inner luminous galaxy.\\

\noindent According to Newton’s Law of Gravitation (Newton 1687), the orbital 
velocity should fall by increasing the distance from the center of the galaxy as,
\begin{equation}
v(r)= \sqrt{G \frac{m(r)}{r}} ,
\end{equation}
\noindent where $v(r)$ is the rotation velocity as a function of radius, and $m(r)$ 
is 
the mass confined within radius, $r$. \\

\noindent From Eq. 1.2, we should expect to observe the fall of orbital velocity as: $v(r)~\propto~1/\sqrt{r}$. But interestingly, the 
galactic rotation curves, as obtained by Rubin and her collaborators, did not 
follow the expected nature. In their publication, Rubin, Kent Ford and Norbert 
Thonnard \cite{Rubin:1980zd} reported their observational results for 21 spiral 
galaxies. Their study showed that with increasing the distance from the center 
of the galaxies the rotational velocity remained constant (or merely increased 
for some galaxies). The rotational velocity of any galaxy can only remain constant if 
the total mass of the system is increasing with radius from the center. The artistic view of their study is shown in 
Fig.~1.2 (a). From this figure, it is 
evident that the radial velocity of the galactic system is much larger than 
what 
would be expected if the gravitational potential of the galaxy came from only 
the luminous matter i.e. from the stars and gas.\\

\begin{figure}[h!]
\subfigure[]
{ \includegraphics[width=0.4\linewidth, height=2.6in]{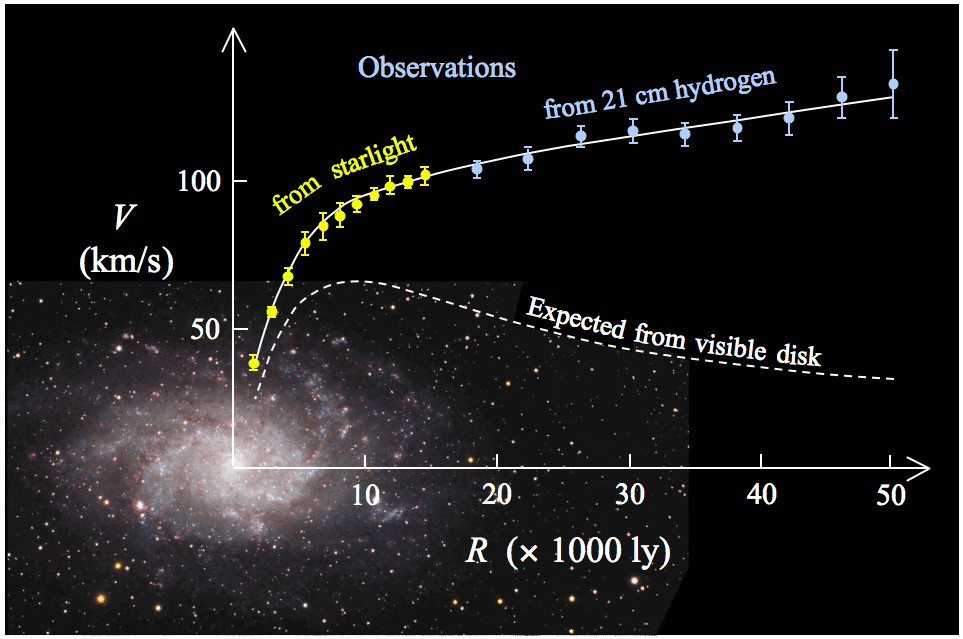}}
\hfill
\subfigure[]
{ \includegraphics[width=0.4\linewidth, height=2.8in]{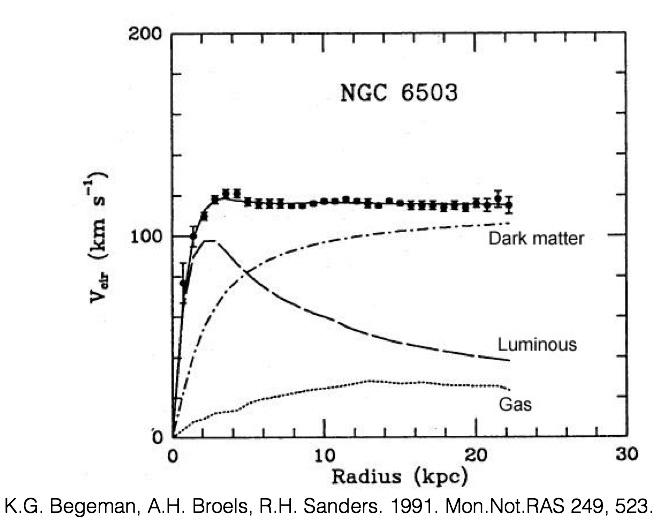}}
\caption{(a) The artistic view of the observed and expected Rotational curve 
from M33 galaxy. (b) The rotational curve of the spiral galaxy NGC6503.}
\end{figure}

\noindent After that, many scientists did similar kinds of studies and all of 
them confirmed the same nature of the galactic rotational curve. The work by 
Begeman, Broeils and Sanders, 1991 \cite{Begeman:1991iy} also reported the same. For their studies, they have chosen the spiral galaxy NGC6503 
(Fig.~1.2 
(b)). They showed the contributions to the rotational 
velocity from luminous disk, gas and dark halo. Their analysis also 
reported the extension of the DM halo beyond the stellar bulge of the 
galaxy.\\

\noindent In the 1970s, scientists tried a new way to understand the 
distribution of DM, `Gravitational lensing'. Einstein's theory of relativity 
postulates that the strong gravitational field can bend the path of light rays, 
i.e., a massive object can bend spacetime and also 
affect the motion of nearby objects. This produces a lensing effect where 
the surrounding objects follow the geodesics of the curved space. This effect 
is called the gravitational lensing \cite{Lynds:1989gry}. For observing the 
effect of gravitational lensing, it requires a very massive object (say the 
cluster of galaxies) and a distant bright light source behind it. If the distant 
object is located directly behind the massive body, the massive object would act 
as a 
gravitational lens and would create numerous images of the distant object. This 
effect would create an Einstein ring structure (see, the blue ring structure from Fig.~1.3) 
\footnote{\tiny{NASA images from Large Synoptic Survey Telescope (LSST); 
http://www.lsst.org/lsst/public}} where the massive object would be at the 
center and the images of the distant object would create the ring (Fig.~1.3). In 
1979, D. Walsh et al. \cite{Walsh:1979nx} was the first to observe this form of 
gravitational lensing. The detailed study of the 
Einstein ring structure allows astronomers to estimate the total mass of any 
massive body, such as: galaxy, cluster of galaxies, etc. Their observational studies show 
that 
only $10\%$ of the total mass of the clusters are in the form of individual 
galaxies, the rest is DM \cite{Walsh:1979nx}.

\begin{figure}
\centering
{ \includegraphics[width=0.5\linewidth]{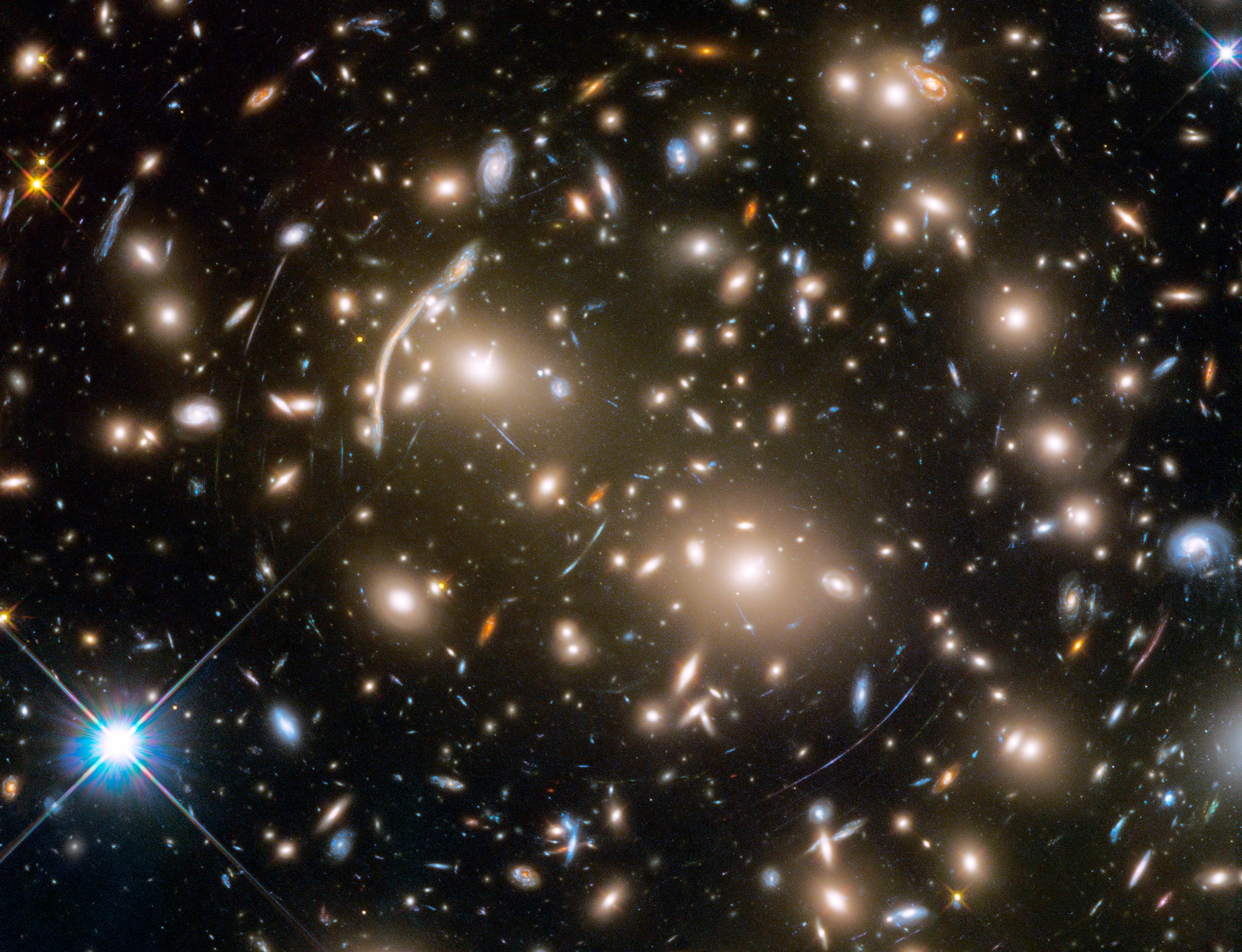}}
\caption{Gravitational lensing of Abell 370 observed by Hubble Space Telescope (HST).}
\end{figure}

\noindent Another very strong evidence for DM is the Bullet cluster 
(1E 
0657-56). This cluster consists of two colliding cluster of galaxies. While the 
galaxies crossed their paths, the stars within the galaxies and other visible 
light passed by each other without being affected much by the collision. But, 
the hot gas clouds which represent most of the baryonic matter merging from 
two colliding galaxies interact electromagnetically and due to friction of 
the gas molecules, the gases of both clusters slowed down much faster than the 
stars. When the gas clouds were slowed down, the visible part of the galaxies 
came into much clearer 
view and that gave the scientists a scope to examine the total mass of the 
Bullet cluster. With the data obtained from the
X-ray telescopes and gravitational lensing observations, the scientists
found that the mass and the gas element do not follow the same 
distribution\cite{Clowe:2006eq}. 
Then by measuring the gravitational lensing effect of the Bullet cluster, the 
scientists determined that the 
cluster bent the path of light more than they could expect from the luminous 
mass. 
This proved that there must be the presence of more mass in the cluster 
than the visible matter. The composite image of bullet cluster (or galaxy 
cluster 1E 0657-56) is shown in Fig. 1.4. The background part of this image is 
showing the visible spectrum of the light stems obtained from Hubble Space and 
Magellan Telescope, while the pink part of this image denotes the X-ray 
emission of the colliding clusters recorded by the Chandra Telescope and lastly 
the blue part shows the mass distribution of the Bullet clusters estimated from 
the gravitational lensing effects 
\footnote{\tiny{Nasa: A matter of fact, August, 2006; 
http://www.nasa.gov/vision/universe/starsgalaxies/dark{\_}matter{\_}proven.html}
, \tiny{X-ray: NASA/CXC/CfA/M.Markevitch el al.}, \tiny{Optical: NASA/STScl, Magellan/U.Arizona/D.Clowe et al.; Lensing 
Map: 
NASA/STScl; EDO WFI; Magellan/U.Arizona/D.Clowe et al.}}.

\begin{figure}
\centering
\includegraphics[width=0.5\linewidth]{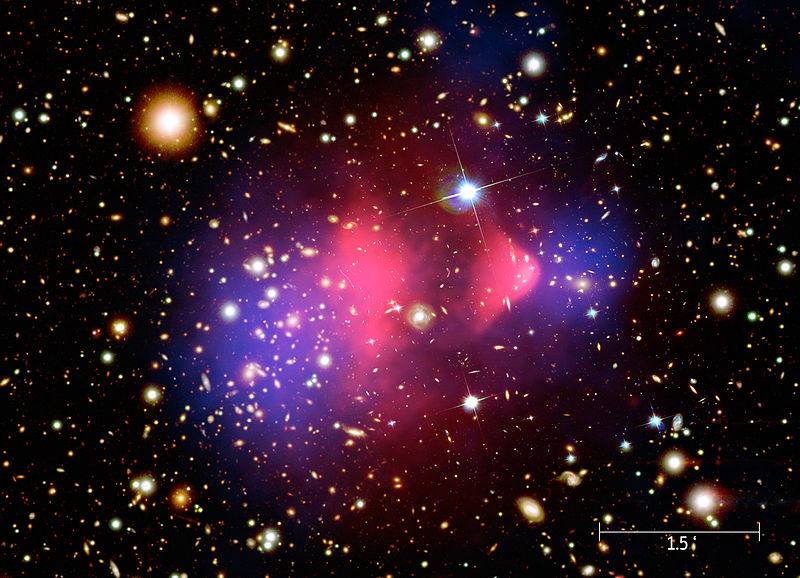}
\caption{X-ray image (pink) of Bullet cluster superimposed over a visible light 
image (blue).}
\end{figure}

\section{Dark Matter Candidates}
In this section, we will discuss the nature of DM and its possible 
candidates. To reveal the nature of the DM particles, the physicists first 
focused on the known astrophysical bodies which are made of ordinary, baryonic 
matters. Later, they have extended the standard model theory to explain the 
non-luminous nature of the DM.\\

\noindent From several observational pieces of evidence, we can summarize the following 
general properties of DM candidates.
\begin{itemize}
\item They do not emit or absorb light, indicating absence of electromagnetic interaction. Hence, they do not carry electric charge. 
\item Majority of them neither participate in strong interaction nor carry colour charge. (A very small fraction of the DM is assumed to be baryonic and only they
can take part in strong interaction.)
\item They do only interact via gravity. The gravitational effect of the DM is very important to form a large structure of the universe.
\end{itemize}

\subsection{Possible DM Candidates:}
\noindent The observational evidence indeed gave us enough hints of the 
existence of the DM, but the true nature of the DM remains unknown. Below we 
will discuss the possible candidates for DM.\\ 

\subsubsection{The Standard Model and the Neutrino and Supersymmetry:}

\noindent The standard model (SM) consists of the following particles - six leptons (electron, muon, tau
and their corresponding neutrinos), six quarks (bup, down, strange, charm, bottom and top)
and five force carriers (photon, gluon, Z, $W^{\pm}$ and the Higgs scalar). Each of the 
above-mentioned leptons and quarks
has their respective antiparticles which are generally denoted with a bar or 
opposite charge sign (for example, the up antiquark's 
symbol is $\bar{u}$). The Higgs Boson with a mass of $\sim$ 125 GeV was discovered in 2012 by the ATLAS \cite{ATLAS:2012yve} and the
CMS experiments performed at the Large Hadron Collider (LHC) of European Organization for Nuclear Research (CERN)\cite{CMS:2012qbp}. \\

\noindent In spite of the success of the SM in explaining behaviour of the elementary particles, it does
not contain any particle which can act as the DM candidate. One of the most stable, neutral and weakly 
interacting particles from SM is the neutrino. But, the recent literature by 
Spergel et al. \cite{Spergel:2003cb} completely ruled out the possibility of neutrinos being the 
entire solution to missing mass of the Universe. From WMAP, they showed the 
neutrino mass to be $m_{v}<~0.23~eV$, which in turn makes the cosmological density 
$\Omega_{v}h^{2}<~0.0072$ \cite{Jarosik:2010iu}. Hence, neutrinos do only 
account for a very small fraction of DM, and cannot be the prime source of 
DM.\\
\noindent Hence, several possible extensions of the SM have been proposed, Supersymmetry (SUSY)
being one of them \cite{Jungman:1995df}. SUSY assumes an additional symmetry between the fermions and the
bosons, i.e., each particle in the SM has its superpartner - fermions have bosons as their
superpartners and vice-versa. The most encouraging finding of the SUSY is that it can 
propose valid DM candidates. Several particles in the SUSY theory are possible DM candidates, like neutralino, sneutrino \cite{Falk:1994es, Hall:1997ah}, and gravitino \cite{Chun:1993vz, Borgani:1996ag}. All of these 
three candidates show a WIMP (Weakly interacting massive particle)-like nature i.e., weakly interacting and 
electrically neutral, but sneutrinos \cite{Falk:1994es, Hall:1997ah} might 
annihilate very rapidly in the early universe and hence its relic densities are 
very low to explain any cosmological phenomenon, whereas, gravitino 
\cite{Chun:1993vz, Borgani:1996ag} would act as a hot DM. In most SUSY models, the lightest neutralino is considered the most promising candidate for
DM.\\

\noindent Several exotic particles are also considered as DM candidates - massive compact halo objects
(MACHOs), black holes, WIMPs, axions, etc. Some
theories also suggest that the DM can be both baryonic and non-baryonic and in that case,
MACHOs are considered as the baryonic type. The dominant part of the DM is mainly
composed of non-baryonic candidates, e.g. neutrinos, WIMPs, axions, etc. Based on the
physical properties, there are different types of DM. We will describe them below.\\

\noindent Kinematically, the DM can be divided into three categories based on their velocity at the
time of its decoupling of universe \cite{silk2000big}. This is important because it has a direct influence on
our galaxy formation and large and small structures of the universe. \\

\begin{itemize}

\item \textbf{Hot Dark Matter (HDM):} The HDM is made of abundant light 
particles. The best candidate for HDM is the normal light neutrino. The mass of 
an HDM is of the order of eV or less, $m_{HDM}$ $<$ 1 eV. \\

\item \textbf{Cold Dark Matter (CDM):} The CDM is at the opposite end of HDM in the mass-velocity spectrum. It is non-relativistic at the time of decoupling. Its 
mass can be in the GeV order or larger. There are many proposed candidates for CDM, 
including weakly interacting massive particles like neutralinos, WIMPZILLAs, 
solitons, etc. \\

\item \textbf{Warm Dark Matter (WDM):} The WMD is something in between the HDM 
and CDM, consisting of particles of $m_{WDM}$ $>$ 1 
KeV which may interact even weaker than a neutrino. It is relativistic at the time of 
decoupling, but non-relativistic at the radiation-to-matter dominance 
transition. There are a few possible candidates for WDM, including sterile 
neutrino, light gravitinos and photino, etc.\\

\end{itemize} 

\noindent The DM can also be classified according to its production mechanism.

\begin{itemize}

\item \textbf{Thermal relics}: The thermal relics particles are assumed to be in 
thermal equilibrium in the early universe and mass of that thermal relic is bound from the above by 340 TeV. Most of the favoured DM candidates are from 
this category.\\

\item \textbf{Non-Thermal relics}: These particles are produced via 
non-thermal mechanism and is believed that
they were never in equilibrium with the thermal bath of the universe. There are 
several favoured DM candidates which are assumed to be non-thermal relics, such 
as axions emitted by cosmic strings, superheavy WIMPZILLAs (masses lie between $10^{12}$ to $10^{16}$ GeV), etc.\\

\end{itemize}

\subsection{Weakly Interacting Massive Particles} 
\noindent One of the leading candidates for DM is weakly interacting massive 
particles (WIMPs) \cite{Steigman:1984ac, Bertone:2004pz}. The most favoured 
DM candidates, like the neutralino from supersymmetry, the lightest 
Kaluza-Klein particle from the superstring theory and theories of extra 
dimensions and some other candidates from beyond the standard model theory are 
assumed to be very massive and only interact via gravitational pull, i.e., 
weakly interacting. These are collectively referred to as WIMPs. They are non-baryonic and well-motivated by
independent considerations of particle physics \cite{Steigman:1984ac}. Systematic theoretical investigations to
understand their properties and experimental searches have to be carried out. \\

\noindent At 
the early universe, say after the Big Bang, the particles were in chemical and 
thermal equilibrium. By chemical equilibrium, we mean that every 
reaction among the particles was reversal (e.g. the creation of WIMPs 
pair-production from SM particles and the WIMP annihilation were in 
equilibrium). Hence, the whole system of universe did not change by any reaction. 
This equilibrium was maintained until the temperature of the universe became 
lower than the particle mass, and as a result, the pair-production of WIMPs 
stopped. When this equilibrium was broken, the abundance of DM candidates 
decayed due to annihilation and this process continued until the annihilation 
rate 
fell below the expansion rate of the universe. This epoch is referred to as 
the ``freeze-out''.\\

\noindent Another class of particles is the Superweakly interacting massive particles (superWIMPs),
which include sterile (right-handed) neutrinos, gravitino. etc. These have annihilation cross-
sections much smaller than that of the weak interaction.

\section{Dark Matter Annihilation}
\noindent In this section, we discuss ways to detect DM candidates of the WIMP type. Generally,
experiments look for the end products of their annihilation or decay channels. One very
popular way to detect the DM candidates is to search for the end products of WIMP
annihilation/decay channels. \\

\noindent We denote the DM and its anti-particles as $\chi$ and 
$\bar{\chi}$, respectively. If DM is a Majorana particle,
$\chi$ and its anti-particle $\bar{\chi}$ would be the same. Several observational evidences as well as theoretical models 
propose that the mass of WIMPs lies in the range of GeV to TeV. If $\chi$ is 
assumed to be thermally relic DM candidates, then $\chi$ and $\bar{\chi}$ should 
participate in the evolution of the universe as other SM particles. \\

\noindent In the early universe, through annihilation and pair-production 
processes, $\chi$ and $\bar{\chi}$ were in equilibrium with ordinary SM particles (i.e equilibrium with fermions (f), quarks (Q), gauge bosons ($W^{\pm}$, Z) etc). The form of the 
annihilation reaction could be described as: \cite{Salati:2014rua}\\
$\chi + \chi \rightarrow Q + \bar{Q} \rightarrow f + \bar{f}, W^{+} + W^{-}, 
Z + Z$,....,
where $Q$ and $\bar{Q}$ denote quark and its antiparticle, respectively. \\

\noindent After the big bang, all of these particles were in equilibrium and 
were at the same temperature. The number density of $\chi$ in equilibrium at a 
given temperature can be described as:

\begin{equation}
\int n_{\chi}^{eq} = \frac{g}{(2\pi)^3} \int f(p) d^{3}p 
\end{equation}

\noindent where, g is the number of internal degree of freedom of $\chi$ and f(p) 
is a function of the three-momentum p of $\chi$. Depending on the spin of the WIMP, $f(p)$ would either follow Dirac-Fermi or Bose-Einstein
distribution. At very high temperatures i.e for $T >> m_{\chi}$, 
$n_{\chi}^{eq} \propto T$, while at lower temperature i.e. for $T << m_{\chi}$, 
$n_{\chi}^{eq} \propto exp(-m_{\chi}/T)$. At $T << m_{\chi}$, the production of 
$\chi$ $\bar{\chi}$ pair from SM particle pair will be suppressed and at the 
same time the annihilation rate will remain the same, hence the number density 
of 
$\chi$ will be exponentially reduced. When the universe expands, the 
temperature of the universe drops to sufficiently low and that would lead to the system out of the equilibrium. \\

\begin{figure}
\centering
\includegraphics[width=0.7\linewidth]{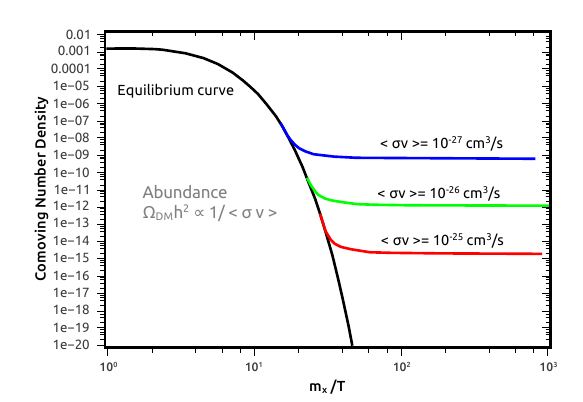}
\caption{Comoving number density evolution as a function of the ratio 
$m_{\chi}/T$ in the context of
the thermal freeze-out.}
\end{figure}

\noindent Due to a significant drop in the number density of $\chi$, it would 
be 
very hard for $\chi$ and $\bar{\chi}$ to find each other to annihilate, or to 
be 
scattered around by ordinary SM particles. Eventually, they would no longer be 
in thermal equilibrium and $\chi$ is decoupled from the rest of the universe. 
Then, except for a very rare occasion, $\chi$ would not annihilate or scatter 
with 
ordinary particles. But $\chi$ continues to expand freely with the Hubble flow. 
The number density of $\chi$ is fixed to the $T^{3}$. In Fig.~1.5, we have 
shown how does the comoving number density of the $\chi$ varies with $m_{\chi}/T$ at the epoch of thermal freeze-out.\\

\noindent The overall geometry of the universe \cite{Gaitskell:2004gd} is 
determined by the density parameter ($\Omega$) of our universe. The expression 
of the density parameter is: $\Omega = \rho/\rho_{c}$. Where $\rho$ is observed 
density and $\rho_{c}$ is the critical density of our universe. The critical 
density is the average of the matter-density that is needed for our universe to 
halt its expansion and it can be expressed as:

\begin{equation}
\rho_{c} = \frac{3H_{0}}{8\pi G}
 \approx 1.88 \times 10^{-26} \, h^{2} \, kg \, m^{-3}
\end{equation} 

\noindent where $H_{0}$ is the Hubble constant and h is the dimensionless form of 
$H_{0}$ in units of 100 km/s/Mpc \cite{Gaitskell:2004gd}. From the density 
parameter of the universe, we can guess the contributions of baryonic matter, 
DM and dark energy, that is, $\Omega = \Omega_{B} + \Omega_{DM} + 
\Omega_{\Lambda}$. Here $\Omega_{B}$, $\Omega_{DM}$ and $\Omega_{\Lambda}$ are 
the relative density parameter for normal baryonic matter, DM and dark 
energy, respectively. The recent observations of the Planck collaboration 
obtained $\Omega_{B}$ = 0.05, $\Omega_{DM}$ = 0.265 and $\Omega_{\Lambda}$ = 
0.685 \cite{Ade:2015xua}. The DM density ($\Omega_{DM}$) depends on 
the 
annihilation cross-section ($\sigma$) weighted by the average velocity ($v$) of 
the particle i.e. on $<\sigma v>$. In order to match the abundance measured by 
the Planck collaboration, the DM relic density would be equal to 
$\Omega_{DM} h^{2} = 0.1197 \pm 0.0022$ \cite{Ade:2015xua}. The expression of 
the 
$\Omega_{DM} h^{2}$ is:

\begin{equation}
\Omega_{DM} h^{2} = 0.11 \frac{3 \times 10^{-26} cm^{3} s^{-1}} {<\sigma v>_{0}}
\end{equation}

\noindent From eq.(1.5), it is evident that DM might have an annihilation 
cross-section, $<\sigma v>_{0} \approx 3 \times 10^{-26} cm^{3} s^{-1}$ at 
thermal freeze-out \cite{Abeysekara:2014ffg}.\\

\noindent Like we already discussed above, WIMPs are thought to be first 
self-annihilate into a quark-antiquark pair and later that pair decays to 
several possible SM particles, as shown in Fig. 1.6. From this image, we can 
observe that as an end product of WIMP annihilation, it can generate 
$\gamma$-ray, lepton pairs such as muon-antimuon pairs ($\mu^{-}\mu^{+}$) or 
electron-positron ($e^{-}e^{+}$) pairs, and also boson pairs like $Z Z$ 
or $W^{+} W^{-}$. Thus, even though the WIMPs are invisible to us, we can
try to probe these SM particles originated from WIMP annihilation. We can start 
our search for DM signature by looking for the areas in the universe that are 
thought to be rich in DM. One of the most popular ways is to 
scan the universe for the end products which might come from the DM 
annihilation/decay. \\

\begin{figure}
\centering
\includegraphics[width=0.8\linewidth]{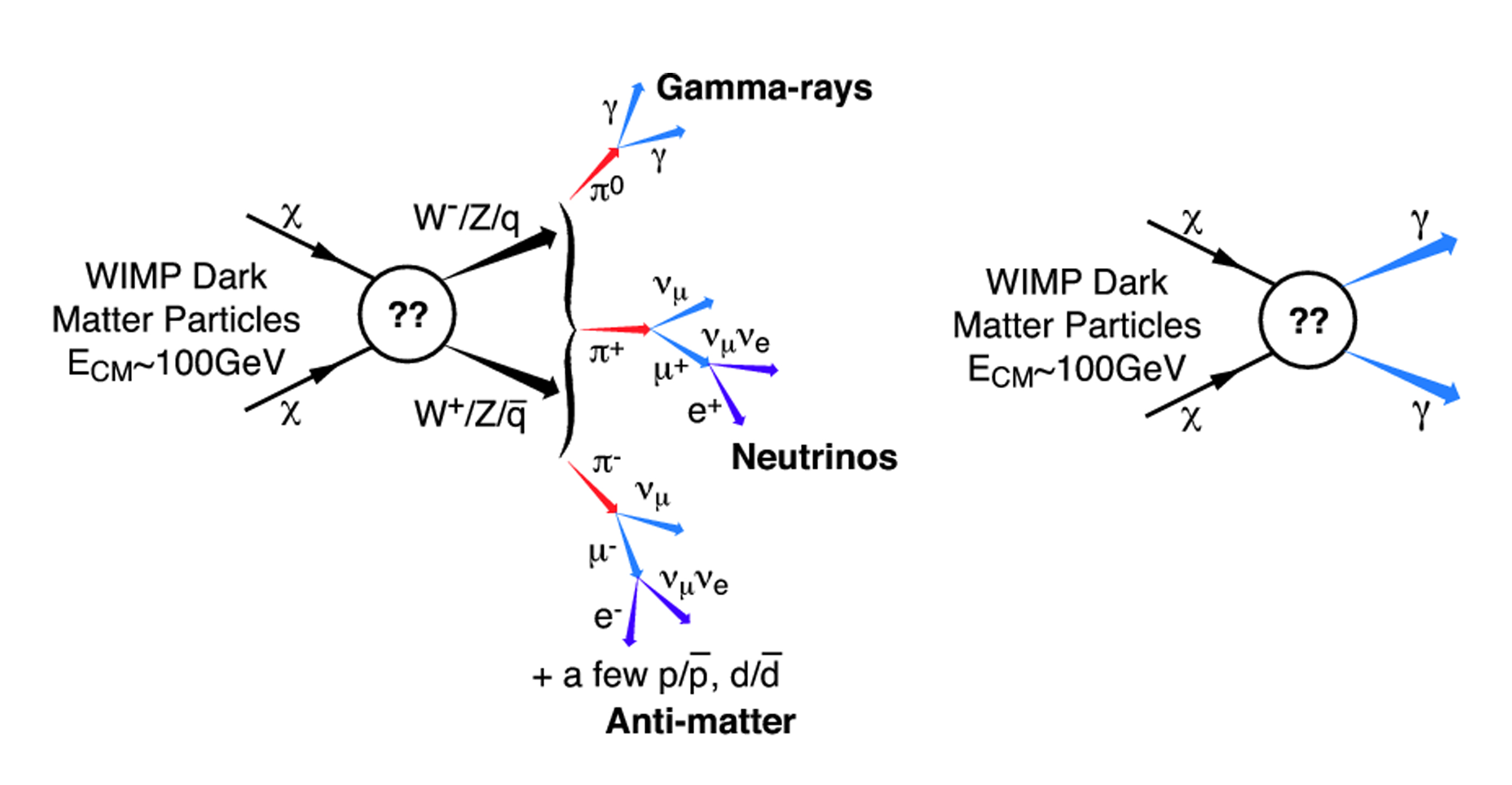}
\caption{WIMP annihilation chain and the end products.}
\end{figure}

\noindent For my thesis, we have focused on five theoretically motivated DM annihilation 
channels (in the later sections we would discuss this in detail). Those channels 
are: $\chi\chi \rightarrow \tau^{+}\tau^{-}$;
$\chi\chi \rightarrow \mu^{+}\mu^{-}$;
$\chi\chi \rightarrow W^{+}W^{-}$;
$\chi\chi \rightarrow b\bar{b}$ and 
$\chi\chi \rightarrow 80\%~b\bar{b} + 20\%~\tau^{+}\tau^{-}$.\\

\noindent The lifetime of tau lepton ($\tau^{-}$) is around $2.9 \times 
10^{-13}$s and its mass is $1776.82 MeV/c^{2}$. $\tau^{-}$ can decay into 
the combination of neutral pions, tau neutrinos and charged pions ($\pi^{\pm}$). 
There are multiple possible decay channels for $\tau^{-}$ and amongst them 
$90\%$ of the decay possibilities are accounting for five channels and the 
remaining around $10\%$ of decay possibilities can be related to twenty five 
different decay modes \cite{Nakamura:2010zzi}. The five dominant $\tau^{-}$ 
decay 
channels are: $\tau^{-} \rightarrow e^{-} + \bar{\nu_{e}} + \nu_{\tau}$ , 
$\tau^{-} \rightarrow \mu^{-} + \bar{\nu_{\mu}} + \nu_{\tau}$, $\tau^{-} 
\rightarrow \pi^{-} + \pi^{0} + \nu_{\tau}$, $\tau^{-} \rightarrow \pi^{+} + 
2\pi^{-} + \nu_{\tau}$, and $\tau^{-} \rightarrow \pi^{-} + 2\pi^{0} + 
\nu_{\tau}$ \cite{Nakamura:2010zzi}. The dominant decay modes of neutral pions are $\pi^{0} \rightarrow 
2\gamma$ (98.82$\%$) and $\pi^{0} \rightarrow \gamma + e^{-} + e^{+}$ (1.17$\%$). Thus decays of taus generate radiation. \\

\noindent The lifetime of muon ($\mu^{-}$) is around $2.2 \times 10^{-6}$ s and 
its mass is 105 $MeV/c^{2}$ and decays via weak interaction: $\mu^{-} \rightarrow e^{-} 
+ 
\bar{\nu_{e}} + \nu_{\mu}$ as an end of final 
state radiation. \\

\noindent The mediator of charged weak interaction W boson, has a mass of 80.4 $MeV/c^{2}$, and decays
into a fermion-antifermion pair. \\

\noindent The two heaviest quarks, top (173210 $MeV/c^{2}$) and bottom (4180 $MeV/c^{2}$) quarks decay via
weak interaction and produce gamma rays as final state.\\

\noindent Four annihilation channels have been chosen for the following reasons. Because of the phase space, we can expect that the DM 
particles would dominantly annihilate into the heaviest possibles channels 
\cite{Abeysekara:2014ffg}. Hence, we consider the $\tau^{+}\tau^{-}$ 
annihilation channels. Several ongoing experiments such as Fermi-LAT, MAGIC, 
etc. have studied $b\bar{b}$ annihilation channel for searching the indirect DM 
signal. Thus, we have chosen the $b\bar{b}$ annihilation channel to check the 
direct comparison of results. We have chosen the bosonic $W^{+}W^{-}$ channel 
because in several experiments this bosonic annihilation channel is widely 
considered. Finally, we have included the $\mu^{+}\mu^{-}$ channel for our 
analysis because this leptonic channel may explain the observed excesses of 
local positrons \cite{Abeysekara:2014ffg}.\\

\section{Dark Matter Detection}

\begin{quote}{Harley White}
\noindent Physicists hunt for DM, to move it\\
With particle accelerators, to prove it\\
Exists as suspected, from data collected\\
With outcome expected, eureka! projected...\\
\end{quote}

\noindent If DM is dominated by WIMPs, then we should have cosmological abundance of WIMPs. Two
different approaches may be used for the detection of DM particles - direct as well as
indirect. A schematic diagram of production and decay of DM is shown in figure 1.7.\\

\begin{figure}
\centering
\includegraphics[width=0.5\linewidth]{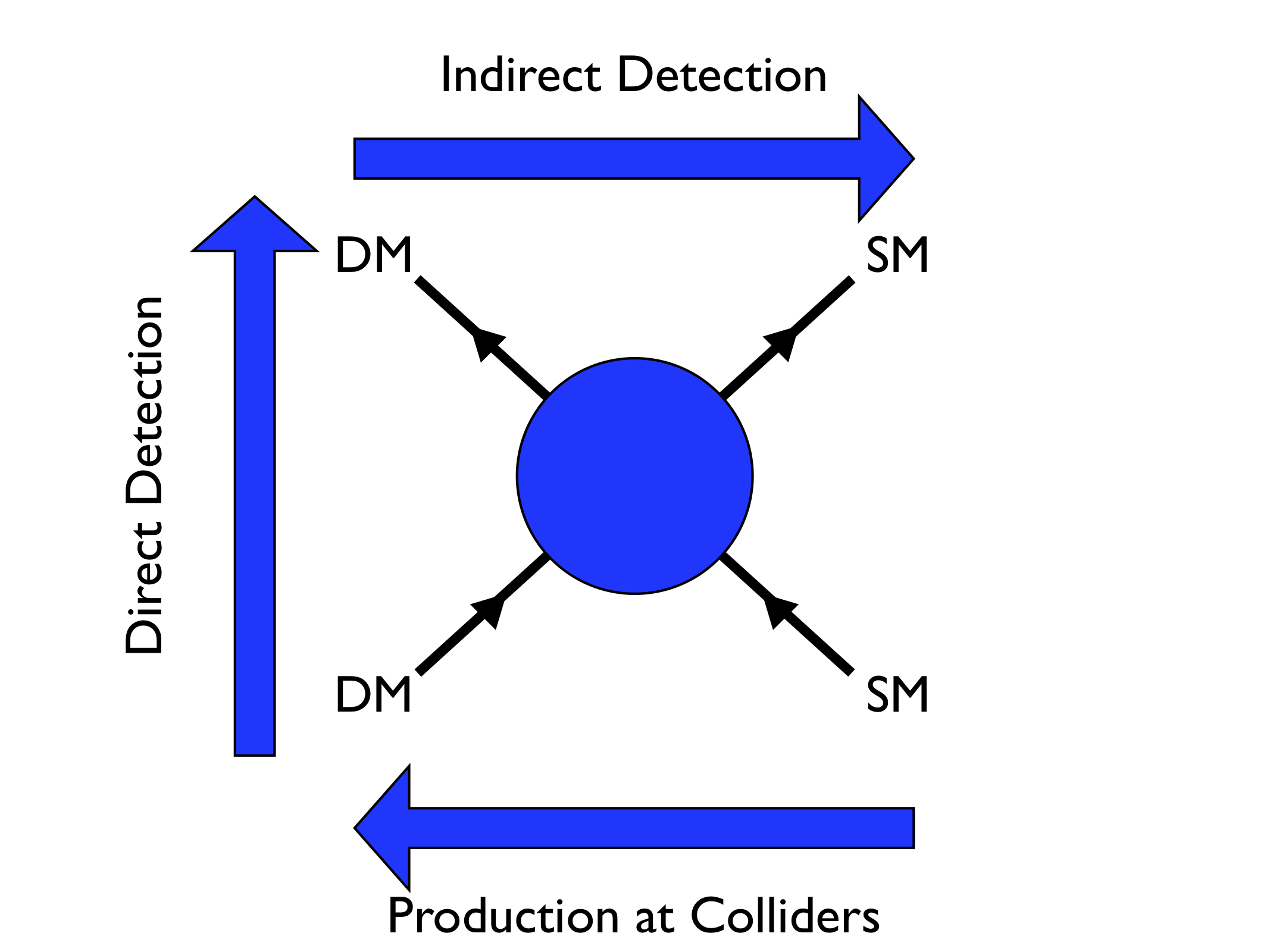}
\caption{Schematic diagram of DM detection through direct, indirect
and its production at Colliders.}
\end{figure}

\noindent It is possible to directly detect DM particles, both from cosmic sources as well as from colliding beam experiments like the LHC. It is assumed that WIMPs 
have weak-scale
scattering cross-section with SM particles and thus it might be possible to 
directly detect the nuclear recoil energy from WIMP-nucleon interactions in 
low-background experiments \cite{Goodman:1984dc, 
Ahmed:2010hw, Aprile:2012nq}. We can also try to generate WIMPs in 
accelerators through
the collision of SM particles. The distinctive signatures (e.g., the missing 
transverse energy)
of these events are expected to be recorded by the collider experiments 
\cite{Baltz:2006fm, Aad:2008zzm, 
Chatrchyan:2008aa}. \\

\noindent Moreover, WIMPs are considered to be a thermal relic and it is 
expected 
that they might
possess a weak-scale self-annihilation cross-section 
\cite{Jungman:1995df, Gunn:1978gr, Stecker:1978du, Bergstrom:1988fp, 
Bergstrom:1997fj}. Thus there might be a fair scope
to indirectly detect the WIMP signature through the SM particles (e.g., photons, neutrinos, positrons, etc.) originating from annihilation.
Different experiments are designed to probe different characteristics of the 
WIMP and all of them have their benefits, difficulties, and uncertainties. But, 
in order to have total knowledge of WIMPs, for example, their eventual 
detection,
identification, and characterization, we need to gather the information from all 
three experimental
techniques.\\

\subsection{Direct Detection}
\noindent The basic assumption for direct detection of DM candidates is that
our Universe is filled with an abundance of WIMPs and many of those WIMPs are 
continuously passing through our terrestrial surface. Thus our terrestrial laboratories 
should 
notice the interaction of WIMPS with the matter by observing the recoil energy 
of nuclei through either ionization, scintillation, or vibrations (phonon). 
This method needs a very clean detector material so that it can detect a 
possible real signal from the background. It is also very important to minimize 
particle backgrounds as much as possible. For the direct experiment, one of the 
most general 
setup is to find 
an underground-site that would effectively reduce the cosmic-ray 
background. The rate of WIMP detection 
depends on various prime factors, such as the mass of WIMPs, the local halo 
density of DM, the velocity distribution in the Milky Way and the cross-section 
on the target nuclei. The detectors generally consist of a very pure crystal as 
in e.g. CDMS\footnote{\tiny{http://cdms.berkeley.edu/}}, DAMA\footnote{\tiny{http://people.roma2.infn.it/ dama/web/home.html}}, CRESST\footnote{\tiny{http://www.cresst.de/}} or a liquid noble gas such as Xenon 
(Xenon100\footnote{\tiny{http://xenon.astro.columbia.edu/xenon100.html}}). 
From the 
theoretical 
prescription, the cross-section of WIMP interactions are predicted to be very 
small, hence very large detectors are needed (e.g. the Xenon100 contains 100 kg 
liquid Xenon) to detect the interactions.\\

\noindent There are several ongoing experiments such as DAMA/LIBRA which are 
designed to detect the DM by using the solid scintillators. For detecting the
particle interactions, the particle detectors of DAMA/LIBRA use the 
thallium-activated sodium 
iodide crystals which are covered in a low radioactivity container with several 
Photomultiplier Tubes (PMTs) \cite{Aad:2012tfa}. These detectors report the annual 
modulation of the signal of confidence level $\approx$ 8.9$\sigma$ 
\cite{Bernabei:2010mq}.\\

\subsection{Indirect Detection}
The indirect detection method is one of the popular ways to identify the 
invisible DM
signals. As we already discussed in the earlier section, WIMP can 
self-annihilate (or
decay) into SM particles. 
With this detection method, we probe the SM particles originated from WIMP
annihilation and then measure the particle spectra generating from them. 
The spectra
would provide us with valuable information about the nature of DM particles. 
There are several
dedicated (ongoing and planned) indirect detection experiments that are designed 
for solving the mystery of the DM. The detection experiments are classified according to the particles they detect.

\begin{itemize}

\item \textbf{Photons:} \\ Gamma-rays, including both direct line photons and 
diffusion photons are one of the most popular methods for indirect DM 
detection. 
WIMP annihilates to a quark and anti-quark products and they later produce a 
jet 
of particles that will generate the gamma-ray spectrum. At high energy, the 
neutral pions decay to a pair of monoenergetic photons that can create the 
prompt line of gamma-rays. When WIMP directly annihilates to the gamma-rays 
i.e., $\chi\chi~\rightarrow~\gamma\gamma$, the energy of the photons is 
proportional to the mass of WIMPs. Since the mass of WIMP is of the order of 
~GeV, it would create a very high energy gamma-rays and detection of any of 
such 
the gamma-ray line would give an obvious indication of the DM annihilation and 
in the indirect detection, it would be referred to as the smoking gun for the 
DM 
search \cite{Bergstrom:2012fi, Weniger:2012tx}. Another source of the 
gamma-rays 
are the internal bremsstrahlung of charged particles produced in the 
annihilation process. The simulation and the observational study suggest that 
Galactic centre, dwarf spheroidal galaxies (dSphs), Low surface brightness 
galaxies (LSBs), cluster of galaxies etc. would be the ideal platform for indirect 
search. The advantage of tracking $\gamma$-ray is that they are electrically 
neutral and do not interact with magnetic fields. Hence, it is possible to track their origin and energy. In the latter 
chapter, we will discuss this method in detail. \\

\noindent For indirect DM detection, there are many dedicated space-based and 
ground-based gamma-ray observatories. The examples of the space-based 
observatories are: Fermi-LAT (Fermi Large Area Telescope), 
AGILE (Astro-rivelatore Gamma a Immagini Leggero \cite{Pittori:2003fgd}),
planned Gamma-400 \cite{Galper:2012fp}, etc. The examples of ground-based Air 
Cherenkov Telescopes (ACTs) are: MAGIC (Mayor Atmospheric Gamma-ray 
Imaging Cherenkov \cite{Aleksic:2011bx}) telescope in La Palma, H.E.S.S. 
(High Energy Stereoscopic System 
\cite{Aharonian:2006pe}) in Namibia, next-generation telescope,
CTA (Cherenkov Telescope Array \cite{Actis:2010bc}), etc. The space-based telescopes 
can directly observe the gamma rays resulting from WIMP pair production within 
their detector, while the ACTs use the atmosphere as part of the detectors and 
detect the Cherenkov light from the air showers which are produced during the 
interaction of gamma rays with the atmosphere. For our thesis work, we have 
considered the space-based gamma-ray telescope but the ACTs have their 
advantages and disadvantages. ACTs can generally observe much higher energy
photons than the space-based telescopes and they have a comparatively large 
collecting area. 
But, 
the advantage of space-based telescopes is that they can cover the whole sky and are more sensitive than ACTs, while 
ACTs need to consider the atmospheric distortions and can not observe the whole 
sky at once.

\item \textbf{Charged particles, positrons, and antiproton:} \\ The possible 
charged particles originating from the WIMP self-annihilation are positrons 
($e^{+}$), electrons ($e^{-}$), antiprotons ($\bar{p}$), and 
anti-deuterons ($\bar{d}$), etc. (see Fig.~1.6). The flux of each charged particles and their anti-particles 
are being estimated from the WIMP mass and the annihilation channels.
There are a few experiments which have prominently reported the excess of 
positrons such as PAMELA (Payload for Antimatter Matter Exploration and 
Light-nuclei Astrophysics), AMS-02 (The Alpha Magnetic Spectrometer), etc. \\

\noindent PAMELA \cite{Adriani:2008zq, Adriani:2008zr} reported an excess in the positron fraction and this 
can be connected to the hint of DM (see e.g. \cite{Bergstrom:2009fa}). 
But there are other existing theories behind such positron excess, some study shows that 
such excess can also be explained by a population of pulsars 
\cite{Malyshev:2009tw}. AMS-02 has also 
observed an excess in the positron fraction and has re-confirmed the findings 
from PAMELA \cite{Accardo:2014lma}. Fermi-LAT, a dedicated 
gamma-ray telescope, can also detect charged particles. Even the results 
obtained from the Fermi-LAT collaboration \cite{Abdo:2009zk} show an excess in 
the electron-positron spectra between 100 to 1000 GeV energy range
and again confirm the positron excess reported by PAMELA 
\cite{FermiLAT:2011ab}. But the problem with Fermi-LAT is that this satellite 
does not have an on-board magnet and so it is not possible for Fermi-LAT to 
distinguish the signal of positrons from electrons.

\item \textbf{Neutrino:} \\ Neutrinos ($\nu$) and anti-neutrinos ($\bar{\nu}$) 
produced in the annihilation of DM particles serve as a good signal for their 
parent particles. The advantage of searching neutrino is that their weak 
interactions lead to the long mean free path. For heavy DM particles, one 
expects 
to see high energy neutrinos coming from the region where the concentration of 
DM is generally high. Unlike photons, neutrinos can be detected in a controlled 
environment of underground laboratories, underwater, or in ice. Presently, the 
active neutrino detectors include Super-Kamiokande (Super-K), ICECUBE 
\cite{Achterberg:2006md}, and ANTARES \cite{Ageron:2011nsa}. The IceCube 
collaboration has looked for muon-neutrino signals from annihilating DM in 
nearby galaxies, galaxy clusters, Galactic centre, Sun, and 
Galactic halo \cite{Aartsen:2012kia, Aartsen:2013dxa, Aartsen:2014hva, 
Abbasi:2012ws}. To date, there are no signals observed in the neutrino channel 
yet from Super-K, ICECUBE, and ANTARES.

\end{itemize} 

\subsection{Collider Searches}

\noindent It is believed that under ideal environment, the DM particles can be produced
in 
the 
colliders. The idea behind the collider searches is to generate DM candidates 
from SM particles i.e., from SM+SM $\rightarrow$ DM+DM. 
Search for DM in colliding beam experiments suffers from several disadvantages. Production
rate of the SM particles is very large compared to the possible DM particles. Also, the DM
candidates may not be directly observed, but the SM particles produced in their decays.
Since it may not be possible to detect all the SM particles produced in the decay of the DM
particles, measurement of their masses in a colliding beam experiment is difficult at the best. \\

\noindent The LHC provided data of proton-proton collisions at centre-of-mass energies 7, 8 and 13
TeV. ATLAS \cite{Aad:2008zzm} and CMS \cite{Chatrchyan:2008aa}, the two major experiments at the LHC have done a number of
analyses and have not seen any signal of DM \cite{Abercrombie:2015wmb}. It is hoped that the next run of the LHC will
reveal evidence of Physics beyond the SM, including the DM candidates.\\

\chapter{Multiwavelength searches for dark matter}
\section{Dark Matter Rich Targets}
\noindent In order to search for the DM signal, we first need to look for the DM 
dominated regions. The observational 
evidences and the optical studies indicate many potential targets. But all of 
these regions have their advantages and 
challenges. We need to explore that in detail. \\

\begin{figure}
\begin{center}
\includegraphics[width=0.8\linewidth]{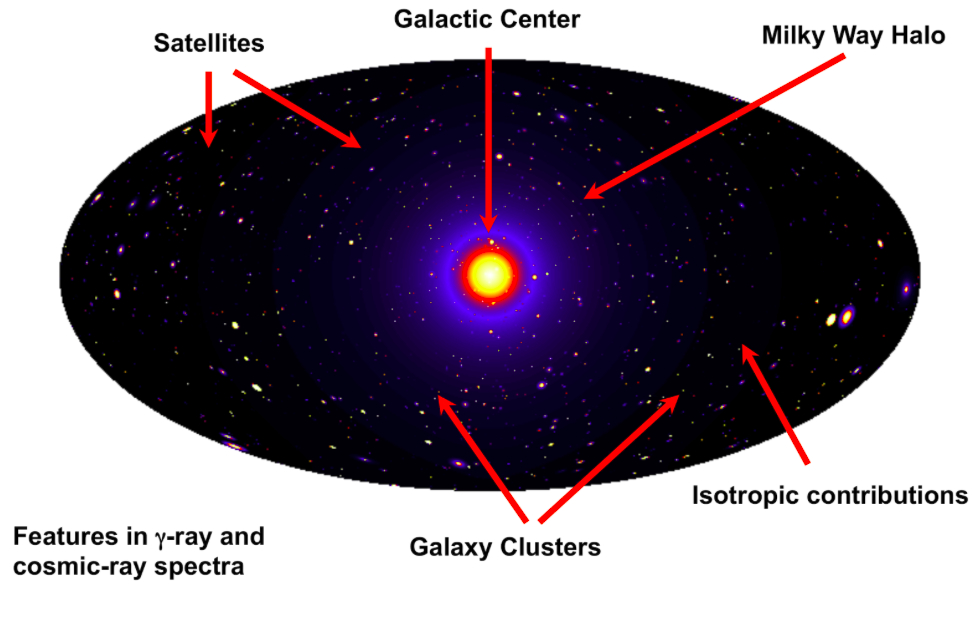}
\caption{DM rich regions.}
\end{center}
\end{figure}

\noindent The Galactic centre (GC) is assumed to be one of the most dense DM 
regions. But this region consists of a large 
number of unknown gamma-ray sources and a very complicated diffuse gamma-ray 
emission resulting from cosmic-ray interacting with interstellar radiation fields 
and gas. Hence, it is very tough to reduce all background emission by modelling 
and that would simultaneously increase the uncertainties in analysis. Looking for DM signal from the GC needs extra caution to avoid background from other
astrophysical sources like pulsars. The spectral line from DM annihilation/decay should be
distinct from the expected background.
But the issue is with the low statistics and inadequate instrumental 
facilities. 
Few studies have observed a hint of a line signal from the GC (e.g. 
\cite{Weniger:2012tx}), but the significance of that emission was very low and 
unfortunately, such significance has decreased over time 
\cite{Weniger:2013dya}. Hence, the true origin of such excess from the GC is 
still under debate \cite{Ackermann:2012qk, Ackermann:2013uma}.\\

\noindent Galactic halo is also believed to be rich in DM. Although we do not have much knowledge
of the mass and shape of the Galactic halo, the background in this region is less
complicated. The resulting upper limits on the annihilation cross-section obtained from the
Galactic halo is comparable to the results from dwarf spheroidal galaxies (dSphs) but they
have much larger uncertainties compared to dSphs \cite{Ackermann:2012rg}.\\

\noindent The high latitude isotropic diffuse emission, i.e., the combination of 
unresolved DM halo and possible Galactic subhalos, can constrain the 
extragalactic isotropic DM signal. The resulting upper limits on the 
annihilation cross-section obtained from the extragalactic diffusion emission 
are comparable to the limits from dSphs for mass above $\approx$ $10^{3}$ GeV 
\cite{Ackermann:2015tah} and for some cases the isotropic diffuse emission can also provide more 
stringent limits than dSphs. But, their DM profile have large uncertainties and 
that makes it difficult to know the true nature of the unresolved sources and 
DM 
halos. Hence, it is hard to distinguish between the positive DM signal with the 
diffuse gamma-ray emission resulting from the unresolved sources 
\cite{Conrad:2015bsa}.\\

\noindent The galaxy clusters are also considered as the DM dominated systems. 
Most of the galaxy clusters are situated at high Galactic latitude and that
significantly reduces the contamination from the galactic diffuse emission. But 
several studies report that in some clusters, the emission might come from their 
cosmic-ray scenarios and DM substructure could have a significant contribution to that. Hence, it also leads to large 
uncertainty in the astrophysical factors \cite{Ackermann:2012rg, 
Ackermann:2013iaq, Pinzke:2011ek}. \\

\noindent In our local universe, low surface brightness (LSB) galaxies would 
also, be considered as
DM dominated galaxies. LSB
galaxies are metal-poor, hardly show any signs of star 
formation\cite{Impey:1997uc} and their stellar disks are generally embedded in 
a rich extended
neutral HI gas disk\cite{deBlok:1996jib, Burkholder:2001sgt, ONeil:2004mqj, 
Du:2015dfg}. The discrepancy between the HI mass estimated from the
rotation curves and the visible baryonic mass (i.e. derived from gas and stars) 
indicates that LSB galaxies are very rich in DM content\cite{deBlok:1997zlw}. 
These galaxies generally do not show any significant emission resulting from their
astrophysical activities i.e., from star-forming regions and hence, important for indirect search
of DM candidates \cite{Honey:2018dfr}. Besides, their extended HI rotation curve and gas
kinematics are used to investigate the distribution of DM at central halo 
and that might help to resolve the much debated
‘cusp-core’ problem in the CDM theory for galaxy 
formation\cite{vandenBosch:2000rza}.
But the problem with the LSB galaxies is that they lie at very large distances 
(e.g. at the order of Mpc) and that would weaken their 
astrophysical factors (at least 3 orders magnitude lower than dSphs). With such a low value of astrophysical factors, it is 
hard 
to provide any strong limit on the DM models. We will discuss this 
in Chapter 7.\\

\noindent The predictions from the cosmological N-body simulation indicate that the structure of the 
CDM halos assumed to be formed by the WIMPs are not even. Recent simulation result indicate that halos contain a large number of bound 
substructures or we can say sub-halos\cite{Abdo:2010ex, Diemand:2005vz, 
Kuhlen:2008qj, Springel:2005nw}. The simulation also hints the 
existence
of a huge number of DM sub-halos around the MW’s 
(MW)\cite{Kuhlen:2009jv, Drlica-Wagner:2013bhh} halo and among all these
predicted sub-halos, few hundreds are assumed to be massive enough to become
the dwarf spheroidal galaxies (dSphs) or the ultra faint dwarf spheroidal 
galaxies (UFDs)\cite{Kuhlen:2009jv}.\\

\begin{enumerate}

\item \textbf{Dwarf Spheroidal Galaxies (dSphs):}\\
The 
dSphs are considered as the largest galactic substructures around the MW. Their 
mass-to-light ratio lies between 100–1000
$M_{\circ} /L_{\circ}$, in which $M_{\circ}$ and $L_{\circ}$ are the solar mass 
and the solar luminosity, respectively. Hence, the dSphs could be the most DM 
dominated structures of the galactic halo. Their large DM content, minimal 
Galactic 
foreground emission, and lack of astrophysical radiation \cite{Mateo:1998wg, 
Grcevich:2009gt} make dSphs promising targets for the indirect detection of DM. 
Since the DM content of each dSphs can be determined from stellar kinematic 
data, it is possible to predict the relative strength and spatial distribution 
of the annihilation signal expected from each galaxy. These characteristics 
provide a mechanism for distinguishing a DM annihilation signal in dSphs from 
conventional astrophysical backgrounds. \\

\item \textbf{Ultra-faint Dwarf Spheroidal Galaxies (UFDs):}\\
Since the last few decades, the Panoramic Survey Telescope and Rapid Response 
System 
(Pan-STARRS) \cite{Kaiser:2002zz, Laevens:2015kla, Laevens:2015una}, the Sloan 
Digital Sky Survey (SDSS) \cite{York:2000gk, 
Belokurov:2010rf}, the Dark 
Energy Survey (DES) \cite{Abbott:2005bi, Bechtol:2015cbp, Koposov:2015cua, 
Kim:2015ila} experiment and the Dark Energy Camera at Cerro Tololo 
\cite{Kim:2015xoa, Martin:2015xla} have detected a set of UFD galaxies. They 
have very low stellar contents and that hints that they could be very rich in DM 
\cite{Grebel:2003zq, 
Evans:2003sc, Bonnivard:2015xpq, York:2000gk, Belokurov:2010rf, Kaiser:2002zz}. 
The UFDs are generally characterised by very old ($\geq$ 12 Gyr) stellar 
populations with large velocity dispersions and inferred mass-to-light ratios 
reaching up to $\approx$ 3000 $M_{\circ} /L_{\circ}$. The high value of 
velocity 
dispersion and mass-to-light ratios support the existence of significant DM in 
UFDs \cite{Koch:2011tb}. Hence, by analysing such UFDs we can aquire a 
substantial knowledge of the nature of the ancient galaxies \cite{Simon:2007dq, 
Kirby:2013wna} that were accreted to form the MW halo \cite{Norris:2010zs, 
Belokurov:2013aga} and the origin of the chemical abundances of the stellar 
population of Milky Way (MW) halo \cite{Starkenburg:2014hca}. Therefore, the UFDs 
are 
considered as the best tracers of early DM sub-halos in the universe 
\cite{Kuhlen:2009jv, Drlica-Wagner:2013bhh, Belokurov:2013aga, Kim:PhDthesis}. 
\\
\end{enumerate}

\section{Dark Matter Density Distributions}
\noindent The exact nature of the DM distribution is still in debate but
several theoretically favoured density profiles can considerably 
fit the N-body simulations data. Amongst all proposed density profiles, the popular 
profiles are the Navarro-Frenk-White (NFW) density profile 
\cite{Navarro:1996gj}, the Burkert (BURK) profile \cite{Burkert:1995yz}, the Pseudo 
Isothermal profile (ISO) \cite{Gunn:1972sv}, the Einasto profile 
\cite{Einasto:1965czb}, etc.

\noindent The NFW profile is defined as
\begin{equation}
\rho_{NFW}(r)=\frac{\rho_{s}r_{s}^{3}}{r(r_{s} + r)^{2}}
\end{equation}
\noindent where, \\
\noindent $\rho_{s}$ = characteristic density of NFW profile and\\
\noindent $r_{s}$ = scale radius of NFW profile.  \\

\noindent The BURK profile is defined as
\begin{equation}
\rho_{BURK}(r)=\frac{\rho_{B}r_{B}^{3}}{(r_{B}+r)(r_{B}^{2} + r^{2})} \\
\end{equation}
\noindent where, \\
\noindent $\rho_{B}$ =central density of BURK profile and\\
\noindent $r_{B}$ = core radius of BURK profile.\\

\noindent The ISO profile is defined as\\
\begin{equation}
\rho_{ISO}(r)=\frac{\rho_{c}}{(1+\frac{r^{2}}{r_{c}^{2}})} \\
\end{equation}
\noindent where, \\
\noindent $\rho_{c}$ = central density of ISO profile and\\
\noindent $r_{c}$ = core radius of ISO profile.\\

\noindent and the Einasto (EINO) profile is defined as

\begin{equation}
\rho_{EINO}(r)= \rho_{e} \exp\Big[\frac{-2((r/r_{e})^\alpha -1)}{\alpha} \Big]
\end{equation}
\noindent where, \\
\noindent $\rho_{e}$ = characteristic density of EINO profile and\\
\noindent $r_{e}$ = scale radius of EINO profile.\\

\noindent For Einasto profile $\alpha$ defines the shape of the distribution. 
In 
Fig. 2.2, we have shown the comparison among NFW, BURK, ISO
and EINO. The NFW profile defines the cuspy 
distribution of DM whereas the BURK and ISO are the cored 
profile with 
constant DM core. The rotation curves of LSB and late-type and gas-rich dwarf 
seem to indicate an approximately
constant DM density in the inner parts of galaxies, while the N-body 
cosmological simulations indicate a steep power-law
type behaviour. This controversy is generally known as the ``core-cusp 
problem'' and till today it remains as one of the unresolved problems in
DM distribution, especially for the small-scale structure \cite{deBlok:2009sp}.

\begin{figure}
\begin{center}
\includegraphics[width=0.8\linewidth]{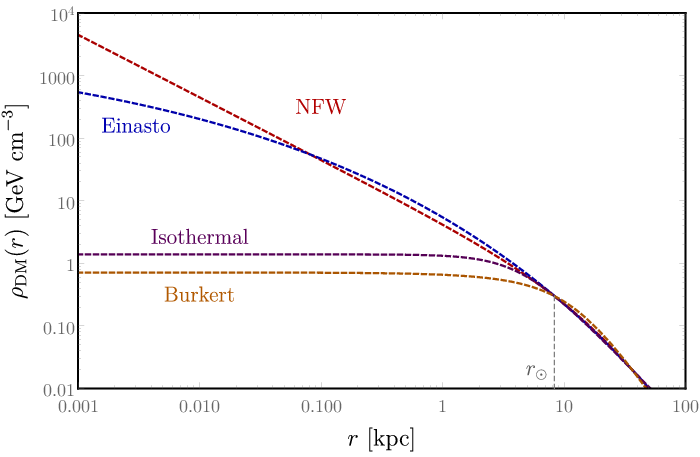}
\caption{NFW, Einasto, Isothermal and Burkert galactic DM density profile. The 
profiles 
are normalized for the Milky-Way such that $\rho_{\odot}$=0.3 GeV $cm^{-3}$ and 
$r_{\odot}$=8.33 Kpc. The diagram is taken from Ref.~Pierre, 2019.}
\end{center}
\end{figure}

\section{The Gamma-ray Signal Resulting from WIMP Annihilation}
\noindent In chapter 1, we have shown how WIMP can annihilate into the SM 
particles and then produce the secondary charged particles and gamma rays as the end 
products of the annihilation chain (see, Fig. 1.6). The gamma-ray resulting 
from the WIMP annihilation is expected to produce a distinct line spectrum 
and 
such a line feature would be completely distinguished from any known
astrophysical phenomena. Thus, it is referred to as the 
``smoking gun''
signature for the indirect search for DM. The nature of the continuum gamma-ray spectra for four different 
annihilation
channels are shown in Fig. 2.3. They are derived from the DMFit package \cite{Jeltema:2008hf} which is 
implemented in the Fermi Science Tools. This DMFit code was first derived using 
the Dark-SUSY \cite{Gondolo:2004sc} package, but later it has been updated by 
Pythia 8.165 \cite{Sjostrand:2007gs} and now this code can consider more 
annihilation channels and cover a wide range of DM masses 
\footnote{\tiny{http://fermi.gsfc.nasa.gov/ssc/data/analysis/software/}}.

\begin{figure}
\begin{center}
\includegraphics[width=0.6\linewidth]{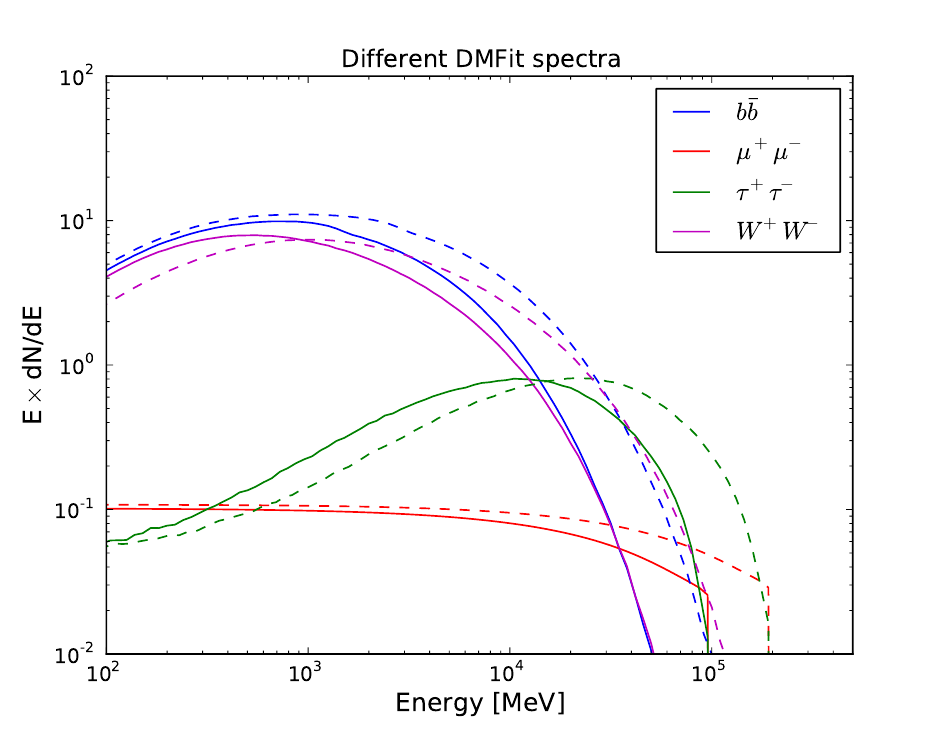}
\caption{Gamma-ray annihilation spectra from WIMPs with masses of 100 GeV 
(solid 
lines) and 200 GeV (dashed lines), annihilating through four different 
annihilation channels (the $b\bar{b}$ channel in blue, $\mu^{+}\mu^{-}$ channel 
in red, $\tau^{+}\tau^{-}$ channel in green, and $W^{+}W^{-}$ channel in 
magenta). Values are obtained from the DMFit package implemented in the Fermi Science Tools.} 

\end{center}
\end{figure}

\noindent The $\gamma$-ray flux originating from the DM annihilation depends on 
both the distribution of DM and the particle physics involving pair 
annihilation. At a specific energy E, the differential $\gamma$-ray flux 
$\phi_{\rm{WIMP}} (E, \Delta \Omega)$ (in units of photons 
$cm^{-2}s^{-1}GeV^{-1}$) arising from the WIMP annihilations of mass $m_{DM}$ 
in a region within a solid angle $\Delta \Omega$ can be expressed as 
\cite{Abdo:2010ex}:
\begin{equation}
\phi_{\rm{WIMP}}(E, \Delta \Omega)~ = ~ \Phi^{pp}(E) \times J(\Delta \Omega),
\end{equation}
\noindent where, $\Phi^{pp}(E)$ is the particle physics factor and $J(\Delta 
\Omega)$ is the astrophysical factor.

\subsection{Particle-Physics Factor}
\noindent The $\Phi^{pp}$ depends on the characteristics of particles 
generating 
through WIMP annihilation. The expression of the particle physics factor can be 
written as \cite{Abdo:2010ex}:
\begin{equation}
\Phi^{pp}(E)~ = ~\frac{<\sigma v>}{8 \pi ~m^{2}_{\rm{DM}}} \sum_{f} 
\frac{dN_{f}}{dE}B_{f}.\\
\end{equation}
\noindent where, $<\sigma v>$ is the thermally-averaged annihilation 
cross-section and $m_{DM}$ is mass of WIMP. $\frac{dN_{f}}{dE}$ denotes the 
differential photon spectrum for each possible pair-annihilation final state and $B_{f}$ is the branching ratio corresponding to the final state, `f'. 
The selection of SM final candidates, through which annihilation would 
occur, is theoretically motivated. Several numerical packages like Pythia 
\cite{Sjostrand:2007gs}, DarkSUSY\cite{Gondolo:2004sc}, DMFit 
\cite{Jeltema:2008hf}, etc. are designed to estimate differential photons yields 
from each annihilation channel.\\

\subsection{Astrophysical Factor (J-factor)}
\noindent The Astrophysical factor (or J-factor) characterizes the astrophysical 
properties of the DM dominated sources. The J-factor 
depends on the spatial distribution of the DM and directly proportional to the 
line-of-sight integral of the squared of DM particle density, i.e. $\propto$ 
$\rho^{2}$. The expression of the J-factor is \cite{Abdo:2010ex}:
\begin{eqnarray}
J (\Delta \Omega) &=& \int \int \rho^{2}(r(\lambda)) d\lambda ~ d\Omega  
\nonumber \\
                  &=& 2 \pi \int_{\theta_{\rm{min}}}^{\theta_{\rm{max}}} 
\rm{sin} \theta \int_{\lambda_{\rm{min}}}^{\lambda_{\rm{max}}} 
\rho^{2}(r(\lambda)) d\lambda ~ d\theta .
\end{eqnarray}
\noindent In Eq. 2.7, $\lambda$ and $r(\lambda)$ are the line-of-sight (l.o.s) and galactocentric distance, respectively. $\theta$ is the angle between the l.o.s and the center of the target. The value of $\theta_{max}$ is the angle over which 
we would average the integration of the J-factor. We generally
use the resolution of the detector as the $\theta_{max}$, for example, if we use 
the data observed by Fermi-LAT, we would consider $0.5^{\circ}$ as $\theta_{max}$ for 
J-factor calculation. For $\theta_{min}$, we generally use $0^{\circ}$.\\

\noindent The expression for $r(\lambda)$ is,
\begin{equation}
r(\lambda) = \sqrt{\lambda^{2} + d^{2} - 2~ \lambda ~d~ \rm{cos \theta}} 
\end{equation}

\noindent where, d is defined as the heliocentric distances of the target.\\

\noindent The maximum and minimum limits of $\lambda$ can be represented as 
\cite{Evans:2016xwx}

\begin{eqnarray}
\lambda_{\rm{max}} = d\rm{cos \theta} + \sqrt{R_{\rm{t}}^{2} - d^{2} 
\rm{sin^{2} \theta}} \\
\end{eqnarray}
\noindent and
\begin{eqnarray}
\lambda_{\rm{min}} = d\rm{cos \theta} - \sqrt{R_{\rm{t}}^{2} - d^{2} 
\rm{sin^{2} \theta}}
\end{eqnarray}
 
\noindent respectively. Here, $R_{t}$ is the tidal radius for DM rich galaxies. For dSphs, 
we generally consider $R_{t}$ for evaluating the maximum
and minimum range of l.o.s distances. The tidal radius of
the dSphs halo in the gravitational potential of
the MW is estimated from the Jacobi
limit \cite{Binney:1987}. But, for comparatively large system, say 
for Low surface brightness galaxies, we use virial radius ($R_{vir}$) in place 
of $R_{t}$.\\

\subsection{Expected Gamma-Ray Flux from WIMP Annihilation}

\noindent Several numerical packages such as 
DarkSUSY\cite{Gondolo:2004sc}, Pythia\cite{Sjostrand:2007gs}, 
DMFit\cite{Jeltema:2008hf}, etc. are generally used for 
producing the spectra of the events resulting from the WIMP annihilation and 
can also be used to model the probable interactions between the incoming and 
the outgoing particles. All of these packages are developed for simulating the 
possible interactions between the self-annihilating DM particles and then to 
estimate the probable number of end products (such as gamma rays, neutrino or 
secondary charged particles) resulting from the DM annihilation. For our work, 
especially for the $\gamma$-ray analysis, we have used 
DMFit tool that is designed to estimate the possible gamma-ray spectrum 
resulting from the DM annihilation for any DM mass in GeV ranges and  
possible annihilation channels. The DMFit tool is developed from several set of 
MC simulations codes used for the hadronization and decay of the DM 
annihilation final products. The same set of MC simulations codes are also used 
by the DarkSUSY package\cite{Gondolo:2004sc} which uses the Pythia 
6.154\cite{Sjostrand:2007gs} as the event-generator. Once such codes estimate the differential flux (i.e. 
$\frac{dN_{f}}{dE}$) originating from pair-annihilation, we can determine the total $\gamma$-ray flux corresponding to the DM signal for any target.\\

\noindent WIMP can self-annihilate to several possible channels, but for our 
analysis, we mostly preferred five combinations, such as: $\chi\chi \rightarrow \tau^{+}\tau^{-}$, $\chi\chi \rightarrow 
\mu^{+}\mu^{-}$, 
$\chi\chi \rightarrow W^{+}W^{-}$, $\chi\chi \rightarrow b\bar{b}$ and 
$\chi\chi 
\rightarrow 80\%~b\bar{b} + 20\%~\tau^{+}\tau^{-}$.
The reasons for 
choosing these annihilation channels are already discussed in section~1.5.

\subsection{The Astrophysical Backgrounds}
\noindent For gamma-ray data analysis, the backgrounds play a very 
significant role. Especially for very faint sources, it is very important to investigate the background region in detail. Otherwise, in due course of analysis, the emission coming from the surroundings can be 
classified as the gamma-ray counts from the source location while they are just the background coming from the 
nearby point sources or the Galactic and the Extragalactic foreground.

\noindent Data collected by the Fermi-LAT has been used for the gamma-ray analysis part of this
thesis, as described in Chapters 3 and 4. 
The space-based telescopes lie above the earth's atmosphere and that help to reduce the possible background contamination at the time of data recording. That enables the detectors to produce much clear images than the ground-based telescope. During the 
analysis, we can again screen our data with event classification process and can 
model 
out the remaining contribution of backgrounds (i.e. gamma-ray emission from diffuse galactic and 
extra-galactic components and nearby gamma-ray sources).

\noindent When Fermi-LAT detects the background particles originating from 
cosmic 
rays (or from cosmic-ray's interactions with the Earth’s atmosphere), initially 
those particles are counted as the events\cite{Ackermann:2012kna}. Fermi-LAT 
has 
the segmented anti-coincidence detector which vetoes most of the passing cosmic rays 
(briefly described in Chapter 3.1) and the remaining cosmic rays are correctly 
deferred in the event selection process. The residual cosmic-ray background 
contamination is included as 
the gamma-ray backgrounds to the source model. The Fermi-LAT 
collaboration provides the necessary background models and source catalogs. The background models consist of a Galactic diffuse emission 
template, an extragalactic isotropic diffuse emission template and the contribution from all the nearby sources that lie within the radius of interest.

\noindent The Galactic diffuse emission template contains both the spatial and 
spectral part of the cosmic emission. The template is derived by fitting the 
inverse Compton radiation maps as predicted by GALPROP \cite{Strong:2007nh} and 
gamma-ray emissivities obtained from gas density maps with known point sources 
and a model for isotropic diffuse 
emission\cite{Nolan:2011sjt}\footnote{\tiny{
http://fermi.gsfc.nasa.gov/ssc/data/access/lat/BackgroundModels.html}}.

\noindent The isotropic template contains the extra-galactic residual 
cosmic-ray contamination and the emission from the unresolved point sources. 
This isotropic diffuse model is 
generated by fitting the extra-galactic residual emission to the high-latitude 
sky with the emission from the Galactic template 
and from other known gamma-ray sources. These two Galactic diffuse emission 
template and  extragalactic isotropic diffuse emission template play an 
important
role to eliminate the possible background emission.

\noindent There is a high possibility that the events detected by the Fermi-LAT 
are 
contaminated by the photons coming from the Earth's albedo effect. The 
Fermi-LAT 
team has recommended 
to 
use the zenith angle cut as 90 degree to reduce the background photons 
resulting 
from the Earth's albedo. Fermi-LAT team has provided the Earth-limb template 
(below 100 MeV) for all available Fermi Gamma-ray LAT (FGL) source catalogs, i.e for 2FGL 
\cite{Nolan:2011sjt}, 3FGL \cite{Acero:2015gva} and 4FGL 
\cite{Fermi-LAT:2019yla}.

\noindent With time, Fermi-LAT team have released several version of source catalogs. The 
first published catalog was 1FGL \cite{Abdo:2010ru} which was 
prepared on 11 months of data. 1FGL catalog contains 1451 sources where 
the sources are modelled with the power-law spectrum. They have implied several 
improvements 
for their second source catalog, i.e. for 2FGL, \cite{Nolan:2011sjt} which have 
24 months of data. In that catalog, the Fermi-LAT team have used the updated
diffuse models. They have also considered extended and non-power-law 
sources and an improved source association process. The 2FGL catalog contains 
1873 sources.

\noindent The third released source catalog by Fermi-LAT team is the 3FGL 
\cite{Acero:2015gva}. This catalog is derived by the first four years 
of 
Fermi-LAT data and for energy range between 100 MeV to 300 GeV. This 
catalog contains 3033 sources (Fig. 2.4). This catalog has included many new 
sources which already have known counterparts in other surveys. The 
new identified or associated sources are from the blazar class (or from active 
galaxies), supernova, pulsar and X-ray binaries \cite{Acero:2015gva}.

\noindent The most recent version of the Fermi-LAT catalog for point sources 
i.e. the 4FGL catalog \cite{Fermi-LAT:2019yla} consists of 5064 sources (Fig. 
2.5). This
catalog is generated from the first eight years of Fermi-LAT data and is 
working between the energy
range from 50 MeV to 1 TeV. Amongst all the Fermi-LAT published source 
catalogs, 4FGL is the deepest one
if we consider the energy range. Relative to the 3FGL catalog, the 4FGL one 
has incorporated the improved analysis method and 
updated models for the Galactic and isotropic diffuse $\gamma$-ray emission. 
The 4FGL catalog consists of
1336 unassociated sources, whereas 239 are the pulsars and more than 3130 of 
the identified or associated sources
are the active galaxies from the blazar class.

\begin{figure}
\begin{center}
\includegraphics[width=1.0\linewidth]{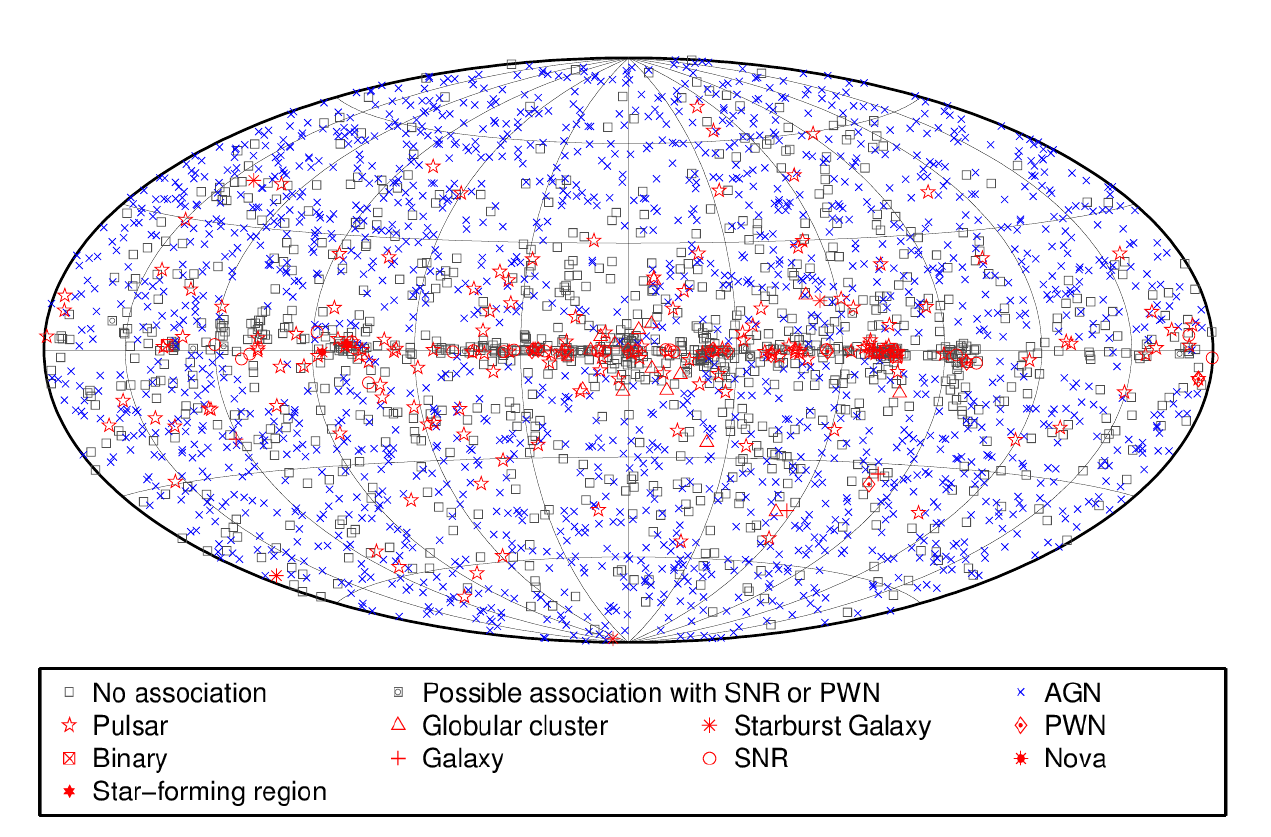}
\caption{Sources from the third Fermi-LAT catalog (3FGL) plotted in Aitoff 
projection.}
\end{center}
\end{figure}

\begin{figure}
\begin{center}
\includegraphics[width=1.0\linewidth]{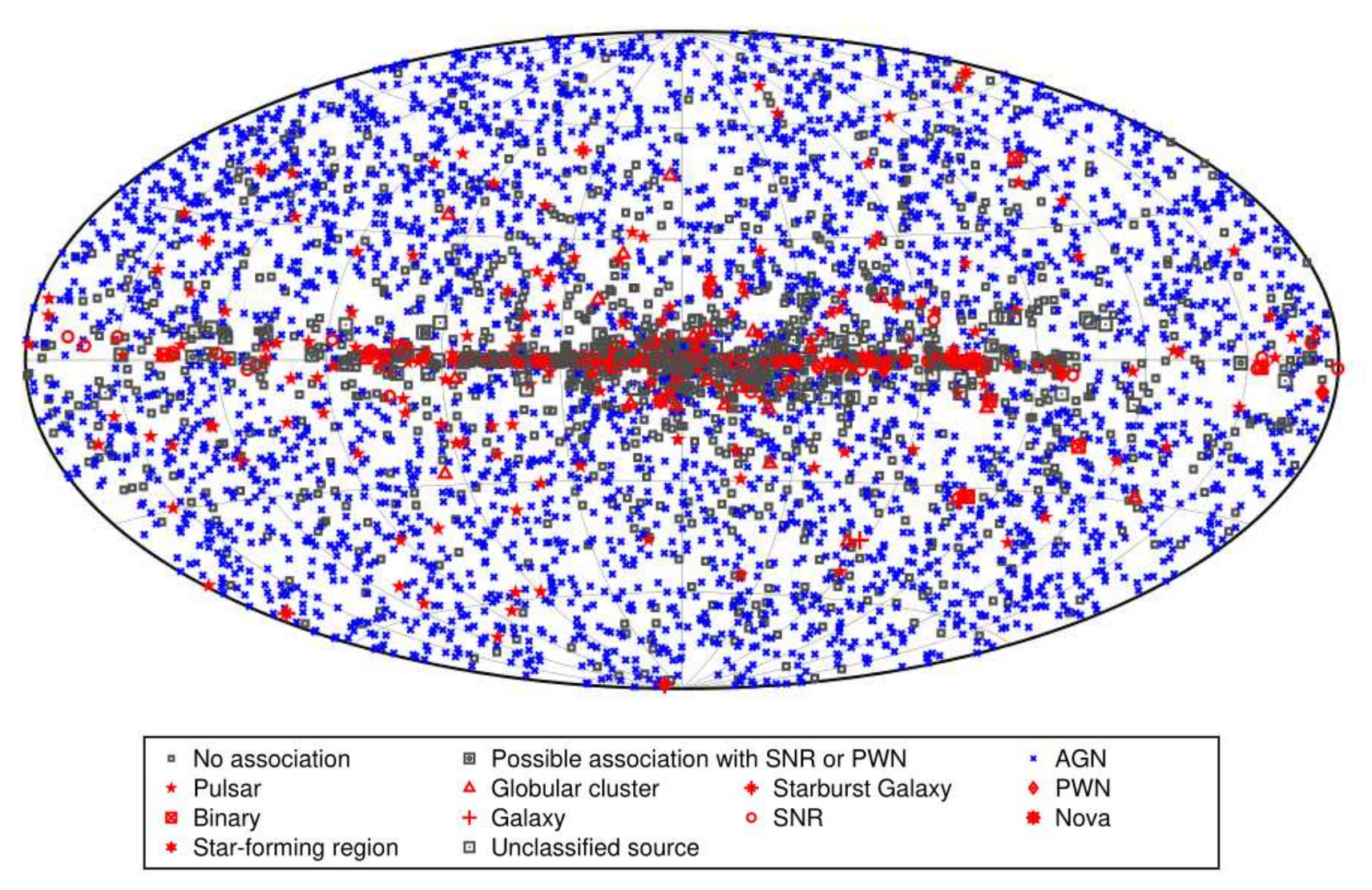}
\caption{Sources from the fourth Fermi-LAT catalog (4FGL) plotted in Aitoff 
projection.}
\end{center}
\end{figure}

\section{The X-ray and Radio Signal Resulting from WIMP Annihilation}
\noindent For DM detection it is important to use multi-wavelength studies complementing the $\gamma$-ray excess with increasing the time period of $\gamma$-ray
analysis. In case of dSphs, it has already been pointed out that the observational limits
obtained from the radio and X-ray data are competitive with $\gamma$-ray \cite{Regis:2017oet,Jeltema:2008ax}.

\begin{figure}
\begin{center}
\includegraphics[width=0.8\linewidth]{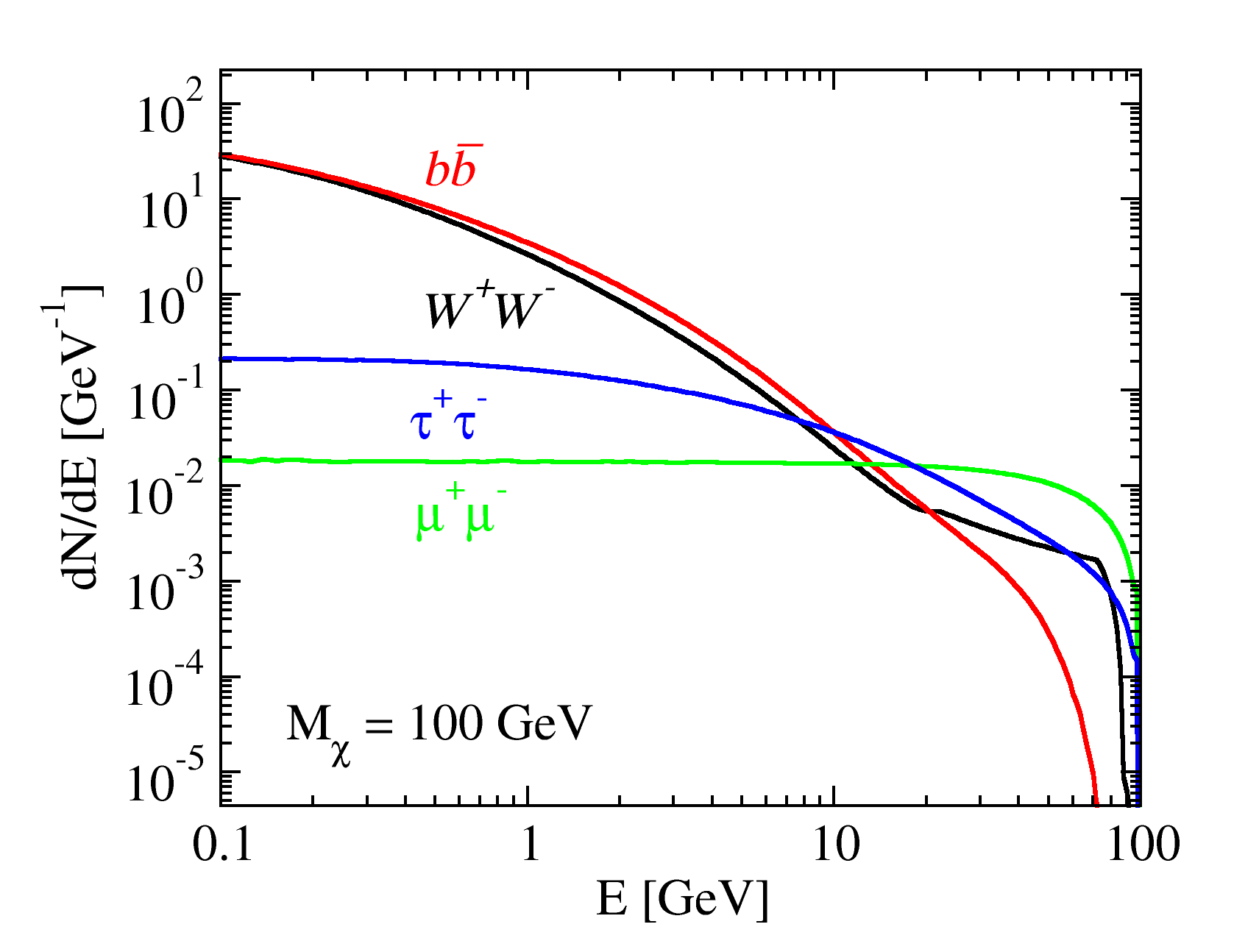}
\caption{The $e^{-}e^{+}$ injection spectra resulting from WIMPs annihilation 
with 100 GeV DM mass
for four different annihilation channels (the $b\bar{b}$ 
channel in red, $\mu^{+}\mu^{-}$ channel in green, $\tau^{+}\tau^{-}$ channel 
in 
blue, and $W^{+}W^{-}$ channel in black). The spectrum is obtained from 
DarkSUSY.}
\end{center}
\end{figure}

\noindent In WIMP annihilation chain, the secondary charged 
particles are generated via various reactions, such as: $\pi^{\pm} 
\rightarrow \mu^{\pm} + \nu_{\mu}(\overline{\nu_{\mu}})$, with $\mu^{\pm} 
\rightarrow e^{\pm} + \overline{\nu_{\mu}}(\nu_{\mu})$. When the charged 
particles are propagating through the interstellar medium, they would lose 
their 
energy through a variety of electromagnetic processes such as inverse Compton 
radiation (IC), synchrotron radiation, Coulomb losses and bremsstrahlung, etc. Charged particles passing through the magnetic field of astrophysical objects produce
electromagnetic emission in the radio frequency range \cite{Ginzburg:1967zja, Longair:2001yy}. For the IC emission, the starlight 
photons and the photons resulting from the 2.7K cosmic microwave background 
(CMB) interact with the relativistic charged particles and produce photons in the X-ray range \cite{Ginzburg:1967zja, 
Longair:2001yy}.
Thus, by various radiation mechanism \cite{Colafrancesco:2005ji, 
Colafrancesco:2006he, 
Ginzburg:1967zja, Longair:2001yy}, especially the synchrotron 
emission and the IC emission at high energies, the charged particles 
originating from WIMP annihilation produce the energy spectrum at 
radio and X-ray frequency range.

\noindent In order to examine the radio and X-ray emission, 
we should consider the diffusion coefficient of the region and the relative energy loss from the charged particles for several possible annihilation channels. In Fig.~2.6, we have shown the $e^{\pm}$ 
injection spectra resulting from the WIMPs of 100 GeV mass that annihilates to $b\bar{b}$, 
$\mu^{+}\mu^{-}$, $\tau^{+}\tau^{-}$ and $W^{+}W^{-}$ final states. 
Formalism for solving the transport equation for the number density ($n_e(r,E)$) of 
$e^\pm$ of a given energy $E$ at the position $\mathbf{r}$ with respect to the 
center of the source has been developed in Refs.~ \cite{Colafrancesco:2005ji, 
Colafrancesco:2006he, McDaniel:2017ppt}. The transport equation looks as

\begin{align}\label{eqn:diffusion}
\frac{\partial}{\partial t} \frac{dn_e}{dE} = \nabla . \Big( D(E,\mathbf{r}) 
\nabla \frac{dn_e}{dE}\Big)
                                                + \frac{\partial}{\partial E} 
\Big( b(E,\mathbf{r}) \frac{dn_e}{dE}\Big)
                                                + Q_e (E,\mathbf{r}).
\end{align}

\noindent Here, $ D(E,\mathbf{r})$ is the space-dependent diffusion coefficient 
and $b(E,\mathbf{r})$ denotes the energy loss term. The Source term ($Q_e$) can 
be defined as
\begin{align}
 Q_e (E,\mathbf{r}) = \frac{\rho^2_\chi(\mathbf{r}) \langle \sigma v\rangle}{2 
m_\chi^2}    \frac{dN_{e}}{dE},
\end{align}

\noindent where, $\frac{dN_e}{dE}$ is the number of $e^{+}/e^{-}$ produced at a 
given energy $E$ per DM annihilation.
The solution of the Eq. 2.11 i.e., $\frac{dn_e}{dE}(r,E)$ would 
give us the number density of $e^{+}/e^{-}$ per unit energy at a distance 
r from the center of the source.\\
\noindent The energy loss term is given by 
\begin{align}
b(E,\mathbf{r})  = & b_{IC}(E) + b_{Syn}(E, \mathbf{r}) + b_{Coul}(E) + b_{brem}(E) \nonumber \\ 
      = & b_{IC}^0 E^2 + b_{syn}^0 B^2 E^2  \nonumber \\ 
        & + b_{Coul}^0 n ( 1 + log(\gamma /n)/75) + b_{brem}^0 n (log(\gamma / 
n) + 0.36).
\end{align}

\noindent where, the magnetic field, $B$ is in unit of $\mu G$, $n$ denotes the 
number density of the thermal electrons in unit of cm$^{-3}$ , $\gamma$ is  $E/m_{e}$ and the energy loss coefficients for all 
radiative mechanisms are
$b_{IC}^0 = 0.25 \times 
10^{-16}$ GeV s$^{-1}$,  $b_{syn}^0 = 0.0254 \times 10^{-16}$ GeV s$^{-1}$, 
$b_{brem}^0 = 1.51 \times 10^{-16}$ GeV s$^{-1}$ and $b_{Coul}^0 = 6.13 \times 
10^{-16}$ GeV s$^{-1}$.

\noindent As we still don't have much detailed knowledge on the structure of DM 
distribution, we have assumed the diffusion coefficient $D(E, \mathbf{r})$ to 
be 
independent of position. So, we can safely consider the Kolmogorov form for 
diffusion coefficient:

\begin{equation}
 D(E) = D_0 \Big(E\Big)^{\gamma_{D}}
\end{equation}
\noindent where $D_{0}$ is defined as the diffusion constant.

\noindent If we assume a uniform magnetic field and stationary state of the 
number density of thermal electrons (i.e. $\frac{\partial,}{\partial t} 
\frac{dn_e}{dE} = 0$), the spherically symmetric solution of the diffusion 
equation is given by

\begin{align}\label{eqn:solutionndifusion}
  \frac{dn_e}{dE}(r,E) = \frac{1}{b(E)} \int_{E}^{M_\chi} dE^\prime G\Big(r, 
v(E)-v(E^\prime)\Big) Q_e(E,r),
\end{align}
\noindent where, the  Green's function is given by 
\begin{align*}
 G(r, \Delta v)  = &\frac{1}{\sqrt{4\pi \Delta v}} \sum_{n=-\infty}^{\infty} 
(-1)^n 
                     \int_{0}^{r_h} dr^\prime \frac{r^\prime}{r_n} 
\Big(\frac{\rho_\chi(r^\prime)}{\rho_\chi(r)}\Big)^2 \\
                  & \times \Big[ exp\Big(-\frac{(r^\prime -r_n)^2}{4 \Delta 
v}\Big) - exp\Big(-\frac{(r^\prime + r_n)^2}{4 \Delta v}\Big)\Big],
\end{align*}
\noindent with $r_n = (-1)^n r + 2nr_h$ and $v(E) = \int_E^{M_\chi} d\tilde{E} 
\frac{D(\tilde{E})}{b(\tilde{E})}$. Here, $r_h$ defines the diffusion zone of 
the 
galaxy. Typically, the value of $r_h$ is taken to be twice the radius of the 
last stellar component of the galaxy (i.e. twice the distance of the outermost 
star from center). The solution is obtained with the free escape boundary 
condition $\frac{dn_e}{dE}(r_h,E) = 0$.  For evaluating Green’s
function, we are considering the average magnetic field strength. So, we express the energy loss term only as a function of E, i.e., $b(E,\mathbf{r})~\approx~b(E)$. \\

\noindent As we already discussed, in the presence of the comparatively strong 
magnetic field, the synchrotron radiation would play the most dominant role.
The synchrotron power spectrum ($P_{\rm synch}(\nu,E,B)$) in the 
presence of $B$ is defined as \cite{Longair:2001yy, Storm:2016bfw}: 

\begin{equation}
 P_{\rm synch}(\nu, E, B) = \pi \sqrt{3} r_0 m_e c \nu_0 \, \int_0^\pi \, 
d\theta \, sin^2\theta \, F\big(\frac{x}{\sin\theta }\big),
\end{equation}
\noindent where, $\theta$ is the pitch angle, $r_0 = e^2/(m_e c^2)$ is the 
classical electron radius and $\nu_0 = eB/(2\pi m_e c)$ is the non-relativistic 
gyro-frequency. While,
\begin{equation}
 F(y) = y \,  \int_y^\infty d\zeta \, K_{5/3}(\zeta) \simeq 1.25 \, y^{1/3}\,
      e^{-y} \, (648 + y^2)^{1/12} \ .
\end{equation}
\noindent The quantity $x$ is given by
\begin{equation}
 x = \frac{2 \, \nu\, m_e^2 \, (1+z)}{3 \, \nu_0\, E^2} 
\end{equation}
\noindent with $z$ being the redshift of the source. 
We can also estimate the local emissivity for the synchrotron radiation i.e., the 
energy radiated at a given $\mathbf{r}$, per unit volume per unit time at a 
given frequency $\nu$ in terms of $P_{\rm synch}$ and $dn_e/dE$, 

\begin{equation}\label{eqn:emissivity}
j_{\rm synch}(\nu, r)
 = \int_{m_e}^{M_{\chi}} dE \left(\frac{dn_{e^+}}{dE}
 + \frac{dn_{e^-}}{dE}\right) P_{\rm synch}(\nu,E,B) = 
 2 \int_{m_e}^{M_{\chi}} dE \, \frac{dn_{e^-}}{dE}\,
 P_{\rm synch}(\nu,E,B) \ .
\end{equation}

\noindent Then the expression for integrated synchrotron flux density would be 
 \begin{equation}\label{eqn:syn_flux}
S_{\rm synch}(\nu) = \frac{1}{4\pi d^2}\int d^3 r \, \,  j_{\rm synch}(\nu,r),  
 \end{equation}
\noindent where, $d$ is the distance to the target galaxy.\\

\noindent For regions with low magnetic fields, the IC radiation process would
play the dominant role. Depending on the mass of DM candidates, the emission 
from the IC mechanism would produce a spectral peak between the soft to hard 
X-ray bands 
\cite{Jeltema:2011bd}. The expression for the IC power ($P_{\rm 
IC}(E_{\gamma},E)$) is:

\begin{equation}
 P_{\rm IC}(E_{\gamma},E) = cE_{\gamma} \int d\epsilon \, n(\epsilon) \, 
\sigma\big(E_{\gamma},\epsilon, E\big),
\end{equation}
\noindent where, $n(\epsilon)$ is photon number density, 
$\sigma(E_{\gamma},\epsilon, E)$ is the IC scattering cross-section and 
$\epsilon$ is the energy of the target CMB photons. E is the energy of the 
relativistic $e^{\pm}$ pair and $E_{\gamma}$ is the energy of the upscattered 
photons. From Klein-Nishina formula, we can define the 
$\sigma(E_{\gamma},\epsilon, E)$ as:

\begin{equation}
\sigma(E_{\gamma},\epsilon, E) = \frac{3\sigma_{T}}{4\epsilon \gamma^{2}} G(q, 
\Gamma)
\end{equation}
\noindent where, $\sigma_{T}$ is the Thomson cross-section and the expression of 
$G(q, \Gamma)$ is:
\begin{equation}
G(q, \Gamma) = \Big[2q \ln q + (1+2q) (1-q) + \frac{(2q)^{2}(1-q)}{2(1+\Gamma 
q)}\Big]
 \end{equation}
\noindent where, $\rm{\Gamma = \frac{4\epsilon \gamma}{m_{e} c^{2}} = \frac{4 
\gamma^{2} \epsilon}{E}}$ and $\rm{q = \frac{E_{\gamma}}{\Gamma 
(E-E_{\gamma})}}$. The range of q lie between $1/(4\gamma^{2}) \geq q \geq 1$.\\

\noindent Similar to the synchrotron emission, we can also find the local 
emissivity for IC emission by folding the power with the electron number 
density 
at equilibrium,
\begin{align}\label{eqn:emissivity}
j_{IC}(\nu, r) = \int_{m_e}^{M_{\chi}} dE \Big(\frac{dn_{e^+}}{dE} + 
\frac{dn_{e^-}}{dE}\Big) P_{IC}(E,E_{\gamma}) \nonumber \\
= 2 \int_{m_e}^{M_{\chi}} dE \, \frac{dn_{e^-}}{dE}\, 
P_{IC}(E,E_{\gamma}),
 \end{align}
  
\noindent The integrated IC flux density spectrum is given by 
 \begin{align}\label{eqn:ic_flux}
S_{IC}(\nu) = \frac{1}{4\pi d^2}\int d^3 r \, \,  j_{IC}(E_{\gamma},r),  
 \end{align}

\noindent where, $d$ is the distance to the target galaxy.\\

\noindent Here, we would like to mention that, unlike the gamma-ray emission, the X-ray and the synchrotron flux
is not directly related to the astrophysical factor (J-factor). They are primarily dependent on the diffusion mechanism and
the energy loss processes of the system. In addition, the magnetic field (B) and the diffusion
coefficient (characterised by $D_{0}$ and $\gamma_{D}$) inside the source would also play a crucial role.\\

\chapter{Fermi Large Area Telescope (Fermi-LAT) Gamma-Ray Observatory}
\section{Instrumental Requirements}
\noindent The possible mass of the DM candidates varies from tens of GeV to a few hundred TeV
depending on the theoretical models \cite{Bertone:2010zza}. Hence, the gamma-ray
detector should have a number of capabilities. The gamma-ray detector for DM searches
should have a good energy resolution and sensitivity over a wide energy range. The instrument should have good angular and energy resolution. 
With good angular resolution, it would be possible to detect a faint gamma-ray 
emission originating from WIMP annihilation. With good energy 
resolution, we can distinguish an annihilation spectrum from 
astrophysical backgrounds. Moreover, 
the instrument should have a large field-of-view (FOV) because that would help 
it to observe a vast region of sky at once. Lastly, the instrument 
should have a good timing resolution and a high observing cadence, so that it can identify variable sources such as pulsars (high frequency) or 
active galactic nuclei (low frequency). 

\noindent In section~1.6.2, we have briefly discussed the
detection methods of various space-based telescopes which are especially dedicated to 
search the indirect signature of WIMP annihilation/decay. The 
telescopes are designed to meet most of the necessary features that we have 
discussed. 
In my thesis, for investigating the DM signature in gamma rays, we have used the data observed by the Fermi Large Area Telescope (LAT). 
In the following sections, we will discuss its working principle in detail.

\section{The Large Area Telescope}

\begin{figure}
\begin{center}
\includegraphics[width=0.5\linewidth]{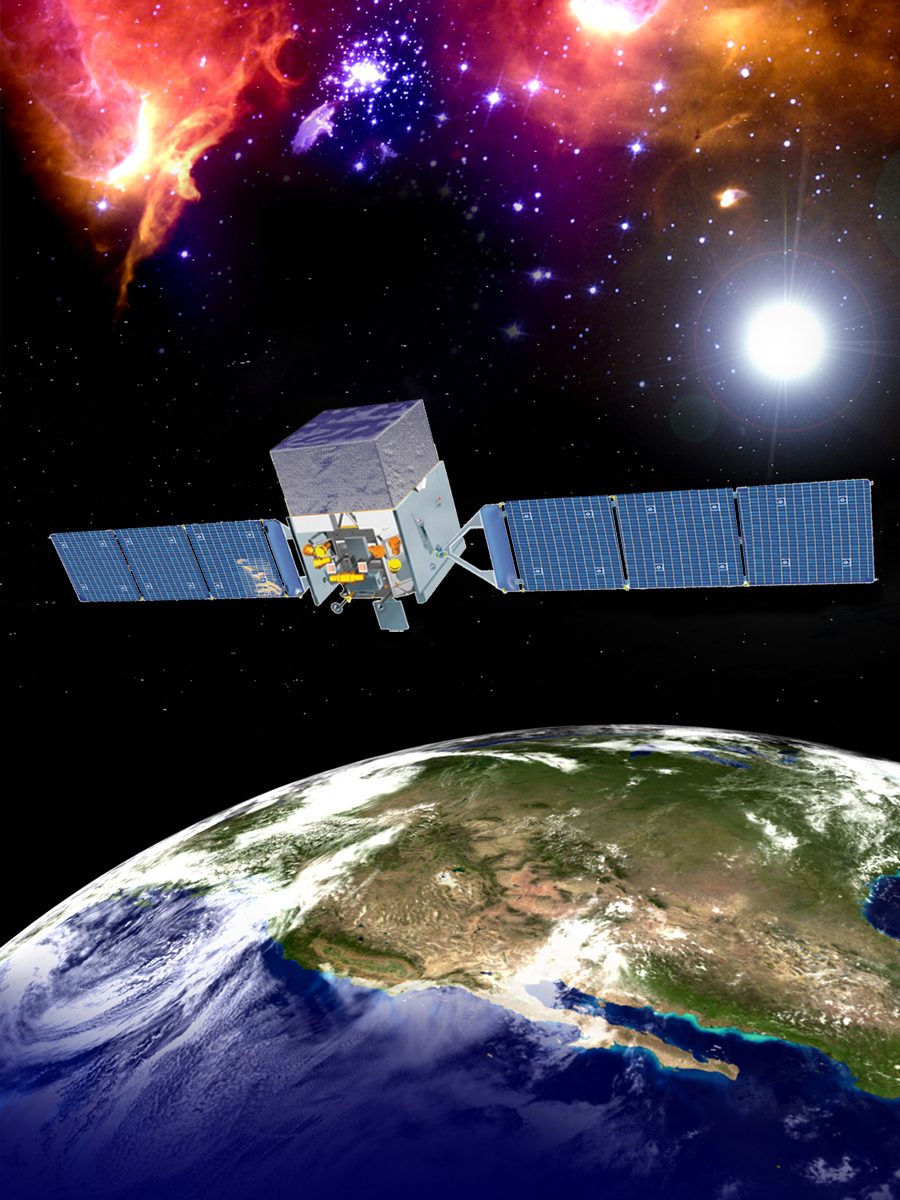}
\caption{A schematic diagram of the Large Area Telescope (LAT).}
\end{center}
\end{figure}

\noindent Fermi-LAT is expected to perform as a brilliant gamma-ray space 
detector over the entire celestial sphere, with comparatively better 
sensitivity 
than other earlier gamma-ray missions. Fermi-LAT team has made
significant improvements in angular resolution, effective area, FOV, energy 
resolution and time resolution of the detector. Such advanced features in
Fermi-LAT can address several unresolved issues in high-energy gamma-ray astrophysics.

\noindent The LAT scans the whole sky for every $\approx$ 192 minute from the 
low-Earth orbit of 565 km altitude at a 25.6-degree inclination with an 
eccentricity $<$0.01 \cite{Atwood:2009ez}. It is launched on June 11, 2008, by 
the Delta II Heavy launch vehicle from Cape Canaveral.

\noindent The principal objective of the Fermi-LAT is to conduct a long term 
high sensitivity observation of the celestial sources for a wide range 
of 
energy band i.e. from $\approx$ 20 MeV to $>$ 500 GeV. It has a large effective 
area combined with good energy, angular and time resolution. Its low deadtime is sufficient enough to study 
transient phenomena. Some key properties of Fermi-LAT are described in Table~3.1
\footnote{\tiny{
https://fermi.gsfc.nasa.gov/ssc/data/analysis/documentation/Cicerone/Cicerone{\_
}Introduction/LAT{\_}overview.html}}.

\begin{center}
\begin{table}
\caption{Properties of Fermi-LAT.} \label{Table-1}
\begin{tabular}{|p{4cm}|p{11cm}|}
\hline
\hline
Parameter & Value or Range  \\ 
\hline
Energy Range & $\approx$ 20 MeV to $>$ 500 GeV \\
\hline
Energy Resolution & $<$ 15 $\%$ at energies $>$ 100 MeV  \\
\hline
Effective Area & $>$ 8,000 $cm^{2}$ maximum effective area at normal incidence 
\\
\hline
Single Photon Angular Resolution & $<~0.15^{\circ}$, on-axis, 68$\%$ space 
angle 
containment radius for E $>$ 10 GeV; $<~3.5^{\circ}$, on-axis, 68$\%$ space 
angle containment radius for E = 100 MeV \\ 
\hline
Field of View & 2.4 sr \\
\hline
Source Location Determination & $<$ 0.5 arcmin for high-latitude source\\ 
\hline
Point Source Sensitivity & $<~6 \times 10^{-9}~ph~cm^{-2}~s^{-1}$ for E $>$ 100 
MeV, 5$\sigma$ detection after 1 year sky survey\\
\hline
Time Accuracy & $<$ 10 microseconds, relative to spacecraft time\\
\hline
Background Rejection (after analysis) & $<~10\%$ residual contamination of a 
high latitude diffuse sample for E = 100 MeV - 500 GeV.\\
\hline
Dead Time & $<$ $<$ 100 microseconds per event\\
\hline
\hline
\end{tabular}
\end{table}
\end{center}

\noindent These features have allowed the LAT to explore the new physics associated with $\gamma$-ray emission.

\subsection{Observational Constraint}

\noindent The LAT has very large FOVs and it can change its direction of 
observation with 
very ease. But its detectors have their observational constraints that need 
to be handled carefully. Fermi-LAT should avoid pointing at or near the Earth 
because 
that can increase the chances of detecting a large number of astrophysical photons. But, at 
low energy Fermi-LAT may sometimes observe the Earth's limb at the time of 
detecting the albedo gamma rays for instrument calibration. Applying a special 
Zenith angle cut can eliminate the photons resulting from the Earth's limb. 
There is another strict precaution for Fermi-LAT observation. Fermi-LAT should 
not record any events when it would transit the South
Atlantic Anomaly (SAA). This region has a high concentration of charged 
particles which are trapped by the Earth's magnetic field.

\subsection{Detection Methodology}

\noindent The Fermi-LAT is a pair conversion detector. During the observation, 
the incident gamma rays penetrate the detector and then interact with a 
high Z converter material. For Fermi-LAT, the tungsten foil is used to convert 
the gamma rays into electron-positron pair. They then pass through the 
silicon strip detectors that track the position of electron-positron pair. As the energy 
of 
gamma-ray is much larger than the rest mass of the electron and positron, the 
daughter products (i.e the charged pair) also predominantly follow the 
direction 
of the incoming gamma-ray. In this process, the new reconstructed direction of 
the gamma rays is restricted by multiple scatterings of the electron-positron 
pair in the tracker and also by the spatial resolution of the tracker material. 
At the bottom of the LAT, the charged particles are deposited into a 
calorimeter 
made of CsI. The calorimeter is thick enough to measure the energy of the 
pairs in the LAT energy band. 

\noindent The charged particles deposit their energy in different parts of the 
tracker and the
calorimeter and then the instrument produces the pulse-height signal as the 
output. In
order to reconstruct the trajectory of the charged particles and their amount 
of energy losses, one needs to combine
the pulse heights with the x-y coordinates from each silicon strip detector 
where the charged
particles hit. 
Both 
on-board and ground analysis reconstruct the tracks of the charged particles 
from their output pulsed data. The Data Acquisition System (DAQ) characterizes 
the interaction that produced the charged particles and also tries to 
distinguish the photons from the background events. Meanwhile, this process 
also 
determines the direction of the incident photon and its estimated energy.
 
\subsection{The LAT Instrument}

\noindent The Fermi-LAT consists of three primary instruments: i) a 
segmented 
anti-coincidence detector (ACD), ii) 16 precision tracker/converter, and iii) 
16 
imaging calorimeter (Figure 3.2). The tracker and the calorimeter form the 
central structure of Fermi-LAT, while the ACD surrounds the tracker 
and the calorimeter. The ACD is again covered by a micrometeorite shield and thermal 
blanket. Trackers and calorimeters act all together to calculate the direction 
of incident particles and their respective energy. The main principle of the 
ACD 
is to identify the incoming charged particles and to distinguish them from 
gamma-rays. The LAT consists of 4$\times$4 arrays of 16 tracker/calorimeter 
modules. The instrument has nearly $10^{6}$ electronic channels operated on 
a power budget of $\approx$ 650 W \cite{Atwood:2009ez}. The working principle pf LAT instrument is depicted in Figure 3.3.

\begin{figure}
\subfigure[]
{\includegraphics[width=0.5\linewidth]{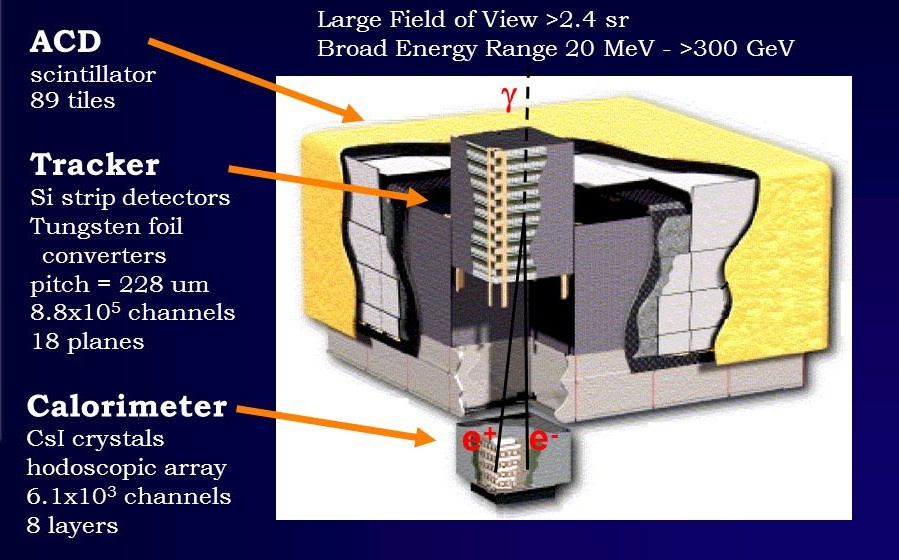}}
\subfigure[]
{\includegraphics[width=0.5\linewidth]{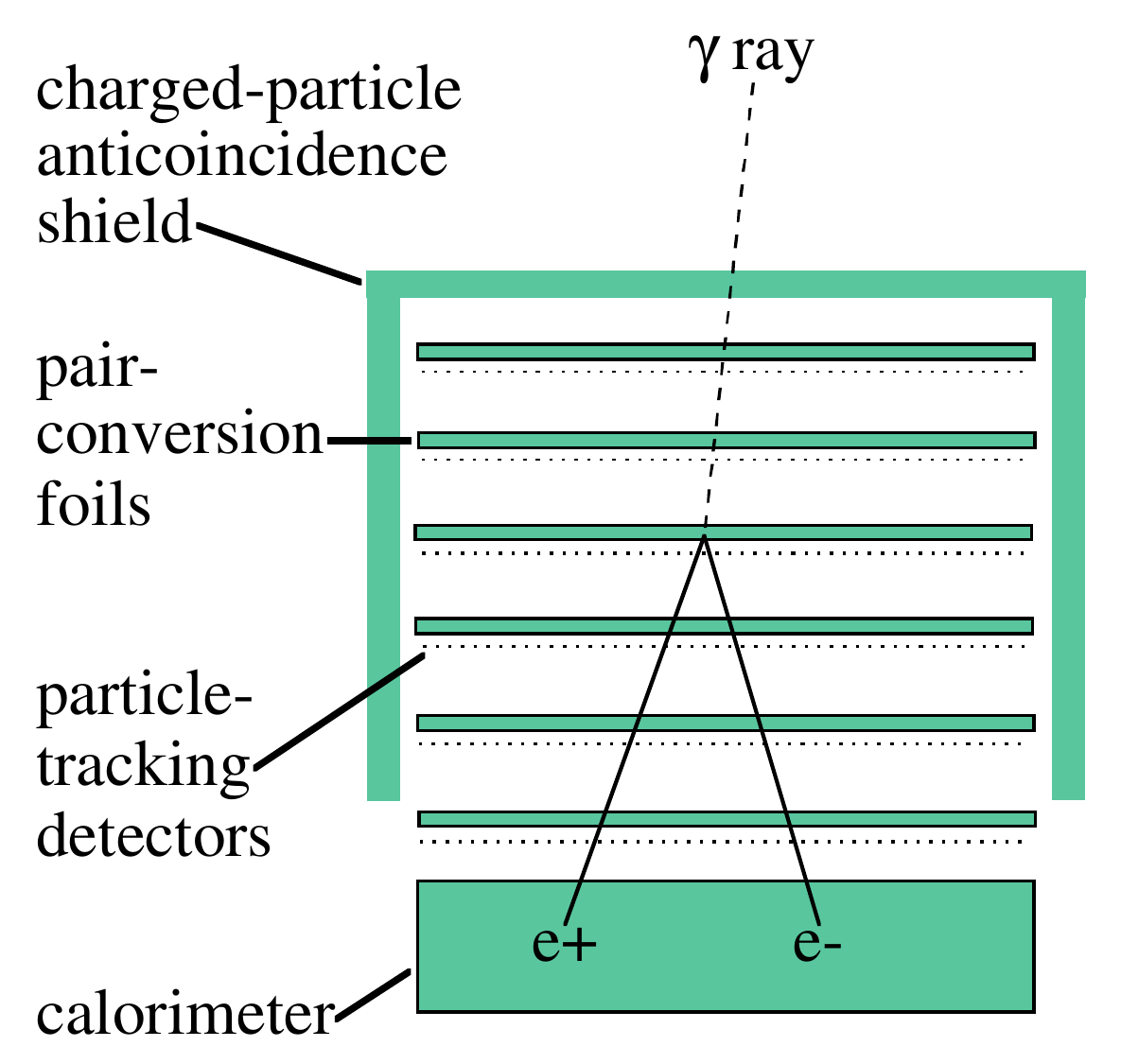}}
\caption{A schematic diagram of the LAT instrument. The dimensions of
the LAT are $1.8m~\times~1.8m~\times~0.72m$. A cutaway image of the LAT module 
shows its
tracker and calorimeter components, while the anticoincidence detector covers 
the tracker and the upper third of the calorimeter. The image is adapted from 
Atwood \textit{et al.}, 2009.}
\end{figure}

\subsubsection{Anti-coincidence Detector (ACD)}

\noindent The cosmic charged particles passing through the LAT can 
generally outnumber the gamma rays by factors of $10^{2}$-$10^{5}$. Those charged particles can be recorded by 
LAT and as a result, the background counts would be increased. In order to 
eliminate such background events resulting from charged particles, the LAT 
instrument is surrounded by an ACD. The
ACD consists of 89 plastic scintillator tiles which are used to identify 
background events and to
issue a veto signal. ACD detects the veto signal through wave-length shifting 
fibers by
two photomultiplier tubes (PMT). In order to detect the charged particles, for 
maximum
ACD efficiency, the tiles are overlapped in one direction and gaps in the other 
direction
are filled by scintillating fiber ribbons.

\noindent ACD covers the entire internal system of the LAT instrument, thus one 
of the responsibilities of the ACD is to identify the charged particles with an 
efficiency of 0.9997 \cite{Moiseev:2007hk}, while ACD also simultaneously needs 
to avoid the ``self-vetoes'' resulting from the backsplash effect. To examine 
the actual energy of the source, it is very advisable to consider the effect of 
backsplash seriously. It is sometimes possible that the secondary charged 
particles generating by an incident high energy photon in the calorimeter 
(potentially a valid event) can again travel back up through the tracker and 
cross the ACD. These particles can Compton scatter and thereby create signals 
from the recoiled electrons. This effect is called the backsplash effect and for 
this effect, the valid gamma rays would be vetoed by ACD. Hence, for reducing 
the effect of backsplash, the LAT team has designed the segmented structure of 
ACD. With the segmented structure, ACD would now only veto those events which 
would trigger an ACD tile in the projected path of the incident photon. The 
segmentation helps to achieve a uniform threshold and also significantly 
increases the sensitivity of Fermi-LAT, especially for high-energy gamma rays.

\noindent There are two types of the output signals generated by the ACD 
photomultiplier: (1) the fast veto pulses that are accessed by on-board LAT 
trigger electronics and (2) the slower pulse-shaped signals that are used for 
charged particle rejection method on the ground. For protecting the ACD from 
the 
space environment, it is covered by a micrometeorite shield and a thermal 
blanket.

\subsubsection{The Tracker (TKR)}

\noindent The principal role of the LAT tracker/converter (TKR) is to convert 
the 
incident $\gamma$ rays into electron-positron pairs and then accurately track 
the resulting particles \cite{Atwood:2007ra}. The TKR consists of 18 XY 
detector planes. Each tracker consists of two orthogonal x-y layers that have 
an 
array of silicon strip detectors (SSDs) for tracking the charged particles. TKR maintains a perfect balance between the thin 
converter for preserving the angular resolution at low energy and the thick 
converter for maximizing $\gamma$-ray conversion efficiency at high energy. For 
this purpose, the TKR is segmented into `FRONT' and `BACK' section. The FRONT 
section consists of 12 planes covering the thin tungsten foil converter of 
0.035 
radiation lengths, while the BACK section consists of 4 planes covering the 
thick tungsten foil converter of 0.18 radiation lengths. For preserving the 
triggering efficiency for $\gamma$ rays that converts in the final thick 
converter, the last 2 final planes that place immediately in front of the 
calorimeter does not have any converter. In order to localize the track of 
charged 
particles, each plane of SSDs has two planes of silicon strips, one is along 
the 
x-direction and the other along the y-direction. In one of the TKR's converting 
tungsten plates, the incoming gamma rays are converting into a pair of electron 
and positron. 

\noindent After the conversion point, SSD planes record the
directions of the incoming electron and positron pair. But the multiple scattering of the 
charged particles in the conversion plane would
affect the angular resolution of the system, especially for low energy range. 
Apart from the electron-positron pair, the cosmic rays also interact inside 
the TKR modules. Thus TKR needs to accurately identify the nature of each 
passing particle 
and their reconstructed energy. The advantage of using the thick converters is 
that it can also
partially shield the FRONT portion of the TKR from the effect of low-energy 
calorimeter backsplash.
The on-axis depth for the TKR module is around 1.5 radiation lengths and that 
increases the probability of the $\gamma$-ray conversion by $\approx 63\%$ 
\cite{Atwood:2007ra}. 

\subsubsection{The Calorimeter (CAL)}
 
\noindent The basic function of the Fermi-LAT calorimeter (CAL) is to estimate 
the energy 
deposited by the electron-positron pair \cite{Atwood:2009ez}. Each CAL 
module contains 96 CsI crystals which are arranged in eight alternating 
orthogonal layers where the total number of crystals is 1536. The output of the 
crystals is recorded on each end by both large and small photodiodes. This 
structure and the segmentation of CAL provide a large dynamic energy range for 
each crystal (2 MeV to 70 GeV) and a precise derivation of the 
three-dimensional 
position of particle shower. The on-axis depth of the CAL is about 8.6 
radiation 
lengths and for the significant amount of gamma rays with energy $\gtrsim$ 100 
GeV, most of the shower fall outside the active region of CAL. But it is 
very interesting to note that the imaging efficiency of the CsI crystals 
provides a precise estimation of the shape of the electromagnetic shower and 
their energy \cite{Bruel:2012bt}.

\subsection{The LAT's Data Acquisition System (DAQ)}

\noindent The Data Acquisition System (DAQ) has a very crucial role in interpreting the signal detected by LAT. In order to control the counts of background events from 
transmitting into the ground, DAQ conducts the onboard filtering on the 
observed 
data. This system converts the detected events into a data stream with a speed 
of around 1.2 Mbps. Apart from that, the DAQ also executes the controlling of 
the system and instrument monitoring such as housekeeping and power switching. 
Sometimes, for improving the performance of the processing, the working
onboard system is modified by uploading new software. 

\noindent Amongst all the penetrated particles through the LAT trackers, the 
astrophysical 
photons only share a very tiny portion. The LAT on-board analysis system 
decreases the raw LAT trigger rate (i.e. from 10 kHz to $\approx$ 400 Hz) and 
then sends the signal to the ground for further analysis. From those 
$\approx$400 Hz counts, only a very small portion (i.e. between $\approx$ 2-5 
Hz) are 
astrophysical photons. When the reprocessed data for an event is passing 
through the on-board analysis, all the conservative cuts, the time stamp and information of 
the signals obtained from various LAT components are saved in a packet.

\noindent As the number of the signal obtained from an event varies, each data 
packets have different
length. The data packets are the primary version of 
the data product. LAT further transfers these data packets to the Solid State 
Recorder (SSR) of
spacecraft
\footnote{\tiny{https://fermi.gsfc.nasa.gov/ssc/data/p7rep/analysis/documentation/Cicerone/Cicerone{\_}Introduction/LAT{\_}overview.html}}.

\begin{figure}
\begin{center}
\includegraphics[width=0.6\linewidth]{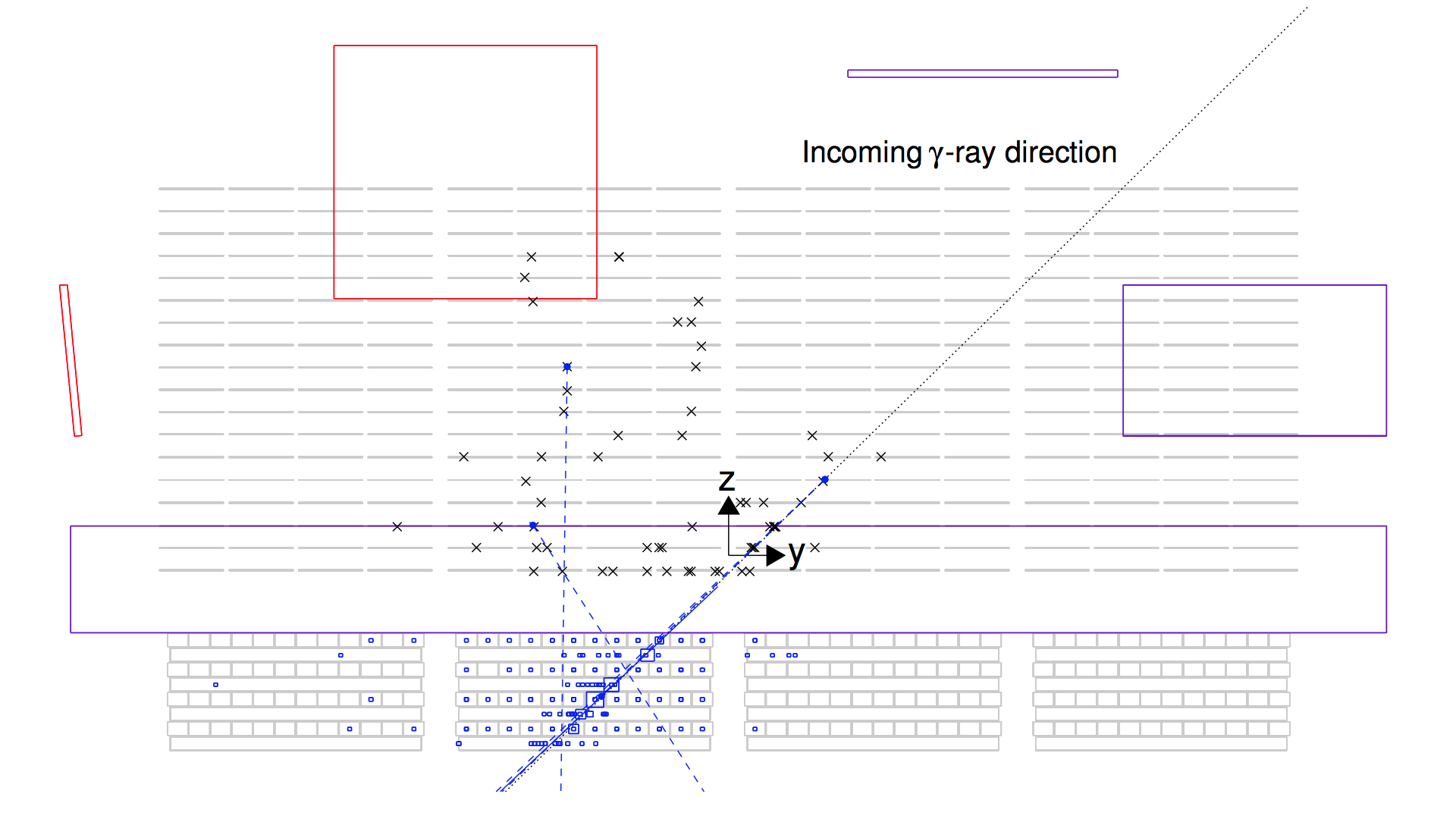}
\caption{Event display of a simulated 27 GeV $\gamma$-ray interacting with the 
LAT instrument. Clusters of hit TKR strips are represented by black crosses, 
while the location and magnitude of energy depositions in the CAL crystals are 
represented by variable-size blue squares. Hit ACD tiles are represented by 
coloured boxes, with a colour corresponding to the amount of energy deposited. 
The 
dotted line represents the true
$\gamma$-ray direction, the dashed lines represent reconstructed TKR tracks, 
and 
the solid line represents the CAL axis. Figure from Ackermann et al., 2012.}
\end{center}
\end{figure}

\subsection{LAT Instrument Performance}

\noindent The instrument response functions (IRF) of any detector is the 
mapping 
between the incoming photon flux and the detected events where the detected 
events depend on the LAT hardware and the analysis process. The analysis 
process 
determines whether the event parameters are from the observables and then 
assigns the probability of the event being a photon. In Fermi-LAT, the IRF is represented by a set of parameters such 
as instrument coordinates, observed event energy ($E^{\prime}$), and incident 
direction ($\widehat{v}^{\prime}$) as a function of true event energy (E), and 
incident direction ($\widehat{v}$).

\noindent The LAT response function is derived by a dedicated GEANT4-based 
Monte 
Carlo simulation of $\gamma$ rays interacting with the LAT detector and Fermi 
spacecraft \cite{Atwood:2009ez}. In order to cover all possible photon 
inclination angles and energies of photons with good statistics, a large number 
of gamma-ray events are being simulated. The Fermi-LAT team designed a separate 
set of IRFs for each event class and event type selection and we need to select 
the correct IRF at the time of performing analysis. 

\noindent In LAT, the performance of the IRF is factorized into three terms: 1) 
efficiency in terms of the detector's effective area (A(E, $\widehat{v}$)), 2) 
resolution as given by the point-spread function (PSF, P($\widehat{v}^\prime|E, 
\widehat{v}$)), and 3) energy dispersion (D($E^{\prime}|E, \widehat{v}$)).\\
\begin{itemize}
\item A(E, $\widehat{v}$) is the product of the geometric 
collection area, $\gamma$-ray conversion probability, and the efficiency of a 
given event selection.\\
\item P($\widehat{v}^\prime|E, \widehat{v}$) is the probability density to reconstruct an event direction $\widehat{v}^\prime$, for a given true
energy (E) and direction $\widehat{v}$.\\
\item D($E^{'}|E, \widehat{v}$) is the probability density to reconstruct an event energy $E^{'}$, for a given true
energy (E) and direction $\widehat{v}$.\\
\end{itemize}

\noindent The Fermi Sciencetools has provided us with multiple IRFs and has allowed the user to choose them according to the preferences of the
analysis types. The most recent version of IRFs released by the LAT team is the
``PASS 8''\footnote{\tiny{https://www.slac.stanford.edu/exp/glast/groups/canda/lat\_Performance.htm}}. In comparison to the earlier version, the ``Pass 8'' improves the LAT analysis by using a completely new set of
event-level reconstruction algorithms and that would effectively decrease the pile-up effects. The performance of the “Pass 8” is shown in Figure 3.4.

\begin{figure}
\subfigure[]
{ \includegraphics[width=0.5\linewidth]{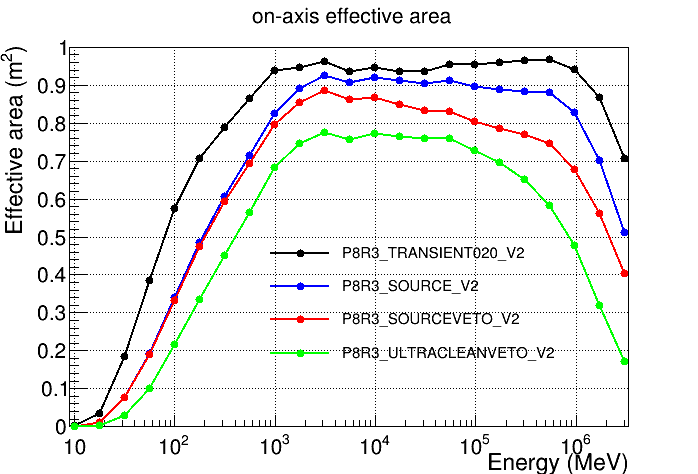}}
\subfigure[]
{ \includegraphics[width=0.5\linewidth]{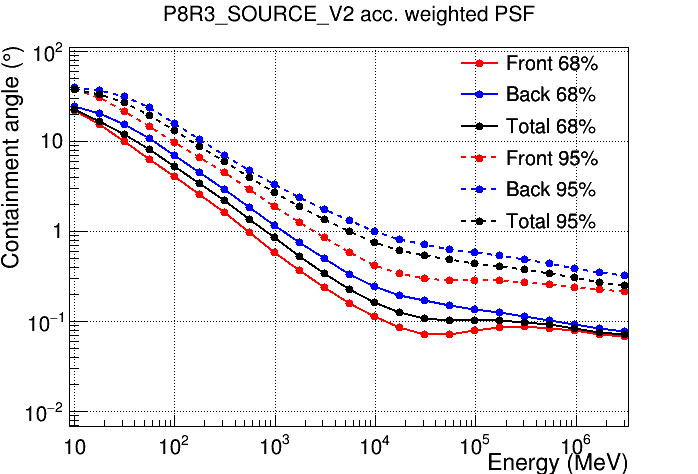}}
\caption{The performance of the Pass 8 at normal incidence as a function of incident photon energy is shown here. 
(a) the effective area and (b) the points spread function. 
The figure is adapted from Bruel et al., 2018.}
\end{figure}

\chapter{Likelihood Analysis of LAT Data}
\section{Overview of Likelihood}
\noindent For analyzing the Fermi-LAT data, it is very important to construct 
the likelihood function. This function would be needed to obtain the best fit 
model parameters during analysis. These best-fitted parameters would describe 
the source's spectrum, its position and its real significance. 

\noindent The likelihood function ($\mathcal{L}$) is the probability 
of obtaining 
the data for the assigned input model if the LAT data is true. The input model 
would include the description of the nature of gamma-ray sources (i.e. whether 
they are point, extended or transient), and also derive the intensity of the 
source and its possible spectra. For this purpose, we would first assume that 
we 
have a handsome knowledge on the mapping of the input model of the gamma-ray 
sky 
to the data.

\section{The LAT Likelihood Function}

\noindent During analysis, it is advisable to distribute the LAT data into a 
large number of bins. The binning is important because the LAT counts are 
dependent on many variables, thus despite having a large number of counts, 
each bin will have a small number of counts. The observed number of counts in 
each bin will follow the Poisson distribution.

\noindent The expression of the Likelihood function is:
\begin{equation}
\mathcal{L(\alpha|D) \equiv P(D|\alpha)}
\end{equation}

\noindent Here, $\mathcal{L}$ is the probability of obtaining the data 
($\mathcal{D}$) for a
given input model with parameters ($\alpha$). 

\noindent For binned LAT likelihood analysis, the function, $\mathcal{L}$, is 
defined as 
the 
a product of Poisson likelihoods i.e. the product of the probabilities of 
observing the detected counts in each bin.
\begin{equation}
\mathcal{L(\alpha|D)} = \prod \limits_{k} \frac{\lambda_{k}^{n_{k}} 
e^{-\lambda_{k}}}{n_{k}!}
\end{equation}
$\mathcal{L(\alpha|D)}$ can also be written as:
\begin{eqnarray}
\mathcal{L(\alpha|D)} &=& \prod \limits_{k} e^{-\lambda_{k}} \prod \limits_{k} 
\frac{\lambda_{k}^{n_{k}}}{n_{k}!}\\
&=& e^{-N_{pred}} \prod \limits_{k} \frac{\lambda_{k}^{n_{k}}}{n_{k}!} ,
\end{eqnarray}

\noindent where, $N_{pred}$ denotes the total number predicted counts from the source model.\\
\noindent 
Instead of $\mathcal{L}$, the logarithm of $\mathcal{L}$ is comparatively easier to handle as this factor is maximized
during the fitting. The log-likelihood can 
be 
expressed in the following form:
\begin{equation}
\mathcal{\log \mathcal{L}} = \sum \limits_{k} n_{k} \log \lambda_{k} - N_{pred}
\end{equation}

\noindent Here, the observed counts in bin, k are $n_{k} = n_{k}(\mathcal{D})$ 
and the counts predicted by the model is $\lambda_{k} = \lambda_{k}(\alpha)$. 
If 
the bin size is infinitely small, then we can assume $n_{k} = 0~\rm{or}~1$ (i.e. for unbinned 
likelihood). In that case, the functional form of $\mathcal{L}$ would be: 
\begin{equation}
\mathcal{L(\alpha|D)} = \prod \limits_{k} \lambda_{k}^{n_{k}} e^{-\lambda_{k}}
\end{equation}

\noindent where k is now representing individual photons. 

\noindent From Eqs. (4.1 - 4.6), we have observed that the likelihood function 
explicitly depends on the predicted counts by the model but it also depends on 
the differential $\gamma$-ray flux of a source. In Eq.~4.7, we express the 
distribution of $\gamma$-ray source as $S(E,\widehat{p}|\alpha$) where 
$\widehat{p}$ denotes the direction of a $\gamma$-ray in celestial coordinates. 
With the help of the spacecraft orientation, direction in celestial 
coordinates can be converted to instrument coordinates, i.e. to 
$\widehat{v}(t|\widehat{p})$. The source model can be expressed as:

\begin{equation}
S(E,\widehat{p}|\alpha) = \sum \limits_{k} s_{k}(E) 
\delta(\widehat{p}-\widehat{p_{k}}) + 
S_{G}(E,\widehat{p}|\alpha)+S_{eg}(E,\widehat{p}|\alpha) + \sum \limits_{l} 
S_{l}(E,\widehat{p}|\alpha)
\end{equation}
\noindent Here, $s_{k}(E), S_{G}(E,\widehat{p}|\alpha)$, 
$S_{eg}(E,\widehat{p}|\alpha)$ and $S_{l}(E,\widehat{p}|\alpha)$ define the 
source model for point source, galactic emission, extragalactic isotropic 
emission and other sources, respectively. In case of faint $\gamma$-ray 
sources, 
it is very important to consider the correct model of the diffuse $\gamma$-ray 
background. In Fermi-LAT, the diffuse background is divided into Galactic and 
Extragalactic components:

\begin{itemize}
 \item \textbf{Galactic Component:} A spatially-structured Galactic component 
corresponding to the $\gamma$-ray emission from the interaction of cosmic rays 
with interstellar gas, dust, and photon fields. \\
 \item \textbf{Isotropic Extragalactic Component:} An isotropic component 
corresponding to the combination of extragalactic $\gamma$-ray emission and 
instrumental charged particle background. 
\end{itemize} 

\noindent At the time of searching for a new $\gamma$-ray source, it is 
advisable to free the normalization parameters of the two diffuse components 
while we can fix the spectral shapes. Now, if we want to calculate the 
predicted counts by a model for a given bin, k, we can then estimate the 
differential flux of each $\gamma$-ray source for their IRFs as.

\begin{equation}
\lambda_{k}(\alpha) = \sum \int\int\int S(E,\widehat{p}|\alpha) 
A(E,\widehat{v}) 
P(\widehat{v}'|E,\widehat{v}) D(E'|E,\widehat{v}) d\Omega dE dt
\end{equation}

\noindent In Eq.~4.8, we have summed over all $\gamma$-ray sources and then 
integrated over the total observing time, the energy range, and the solid angle 
with respect to the LAT frame.

\noindent Several simplifying assumptions can be considered for lowering the 
total computational cost of likelihood calculation. The region of interest 
(ROI) 
defines centred region around the location of our source of interest and during 
the 
likelihood process the $\gamma$-ray emission model is generated for only those 
sources which are situated within a few PSF-widths of this region. The duration 
of observation and exposure are precomputed and that helps to discard the IRF's 
dependence
on the azimuthal angle. When the bin size of the 
analysis is comparatively larger (i.e. for binned analysis) than the scale of 
the energy dispersion, we 
can ignore the effects of energy dispersion.

\section{The Profile Likelihood}

\noindent For estimating the best-fit parameters for a given model, we need to 
maximize 
the likelihood with respect to the parameters of interest, i.e.,

\begin{equation}
\widehat{\alpha} = arg_{\alpha} max \mathcal{L(\alpha|D)} 
\end{equation}

\noindent where $\widehat{\alpha}$ represents the estimator of maximum 
likelihood (MLE) for the parameters, $\alpha$. Practically, performing a 
non-linear maximum likelihood for a large set of parameters is not 
computationally possible. Hence, a conventional solution is to partition the 
set 
of parameters i.e., $\alpha$, into a set of parameters of interest i.e., $\mu$, 
and a set of nuisance parameters i.e., $\theta$, such that $\alpha = 
\{\mu,\theta\}$. For instance, when we are trying to determine the possible 
spectra 
for gamma-ray source, the spectral index or flux of a specific $\gamma$-ray 
source can be the parameters of interest while the background $\gamma$-ray 
sources or constraints on source characteristics derived from independent 
analysis could be the the nuisance parameters. The expression of the Profile 
likelihood is:

\begin{equation}
\mathcal{L}_{p}(\mu|\mathcal{D}) = sup_{\theta} \mathcal{L(\mu,\theta|D)}
\end{equation}

\noindent The advantage of the profile likelihood is that by maximizing the 
likelihood function concerning the 
nuisance parameters, it decreases the dimensionality of likelihood. The 
profile likelihood function does not disclose the full 
distribution of the nuisance parameters of the system but it still maintains 
the 
statistical properties of the likelihood function\cite{Bartlett:1953gfw}.

\section{The Joint Likelihood}

\noindent The sources which belong to the same class generally
share a common set of physical characteristics i.e., they have the same 
range of luminosity or can be described by the same spectral models, etc. For 
such cases, the sensitivity to the characteristic of the sources can be 
increased by combining them. This formulation would 
follow the likelihood-based analysis and would lead to the concept of joint 
likelihood. The joint likelihood is the function of the product of the 
individual likelihoods where the function combines the parameters of all individual sources.
For each source, i, the expression of the joint likelihood is:

\begin{equation}
\mathcal{L}(\mu, \{\alpha_{i}\}|\mathcal{D})=\prod \limits_{i=1} 
\mathcal{L}_{i}(\mu, \alpha_{i}|\mathcal{D})
\end{equation}

\noindent Here, the parameters are divided into a set of parameters shared by 
all 
sources i.e., $\mu$, and a set of parameters depending on each individual 
source i.e., {$\alpha_{i}$}. 

\noindent For example, when we are considering a set of dSphs for DM study, the 
intrinsic luminosity or spectral model can be treated as the set of shared 
parameters, while the distance to each source, point-like background sources 
within each ROI, or the normalization of a diffuse background near each source 
will be considered as the independent parameters. The joint likelihood 
would 
then act as the likelihood of a single source, including the construction of a 
profile joint likelihood. In order to obtain the combined limits for DM 
annihilation signal, the individual likelihood function is weighted with their respective J-factor.

\section{Hypothesis Testing}

\noindent The likelihood formalism allows a robust statistical framework for 
hypothesis testing. In hypothesis testing we check how much the parameters of 
interest ($\mu$) deviate from their nominal expected value ($\mu_{0}$). From 
the ratio of the maximum likelihood test assuming for two hypothesis 
\cite{Neyman:1933wgr}, we can derive the ``test statistic'' (TS). The 
expression 
of the TS is:

\begin{equation}
TS = 2~ln \Big( 
\frac{\mathcal{L(\widehat{\mu}|D)}}{\mathcal{L}(\mu_{0}|\mathcal{D})} \Big) = 
2~(ln \mathcal{L(\widehat{\mu}|D)} - ln\mathcal{L}(\mu_{0}|\mathcal{D}))
\end{equation}

\noindent where, $\mathcal{L}(\mu_{0}|\mathcal{D})$ is the maximum likelihood 
value for a model without an additional source (i.e., the `null hypothesis') and 
$\mathcal{L(\widehat{\mu}|D)}$ is the maximum likelihood value for a model with 
the source at a specified location.

\noindent Wilks theorem \cite{Wilks:1938dza} and Chernoff 
\cite{Chernoff:1954eli} theorem state that the asymptotic distribution of TS 
values under the null hypothesis (i.e., $\mu = \mu_{0}$) should follow a 
$\chi^{2}_{n}$-distribution, where n represents the dimensionality of $\mu$. 
Hence, it signifies that the TS value can be drawn from this 
$\chi^{2}_{n}$-distribution if the null hypothesis holds. The large TS 
indicates that the source is 
present 
in the location i.e., the null hypothesis is incorrect.

\noindent The most general application of the likelihood ratio test is to check 
the significance of a $\gamma$-ray source. The detection significance 
($\sigma$) 
of any source is approximately equal to the square root of the TS value, i.e., 
$\sigma \approx \sqrt{TS}$. In order to check the significance of the source, 
the parameter of interest is the flux of the gamma-ray source whereas the null 
hypothesis assumes that the gamma-ray flux from the source location is zero. 
From Eq.~4.12, the TS value can be defined by maximizing the likelihood 
function for both the putative source flux that is free to vary and the 
putative source flux that is fixed to zero. As a thumb rule, the threshold for 
detection of the real signal is set at TS $\geq$ 25 i.e., corresponds to 5$\sigma$. 
However, for some cases, the spectral index of the source is left free during 
the model fitting, and that can decrease the detection significance of the 
source from 5$\sigma$ to $\approx$ 4.2$\sigma$ \cite{Abdo:2010ru}.

\noindent Apart from finding the best-fitted parameter values for a 
source-model, the likelihood algorithm can also estimate the uncertainty of 
those parameters of interest \cite{Neyman:1937uhy}. From the shape of the 
likelihood function, we can determine the uncertainty in the best-fit parameters 
of interest. For high significant sources where the null hypothesis is not 
valid, we choose the two-sided confidence interval for the estimation of maximum 
likelihood, while for the faint sources when we could not strongly eliminate the 
null hypothesis, we set the one-sided confidence interval on the maximum 
likelihood estimate. But unfortunately, the calculation is not straight forward 
for the system with low-counts where we could not directly use the asymptotic 
formula to estimate the significance of the sources. Many literatures have 
already provided the solutions for this issue\cite{Helene:1990yi, 
Feldman:1997qc, Roe:1998zi, Rolke:2004mj} and for our analysis, we have 
particularly used the delta-log-likelihood method provided by the Rolke et al. 
\cite{Rolke:2004mj}.

\section{Derivation of Flux Upper limits}
\noindent From section 4.5, we find that for deriving the significance of any source, we need
to check the likelihood ratio test. In Wilks’s theorem, the
null hypothesis means no source exists, while the alternative hypothesis
assumes that the source exists. 
For the analysis of the faint sources (or we can consider the case for DM searches), we observe a little to no gamma-ray emission from 
the direction of the target. Hence, for that scenario, null hypothesis is a good approximation and it 
is advisable to estimate upper bound on the gamma-ray flux.

\noindent In order to estimate the flux upper limit, we generally prefer to use the profile likelihood method 
\cite{Rolke:2004mj, Barbieri:1982eh}. If we
assume that the delta-log-likelihood behaves asymptotically i.e., as $\chi^{2}$, then the 90$\%$ confidence region
would be relative to the change in log-likelihood by 2.71/2. Here, we need to mention that such changes in log-
likelihood function corresponds to the two-sided confidence interval. 
If we would like to derive the
upper-limit corresponds to a one-sided 95$\%$ C.L., during likelihood, all the normalization parameters
along with the two diffuse components would be fitted with the entire dataset until
the logarithmic difference between two likelihood functions reach at 1.35 \cite{Abdo:2010ex}.

\section{P-value}

\noindent In statistics, the p-value or the calculated probability is the probability of obtaining results at least as extreme, when the null hypothesis ($H_{0}$) is assumed to be true \cite{pvalue}. However, in statistics, the term ‘extreme’ depends on how we are dealing with the hypothesis test. The null hypothesis ($H_{0}$) is usually an hypothesis of ``no difference'' whereas the alternative hypothesis ($H_{1}$) is the opposite of the null hypothesis. A smaller p-value means that there is stronger evidence in favor of the alternative hypothesis. \\

\noindent P-values are usually found using p-value tables or spreadsheets/statistical software. These calculations are based on the assumed or known probability distribution of the specific statistic being tested. P-values are calculated from the deviation between the observed value and a chosen reference value, given the probability distribution of the statistic, with a greater difference between the two values corresponding to a lower p-value. \\

\noindent In the contest of null-hypothesis testing, if we consider that the observed test-statistic (t) is drawn from the probability distribution (T), then the p-value would be of observing a test-statistic value at least as ``extreme'' as t when the null hypothesis ($H_{0}$) is true. In a formal significance test, the null hypothesis ($H_{0}$) is rejected if the p-value is less than a threshold alpha level ($\alpha$) or significance level. The term significance level ($\alpha$) is used to refer to a pre-chosen probability whereas we use the p-value to indicate a probability after a given study. The value of $\alpha$ is instead set by the researcher before examining the data. By convention, $\alpha$ is commonly set to 0.05, though lower alpha levels are sometimes used. \\

\chapter{Constraints on dark matter models from the Fermi-LAT observation of Triangulum-II}
\section{Triangulum-II}
\noindent Since last two decades, the Sloan Digital Sky Survey (SDSS) 
\cite{York:2000gk} 
discovered a new member of Milky Way satellites. They are ultra-faint and have 
a 
very
high mass to light ratio.Thus we can assume that they might be very rich in 
DM contents
\cite{Willman:2004kk, Zucker:2006bf, Belokurov:2006ph, Irwin:2007jz, 
Walsh:2007tm,
Strigari:2007at}. Over the past few years, the Panoramic Survey 
Telescope and Rapid Response System (Pan-STARRS) \cite{Kaiser:2002zz} and the 
Dark Energy Survey (DES) \cite{Abbott:2005bi} have observed a new population of 
dSphs 
\cite{Laevens:2015una,Bechtol:2015cbp,Kim:2015ila,Drlica-Wagner:2015ufc} around 
our 
Milky Way. Triangulum-II (hereafter, we would refer to as Tri-II) is one of the 
newly discovered dSphs \cite{Biswas:2017meq},
which has been detected by the Pan-STARRS Survey \cite{Laevens:2015una}. This 
survey has concluded that Tri-II is either an ultra-faint and DM dominated 
dwarf 
galaxy or a globular cluster. There are several pieces of studies 
\cite{Genina:2016kzg, Hayashi:2016kcy, Biswas:2017meq} which have claimed that 
Tri-II may come as a very potential target for indirect DM detection. 
In this chapter, we would describe our findings for Tri-II \cite{Biswas:2017meq}.\\

\noindent For our study, we have considered Tri-II as a metal-poor galaxy with 
large
mass to light ratio\cite{Biswas:2017meq}. But so far very few numbers of member stars of Tri-II have
been detected and its exact number is still uncertain \cite{Kirby:2015bxa, 
Kirby:2017cyb}. Ref.~\cite{Kirby:2015bxa,
Martin:2016cyb} had observed nearly $6$ member stars in Tri-II, while their 
most 
recent 
study \cite{Kirby:2017cyb} have discovered the existence of $13$ stars along 
with 
very velocity dispersion $\approx~\sigma_{\rm v}<4.2~{\rm {km}}~{\rm s}^{-1}$ 
and 
$<3.4~{\rm {km}}~{\rm s}^{-1}$ for $95\%$ and $90\%$ C.L., 
respectively. 
In Table~5.1, we have described some important properties of Tri-II.

\begin{table}[h!]
\begin{center}
\caption{Properties of Triangulum-II.}
\begin{tabular}{|p{4 cm}|p{6 cm}|p{4 cm}|}
\hline
\hline
Property &  Value   & Reference  \\ 
\hline
Galactic longitude & $\rm{141.4^{\circ}}$  & \cite{Laevens:2015una} \\
\hline
Galactic latitude & $\rm{-23.4^{\circ}}$ & \cite{Laevens:2015una} \\
\hline
Galactocentric distance & $\rm{36_{-2}^{+2}~kpc}$ & 
\cite{Genina:2016kzg,Kirby:2015bxa}\\
\hline
2D half light radius ($\rm{r_{h}}$) & $\rm{34_{-8}^{+9}~pc}$ & 
\cite{Kirby:2015bxa,Kirby:2017cyb} \\
\hline
Velocity relative to galactic standard of rest (GSR) ($\rm{v_{GSR}}$) & -261.7 
km $\rm{s^{-1}}$ & \cite{Kirby:2017cyb}\\
\hline
Mean heliocentric velocity  $~<\rm{v_{helio}}>$ & $\rm{-381.7\pm2.9~km~s^{-1}}$ 
& \cite{Kirby:2017cyb}\\
\hline
Stellar Velocity Dispersion ($\rm{\sigma_{v}}$) & 
$<~\rm{3.4~km~s^{-1}~(90\%~C.L.)}$ & \cite{Kirby:2017cyb}\\
 & $<~\rm{4.2~km~s^{-1}~(95\%~C.L.)}$ & \cite{Kirby:2017cyb}\\
\hline
Mass within 3D half-light radius \Big($\rm{\frac{M_{1/2}}{M_{\odot}}}$\Big) & 
$\rm{<~3.7~\times~10^{5}~(90\%~C.L.)}$ & \cite{Kirby:2017cyb}\\
& $\rm{<~5.6~\times~10^{5}~(95\%~C.L.)}$ & \cite{Kirby:2017cyb}\\
\hline
Mass-to-light ratio within 3D half-light radius 
\Big($\rm{(M/L_{v})_{1/2}}$\Big) 
& $\rm{<~1640~M_{\odot}~L_{\odot}^{-1}~(90\%~C.L.)}$ & \cite{Kirby:2017cyb}\\
& $\rm{<~2510~M_{\odot}~L_{\odot}^{-1}~(95\%~C.L.)}$ & \cite{Kirby:2017cyb}\\
\hline
Density within 3D half-light radius $\rm{\rho_{1/2}}$ & 
$\rm{<~2.2~M_{\odot}~pc^{-3}~(90\%~C.L.)}$ & \cite{Kirby:2017cyb}\\
& $\rm{<~3.3~M_{\odot}~pc^{-3}~(95\%~C.L.)}$ & \cite{Kirby:2017cyb}\\
\hline
Metallicity ([$\rm{Fe/H}$]) & $\rm{-2.24\pm0.05}$ & \cite{Kirby:2017cyb}\\

\hline
\hline
\end{tabular}
\end{center}
\end{table}
\begin{center}
\end{center}

\noindent In Table~5.1, $M_{\odot}$ and $L_{\odot}$ denote the mass and the 
luminosity of 
the Sun, respectively. The values of $\rm{M_{1/2}}$, $\rm{(M/L_{v})_{1/2}}$ 
and $\rm{\rho_{1/2}}$ have been taken from~\cite{Kirby:2015bxa,Kirby:2017cyb}.
For our study, we have assumed Tri-II as spherically symmetric (because of its 
low ellipticity)
 and in a state of dynamical equilibrium \cite{Biswas:2017meq}.
From the observational study, ref.~\cite{Kirby:2017cyb} had obtained that 
Tri-II 
has large velocity dispersion concerning the galactic standard of rest (GSR). 
All the observational studies (from \cite{Kirby:2015bxa, Kirby:2017cyb}) 
suggested that Tri-II might be
affected by the total tidal effect from the Milky Way \cite{Kirby:2015bxa}. 
Several studies have also suspected the association of Tri-II with the 
Triangulum-Andromeda halo sub-structures \cite{Laevens:2015una, 
Majewski:2004zi} 
and with the PAndAS stream 
\cite{Martin:2014dja}, that might cause the effect of tidal disruption of 
Tri-II.
Indeed, the above-mentioned observations did not provide any concrete proof 
that 
Tri-II is in dynamical equilibrium \cite{Biswas:2017meq}. But, any tidally disrupting galaxy would 
show a high ellipticity, whereas Tri-II has low ellipticity \cite{Biswas:2017meq}. Moreover, the 
tidal 
radius of Tri-II is nearly three times of the 3D half-light radius of Tri-II 
and 
from that observational data, we can also predict that
the shape of Tri-II is insulated from Galactic tides. Thus, high mass to 
light 
ratio and large velocity dispersion value indicate a high concentration of DM \cite{Biswas:2017meq}. 

\section{The \textit{Fermi}-LAT Data Analysis of Tri-II}

\noindent For examining the possible signal from the Tri-II, we have analysed 
the 
gamma-ray observed by the Fermi-LAT \cite{Biswas:2017meq}.
For our analysis, we have used the \texttt{Fermi 
ScienceTools} version \texttt{v10r0p5} (released on June~24, 
2015)\footnote{\tiny{https://fermi.gsfc.nasa.gov/ssc/data/analysis/software/}}. 
Here, we have used the fully reprocessed Pass8 dataset 
\footnote{\tiny{
https://fermi.gsfc.nasa.gov/ssc/data/analysis/documentation/Pass8{\_}usage.html}
}. 
Such Pass8 processed data provides an improved event reconstruction, a wider 
energy range, a better energy resolution and a significantly increased 
effective 
area, especially for energy below $<$ 100 MeV. We have chosen 
$10^{\circ}$ as 
ROI\footnote{\tiny{https://fermi.gsfc.nasa.gov/ssc/data/analysis/scitools/data{
\_}preparation.html}} \cite{Biswas:2017meq}.
With `gtselect', we have applied $0.1 \le E \le 50$~GeV energy cut on the 
photon 
events \cite{Biswas:2017meq}. We have chosen this energy range to avoid the possible calibration 
uncertainties for energy below 100 MeV and background contamination at 
high 
energy. To avoid the albedo contamination, we have used the zenith angle cut at 
$\theta~=~ 
90^{\circ}$ \cite{Biswas:2017meq}. Next, we have performed the `binned likelihood', which is 
implemented in the \texttt{ScienceTools} \cite{Cash:1979vz,Mattox:1996zz}, on 
our reconstructed 
dataset\footnote{\tiny{
https://fermi.gsfc.nasa.gov/ssc/data/analysis/scitools/binned{\_}likelihood{\_}
tutorial.html}}\cite{Biswas:2017meq}. For our analysis, we have used the \textit{event class 128} and \textit{event type 3} \cite{Biswas:2017meq}. \textit{Event class 128} provids a good sensitivity to the point
sources and the moderately extended sources. The \textit{event type 3} is preferred for the $e^{+}$,~$e^{-}$ pair conversion that 
is supposed to occur both at the FRONT and the BACK tracker layers of the 
Fermi-LAT. Along with the above-mentioned selections, we have adopted 
$P8R{\_SOURCE\_V6}$ as IRF\footnote{\tiny{
https://fermi.gsfc.nasa.gov/ssc/data/analysis/documentation/Cicerone/Cicerone{\_
}LAT{\_}IRFs/IRF{\_}overview.html}},
\footnote{\tiny{
https://fermi.gsfc.nasa.gov/ssc/data/analysis/documentation/Cicerone/}} \cite{Biswas:2017meq}.\\

\noindent Along with all the sources within $10^{\circ}$ ROI, we have added the 
model for 
Tri-II in our source model \cite{Biswas:2017meq}.
As there was no pre-existing study of Tri-II by the 
Fermi-LAT collaboration team, we have modelled Tri-II with a power-law 
spectrum. 
The spectral and spatial models of all the remaining sources in the ROI have 
been taken from the 3FGL catalog. Moreover, to eliminate the possible galactic 
and extragalactic diffuse emission, we have also included the galactic diffuse 
model and its possible isotropic component to the source model ($gll{\_iem\_v05}.fits$ $\&$ $iso{\_source\_v05}.txt$) 
\footnote{\tiny{
https://fermi.gsfc.nasa.gov/ssc/data/access/lat/BackgroundModels.html}}.
During maximum likelihood fitting procedure, the normalisation parameter of 
those diffuse components are kept free. In that process, the spectral parameters 
for all the sources within $5^{\circ}$ from the location of Tri-II are left 
free 
but the parameters of other remaining sources in the ROI are kept fixed to 
their 
preferred values from 3FGL catalog. We have also fixed the localisation of 
Tri-II \cite{Biswas:2017meq}.

\subsection{Results from the Power-law Modelling} 

\begin{figure}
\centering
\subfigure[]
{\includegraphics[width=0.8\linewidth]{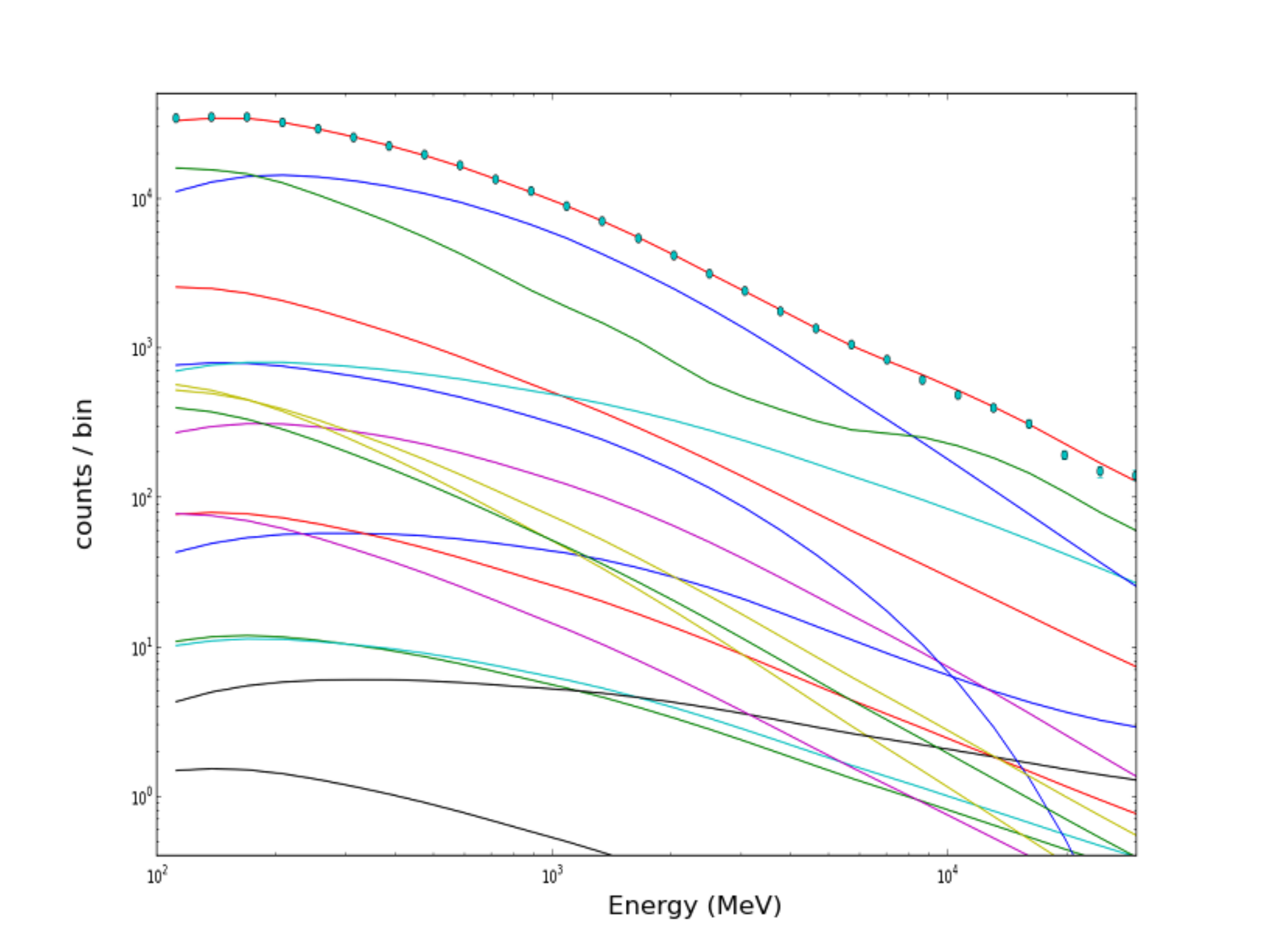}}
\subfigure[]
{\includegraphics[width=1.0\linewidth]{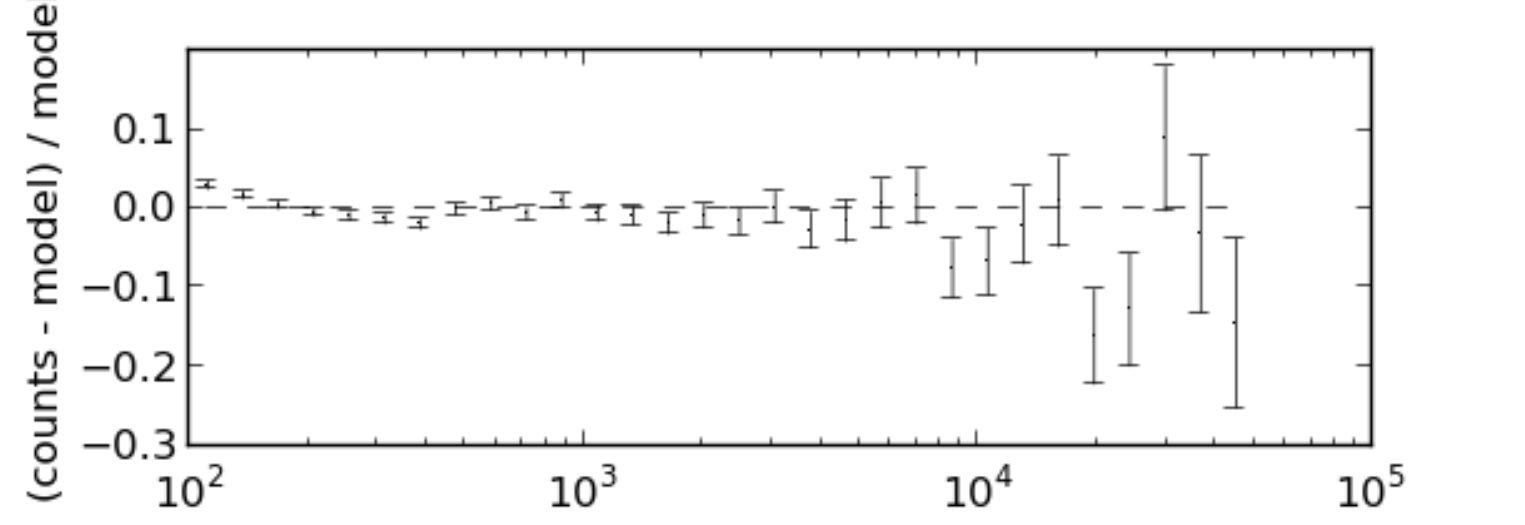}}
\caption{(a) The spectral fit to the observed data counts and (b) the
residual 
plot for the location of Tri-II. We have modelled the Tri-II with
the power-law spectrum for $\Gamma = 2$. In figure 5.1(a), the solid red curve displays the best-fit total spectrum, along with the corresponding LAT-observed data
points (in green); the upper-most solid blue and green curves display the galactic diffuse background and
the isotropic background component, respectively.}
\end{figure}

\noindent We have modelled Tri-II with power-law spectrum and the differential 
photon flux 
obtained from the Tri-II would be \cite{Abdo:2010ex, Biswas:2017meq}:
\begin{equation}
\rm{\frac{dN}{dA dE dt} = N_{0} \Big(\frac{E}{E_{0}}\Big)^{-\Gamma}},
\end{equation} 

\noindent Here $dN$ is the number of photons and their reconstructed energies 
vary between 
$E$ to $E + dE$. The photons are incident on an elemental area $dA$ of the 
detector with an elemental time interval of $dt$. In Eq.~(5.1), 
$\Gamma$ is the spectral index and $N_{0}$ is a normalisation parameter. We have fixed 
the 
energy scale $E_{0}$ at $100~\rm{MeV}$ \cite{Biswas:2017meq}. 
During the model fitting of Tri-II, we have considered five different values of 
the $\Gamma$ i.e., 1, 1.8, 2, 2.2 and 2.4 and have repeated the binned 
likelihood 
analysis for each $\Gamma$ value \cite{Biswas:2017meq}.
We have used $\Gamma = 1$ because of its connection with DM annihilation model 
(\cite{Essig:2009jx, Biswas:2017meq}),
while the other four $\Gamma$ is considered to check constraints on the general 
astrophysical source spectra.\\

\noindent In our power-law modelling, we have determined the best-fit values of 
$N_{0}$ 
along with the normalisation parameter of
isotropic and the galactic diffuse model for each $\Gamma$ \cite{Biswas:2017meq}. In fig.~5.1(a), we 
have displayed the spectral fits to the data from all the sources within ROI, 
along with two diffuse background models for $\Gamma~=~2$ \cite{Biswas:2017meq}. In this figure, the sum of the best fit spectrum along with the LAT counts within ROI $10^{\circ}$ is denoted by the top red curve, while the best fit spectrum for the galactic and isotropic components are defined by the top blue and the top green curves, respectively. On the other hand, 
in 
Fig.~5.1(b), we have shown the residual plot of Tri-II for $\Gamma~=~2$ \cite{Biswas:2017meq}. \\
 
\noindent In Table~5.2, we have mentioned our obtained best-fit values of 
$N_{0}$, their 
statistical errors and 
the $\rm{TS}$ values for all spectral indices \cite{Biswas:2017meq}. \\

\noindent From Table~5.2, we can find that for each of spectral indices, 
the normalisation constant, $N_{0}$ of Tri-II is less than the statistical 
errors 
obtained from fitting procedure \cite{Biswas:2017meq}. The corresponding TS value is less than unity. 
Hence, from Table~5.2, we could conclude that
the LAT could not detect any $\gamma$-ray signal from Tri-II \cite{Biswas:2017meq}.\\

\noindent As no significant emission has been detected by Fermi-LAT from the 
direction 
Tri-II, we have then 
derived the upper limit of the possible $\gamma$-ray flux from the location of
Tri-II \cite{Biswas:2017meq}. With profile likelihood method \cite{Barbieri:1982eh, Rolke:2004mj} we 
have determined the $\gamma$-ray flux upper limits for a full dataset where, we 
have considered the full range of reconstructed energy of photons for our 
analysis. 
The upper limits of $N_{0}$ are evaluated with $95 \%$ confidence level (C.L.) 
and during this procedure, $N_{0}$ and the normalisation parameters of the 
galactic and isotropic model are fitted with the LAT-obtained spectrum at each 
step. This analysis would continue until the difference of the logarithm of the 
likelihood function reaches the value $1.35$ 
\cite{Abdo:2010ex} corresponding to a one-sided  $95\%$ C.L. Next we have 
applied the 
Bayesian method to our dataset (\cite{Helene:1990yi}). This method is already 
implemented in the \texttt{Fermi ScienceTools} 
\cite{Abdo:2010ex} and is used for obtaining a more appropriate value of the upper 
limit of the 
$\gamma$-ray flux in $95 \%$ C.L. \cite{Biswas:2017meq}. In Table 5.3, we have displayed the upper 
limits of the $\gamma$-ray flux for all five spectral indices \cite{Biswas:2017meq}. From Table 5.3, 
we can observe that the gamma-ray flux upper limits
for $\Gamma = 1$ is about $16$ times lower than the one for $\Gamma = 2.4$ \cite{Biswas:2017meq}. 
This result is in agreement with the results derived by Ref.~\cite{Abdo:2010ex}.\\

\begin{table}[!h]
\begin{center}
\caption{Best fit value of the normalisation parameter of Tri-II and the TS values for five $\Gamma$s.}
\begin{tabular}{|p{3cm}|p{5cm}|p{5cm}|}
\hline 
\hline
Spectral~Index~($\Gamma$) & $\rm{N_{0} \times 10^{-5}}$  & Test Statistic (TS) 
value \\
\hline 
$1$  & $(1.41\pm2.75)\times10^{-9}$ & 0.41   \\
\hline 
$1.8$   & $(6.66\pm11.49)\times10^{-8}$ & 0.44   \\
\hline 
$2$  & $(1.06\pm2.41)\times10^{-7}$ & 0.23   \\
\hline 
$2.2$   & $(1.88\pm5.53)\times10^{-7}$ & 0.02   \\
\hline 
$2.4$   & $(1.41\pm2.75)\times10^{-11}$ & $-7.45\times 10^{-8}$   \\
\hline 
\hline
\end{tabular}
\end{center}
\end{table}

\begin{table}[!h]
\caption{Estimated $\gamma$-ray flux upper limits in $95\%$ C.L..}
\begin{center}
\begin{tabular}{|p{3cm}|p{8cm}|}
\hline 
\hline
Spectral~Index~($\Gamma$) & 
Flux~upper~limits~at~$\rm{95\%~C.L.~(cm^{-2}~s^{-1})}$ \\
\hline 
1 & $8.29\times10^{-11}$ \\
\hline 
1.8 & $4.55\times10^{-10}$ \\
\hline 
2 & $7.14\times10^{-10}$ \\
\hline 
2.2 & $1.04\times10^{-9}$ \\
\hline 
2.4 & $1.37\times10^{-9}$ \\
\hline
 \hline
\end{tabular}
\end{center}
\end{table}

\section{J-factor for Tri-II}

\noindent For this work, we have modelled the DM distribution of Tri-II with 
NFW 
density 
profile 
\cite{Navarro:1996gj}. For estimating the J-factor value of Tri-II, we have 
used 
the following simple analytical 
relation derived by ref.~\cite{Evans:2016xwx}. The J-factor estimated by this 
analytical formula is in good agreement 
with the numerically estimated values of J-factor \cite{Biswas:2017meq}.
\begin{equation}
J \approx\frac{25}{8G^{2}} \frac{\sigma_{{\rm{v}}}^{4}\theta}{dr_h^{2}} .
\end{equation} 
here, $G$ is the gravitational constant, $\sigma_{{\rm{v}}}$ is the velocity 
dispersion and $r_{h}$ is the 2D projected 
half-light radius. For Tri-II, we have considered $\theta  = 0.15^{\circ}$ 
\cite{Genina:2016kzg}.\\

\noindent In Table 5.4, we have shown two different values of J-factor \cite{Biswas:2017meq} for two 
different 
$\sigma_{{\rm{v}}}$ values of Tri-II \cite{Kirby:2017cyb}. 
 
\begin{table}[!h]
\begin{center}
\caption{Parameters to calculate the J-factors.}
\begin{tabular}{|p{2cm}|p{3cm}|p{2cm}|p{2cm}|p{4cm}|}
\hline 
\hline
d (kpc) \cite{Laevens:2015una} & $\sigma_{{\rm{v}}}$ ($\rm{km~s^{-1}}$) 
\cite{Kirby:2017cyb} 
& $r_{h}$ (pc) \cite{Kirby:2017cyb}& $\theta$ (deg) \cite{Genina:2016kzg}& 
J-factor from 
Eq.~(5.2) ($\rm{{GeV^{2}~cm^{-5}}}$)\\
\hline 

$\rm{30\pm2}$  & $4.2~(95\%~\rm{C.L.})$ & 34 & $0.15^{\circ}$ & 
$\rm{0.17\times10^{20}}$ \\
\hline
$\rm{30\pm2}$ &  $3.4~(90\%~\rm{C.L.})$ & 34  & $0.15^{\circ}$ & 
$\rm{0.75\times10^{19}}$ \\
\hline
\hline
\end{tabular}
\end{center}
\end{table}

\noindent For our J-factor calculation, we did not consider the contribution 
from the 
substructures in Tri-II \cite{Biswas:2017meq}
which can increase the value of J-factor by few factors 
\cite{Martinez:2009jh,Abdo:2010ex}.\\

\noindent We would also like to point out that, in our present calculation, we 
did not 
take into account the effect of
Sommerfeld enhancement \cite{ArkaniHamed:2008qn,Abdo:2010ex, 
Feng:2010zp}. Such enhancement can increase the $\gamma$-ray flux due to the
dependence of annihilation cross-section (i.e., $<\sigma~v>$) on the relative velocity of particles. 
This enhancement effects the relative velocity of thermal relics cross section 
at freeze-out. Thus the value for
$<\sigma~v>$ would differ by few factors if we consider the 
Sommerfeld and can almost maximize the cross section by a factor of
$7$ to $90$ for WIMP masses between $100$ GeV to $3$ TeV \cite{Feng:2010zp}. 
To make a conservative approach, we have not included any such effect for our 
calculation.

\section{Constraints on Annihilation Cross-section} 
\noindent In this section, we would try to examine the possible $\gamma$-ray 
emission 
resulting
from the DM annihilation in Tri-II. For this purpose, we have determined the 
$95 
\%$ C.L. 
upper limits on $\gamma$-ray flux as a function of the WIMP mass for some 
specific annihilation channels \cite{Biswas:2017meq}. \\

\noindent To estimate the flux upper limits in 95$\%$ C.L. and the 
corresponding 
upper 
limits to 
the thermally averaged pair-annihilation $<\sigma v>$ of the WIMPs with the 
variation of 
the plausible WIMP masses ($\rm{m_{DM}}$), we have used the Bayesian approach 
(\cite{Helene:1990yi}) 
as this is more sensitive~\cite{Rolke:2004mj, Barbieri:1982eh} than the profile
likelihood method for low statistics \cite{Biswas:2017meq}. \\

\noindent For estimating the plausible flux upper limits and limits to the
$<\sigma v>$ of WIMP pair-annihilation, we have fitted the
$\gamma$-ray spectrum arising from the DM-dominated dSphs \cite{Biswas:2017meq} with an
MC-simulated DM self-annihilation spectrum, DMFitFunction
\cite{{Jeltema:2008hf}}.\\

\noindent The functional form of DMFitFunction
(a modified form of Eq.~2.5) can be written as \footnote{https://fermi.gsfc.nasa.gov/ssc/data/analysis/scitools/source{\_}models.html}:

\begin{equation}
\label{eqn:dm_spectrum}
\frac{dN}{dE} (E,\Delta \Omega) = <\sigma v> J(\Delta \Omega) (B~F(M_{DM},C_{0}) 
+ (1 - B~F(M_{DM},C_{1}))
\end{equation}

\noindent From Eq. 5.3, B, $C_{0}$ and $C_{1}$ define the branching ration, 
primary decay channel and secondary decay channel,
respectively. The DMFitFunction is implemented in Fermi
\texttt{ScienceTools} as a DMFit package
\cite{Jeltema:2008hf} and the values of F($M_{DM}$,C) are provided by
the Fermi-LAT team. For J-factor, we have taken the values from Table~5.4 \cite{Biswas:2017meq}.\\

\noindent For this work, we have considered five supersymmetry-motivated pair
annihilation final states, such as  $100\%$ b$\rm{\bar{b}}$, $100\%$ 
$\rm{\tau^{+} \tau^{-}}$, 
$80\%$ b$\rm{\bar{b}} + 20\%$ $\rm{\tau^{+} \tau^{-}}$, $100\%$ $\rm{\mu^{+} 
\mu^{-}}$ and $100\%$ $\rm{W^{+} W^{-}}$, respectively
\cite{Jungman:1995df}. These annihilation channels are
particularly favored when we consider the neutralino as the WIMP
candidate \cite{Biswas:2017meq}. Though the supersymmetry theory prefers neutralino as a
valid cold DM candidate, our obtained result would be generic
to all theoretical WIMP models \cite{Biswas:2017meq}. \\

\begin{figure}
\begin{center}
\includegraphics[width=0.5\textwidth,clip,angle=0]{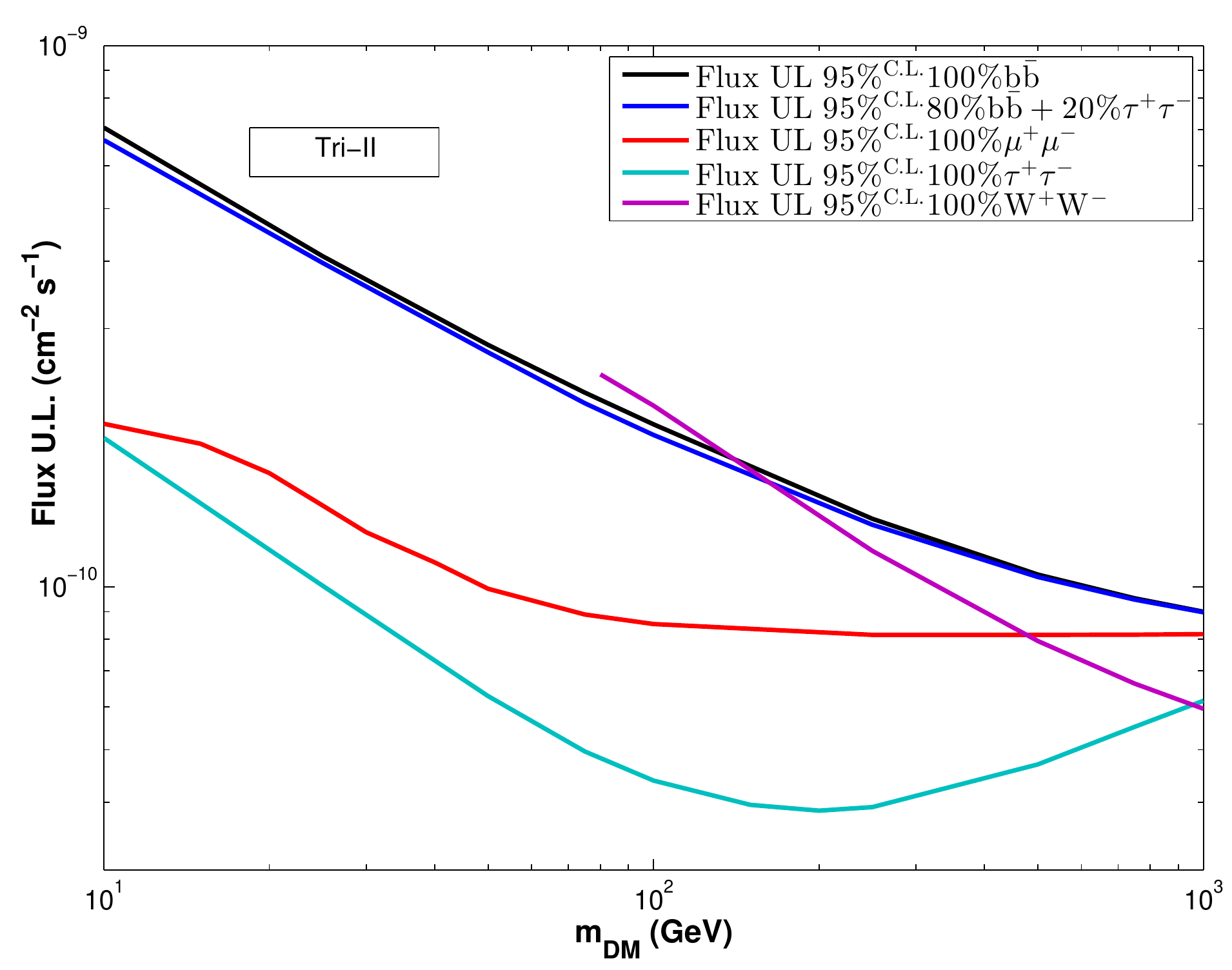}
\caption{The $\gamma$-ray lux upper limits of Tri-II for 
five WIMP annihilation final states.}
\end{center}
\end{figure}

\noindent In Fig.~5.2, we have shown the variation
of flux upper limits in $95\%$ C.L. with DM mass for all five annihilation 
channels.
From Fig.~5.2, we can observe that for all five final annihilation states, with 
increasing mass, the spectrum
from WIMP annihilation always shifts to higher energies
\cite{Serpico:2009vz} and so, we can expect that the variation of flux
upper limits would be comparatively less for high $m_{\rm{DM}}$ \cite{Biswas:2017meq}. 
Among all five final annihilation states producing hard $\gamma$-ray spectrum, 
$100\%$ $\tau^{+}\tau^{-}$ and $100\%$ $\mu^{+} \mu^{-}$ final states produce the abundant photon
fluxes especially at high energies where, the diffuse background is comparatively clear \cite{Biswas:2017meq}. From Fig.~5.2, we can also find that at 
$m_{\rm{DM}} \sim 1$~TeV, 
the flux upper limit for all five channels varies within a factor of $3$ 
but for low mass WIMP, this variation is more than an order of magnitude \cite{Biswas:2017meq}. \\

\noindent The results that we have shown in Fig.~5.2 are only dependent on the 
annihilation final state and DM mass \cite{Biswas:2017meq}. Thus, the flux upper limits from 
Fig.~5.2 are generic to all DM theoretical models and do not depend on any 
particular theory. Next, we have considered a few specific models to study the 
annihilation cross-section of WIMPs \cite{Biswas:2017meq}.\\

\begin{figure}
\centering
\subfigure[]
 { \includegraphics[width=0.48\linewidth]{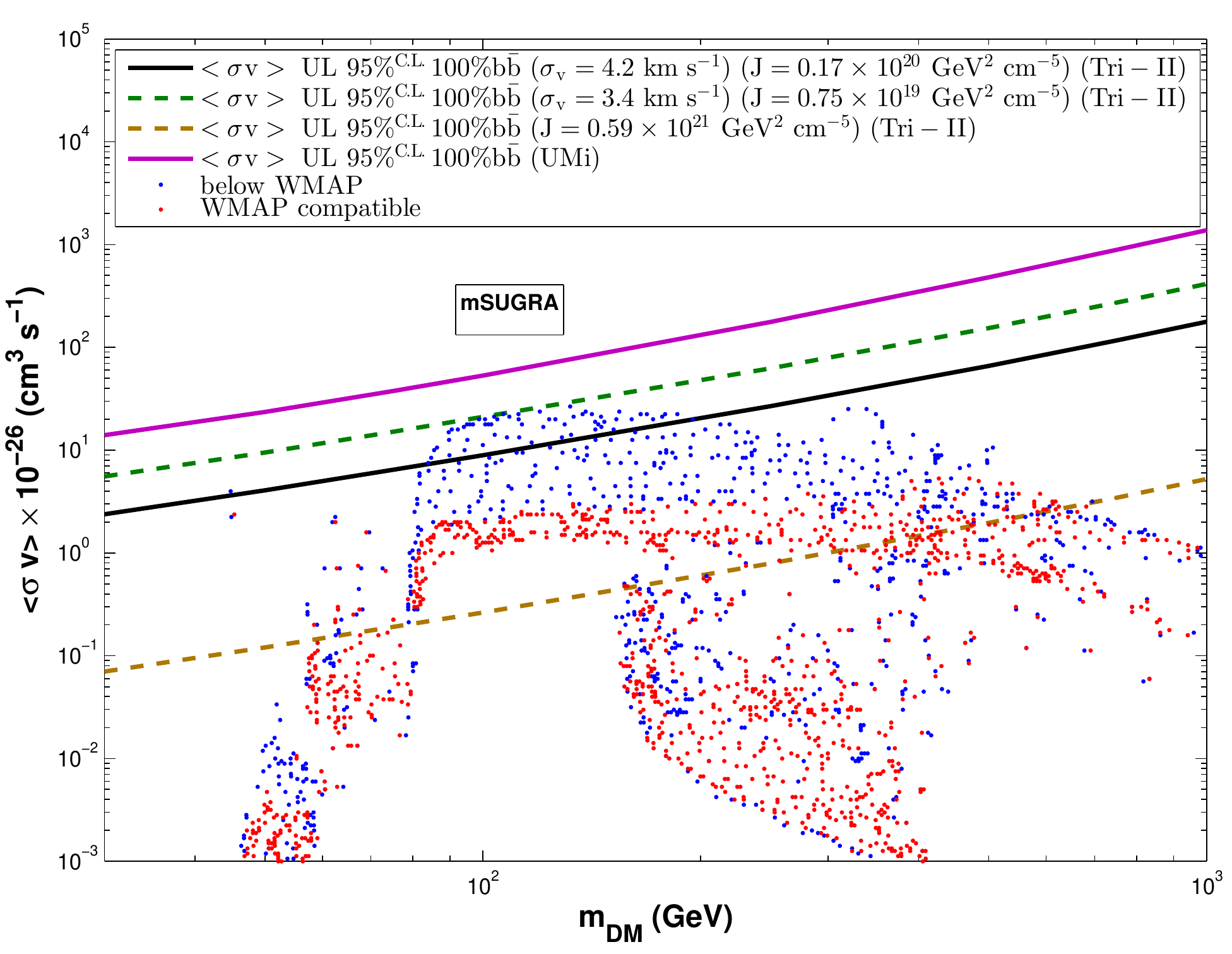}}
\subfigure[]
 { \includegraphics[width=0.48\linewidth]{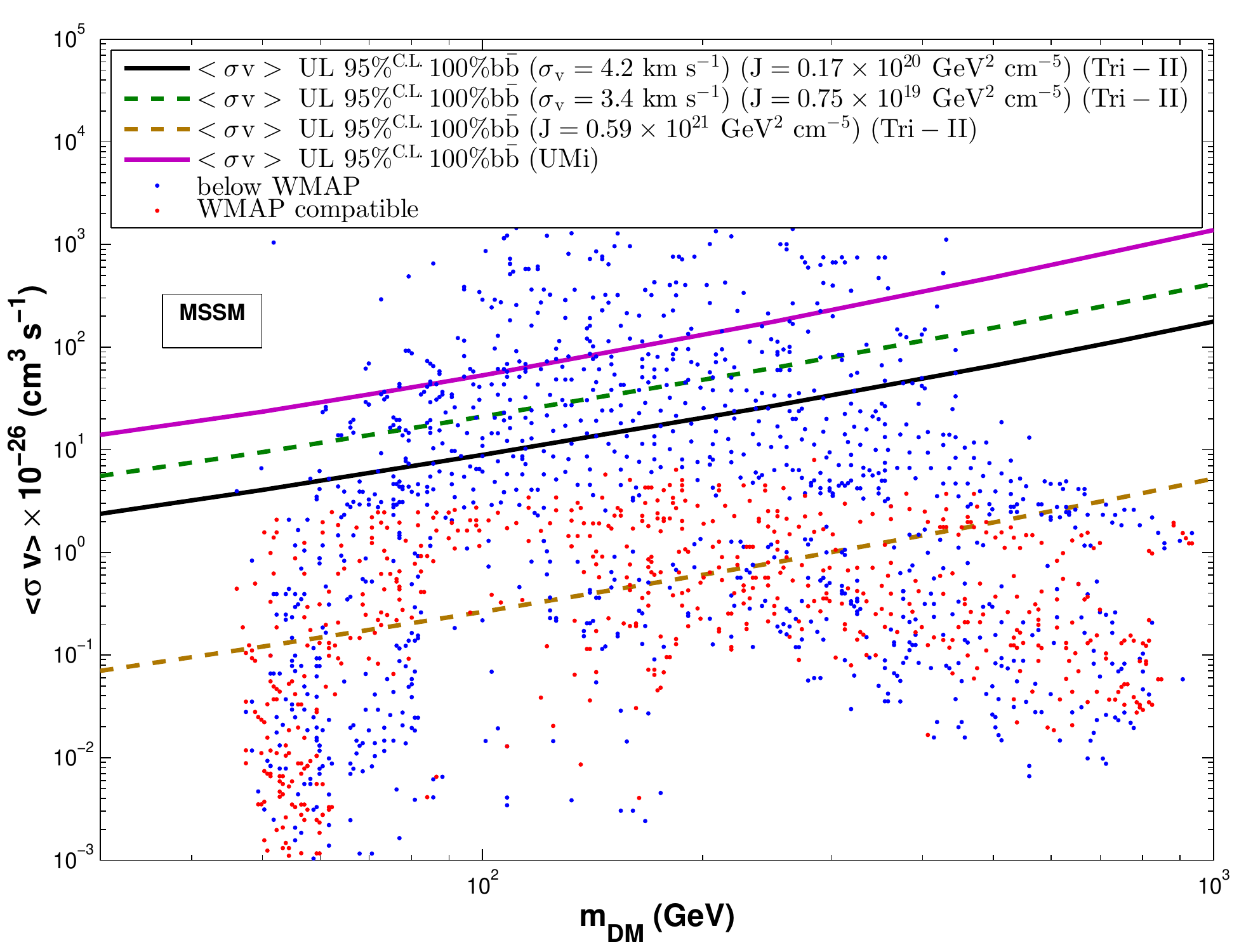}}
 \caption{Predictions from (a) the mSUGRA and (b) the MSSM models are shown in the parameter plane of
($m_{DM},<\sigma v>$). The red points denote the thermal relic abundance which is related to the DM 
density, while the blue points denote the lower thermal relic DM density. The red and blues points have been taken from the Abdo \textit{et al.}, 2010. In both figures, the $<\sigma~v>$ upper limits for two 
velocity dispersion ($\sigma$) values of Tri-II have been estimated for
$95\%$ C.L. The $<\sigma~v>$ upper limits for UMi have been similarly
estimated. The $<\sigma~v>$ upper limits of Tri-II for the higher value of J-factor (Genina \textit{et al.}, 2016) is denoted by the yellow curves. For UMi, we have used the parameter set mentioned in Abdo \textit{et al.}, 2010.}
\end{figure}

 \begin{figure}
 \centering
{\includegraphics[width=0.5\linewidth]{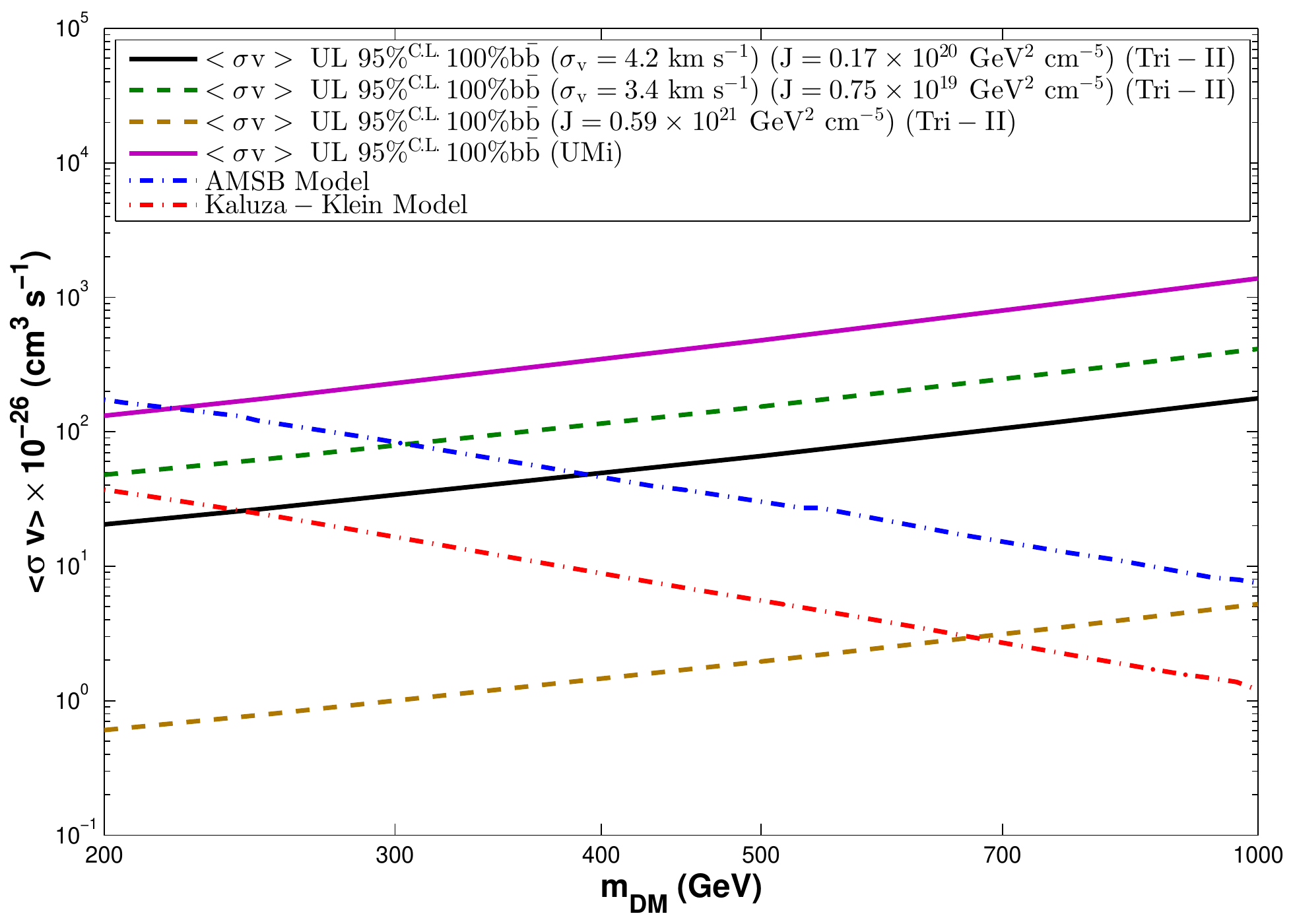}}
\caption{Predictions from the AMSB and the Kaluza-Klein UED models are shown in the parameter plane of
($m_{DM},<\sigma v>$). The $<\sigma~v>$ upper limits for two 
velocity dispersion ($\sigma$) values of Tri-II have been estimated for
$95\%$ C.L. The $<\sigma~v>$ upper limits for UMi have been similarly
estimated. The $<\sigma~v>$ upper limits of Tri-II for the higher value of J-factor (Genina \textit{et al.}, 2016) is denoted by the yellow curves. For UMi, we have used the parameter set mentioned in Abdo \textit{et al.}, 2010.}
\end{figure}

\noindent In Figs.~5.3(a,b) and Fig.~5.4, we have compared the resulting LAT 
sensitivity 
for 
Tri-II for three J values with a classical dSph, Ursa Minor (UMi) \cite{Biswas:2017meq}. 
In Figs.~5.3(a,b), we have considered two theoretically preferred DM models 
namely, minimal supergravity 
(mSUGRA) \cite{Chamseddine:1982jx} and Minimal Supersymmetric Standard Model 
(MSSM) 
\cite{Chung:2003fi}, respectively. In the mSUGRA model, the supersymmetry 
breaking 
parameters are defined at a high energy scale; generally of the order of 
grand unification scale $\sim 2 \times 10^{16}$~GeV. 
In the MSSM model, all the supersymmetry breaking parameters are specified 
at low energy scale i.e., in the electro-weak energy range. 
In Fig.~5.4, we have considered two other DM models, namely,
anomaly mediated supersymmetry breaking (AMSB) model \cite{Giudice:1998xp} and
Kaluza-Klein particle of universal extra dimensions (UED) 
\cite{Cheng:2002ej,Servant:2002aq,Hooper:2007qk}.
In the AMSB model, the supersymmetry breaking scenario might lead to the 
production of 
the wino-like neutralinos or the winos.
Winos are the mass eigenstate of neutralino that corresponds to the 
supersymmetric fermionic partners of the SU(2) gauge bosons. At about 
$2$ TeV mass of wino, the universal DM density agrees with the thermal relic DM
density generated by the winos. Several non-thermal production scenarios can 
interpret the connection of wino with the lighter DM candidates for masses less 
than 1 TeV\cite{Abdo:2010ex}.
In the Kaluza-Klein model, with very minimum setup, the first order excitation 
of the U(1) 
hypercharge gauge boson is also known as $\rm{B}^{(1)}$, might have a 
connection 
with the DM candidate. For Kaluza-Klein model, there exists a nearly exact 
relationship between $m_{DM}$ and the pair annihilation $<\sigma v>$. Moreover, 
from this model, we can obtain the thermal relic rate corresponding to the DM density
for DM masses above $700$~GeV \cite{Servant:2002aq}. \\

\noindent In Figs.~5.3(a,b) and Fig.~5.4, we have compared the LAT sensitivity 
obtained 
for Tri-II in the ($\rm{m_{DM}}$, $<\sigma v>$) plane with the $<\sigma v>$ 
limits predicted from the above mentioned four DM models namely mSUGRA, MSSM, 
Kaluza-Klein DM in UED and wino-like DM 
in AMSB \cite{Biswas:2017meq}. In figs.~5.3(a,b), the red points are associated with the thermal 
relic 
production while the blue points are related to the lower thermal relic density 
\cite{Servant:2002aq}. All these assumptions have been taken from the 
Ref.~\cite{Servant:2002aq}. 

\noindent In figs.~5.3(a,b) and 5.4, we have showed the upper limits on 
$<\sigma 
v>$ 
obtained for Tri-II for its two 
different values of velocity dispersion \cite{Kirby:2017cyb} and the 
predictions 
obtained 
from mSUGRA, MSSM, AMSB and Kaluza-Klein UED models 
respectively \cite{Biswas:2017meq}. The study by ref.~\cite{Kirby:2017cyb} has derived one optimistic 
value 
of $\sigma_{{\rm{v}}}$ $<$ 4.2~km~s$^{-1}$ in $95\%$ C.L. and another 
conservative value 
of $\sigma_{{\rm{v}}}$ $<$ 3.4~km~s$^{-1}$ in $90\%$ C.L. \cite{Biswas:2017meq}. In addition, we have 
also compared 
limits obtained for Tri-II with UMi \cite{Biswas:2017meq}. From Figs.~5.3(a,b), we can observe that 
even for 
the velocity dispersion value of 3.4~km~s$^{-1}$, at $m_{DM}= 100$ GeV,
the constraints obtained from mSUGRA and MSSM models for low thermal densities 
are nearly a factor 2.5 lower than 
the limits obtained from UMi \cite{Biswas:2017meq}. For velocity dispersion value of 4.2~km~s$^{-1}$, 
the constraints has further improved by 
a factor of $\sim$ 6 \cite{Biswas:2017meq}. Moreover, the Fig~5.4 has also indicated that for 
$\sigma_{{\rm{v}}}$ = 4.2~km~s$^{-1}$,  $<\sigma v>$ upper limits obtained from 
Tri-II
has disfavored the Kaluza-Klein in UED and AMSB models for masses 
$\lesssim 230$~GeV and $\lesssim 375$~GeV respectively \cite{Biswas:2017meq}. For $\sigma_{{\rm{v}}}$ 
= 3.4~km~s$^{-1}$, the limits obtained from Tri-II could not provide any 
effective constraints on Kaluza-Klein in UED models, whereas it disfavors the 
AMSB models for masses $\lesssim 300$~GeV \cite{Biswas:2017meq}. 
Here we also want to mention that for $\gamma$-ray observation, $100\%$ 
$b\bar{b}$ channel provides the more stringent limits. Thus, in 
Figs.~5.3(a,b) and 5.4, 
we have only shown the results for $100\%$ b$\rm{\bar{b}}$ channel \cite{Biswas:2017meq}.\\

\noindent From Figs.~5.3(a,b) and 5.4, we want to note the fact that for even higher value of J-factor, 
say for $\rm{0.59\times10^{21}}~\rm{{GeV^{2}~cm^{-5}}}$ (obtained from 
ref.~\cite{Genina:2016kzg}), Tri-II 
would put more stringent limits on the theoretical DM models than we have 
obtained by 
from the J values associated with the velocity dispersion of 
Tri-II. At $m_{\rm{DM}}= 100$~GeV, the predicted value of 
$<\sigma~v>$ corresponding to the J~$= \rm{0.59\times10^{21}}~\rm{{GeV^{2}~cm^{-5}}}$ is nearly 
$\sim 30$ factor lower than 
the $<\sigma~v>$ limits obtained for J~$= 
\rm{0.17\times10^{20}}~\rm{{GeV^{2}~cm^{-5}}}$ \cite{Biswas:2017meq}. In addition, 
this high J value disfavors the Kaluza-Klein in UED and AMSB model for mass 
$<~700$~GeV $<~1000$~GeV, respectively \cite{Biswas:2017meq}.

\begin{figure}
\centering
 {\includegraphics[width=0.5\linewidth]{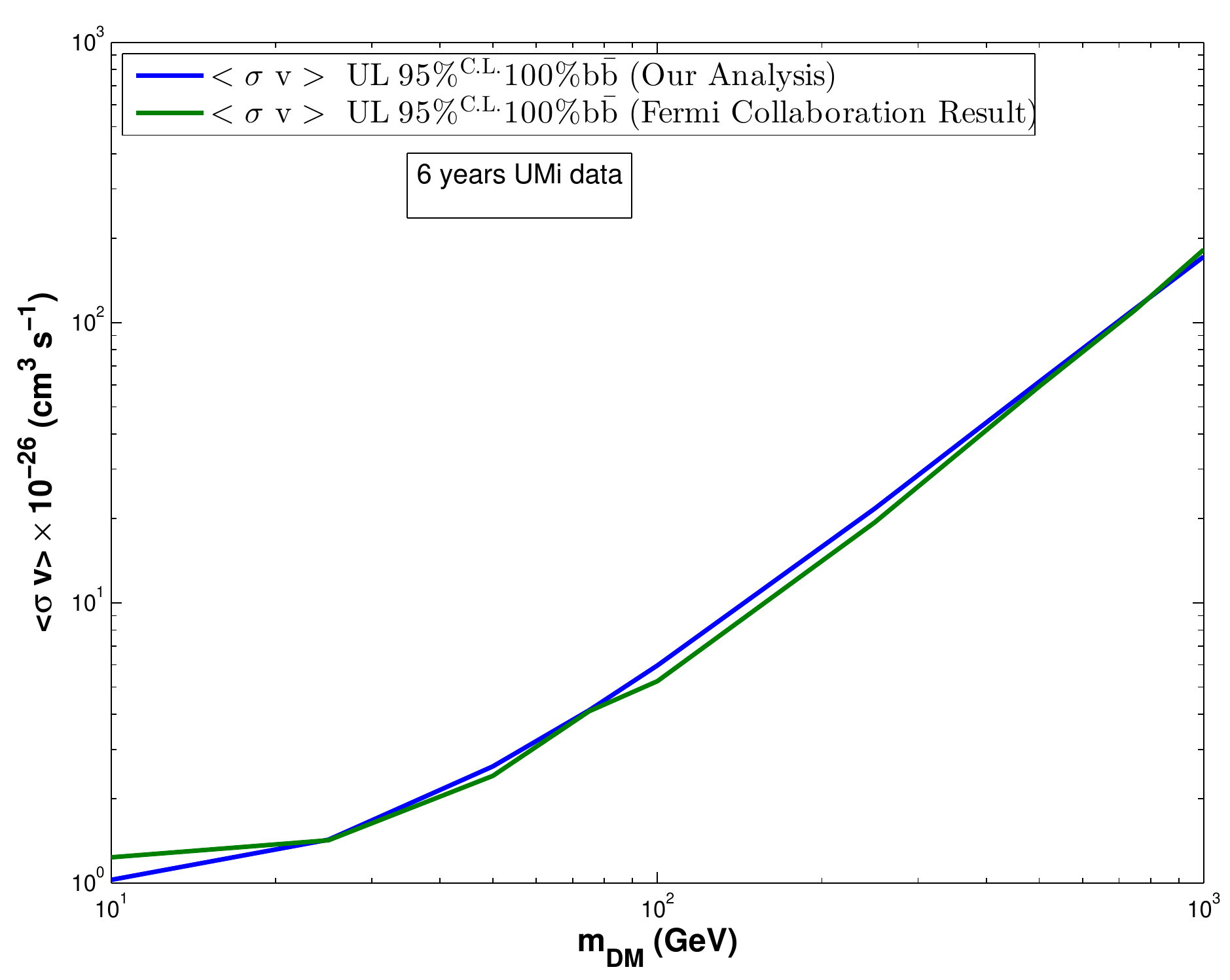}}
\caption{Comparison between the $<\sigma~v>$ upper limits for the b$\rm{\bar{b}}$ 
annihilation channel obtained from our analysis and obtained by the
\textit{Fermi} collaboration for UMi (Ackermann \textit{et al.}, 2015).}
\end{figure}

\noindent We would also like to point out that to check the reliability of our 
analysis 
method, we have compared our analysis result for UMi with the result obtained 
by 
\textit{Fermi} collaboration \cite{Ackermann:2015zua}. For this comparison, we 
have followed the
same data selection and analysis procedure as \textit{Fermi} collaboration by 
ref.~\cite{Ackermann:2015zua}. In Fig.~5.5, we have shown the comparison. From 
Fig.~5.5, we can conclude that \cite{Biswas:2017meq} our result is in good agreement with the result 
obtained by ref.~\cite{Ackermann:2015zua}. This study supports the reliability 
of the analysis procedure that we followed in this work \cite{Biswas:2017meq}.

\section{Conclusions $\&$ Discussions}
\noindent In this work, we have analysed nearly seven years of $\gamma$-ray 
data 
from the 
direction of Tri-II by \texttt{Fermi ScienceTools} but
could 
not observe any $\gamma$-ray excess
from the location of Tri-II. Thus, we have derived the upper limit of 
$\gamma$-ray 
flux for two possible scenarios.\\

\noindent Using the DM annihilation spectra, we have estimated the upper limits 
of 
$<\sigma v>$ where, we consider that DM candidates are in form of 
neutralinos. From our analysis, we have shown that for $\rm{\sigma_{v}= 
4.2~km~s^{-1}}$ with $100\%$ b$\rm{\bar{b}}$ channel, $<\sigma v>$ limits 
obtained from 
Tri-II 
constrain the mSUGRA and the MSSM models with low thermal relic densities, 
whereas the limits 
constraint the Kaluza-Klein DM in UED and the AMSB models for masses $\lesssim 
230$~GeV 
and $\lesssim 375$~GeV, 
respectively. \\

\noindent Even for the velocity dispersion with 90 $\%$ C.L. i.e., for
 $\rm{\sigma_{v} = 3.4~km~s^{-1}}$, Tri-II can constrain the MSSM model with 
low 
thermal relic densities and the AMSB model for masses $\lesssim 300$~GeV. 
Besides, from our work, we have found that $\gamma$-ray data from Tri-II 
can even put stronger limits on the theoretical DM models than UMi. 
We would also like to point out that our results are entirely based on 
the standard NFW profile and we do not consider any effects of boost factor 
related to substructures in Tri-II or the Sommerfeld effect in accord to the 
annihilation cross-section. Finally, from our work, we can state that with more 
precise observations of Tri-II, in future we can establish Tri-II as a very 
strong 
DM candidate for indirect DM searching.

\chapter{Analysis of Fermi-LAT data from Tucana-II: an intriguing hint of a signal}
\section{Tucana-II}
\noindent Inspired by the ongoing research interest in the indirect search for 
DM signal from
the UFDs or dSphs, in this work, we have studied a recently discovered UFD, 
namely Tucana-II (Tuc-II) \cite{Bhattacharjee:2018xem}. It is also referred to as DES~J2251.2-5836.
\cite{Koposov:2015cua, Bechtol:2015cbp, Drlica-Wagner:2015ufc}. The observation 
by ref.~\cite{Walker:2016mcs} has confirmed its status. Their study suggested 
that Tuc-II is a UFD and not a member of any globular cluster. Its high mass to 
luminosity ratio, large half light radius, large velocity dispersion value and 
luminosity-metallicity relation, all of these qualities have well-established 
Tuc-II as a confirmed candidate of UFD galaxies\cite{Walker:2016mcs, 
Gilmore:2006iy, Kirby:2013wna, 
Straizys:1974jf, Gallagher:2003nx, Grcevich:2009gt, Ji:2016cvd} and make Tuc-II 
a very promising source for the indirect search of DM 
signal\cite{Walker:2016mcs, Drlica-Wagner:2015xua, Hooper:2015ula, 
Fermi-LAT:2016uux, Calore:2018sdx}. The shape of the Tuc-II is a bit distorted 
and its outer region appears to be little elongated but the observational noise 
could be the reason for the distortion in Tuc-II \cite{Bechtol:2015cbp, 
Martin:2008wj, Munoz:2009hj}. \\

\noindent By using the Michigan Magellan Fibre System (M2FS) \cite{Mateo:2012mn}, the spectroscopic 
study done by Walker \textit{et al.}, 2016 \cite{Walker:2016mcs} has identified 
some of the member stars in the direction of Tuc~II. The study by 
ref.~\cite{Walker:2016mcs} and other previous photometric observation of 
Tuc-II~\cite{Koposov:2015cua, Drlica-Wagner:2015ufc}, have identified
eight probable member stars of Tuc-II. Those member stars are well-resolved 
enough to 
estimate the internal velocity dispersion of Tuc-II but they also lead to a 
large asymmetrical 
uncertainties in velocity dispersion (i.e., 
$\rm{\sigma_{v}}~=~8.6_{-2.7}^{+4.4}~{\rm {km}~{s}}^{-1}$) about a 
mean velocity of $-129.1_{-3.5}^{+3.5}~{\rm {km}~{s}}^{-1}$. Some important 
properties of Tuc-II obtained by several studies \cite{Koposov:2015cua, 
Walker:2016mcs, Chiti:2018cds} have been mentioned in Table~6.1 \cite{Bhattacharjee:2018xem}.

\begin{table}
\begin{center}                                                                                                                                                                                                         
\caption{Properties of Tucana-II.}
\label{Table-1}
\begin{tabular}{|p{4 cm}|p{4 cm}|p{2 cm}|}
\hline
\hline
Property &  Value   & Reference  \\ 
\hline
Galactic longitude & $\rm{328.0863^{\circ}}$  & \cite{Koposov:2015cua} \\
\hline
Galactic latitude & $\rm{-52.3248^{\circ}}$ & \cite{Koposov:2015cua} \\
\hline
Heliocentric distance ([d]) & $\rm{57_{-5}^{+5}~\rm {kpc}}$ & 
\cite{Koposov:2015cua}\\
\hline
Metallicity ([$\rm{Fe/H}$]) & $\rm{<0.4}$ & \cite{Walker:2016mcs}\\ 
\hline
Projected half light radius ($\rm{R_{h}}$) & $\rm{165^{+27.8}_{-18.5}~pc}$ & 
\cite{Koposov:2015cua} \\
\hline
Maximum galactocentric angular distance in the sample of the observed member 
stars in Tuc-II, as measured from the observer's position ([$\theta_{\rm 
{max}}$]) & $0.30^{\circ}$ & \cite{Chiti:2018cds}\\ 
\hline
Square-root of the luminosity-weighted square of the line-of-sight stellar 
velocity dispersion ($\rm{\sigma_{v}}$) & $~\rm{8.6_{-2.7}^{+4.4}~km~s^{-1}}$ & 
\cite{Walker:2016mcs}\\
\hline
Mass within the projected half-light radius 
\Big($\rm{\frac{M_{1/2}}{M_{\odot}}}$\Big) & 
$\rm{~2.7_{-1.3}^{+3.1}~\times~10^{6}}$ & \cite{Walker:2016mcs}\\
\hline
Dynamical mass-to-light ratio \Big($\rm{(M/L_{v})_{1/2}}$\Big) & 
$\rm{~1913_{-950}^{+2234}~M_{\odot}~L_{\odot}^{-1}}$ & \cite{Walker:2016mcs}\\
\hline
\hline
\end{tabular}
\end{center}
\end{table}

\section{The \textit{Fermi}-LAT Data Analysis of Tuc-II}

\begin{table}
    \caption{The parameter set that we used for our \textit{Fermi}-LAT 
analysis.}
    \begin{tabular}{||p{7 cm}p{8 cm}||}
        \hline \hline
        {\bf Parameter for data extraction } &\\        
        \hline\hline
        Parameter & Value \\
        \hline \hline
        Source & Tucana-II \\
        Right Ascension (RA) & 342.9796 \\
        Declination (DEC) & -58.5689 \\
        Radius of interest (ROI) &  $10^{\circ}$ \\
        TSTART (MET) & 239557418 (2008-08-04 15:43:37.000 UTC) \\
        TSTOP (MET) & 530362359 (2017-10-22 10:52:34.000 UTC) \\
        Energy Range & 100 MeV - 300 GeV \\
        \hline \hline
        \texttt{gtselect} for event selection & \\ 
        \hline \hline
        Event class & Source type (128)\\ 
        Event type & Front+Back (3)\\
        Maximum zenith angle cut & $90^{\circ}$\\
        \hline \hline
        \texttt{gtmktime} for time selection &\\ 
        \hline \hline
        Filter applied & $\textit{(DATA\_QUAL>0)\&\&(LAT\_CONFIG==1)}$\\ 
        ROI-based zenith angle cut & No\\
        \hline \hline
       \texttt{gtltcube} for livetime cube &\\ 
        \hline \hline
        Maximum zenith angle cut ($z_{cut}$) & $90^{\circ}$\\  
        Step size in $cos(\theta)$ & 0.025\\
        Pixel size (degrees) & 1\\
        \hline \hline
        \texttt{gtbin} for 3-D (binned) counts map &\\ 
        \hline \hline
        Size of the X $\&$ Y axis (pixels) & 140\\
        Image scale (degrees/pixel) & 0.1\\
        Coordinate system & Celestial (CEL)\\
        Projection method & AIT\\
        Number of logarithmically uniform energy bins & 24\\ 
        \hline \hline
        \texttt{gtexpcube2} for exposure map &\\ 
        \hline \hline
        Instrument Response Function (IRF) & $\rm{P8R2\_SOURCE\_V6}$ \\ 
        Size of the X $\&$ Y axis (pixels) & 400\\
        Image scale (degrees/pixel) & 0.1\\
        Coordinate system & Celestial (CEL)\\
        Projection method & AIT\\
        Number of logarithmically uniform energy bins & 24\\ 
        \hline \hline
        \texttt{gtlike} for likelihood analysis &  \\ 
        \hline \hline
        Galactic diffuse emission model & $\rm{gll\_iem\_v06.fits}$ \\        
        Extragalactic isotropic diffuse emission model & 
$\rm{iso\_P8R2\_SOURCE\_V6\_v06.txt}$ \\ 
        Source catalog & 3FGL \\
        Extra radius of interest &  $5^{\circ}$ \\
        Response functions & $\rm{P8R2\_SOURCE\_V6}$\\
        Optimizer & NEWMINUIT\\
        Spectral model of Tucana-II & Power law (in Section-6.2.1) $\&$ DMFit 
Function (in Section-6.4) \\ 
        \hline \hline
       \end{tabular}
    \label{table_1}
\end{table} 

\noindent Like Tri-II, for Tuc-II we have analysed the gamma-ray data observed 
by the Fermi-LAT and have mostly followed the same analysis method that we have 
applied for Tri-II \cite{Bhattacharjee:2018xem}.\\

\noindent We have used the Fermi ScienceTools version \texttt{v10r0p5} and the 
dataset was 
pre-processed with an 
improved IRF, P8R2\_SOURCE\_V6 of the Fermi-LAT \cite{Bhattacharjee:2018xem}.\\

\noindent For analysing the possible signal coming from the direction of 
Tuc-II, 
we have extracted about nine years of Fermi-LAT data in between 100 MeV to 300 
GeV energy range and have selected  
a $10^{\circ}~\times~10^{\circ}$ ROI centred on the 
location of Tuc-II \cite{Bhattacharjee:2018xem}.\\

\noindent In the source model, we have included our source of interest, Tuc-II 
along with 
all the sources from 3FGL catalog~\cite{Acero:2015gva} within 15$^{\circ}$ ROI 
from the location of Tuc-II \cite{Bhattacharjee:2018xem}. Then with `gtlike' tool, we have run the binned 
likelihood on our 
dataset~\cite{Cash:1979vz, Mattox:1996zz}. During the likelihood process, the 
spectral parameters of 
all the sources within $10^{\circ}~\times~10^{\circ}$ ROI and the 
normalization parameters of two diffuse backgrounds models (i.e., 
$\rm{gll\_iem\_v06.fits}$ 
and  $\rm{iso\_P8R2\_SOURCE\_V6\_v06.txt}$) have been left free \cite{Bhattacharjee:2018xem}. The remaining 
all the background sources within the 
$15^{\circ}~\times~15^{\circ}$ ROI have been kept fixed to their 3FGL catalog 
\cite{Acero:2015gva} mentioned values. All the necessary information for 
performing the  
\textit{Fermi}-LAT analysis is mentioned in TABLE~6.2 \cite{Bhattacharjee:2018xem}. In TABLE~6.2, TSTART and TSTOP define the start and the end of observation in unit Mission Elapsed Time (MET), respectively.\\

\noindent In the following section, to check any possible emission from Tuc-II 
location, we would first model our source with a power-law spectrum for 
different spectral indices.

\subsection{Results of the Power-law Modelling} 

\begin{figure}
\centering
\subfigure[]
{ \includegraphics[width=0.8\linewidth]{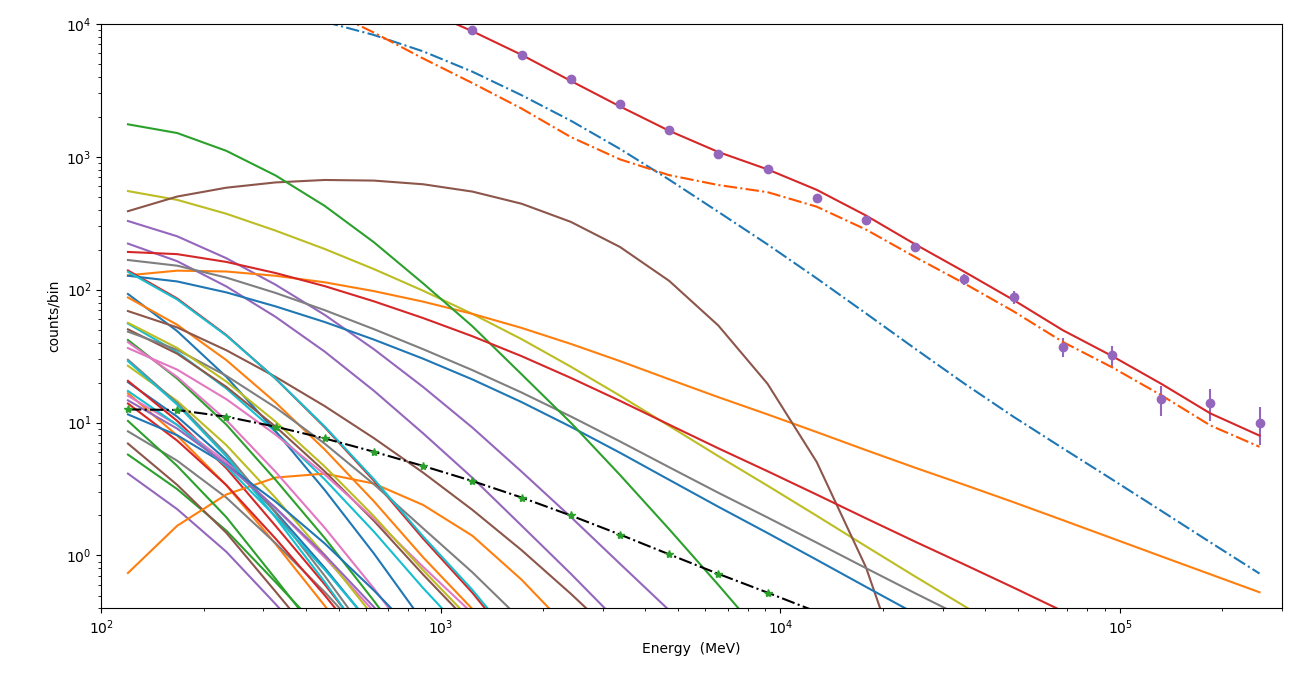}}
\subfigure[]
{ \includegraphics[width=0.9\linewidth]{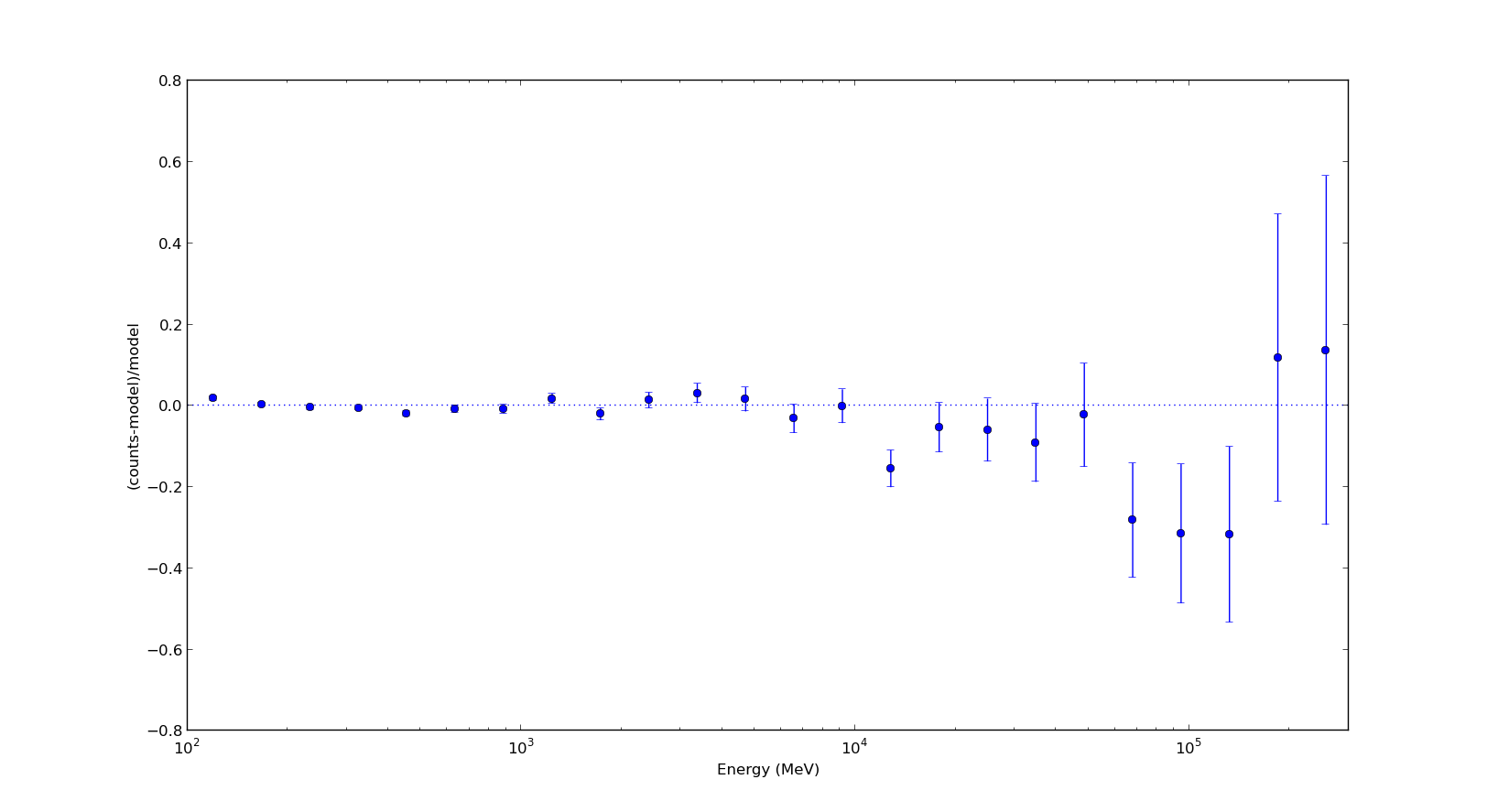}}
\caption{(a) The spectral fit to the observed data counts per energy bin and (b) the
residual 
plot for the location of Tuc-II has been shown here. We have modelled the Tuc-II with
the power-law spectrum for $\Gamma = 2$. In figure 6.1(a), the solid dark reddish-
brown curve displays the best-fit total spectrum, along with the corresponding LAT-observed data
points (in purple); the dot-dashed sky-blue and orange curves display the galactic diffuse background and
the isotropic background component, respectively; the dot-dashed black curve along with green points
denotes the spectral fit of Tuc-II. The rest of the curves correspond to various point sources other
than Tuc-II, lying within the ROI that are not distinctly labeled in figure 6.1(a).}
\end{figure}

\noindent Like our previous chapter, we have modelled the Tuc-II with power-law 
spectrum (Eq.~5.1) and have performed the fitting for five spectral indices 
($\Gamma$). In Fig.~6.1, we have shown the fitting results of Tuc-II for the spectral index $\Gamma =2$.\\

\noindent In Fig.~6.1(a), we have shown the spectral fit of all the sources that lie within the ROI \cite{Bhattacharjee:2018xem}. 
In this figure, the sum of the best fit spectrum along with the LAT counts (in purple) is denoted by the solid
dark reddish-brown curve, while the best fit spectrum for the galactic and isotropic components are defined by the `dot-dashed' sky-blue and
orange curves, respectively. The black `dot-dashed' curve along with the green points refer to the best-fit spectra of Tuc-II and
the remaining curves are related to other sources within ROI. In Fig.~6.1(b) we have displayed the residual plot of Tuc-II for the
spectral index, $\Gamma$=2 \cite{Bhattacharjee:2018xem}.\\

\noindent The best-fitted value of the normalisation parameter, $N_{\rm {0}}$ 
and the TS value obtained from 
Tuc-II for all five spectral indices ($\Gamma$) is shown in TABLE~6.3 \cite{Bhattacharjee:2018xem}. 
Among all five spectral indices, $\Gamma = 1$ is 
assumed to have the connection with DM annihilation
(Ref.~\cite{Essig:2009jx}) and we have chosen other our $\Gamma$'s values to 
examine the astrophysical spectrum of Tuc-II \cite{Bhattacharjee:2018xem}. From TABLE~6.3 we can observe 
that 
for $\Gamma = 1$, the value of statistical error on $N_{\rm {0}}$ is slightly 
higher than the value of $N_{\rm {0}}$ itself and the TS values of Tuc-II for 
all $\Gamma$s are much less than the threshold limit for detection (i.e., 
TS$\ge$25) \cite{Bhattacharjee:2018xem}.

\begin{table}
\begin{center}
\caption{The best-fit normalization parameters ($N_{0}$) of Tuc-II and the TS values for
five spectral indices ($\Gamma$).}
\begin{tabular}{|p{3cm}|p{5cm}|p{3cm}|}
\hline 
\hline
Spectral~Index~($\Gamma$) & $\rm{N_{0} \times 
10^{-5}~(cm^{-2}~s^{-1}~MeV^{-1})}$  & Test Statistic (TS) value \\
\hline 
$1$  & $(2.457\pm11.17)\times10^{-10}$ & 0.056   \\
\hline 
$1.8$  & $(1.173\pm1.126)\times10^{-7}$ & 1.215   \\
\hline 
$2$  & $(3.146\pm 2.565)\times10^{-7}$ & 2.077   \\
\hline 
$2.2$  & $(7.458\pm4.923)\times10^{-7}$ & 2.973   \\
\hline 
$2.4$ & $(1.433\pm0.839)\times10^{-6}$ & 3.592   \\
\hline 
\hline
\end{tabular}
\end{center}
\end{table}

\begin{table}
\caption{The $\gamma$-ray flux upper limits in $95\%$ C.L. obtained from Tuc-II for five spectral indices ($\Gamma$).}
\begin{center}
\begin{tabular}{|p{3cm}|p{8cm}|}
\hline 
\hline
Spectral~Index~($\Gamma$) & 
Flux~upper~limits~in~$\rm{95\%~C.L.~(cm^{-2}~s^{-1})}$ \\
\hline 
1 & $3.248\times10^{-11}$ \\
\hline 
1.8 & $4.484\times10^{-10}$ \\
\hline 
2 & $8.362\times10^{-10}$ \\
\hline 
2.2 & $1.401\times10^{-9}$ \\
\hline 
2.4 & $2.113\times10^{-9}$ \\
\hline
 \hline
\end{tabular}
\end{center}
\end{table}

\noindent As we have not detected any excess emission from Tuc-II location, 
we have determined the flux upper limit in $95\%$ C.L. by the profile 
likelihood 
method 
\cite{Barbieri:1982eh, Rolke:2004mj}.\\

\noindent We have next derived the flux upper limits in 95$\%$ C.L. by using 
the 
semi-Bayesian method with flat prior \cite{Bhattacharjee:2018xem}. This semi-Bayesian method is developed 
from Helene’s approach \cite{Helene:1990yi} and is 
already implemented in the \texttt{ScienceTools}.\\

\noindent In Table~6.4, we have shown the flux upper limits in 95$\%$ C.L. 
derived from the semi-Bayesian method \cite{Bhattacharjee:2018xem}.
From this table~6.4, we can note that, the $\gamma$-flux upper limit for 
$\Gamma=1$ is almost 2 orders lower than the flux upper limits corresponding to 
$\Gamma=2.4$ \cite{Bhattacharjee:2018xem}. This result is consistent with our finding for 
Tri-II~\cite{Biswas:2017meq, Bhattacharjee:2018xem}. Here, we would like to point out that flux upper 
limits developed from the semi-Bayesian method and the profile likelihood 
method 
are hardly differed by 1.2 to 1.3 factor \cite{Bhattacharjee:2018xem}.\\

\noindent In the next section, we have attempted to study the possible DM 
signature coming from the location of Tuc-II \cite{Bhattacharjee:2018xem}. Thus, now we would model Tuc-II 
with the $\gamma$-ray spectrum from DM annihilation 
(i.e., with DMFit function) that is implemented in \textit{Fermi} 
\texttt{ScienceTools}. For comparison, along with Tuc-II, we would also 
introduce two newly discovered dSphs, namely, 
Reticulum-II (Ret-II) and Ursa Minor (UMi) \cite{Bhattacharjee:2018xem}.\\

\section{Estimation of Astrophysical Factor (J-factor) for Tuc-II}

\noindent The main difficulties in studying the newly discovered UFDs is their 
insufficient kinematics data. That also questions the reliability of J-factors 
of
the dSphs and UFDs \cite{Bhattacharjee:2018xem}. For our work, we have taken the J-factors of Tuc-II and other two dSphs (Ret-II and UMi) from Evans 
\textit{et al.}, 2016 
\cite{Evans:2016xwx}. Their study suggests that the 
analytical formula for J-factor 
can estimate more or less accurate results if we compare it to the spherical 
Jeans formula for J-factor calculation driven by Markov Chain Monte Carlo 
techniques. Evans \textit{et 
al.} \cite{Evans:2016xwx} argued that their derived formula for J-factors can 
even reproduce the computational results.

\section{DM Annihilation Constraints from Tuc-II}
\subsection{Searching for $\gamma-ray$ Emission due to DM 
Annihilation from Tuc-II}

\begin{figure}
\centering
\subfigure[]
 {\includegraphics[width=0.49\linewidth]{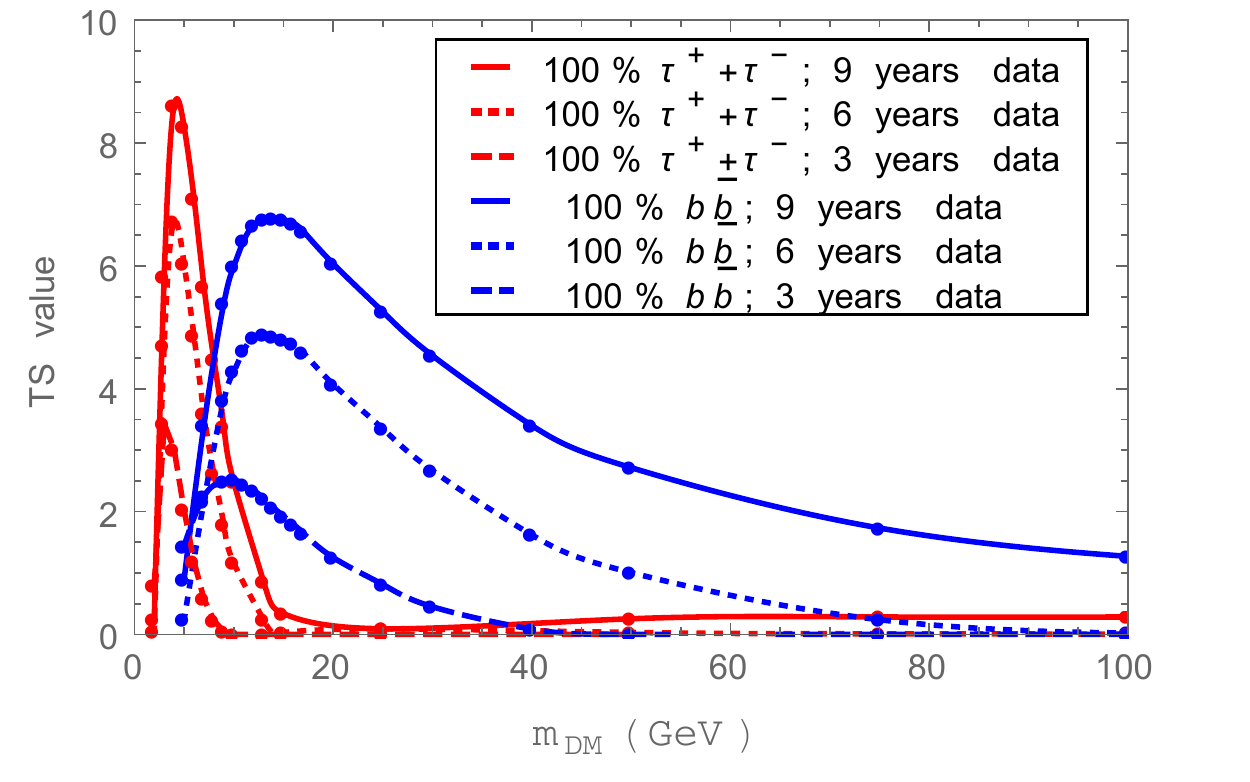}}
\subfigure[]
 {\includegraphics[width=0.49\linewidth]{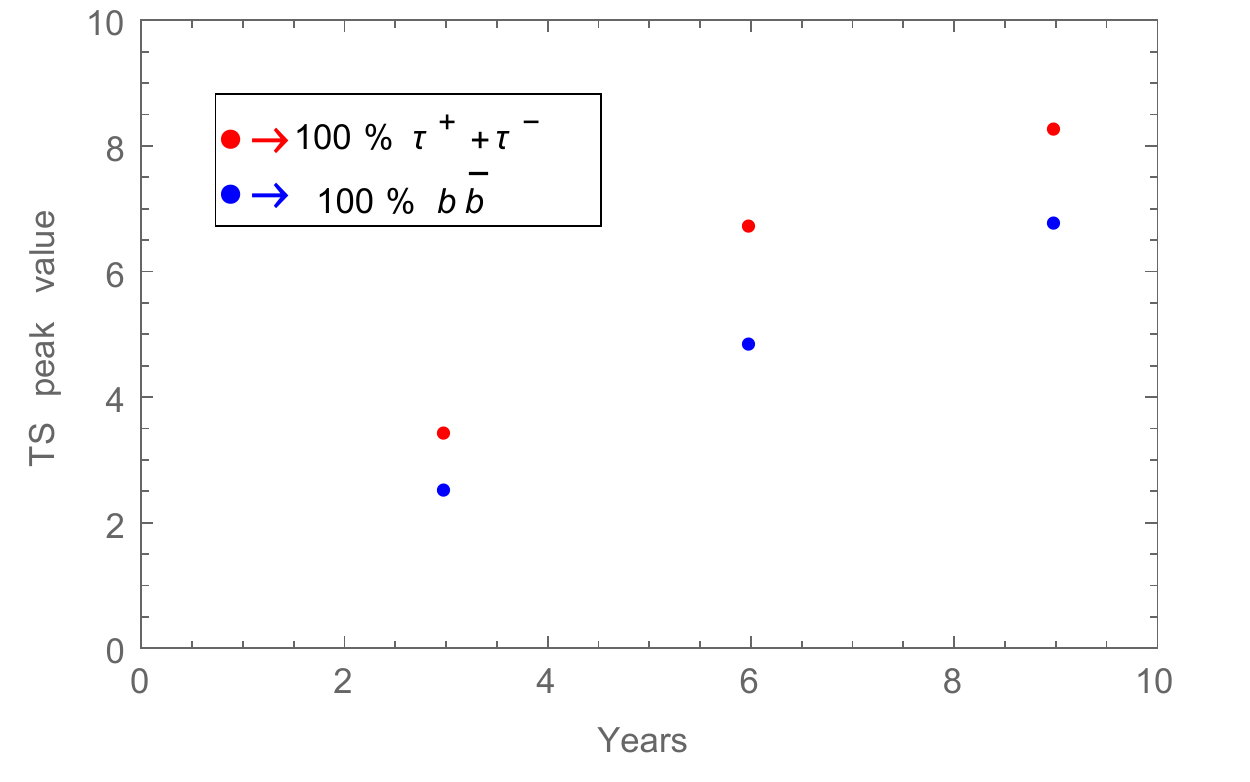}}
\caption{(a) The variation of the TS values of Tuc-II with $m_{DM}$ for two WIMP annihilation channels; i) $100\%$ $b\bar{b}$ 
(blue) and ii) $100\%$ $\tau^{+}\tau^{-}$ (red). We have also shown the results for three different periods 
of LAT data. (b) The peak TS value observed from the location of Tuc-II for three periods of LAT data, while the red and the blue markers refer to the peak value of TS 
for $b\bar{b}$ and $\tau^{+}\tau^{-}$ WIMP annihilation final states, 
respectively.}
\end{figure}

\noindent Here first we have fitted the possible $\gamma$-ray flux arising from 
the
Tuc-II location with the $\gamma$-ray spectrum for DM pair-annihilation \cite{Bhattacharjee:2018xem}. For 
this calculation, we have employed 
the MC simulation package DMFit~\cite{Jeltema:2008hf, Gondolo:2004sc} which is 
implemented in the Fermi-\texttt{ScienceTools}. 
We have defined the Tuc-II as a point source and its significance is derived by 
the $\Delta TS$ method that we have followed in section 6.2.1. \\

\noindent In this section, we would try to examine after modelling the Tuc-II 
with $\gamma$-ray annihilation spectrum,
whether we can obtain any excess from the location of Tuc-II. Interestingly, we 
have detected a very faint emission from Tuc-II \cite{Bhattacharjee:2018xem}.\\

\noindent From Fig.~6.2(a), we can observe the variation of the detection 
significance of 
$\gamma$-ray 
excess (i.e the TS values) from the location of Tuc-II as a function of WIMP 
mass ($m_{DM}$) and for two pair annihilation final states, i.e., $100\%$ 
$b\bar{b}$ and 
$100\%$ $\tau^{+}\tau^{-}$ \cite{Bhattacharjee:2018xem}.  In Fig. 6.2(b), we
have also shown the variation of TS values for three, six and nine years of LAT 
data \cite{Bhattacharjee:2018xem}.
For this comparison, we have performed the same 
analysis method in all three years of the dataset. From Fig.~6.2(b), we can 
observe that the peak value of TS is increased with increasing the dataset and both annihilation channels
have followed the same nature. The observed emission from Tuc-II location is 
indeed too faint (i.e., 
less than TS=25) to claim anything precisely, but the most interesting finding 
of this analysis is that the peak value of TS is gradually increasing with time 
period \cite{Bhattacharjee:2018xem}. From this signature, we can expect that in future we can possibly 
detect 
a real signal from Tuc-II either due to its connection with
any astrophysical source or resulting from DM annihilation \cite{Bhattacharjee:2018xem}. In Fig.~6.2(a), 
with 
nine years of \textit{Fermi}-LAT data, 
the TS value peaks at $m_{DM}$~=~14 GeV for $100\%$ $b\bar{b}$ channel, while for $100\%$ $\tau^{+}\tau^{-}$ it
peaks at $m_{DM}$~=~4 GeV  \cite{Bhattacharjee:2018xem}.\\

\noindent There were many earlier studies that already analyzed 
Tuc-II 
with six or 
seven years of \textit{Fermi}-LAT data \cite{Drlica-Wagner:2015xua, 
Hooper:2015ula, 
Fermi-LAT:2016uux, Calore:2018sdx}. But in our analysis, we have studied Tuc-II 
with nine years of \textit{Fermi}-LAT data and thus the increase in TS peak 
values possibly originate from the larger dataset \cite{Bhattacharjee:2018xem}. Hence, such
an increase in $\gamma$-ray excess with increasing the time period of analysis 
seems to encourage the indirect detection of DM annihilation signal \cite{Bhattacharjee:2018xem}.\\

\begin{table}
\caption{The overview of the TS and the $\Delta$ TS values for two spectrum 
models that we have considered for this work: 1) the power-law (PL) for the spectral index, $\Gamma=~2.4$ and 2) the 
best-fit DM model corresponds to the highest TS values (for
our case, it is $\rm{100\%~\tau^{+}\tau^{-}}$ final state at $m_{DM}$= 4 GeV). The p-value 
is estimated by assuming the $\chi^{2}$ distribution for the 1 degree of freedom.}.
\label{Tab-3}
\centering
\begin{tabular}{|p{1.5cm}|p{1.5cm}|p{1.5cm}|p{1.5cm}|p{1.5cm}|p{1.5cm}|p{1.2cm}
|p{2cm}|}
\hline 
Our source & TS for PL & $\sigma~(= \sqrt{TS})$ for PL & p-value for PL & TS 
for 
DM & $\sigma~(= \sqrt{TS})$ for DM & p-value for DM & $\Delta$ TS (DM-PL)\\
\hline
Tucana-II & 3.59 & 1.89 & 0.05 & 8.61 & 2.93 & 0.003 & 5.02  \\
\hline
\end{tabular}
\end{table}

\begin{figure}
\centering
\subfigure[]
 { \includegraphics[width=0.48\textwidth, height=0.35\textwidth]{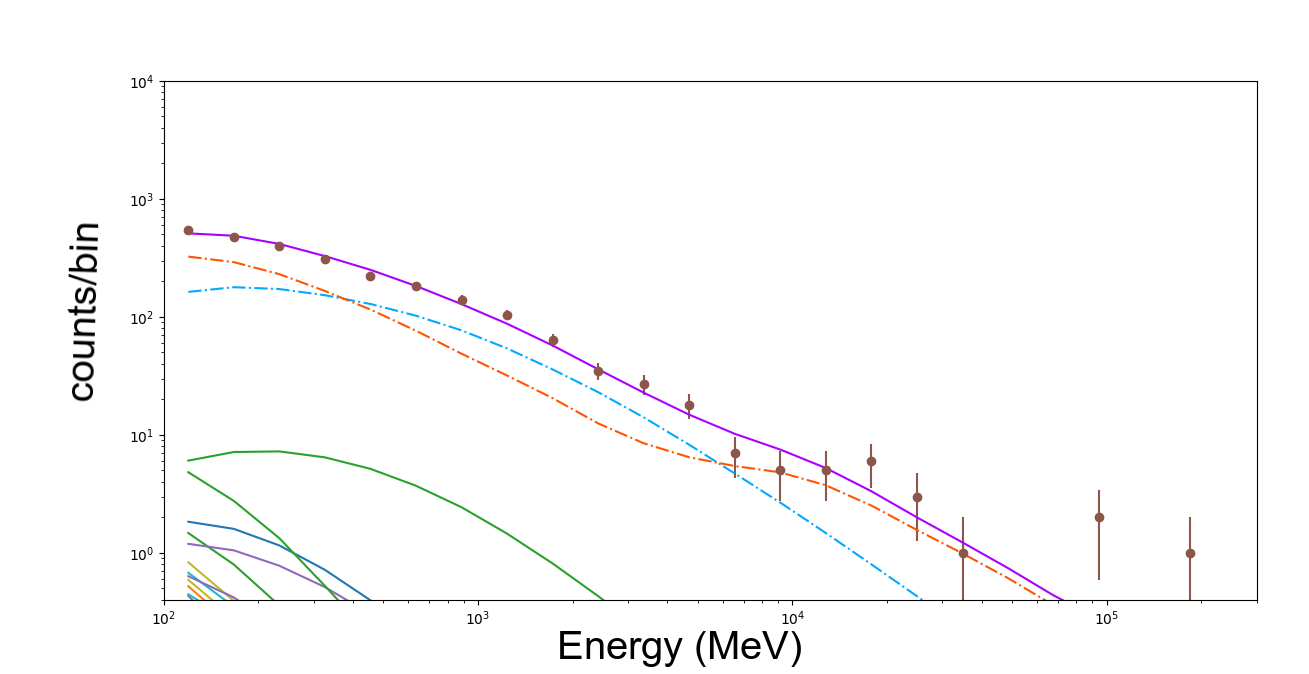}}
\subfigure[]
 { \includegraphics[width=0.48\linewidth, height=0.34\textwidth]{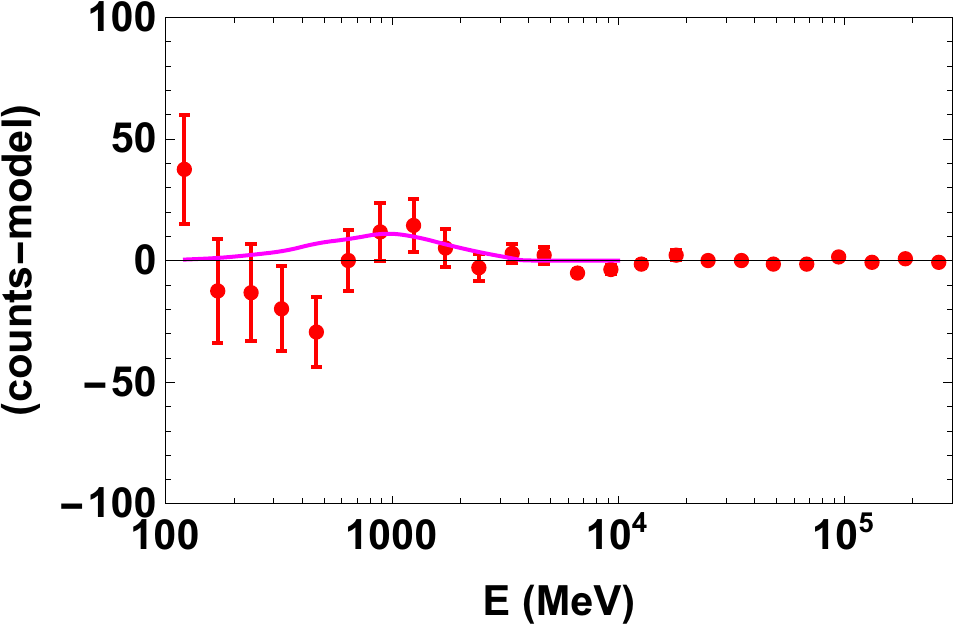}}
\caption{(a) The spectral fit to the observed counts and (b) the residual 
plot for $1^{\circ}$ $\times$ $1^{\circ}$ of ROI centred on the location of
Tuc-II. 
From Fig. 6.3(a), the sum of the best-fit spectrum 
along with the Fermi-LAT detected counts (in brown) is shown with the solid purple curve, while the diffuse galactic and isotropic components are displayed by the `dot-dashed' sky-blue and orange curves. For Fig. 6.3(b), with the magenta solid cure, we have represented the best-fit DM annihilation spectra for 100$\%$ $\tau^{+}\tau^{-}$ channel at DM Mass,$m_{DM}$ = 4 GeV. The corresponding residual points (in red) within 100 MeV to 300 GeV energy ranges are overplotted with error bars.}
\end{figure}

\noindent If we compare between the power-law and the DM annihilation spectra, 
we would
find that the peak value of TS value is significantly improved for DM 
annihilation hypothesis \cite{Bhattacharjee:2018xem}. Besides, the p-value (p-value is defined as the 
probability of getting the signal-like data obtaining from background excess) 
related to local significance is reduced for DM annihilation spectra. We have 
derived the p-value by assuming a $\chi^{2}$ distribution for a degree of 
freedom~=~1 \cite{Bhattacharjee:2018xem}. All the necessary details are displayed in TABLE~6.5 \cite{Bhattacharjee:2018xem}. From this 
table, we can find that excess obtained from the Tuc-II location might favour 
the DM annihilation scenario over its connection with the astrophysical 
phenomenon \cite{Bhattacharjee:2018xem}. But here we also want to mention that for both DM annihilation 
hypothesis and power-law, we have obtained the comparable value of 
-log(Likelihood). 
Thus, we could not firmly rule out the possibility of having an astrophysical 
connection with the faint excess from Tuc-II \cite{Bhattacharjee:2018xem}. Hence, from our analysis, at 
present, we can conclude that our results at best show a hint of a DM signal 
from the location of Tuc-II \cite{Bhattacharjee:2018xem}. For DM annihilation spectra, we have
obtained $\sigma$ = 2.93 and next, we will examine the 
effect of surrounding unresolved sources that have not been detected by the 
Fermi-LAT and will review whether their effects could decrease the local
significance ($\sigma$) for DM annihilation model.\\

\noindent Till now, we have executed the likelihood analysis over 
$10^{\circ}~\times~10^{\circ}$ ROI but it is quite impossible to distinguish 
any 
special features of Tuc-II from that large region of the sky. Hence, in 
Fig.~6.3(a,b), we have displayed the best-fitted spectra and corresponding 
residual plot of Tuc-II for $1^{\circ}~\times~1^{\circ}$ ROI region \cite{Bhattacharjee:2018xem}. For 
obtaining the best-fitting spectra for $1^{\circ}~\times~1^{\circ}$ ROI region, 
we have fixed all the background sources to the best-fitted 
values obtained from the $10^{\circ}~\times~10^{\circ}$ ROI fitting \cite{Bhattacharjee:2018xem}. Now, for 
investigating any interesting signature originating from the location of 
Tuc-II, 
in Fig. 
6.3(a) we have not included Tuc-II in the source model. \\

\noindent From Fig.~6.3(a), we can check the spectral fit per energy bin of all 
the sources within $1^{\circ}~\times~1^{\circ}$ ROI along with the isotropic 
and 
the galactic diffuse background model except for Tuc-II. In Fig.~6.3(b), we 
have 
over-plotted the best-fitted spectra of Tuc-II with a magenta solid line and 
the 
corresponding residual plot
between 100 MeV to 300 GeV energy range are shown with the red points.  \\

\noindent In Fig.~6.3(b), we have considered the best-fitted DM spectra for 
100$\%$ $\tau^{+}\tau^{-}$ annihilation channel at DM mass 4 GeV as this 
channel produces the highest TS peak value of Tuc-II (check Fig.~6.2(a,b)) \cite{Bhattacharjee:2018xem}. 
Now, 
to check the goodness of fitting between the DM annihilation spectra for 
$\tau^{+}\tau^{-}$ annihilation channel and the data derived from the residual 
energy spectrum (Fig.~6.3(b)), we have applied the T-TEST method \cite{Welch:1947df, Ruxton:2006dh}. This method 
is generally favoured for the system which is dealing with the small number of 
events. T-TEST is the test for the statistical hypothesis that examines whether 
there exists any considerable deviation between the means of two samples. Under 
the null hypothesis, T-TEST expects that both the samples are drawn from the 
same populations (Appendix A and B from Chapter 9). For our case, 
with the T-TEST method, we have tried to check whether our selected DM model 
spectrum can provide an acceptable fit to the data obtained from the residual 
energy spectrum (Fig.~6.3(b)) \cite{Bhattacharjee:2018xem}. In Fig.~6.3(b),
we have combined the residuals from all pixels into the energy bins. From this 
figure (Fig.~6.3(b)), we can observe that even for the full energy range (i.e., 
including both the positive bump for energy above 500 MeV and the negative bump for 
energy below 500 MeV) the spectrum for DM annihilation model with 
$\tau^{+}\tau^{-}$ final state can produce an acceptable fit to the residual 
energy spectrum with a p-value of $\approx$ 0.112 (p-value is related to the 
goodness of fitting of the T-Test) \cite{Bhattacharjee:2018xem}. P-value of $>0.05$ implies that we are not 
in a position to reject the assumption for the null hypothesis. Thus, we could 
not reject the idea that the DM annihilation spectrum for its 
$\tau^{+}\tau^{-}$ final state (for both the positive and the negative bumps) 
is 
consistent with the residual spectrum. Besides, if we only focus on the 
positive 
residual energy bump above 500 MeV, we would find that the DM annihilation 
model 
for $\tau^{+}\tau^{-}$ final state provides a fit to the residual energy 
spectrum with a p-value of $\approx$ 0.782 \cite{Bhattacharjee:2018xem}. This positive bump from the 
residual 
energy spectrum indicates an intriguing hint of the DM annihilation signal from 
Tuc-II \cite{Bhattacharjee:2018xem}. Here, we would like to mention that from Fig 6.3(b) the negative bump 
for energy below 500 MeV is nearly the same significant as the positive bump 
for 
energy above 500 MeV and this negative bump at lower energies might be 
connected 
to the poor modelling of the diffuse background templates. The TS peak values 
that we have obtained from the Tuc-II location is much lower than the detection 
threshold limit for
the \textit{Fermi}-LAT. Hence, we could not completely rule out the possibility 
that such excess might come from the statistical fluctuations or it might have 
a connection with some nearby unassociated sources. In our next section, we 
would investigate this in detail.

\subsection{Distribution of the Excess Obtained from $\gamma$-ray 
Spectra of DM Annihilation}
\begin{table}
\caption{The list of CRATES and BZCAT sources within the 1$^{\circ}$ ROI of the
Tuc-II. The J225455-592606 is listed in both catalogs. Thus for this source, we have used its 
CRATES coordinates.}
\label{Tab-3}
\centering
\begin{tabular}{|p{2cm}|p{6cm}|p{5cm}|}
\hline 
Our source & Nearby sources from BZCAT and CRATES catalog & Distance to the 
Tuc-II ($^{\circ}$)\\
\hline
Tucana-II & J~225134-580103 & 0.55 \\
\hline
& J~225008-591029 & 0.66 \\
\hline
& J~225455-592606 & 0.95 \\
\hline
\end{tabular}
\end{table}

\begin{table}
\caption{The TS values for Tuc-II, 4FGL 2247.7-5857, and three sources from the BZCAT and the CRATES catalog that lie within $1^{\circ}$ of Tuc-II are mentioned. For 
Tuc-II, we have shown its TS peak value for $100\%~\tau^{+}\tau^{-}$ 
annihilation channel at $m_{DM}$=4 GeV. The three nearby CRATES 
sources are modelled with the power-law spectra for 
$\Gamma=2.2$. In case of 4FGL 2247.7-5857, we have modelled it with
power-law spectra and have used the parameter values the 4FGL catalog of Fermi-LAT.}
\begin{tabular}{|p{0.8cm}|p{1.2cm}|p{1.2cm}|p{1.2cm}|p{1.2cm}|p{1.2cm}|p{3cm}
|p{2.5cm}|}
\hline 
Year & Tuc-II from by $\Delta~TS$ method &  TS value of J225134-580103 & TS 
value of 
J225008-591029 & TS value of J225455-592606 & TS value of 4FGL 2247.7-5857 & TS 
value of Tuc-II after including three CRATES sources and 4FGL 2247.7-5857 to 
source model &  Rescaled TS value of Tuc-II due to all possible background 
fluctuation.\\
\hline
3 & 3.0868 & 0.05 & 0.027 & 0.49 & 5.61 & 3.04 & $\approx$ 1.7167\\
\hline
6 & 6.8802 & 0.66 & 1.22 & 0.98  & 10.45 & 5.24 & $\approx$ 3.8265\\
\hline
9 & 8.61  & 2.043 & 3.82 & 2.01 & 21.67 & 7.05 & $\approx$ 4.7885\\
\hline
\end{tabular}
\end{table}

\noindent In subsections 6.2.1 and 6.4.1, we have determined the TS value for 
Tuc-II but we have not examined whether there is any nearby background 
fluctuation. Such surrounding fluctuation can influence the 
significance that we have earlier obtained for 
Tuc-II. Apart from this, we have obtained a very faint emission from the 
location of Tuc-II (i.e., TS value of 8.61) \cite{Bhattacharjee:2018xem}. Thus, before claiming its 
connection with the spectrum resulting from DM annihilation, next, we would try 
to carefully examine
the origin and the reliability of such faint excess \cite{Bhattacharjee:2018xem}. \\

\noindent There is a fair chance that the excess that we obtained from Tuc-II 
location
could be the result of either any surrounding unresolved sources or the 
deficiency of background models \cite{Bhattacharjee:2018xem}. Carlson \textit{et al.}, 
2015~\cite{Carlson:2014nra} have suggested 
that such faint $\gamma$-ray emission from dSphs can plausibly come from 
several 
nearby unresolved faint $\gamma$-ray sources such as radio galaxies 
\cite{Inoue:2011bm}, blazars 
\cite{Abdo:2010gqa}, star-forming galaxies 
\cite{Ackermann:2012vca, Linden:2016fdd} and millisecond pulsars 
\cite{Hooper:2013nhl}. Among these, 
blazars are the most responsible candidates for such background fluctuations 
\cite{Carlson:2014nra}. At high-latitude, blazars are the most numerous point 
sources 
and thus they are assumed to be the prime source of anisotropy in extragalactic 
gamma-ray sky \cite{Ackermann:2012uf, Abazajian:2010pc, Venters:2010bq, 
Venters:2011gg, Cuoco:2012yf, 
Harding:2012gk}. A non-negligible amount of 
$\gamma$-ray emission can also arise from the star-forming and the radio 
galaxies.\\

\noindent Motivated by the work of Carlson \textit{et al.}, 
2015~\cite{Carlson:2014nra}, in this section, we have performed a detailed 
analysis to examine the possible reason for obtaining a faint excess from the 
location of Tuc-II. For our purpose, we have used two multiwavelength catalogs 
for blazar such as CRATES 
\cite{Healey:2007by} and BZCAT \cite{Massaro:2008ye}. BZCAT catalog consists of 
nearly 3149 blazars and among them, 2274 are located at high galactic latitude 
i.e., $|b|>30^{\circ}$. CRATES catalog contains
nearly 11,000 bright flat-spectrum radio sources. Within $1^{\circ}$ ROI of 
Tuc-II, we have observed one blazar 
from the BZCAT catalog and three radio sources from the CRATES catalog \cite{Bhattacharjee:2018xem}. The 
source that is included in the BZCAT catalog has also been detected by the 
CRATES catalog. For our examination, we have considered three CRATES sources 
such as J225134-580103, J225008-591029, and J225455-592606. All three sources 
are located within a 
1$^{\circ}$ ROI of Tuc-II \cite{Bhattacharjee:2018xem}. We have not considered any other radio sources 
beyond 
1$^{\circ}$ because any source beyond 
1$^{\circ}$ might would not produce any significant changes to the local 
emission of dSphs \cite{Carlson:2014nra}. In Table~6.6, we have mentioned the 
list of CRATES sources within 
1$^{\circ}$ ROI of Tuc-II \cite{Bhattacharjee:2018xem}.\\

\noindent Inspired by ref.~Carlson et al.\cite{Carlson:2014nra}, we have modelled all three 
radio sources with the power-law spectrum of the index ($\Gamma$)=2.2 and then 
have derived the TS values of these three radio sources for different time 
periods of Fermi-LAT dataset\cite{Bhattacharjee:2018xem}. In Table~6.7, we have mentioned our result. After 
the inclusion of these three sources, we have observed that the significance of 
Tuc-II is only reduced by $\sim$ 10$\%$ \cite{Bhattacharjee:2018xem}. Here we would like to mention that 
Carlson 
\textit{et al.}, 2015~\cite{Carlson:2014nra}, has also observed the same 
reductions. They have 
again concluded that the blazars are responsible for only 10$\%$ of local 
TS value of the source and the large part of the excess from dSphs is not
related to the nearby radio sources.\\

\noindent To investigate the distribution of local excess obtained from the 
location of Tuc-II, we have generated the $2^{\circ}$ x $2^{\circ}$ residual TS 
map around Tuc-II with \textit{`gttsmap'} for energy range between 100 MeV to 
300 GeV \cite{Bhattacharjee:2018xem}. During this process, the spectral parameters of all the sources within 
$10^{\circ}$ ROI were kept fixed to their values obtained from their fittings 
performed on nine years of Fermi-LAT data \cite{Bhattacharjee:2018xem}. But the normalization values for the 
galactic and the isotropic models were left free. We have generated the TS map 
for three 
cases \cite{Bhattacharjee:2018xem}: 1) Fig.~6.4 (extreme left) Tuc-II and three sources from BZCAT and 
CRATES catalog that lies within a $1^{\circ}$ x $1^{\circ}$ ROI of Tuc-II were 
not included to the source model, 
2) Fig.~6.4 (middle); The three radio sources from BZCAT and CRATES catalog 
that lie within a $1^{\circ}$ x $1^{\circ}$ ROI of Tuc-II were included to 
source model but Tuc-II were not included, 3)
Fig.~6.4 (extreme right); Tuc-II and three sources from BZCAT and 
CRATES catalog that lies within a $1^{\circ}$ x $1^{\circ}$ ROI of Tuc-II were 
included to the source model. For generating the residual TS map (for right 
image of Fig.~6.4), we have taken the best-fitted parameters of Tuc-II obtained 
from its DM annihilation spectra for $100\%~\tau^{+}\tau^{-}$ annihilation 
channel at $m_{DM}$ = 4 GeV.

\begin{figure}
\centering
 { \includegraphics[width=1.0\linewidth]{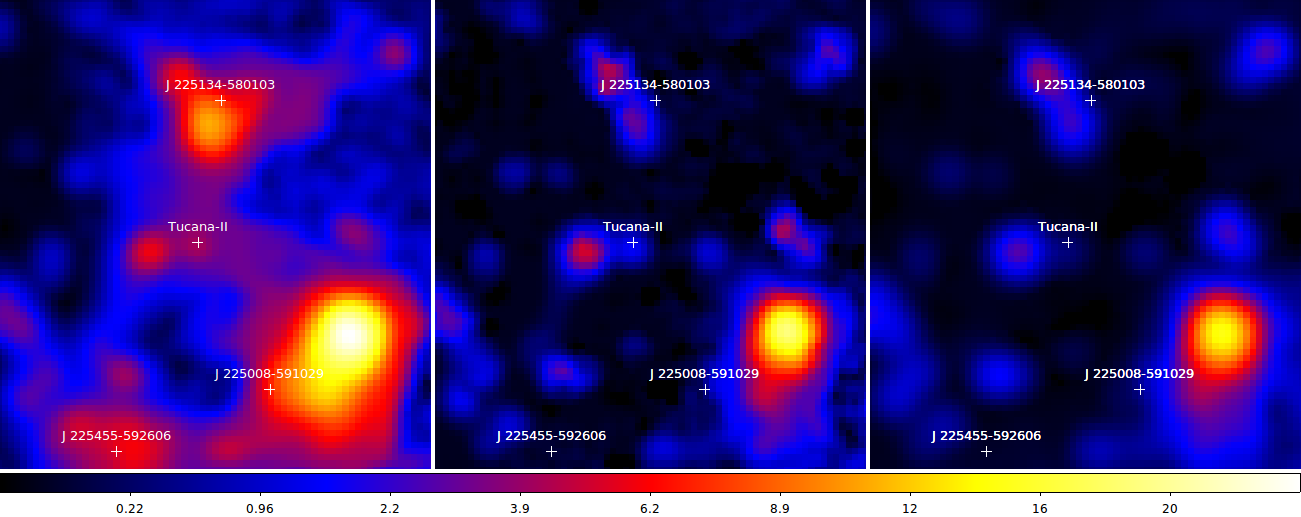}}
\caption{The residual TS maps (between 100 MeV to 300 GeV) for $2^{\circ}~\times~2^{\circ}$ 
ROI centred at Tuc-II extracted from the $10^{\circ}$ ROI. The image scale for TS map is 0.025 $pixel^{-1}$. In 
the left Fig., the three CRATES sources and Tuc-II are not added to our source model; For middle Fig., Tuc-II is not added to the source model but we have included three CRATES sources in our source model; in right Fig., the three CRATES sources and Tuc-II are 
included to our source model. We have mentioned the name of Tuc-II and three CRATES sources that lie within $1^{\circ}$ with the white cross.} 
\end{figure}

\noindent From Fig.~6.4 (extreme left, middle), we can observe a hint of a 
localized-emission of TS value $\approx$ 6.5 and that region is very close to 
the location of Tuc-II \cite{Bhattacharjee:2018xem}. This is true that the region of faint emission is not 
directly localized to the position of Tuc-II, 
but that is mere $0.18^{\circ}$ away from the position of Tuc-II \cite{Bhattacharjee:2018xem}. 
Interestingly, from Fig.~6.4(right), we can find that just after including 
three radio sources from CRATES catalog and Tuc-II to the source model, the 
significance of the nearby localized-excess is considerably reduced \cite{Bhattacharjee:2018xem}. Thus, from 
Fig.~6.4, we can conclude that there is a fair possibility that Tuc-II is 
associated with the nearby localized emission \cite{Bhattacharjee:2018xem}.\\

\noindent Apart from that nearby localized-excess region of Tuc-II, from Fig. 
6.4, we can notice a very bright emission of TS value $\approx$5$\sigma$ at the 
bottom of the right corner and after investigating the above three TS maps, we 
can safely state that such bright excess is not associated with Tuc-II \cite{Bhattacharjee:2018xem}. Thus, 
we 
have checked 4FGL 
catalog of Fermi-LAT (\cite{Fermi-LAT:2019yla}) and have noticed that source,  
4FGL 2247.7-5857 is exactly overlapping with that bright region \cite{Bhattacharjee:2018xem}. Next, we have 
produced the residual TS map of $2^{\circ}~\times~2^{\circ}$ for four 
cases (see Fig. 6.5) \cite{Bhattacharjee:2018xem}; First and second residual TS maps of Fig. 6.5 
are same 
as first two TS maps of Fig. 6.4. For the third TS map of Fig. 6.5, the three 
radio sources from BZCAT and CRATES catalog 
that lie within a $1^{\circ}$ x $1^{\circ}$ ROI of Tuc-II and 4FGL 2247.7-5857 
were included to source model but Tuc-II were not included. For the last TS 
map 
(extreme right of Fig.~6.5); 4FGL 2247.7-5857, Tuc-II and three sources from 
BZCAT 
and 
CRATES catalog that lies within a $1^{\circ}$ x $1^{\circ}$ ROI of Tuc-II were 
included to the source model.

\begin{figure}
\centering
{ \includegraphics[width=1.0\linewidth]{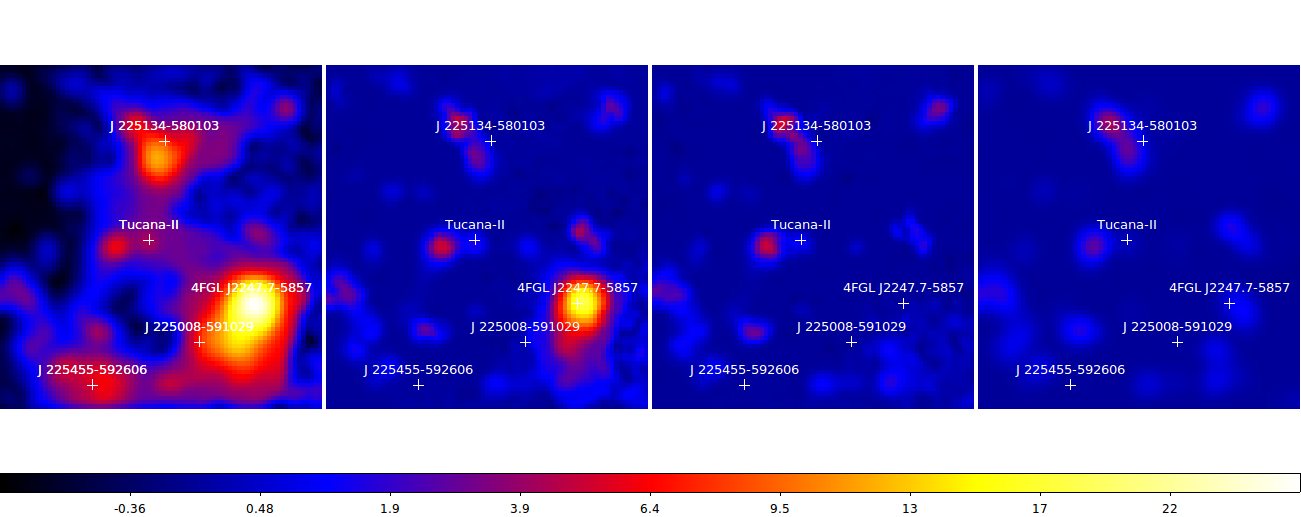}}
\caption{The residual TS maps (between 100 MeV to 300 GeV) for $2^{\circ}~\times~2^{\circ}$ 
ROI centred at Tuc-II extracted from the $10^{\circ}$ ROI. The image scale for TS map is 0.025 $pixel^{-1}$. In 
the extreme left Fig., the three CRATES sources, 4FGL 2247.7-5857 and Tuc-II are not added to our source model; For Second left Fig., Tuc-II and 4FGL 2247.7-5857 are not added to the source model but we have included three CRATES sources in our source model; for Third Fig., Tuc-II in not added to the source model but we have included 4FGL 2247.7-5857 and three CRATES sources in our source model; in extreme right Fig., the three CRATES sources, 4FGL 2247.7-5857 and Tuc-II are 
included to our source model. We have mentioned the name of Tuc-II, 4FGL 2247.7-5857 and three CRATES sources that lie within $1^{\circ}$ with the white cross.} 
\end{figure}

\noindent Now if we check the extreme right image of Fig. 6.5, we can observe 
that after inclusion of 4FGL 2247.7-5857 to source model, the emission from 
that 
bright region at the bottom of the right corner is greatly decreased. Hence, 
this result shows that the bright excess from residual TS map has an 
astrophysical connection and primarily originates from the source 4FGL 
2247.7-5857 \cite{Bhattacharjee:2018xem}.\\

\noindent From our analysis, we would like to mention that, even after 
including 
4FGL 2247.7-5857 and three radio sources from CRATES catalog, from Fig.~6.4 and 
Fig.~6.5 we can still detect plenty of delocalized excesses \cite{Bhattacharjee:2018xem}. The deficiency in 
background models for Fermi-LAT can be the reason for leakage \cite{Bhattacharjee:2018xem}. There is also a 
possibility 
that these delocalized excess regions are associated with some unresolved 
astrophysical 
sources as well as the DM 
subhalos can also be linked with such emissions. There are some studies which 
argue that even if we try to accurately model all the astrophysical sources to an extent, the DM subhalos will still be accountable for an 
irreducible background, say$\approx5\%-10\%$, for the gamma-ray sky 
\cite{Carlson:2014nra, 
Ackermann:2012uf, Lee:2008fm, SiegalGaskins:2009ux}. But with detailed 
multiwavelength study, we can positively reduce the contamination from 
most of the unresolved sources in our blank sky.\\

\noindent In this work, for calculating the TS value, we have considered the 
background models provided by Fermi-LAT and not the blank sky. Hence, 
there is a high chance that we have overestimated the significance of the 
source 
even after including all possible nearby sources to our source model \cite{Bhattacharjee:2018xem}. There are 
several works by Fermi collaboration 
\cite{Ackermann:2013yva, Fermi-LAT:2016uux} which have reported that from a 
vast 
region of the blank sky, we might observe an excess of TS $>$ 8.7. Such 
emission 
would decrease the source
significance from 2.95$\sigma$ to 2.2$\sigma$ \cite{Ackermann:2013yva, 
Fermi-LAT:2016uux}. 
Following this prescription \cite{Ackermann:2013yva, Fermi-LAT:2016uux}, we 
have 
also re-calibrated 
the TS value estimation and this effect reduces the TS value of Tuc-II from 
8.61 
to 4.79 i.e., p-value from 
0.003 to 0.029 \cite{Bhattacharjee:2018xem}. All our obtained results are mentioned in Table~6.7 \cite{Bhattacharjee:2018xem}. In 
column 2, we have shown the TS value from $\Delta TS$ method; in 
columns 3, 4, 5 and 6, we have given the TS value of all three radio sources 
from CRATES catalog and 4FGL 
2247.7-5857; in column 7, we have provided the revised TS value of 
Tuc-II after including 4FGL 2247.7-5857 and three radio sources from CRATES 
catalog to the source model; 
and in column 8, we have shown the re-scalled TS value of Tuc-II by considering 
all probable background fluctuations.

\subsection{Possible DM Annihilation Constraint on Theoretical DM 
Models with 9 Years of Tuc-II \textit{Fermi}-LAT Data}

\noindent In our earlier section, we have already discussed that peak value of TS
for $\tau^{+}\tau^{-}$ annihilation channel is lower than the detection 
threshold limit of Fermi-LAT
(i.e., TS~$<$~25). Thus, in this section, we would estimate 
$\gamma$-ray flux upper limit in $95~\%$ C.L. for Tuc-II by employing the 
$\gamma$-ray spectrum from DM annihilation. For this purpose, we have used the 
semi-Bayesian method~\cite{Helene:1990yi}, as described in section 6.2.1. With DMFIt 
Function, 
we have also
determined the upper limits to the $<\sigma 
v>$ as a function of DM mass ($\rm{m_{DM}}$), for five pair-annihilation final 
states \cite{Jungman:1995df}. 
Like Tri-II, in this analysis, we have again considered these five 
supersymmetry-favoured pair annihilation final states \cite{Jungman:1995df}, 
such as
$100\%$ $b\bar{b}$, $80\%$ $b\bar{b}$+$20\%$ $\tau^{+}\tau^{-}$, $100\%$ 
$\tau^{+}\tau^{-}$, $100\%$ $\rm{\mu^{+} \mu^{-}}$ and $100\%$ $W^{+}W^{-}$, 
respectively. In Fig.~6.6(a,b), we have shown the variation of $\gamma$-ray 
flux 
upper limits of Tuc-II in 95 $\%$ C.L. and the relative upper limits to 
$<\sigma 
v>$ as a function of $\rm{m_{DM}}$ and annihilation final states \cite{Bhattacharjee:2018xem}.

\begin{figure}
\centering
\subfigure[]
{\includegraphics[width=0.49\textwidth,clip,angle=0]{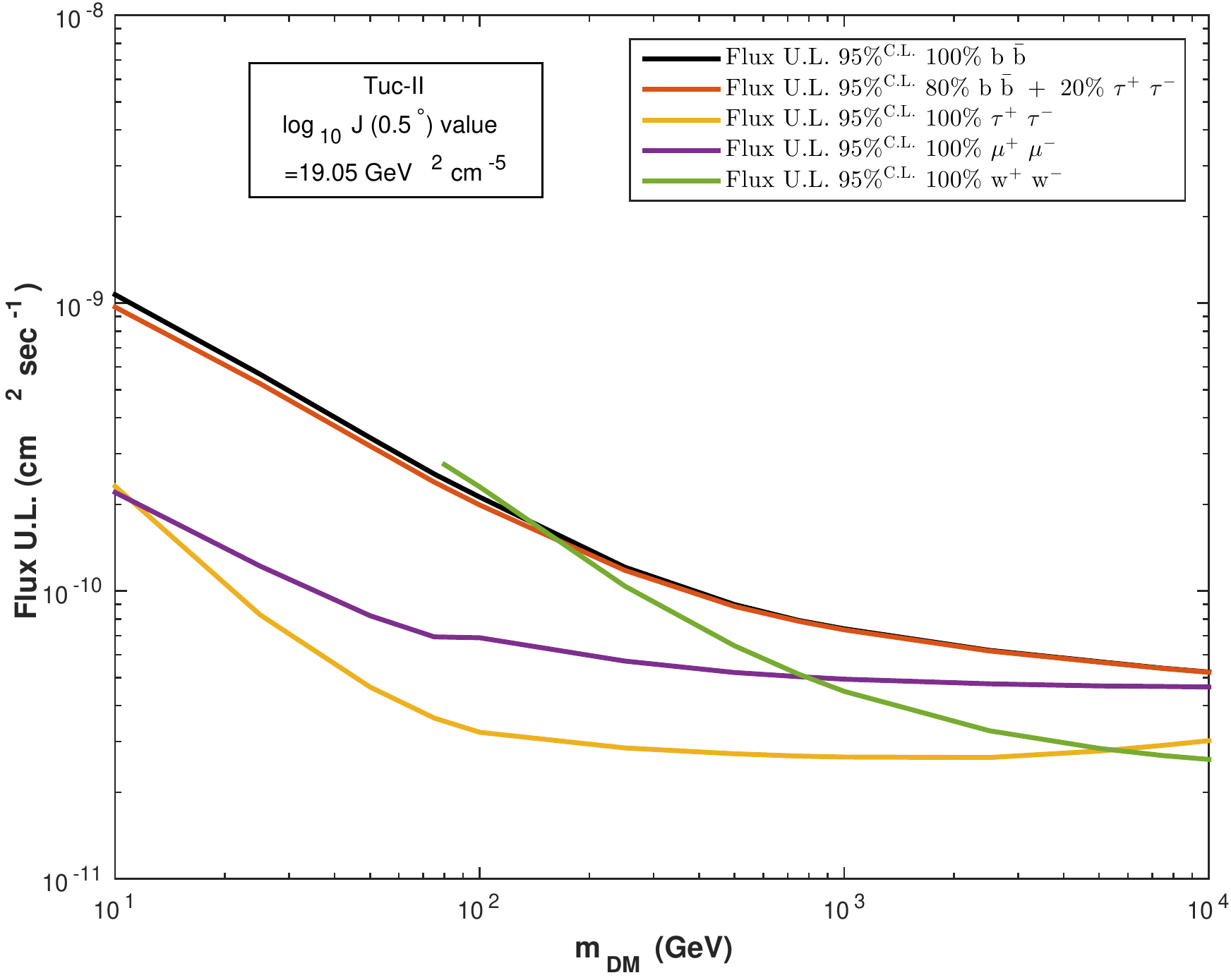}}
\subfigure[]
{\includegraphics[width=0.49\linewidth]{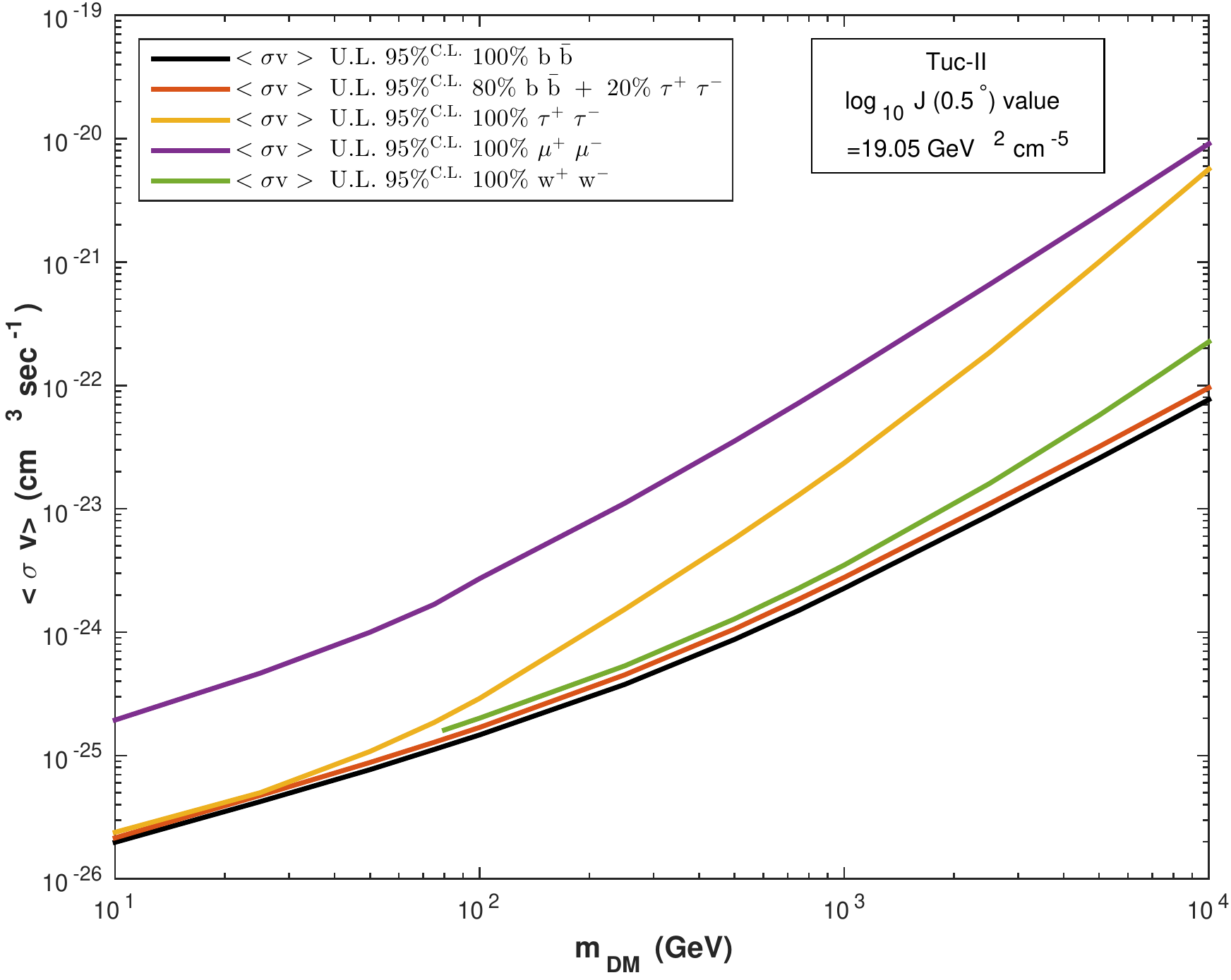}}
\caption{The variations of (a) $\gamma$-ray flux upper 
limits and (b) the respective WIMP pair annihilation $<\sigma~v>$ in $95\%$ C.L. with DM mass, $\rm{m_{DM}}$ estimated for five 
annihilation channels, ``f". The results are produced by considering the median value of $\rm{J(0.5^{\circ})}$-factor value for 
Tuc-II.}
\end{figure}

\begin{figure}
\centering
\subfigure[]
{\includegraphics[width=0.49\linewidth]{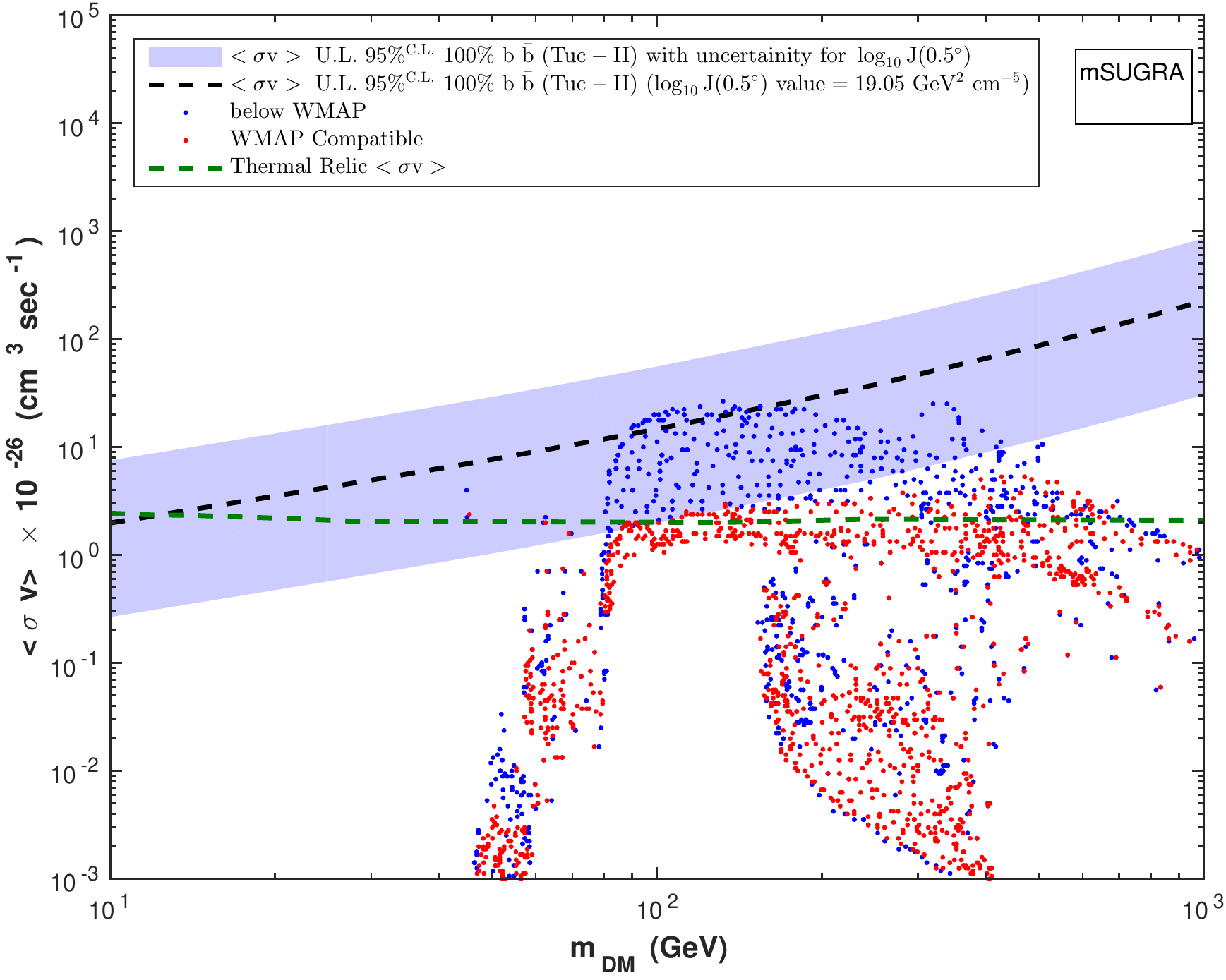}}
\subfigure[]
{\includegraphics[width=0.49\linewidth]{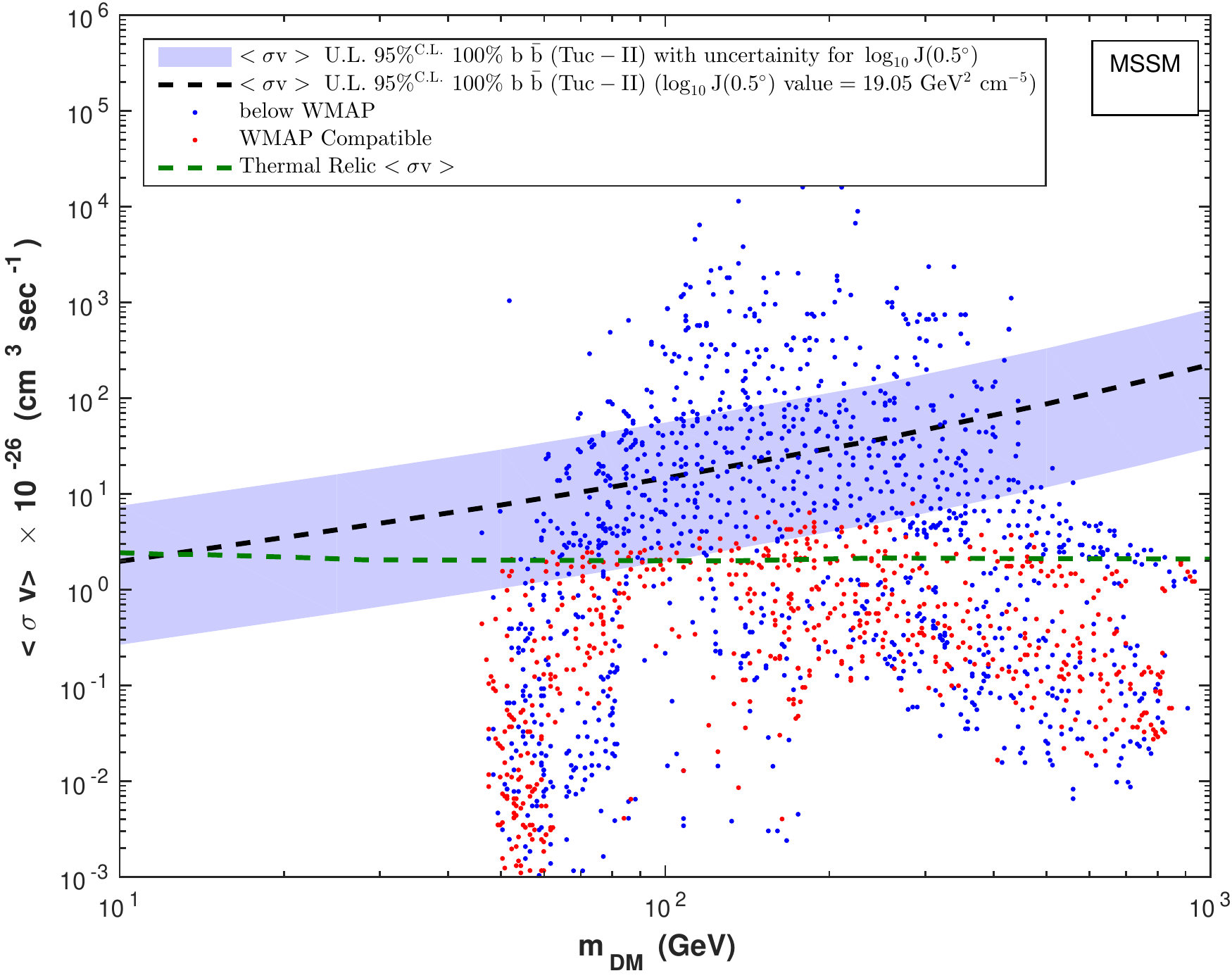}}
\caption{The variation of $<\sigma v>$ upper limits of Tuc-II with $\rm{m_{DM}}$ for $b\bar{b}$ annihilation channels is shown in the parameter plane of ($\rm{m_{DM},<\sigma v>}$) for the median value of J-factor 
with its associated uncertainties. The shaded region denotes the uncertainty associated with the DM profiles for Tuc-II. The $<\sigma~v>$ limits obtained from Tuc-II are compared with the limits predicted by (a) the mSUGRA and (b) the MSSM 
DM models. In both (a) and (b), the red points are related to the thermal relic DM density, while the blue points correspond to the higher $<\sigma~v>$ and low thermal relic DM density. The thermal-relic cross section rate ($\rm{2.2\times10^{-26}~cm^{3}~s^{-1}}$) estimated by the Steigman \textit{et al.}, 2012 is displayed by a green dashed line.}
\end{figure} 

\begin{figure}
\centering
{\includegraphics[width=0.5\linewidth]{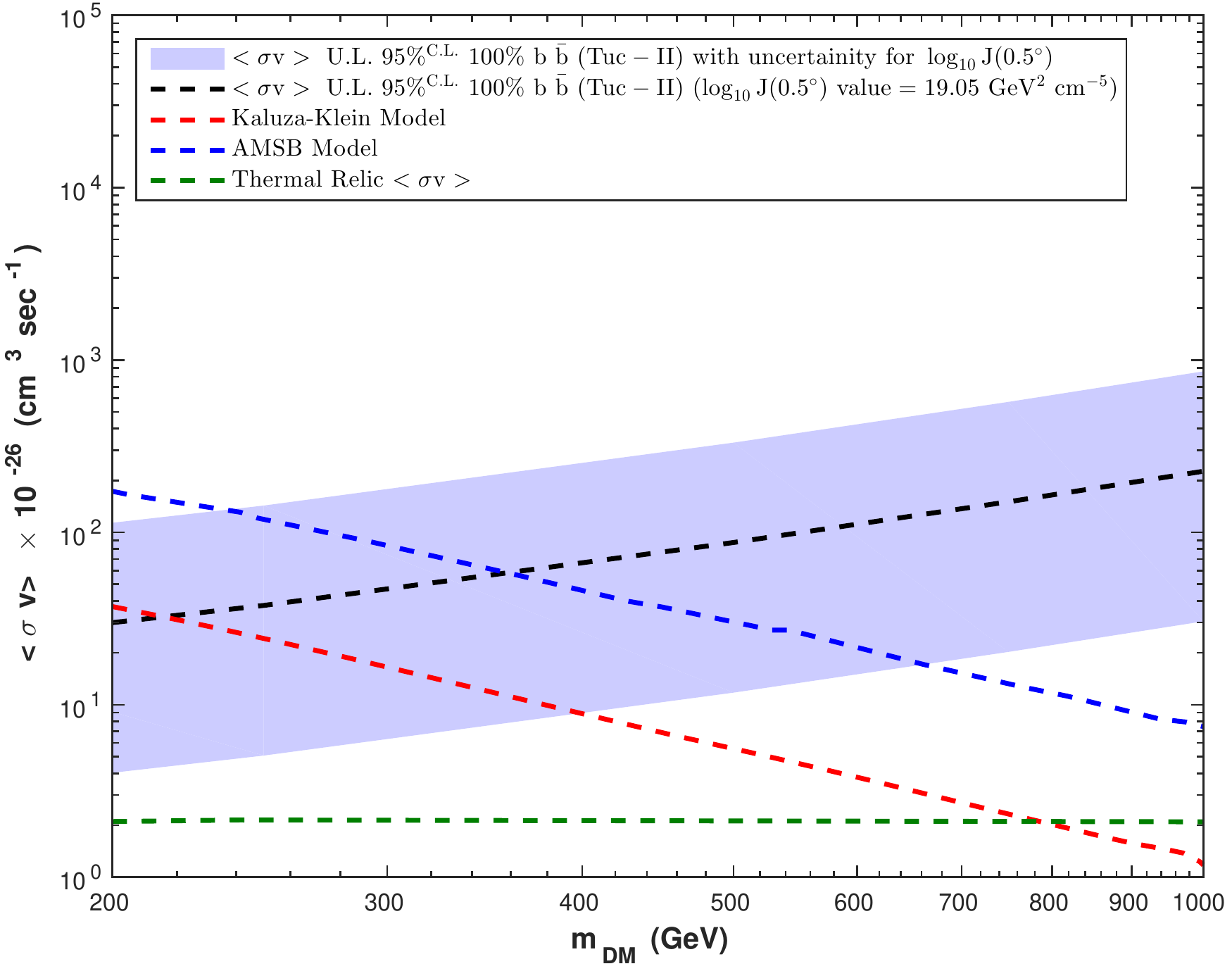}}
\caption{The comparison of the $<\sigma~v>$ upper limits obtained from the Tuc-II with the $<\sigma~v>$ limits predicted by the AMSB and the Kaluza-Klein UED DM models is displayed in this figure. The shaded region denotes the uncertainty associated with the DM profiles for Tuc-II. The thermal-relic cross section rate ($\rm{2.2\times10^{-26}~cm^{3}~s^{-1}}$) estimated by the Steigman \textit{et al.}, 2012 is displayed by a green dashed line.}
\label{fig:fig}
\end{figure}

\noindent In this work, we have also tried to check whether Tuc-II can impose 
any strong constraint on theoretically favored DM models \cite{Bhattacharjee:2018xem} and for that purpose, 
we have again considered the mSugra~\cite{Chamseddine:1982jx} model, the 
MSSM~\cite{Chung:2003fi}, the Kaluza-Klein in UED~\cite{Cheng:2002ej, 
Servant:2002aq, Hooper:2007qk} and the AMSB model~\cite{Giudice:1998xp}, 
respectively. In Figs.6.7~(a,b) and 6.8, we have shown $<\sigma v>$ upper limits of 
Tuc-II for $100\%$ b$\rm{\bar{b}}$ annihilation channel, as a function of 
$\rm{m_{DM}}$, 
for its median J value and uncertainties in J-factor \cite{Evans:2016xwx}. 
Here, 
we have only considered the $100\%$ b$\rm{\bar{b}}$ annihilation channel 
because 
for $\gamma$-ray analysis, this channel provides the most stringent limits on 
theoretical model \cite{Bhattacharjee:2018xem}. 
In Figs.~6.7(a,b) and 6.8, we have denoted the relic thermal cross 
section rate derived by Steigman et. al.~\cite{Steigman:2012nb} with
a horizontal dashed green line. These results are 
then compared with the $<\sigma~v>$ values obtained from the mSugra (in 
Fig.~6.7(a))~\cite{Chamseddine:1982jx} model, the MSSM (in 
Fig.~6.7(b))~\cite{Chung:2003fi}, the Kaluza-Klein in UED 
(Fig.~6.8)~\cite{Cheng:2002ej, Servant:2002aq, Hooper:2007qk} and the AMSB 
model 
(Fig.~6.8)~\cite{Giudice:1998xp}, respectively.\\

\noindent From Figs. 6.7~(a,b) and 6.8, we can immediately observe that for its 
lowest 
limit of the shaded band, Tuc-II could provide a very strong limit on the 
parameter space of all four theoretical DM models \cite{Bhattacharjee:2018xem}. From Figs.~6.7(a,b), it is 
very encouraging to mention that, even for the median of 
J($0.5^{\circ}$)-factor 
of Tuc-II (i.e., 
$\rm{\log_{10} J(0.5^{\circ})}$=19.05~$\rm{GeV^{2}~cm^{-5}}$), the upper 
limits of $<\sigma~v>$ can significantly constrain the blue points in both the 
MSSM and 
the mSUGRA model, while the uncertainty band of J-factor of Tuc-II have already 
started 
to limit the red points for both the models \cite{Bhattacharjee:2018xem}. From Fig.~6.8, it is interesting 
to 
note that the $<\sigma v>$ upper 
limit from Tuc-II for the median value of J($0.5^{\circ}$)-factor (i.e., 
$\rm{\log_{10} J(0.5^{\circ})}$=19.05~$\rm{GeV^{2}~cm^{-5}}$), disfavors the 
Kaluza-Klein in UED model and the AMSB model for masses 
$\approx<220$~GeV and $\approx<400$~GeV, respectively \cite{Bhattacharjee:2018xem}.\\

\noindent For Tuc-II, the insufficient kinematics data is the main reason 
behind 
its large uncertainties in J-factor. But, in future, with more detailed 
observation of the structure of Tuc-II, we should positively reduce such large 
uncertainty band 
to a single upper limit curve for $<\sigma~v>$ and that would definitely 
improve 
the $<\sigma~v>$ limit on beyond SM \cite{Bhattacharjee:2018xem}.

\subsection{Comparison of the Constraints on the DM Annihilation 
Cross-section ($b\bar{b}$ Channel) Obtained from Tuc-II, Ret-II and UMi}

\noindent In this section, we have introduced two newly discovered dSphs, 
Ret-II 
and UMi. In Fig.~6.9, we have shown the comparison between the Tuc-II, Ret-II 
and UMi in space of ($<\sigma v>$, $m_{DM}$) and for this comparison, we have 
again chosen the $b\bar{b}$ annihilation channel \cite{Bhattacharjee:2018xem}.  
For obtaining the $<\sigma~v>$ upper limit in 95$\%$ C.L. of Ret-II and UMi, we 
have analysed the nine years of Fermi-LAT and followed the same method that we 
have used for Tuc-II (check Table~6.2).\\

\noindent In Fig.~6.9, the median value of 
J-factor is denoted by the dashed lines, while the shaded band represents the 
range of uncertainties in J-factor 
for all three UFDs \cite{Bhattacharjee:2018xem}. In case of newly discovered UFDs, a very few numbers of 
member stars have been observed that leads to the main difficulties in 
understanding the DM distribution in UFDs. The large uncertainty bands of UFDs 
actually represent our insufficient knowledge of their internal structures. 

\begin{figure}
\begin{center}
\subfigure[]
{\includegraphics[width=0.5\textwidth,clip,angle=0]{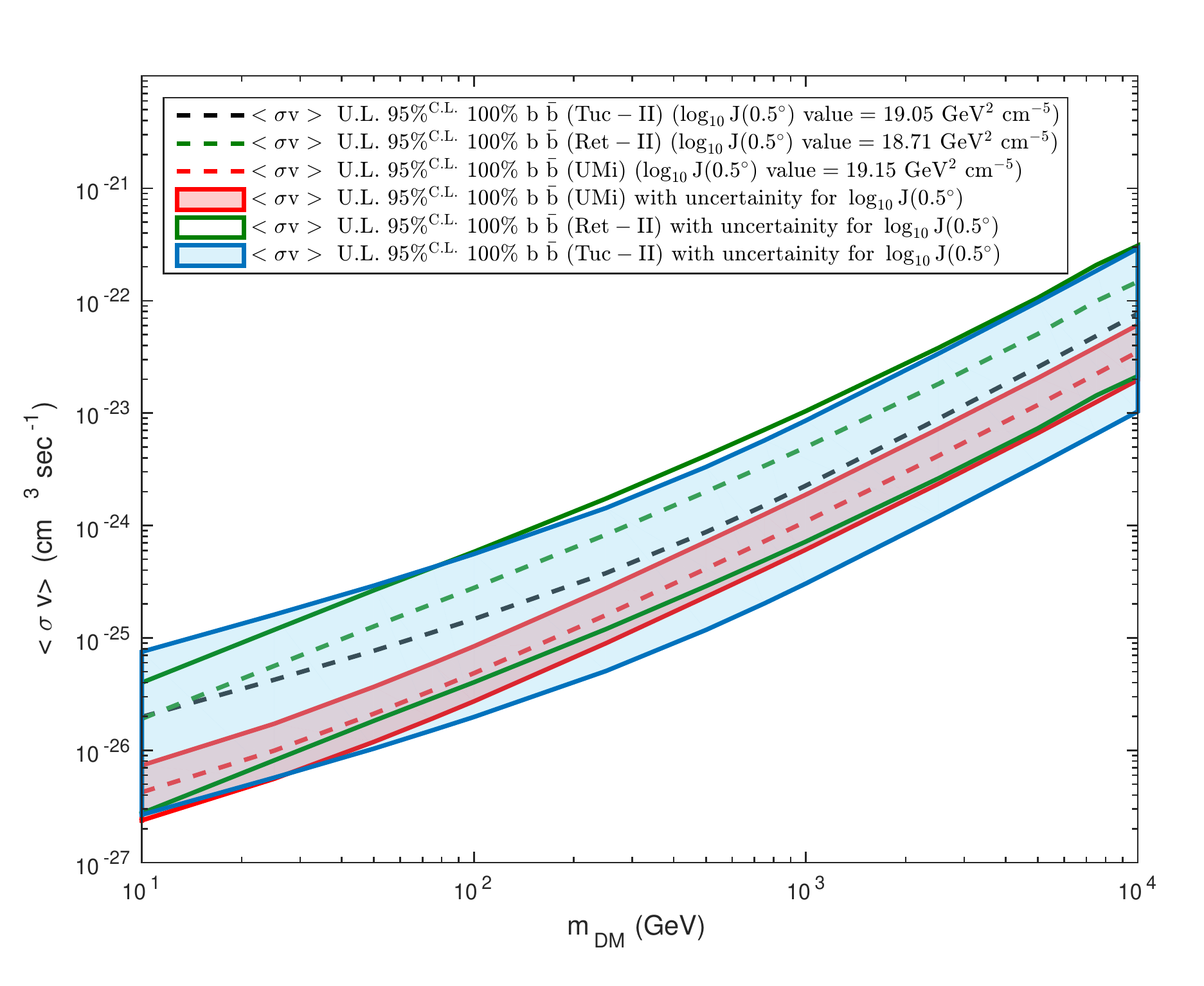}}
\caption{The variations of $<\sigma v>$ with $\rm{m_{DM}}$ for the $b\bar{b}$ annihilation channel of Tuc-II, UMi and Ret-II is shown in the parameter plane of ($\rm{m_{DM},<\sigma v>}$). The 
shaded region denotes the uncertainty associated with the DM profiles of UFDs, while the dashed line represents the $<\sigma v>$ upper limits in 95 $\%$ C.L. for their median value of J-factor.}
\end{center}
\end{figure}

\noindent From Fig.~6.9, we can notice that compared to UMi and Ret-II, Tuc-II 
shows larger uncertainty in DM 
density profile \cite{Bhattacharjee:2018xem}. We can also observe an overlapping region between the 
uncertainties band of the Ret-II, Tuc-II and UMi in parameter space of 
($<\sigma~v>$, 
$m_{DM}$). So, from this scenario, we could not favour 
Tuc-II over other two UFDs \cite{Bhattacharjee:2018xem}. But from Fig.~6.9, it is also important to note 
that 
above 
$m_{DM}~\sim~100~GeV$, Tuc-II has provided a better constraint on 
($<\sigma~v>$, 
$m_{DM}$) space than Ret-II for its median value of J($0.5^{\circ}$)-factor \cite{Bhattacharjee:2018xem}.

\subsection{Comparative Study between the Limits Obtained from Tuc-II and the Limits Obtained from 
Several Collaboration Works on dSphs/UFDs}
\noindent Here, we have performed a comparative study between the upper limits 
of 
$<\sigma~v>$ obtained from Tuc-II and the $<\sigma~v>$ limits obtained from 
several collaboration works on dSphs/UFDs and the related plot is shown in 
Fig.~6.10 \cite{Bhattacharjee:2018xem}. For comparison, we have included the
results from the combined analysis~\cite{Ackermann:2015zua} of 15 dSphs with 
six 
years of 
\textit{Fermi}-LAT data, the results obtained by the High 
Energy Stereoscopic System (H.E.S.S.) telescope from a combined analysis of $5$ 
dSphs~\cite{Abramowski:2014tra}, the results obtained by the High Altitude 
Water 
Cherenkov (HAWC) gamma-ray observatory from the combined analysis of $15$ 
dSphs~\cite{Proper:2015xya}, the results obtained by the Very Energetic 
Radiation Imaging Telescope Array System (VERITAS) from $4$ 
dSphs~\cite{Archambault:2017wyh}, the results obtained by the Major Atmospheric 
Gamma-ray Imaging Cherenkov Telescopes (MAGIC) \cite{Ahnen:2016qkx} for 
Segue-I, as well as the results obtained for Segue-I by the combined analysis 
from the \textit{Fermi}+the MAGIC~\cite{Ahnen:2016qkx} 
collaboration. From Fig.~6.10, it is evident that among all observational 
results, the combined \textit{Fermi} $\&$ MAGIC analysis for Segue-I imposes 
the 
best limit on the WIMP pair-annihilation $<\sigma~v>$ for a very wide range of 
DM masses. The combined limits obtained from 15 dSph performed by the 
\textit{Fermi}-LAT collaboration also provides a strong constraint up to around 
DM mass $1$~TeV, 
and beyond that DM mass, because of the low statistics, Fermi-LAT could not 
perform well. It is also interesting to mention that the $<\sigma~v>$ 
upper-limits obtained from both HAWC and \textit{Fermi}+MAGIC collaboration 
tend 
to converge for the mass range 
$\approx 100$~TeV and that signature indicates that they are competitive in 
place of searching the DM signal from
dSphs/UFDs. Thus, from Fig.~6.10, we can conclude that the combined data are 
taken 
from several ground and space-based $\gamma$-ray telescopes can improve the 
present limits of WIMP annihilation $<\sigma~v>$.
\begin{figure}
\centering
{\includegraphics[width=0.6\linewidth]{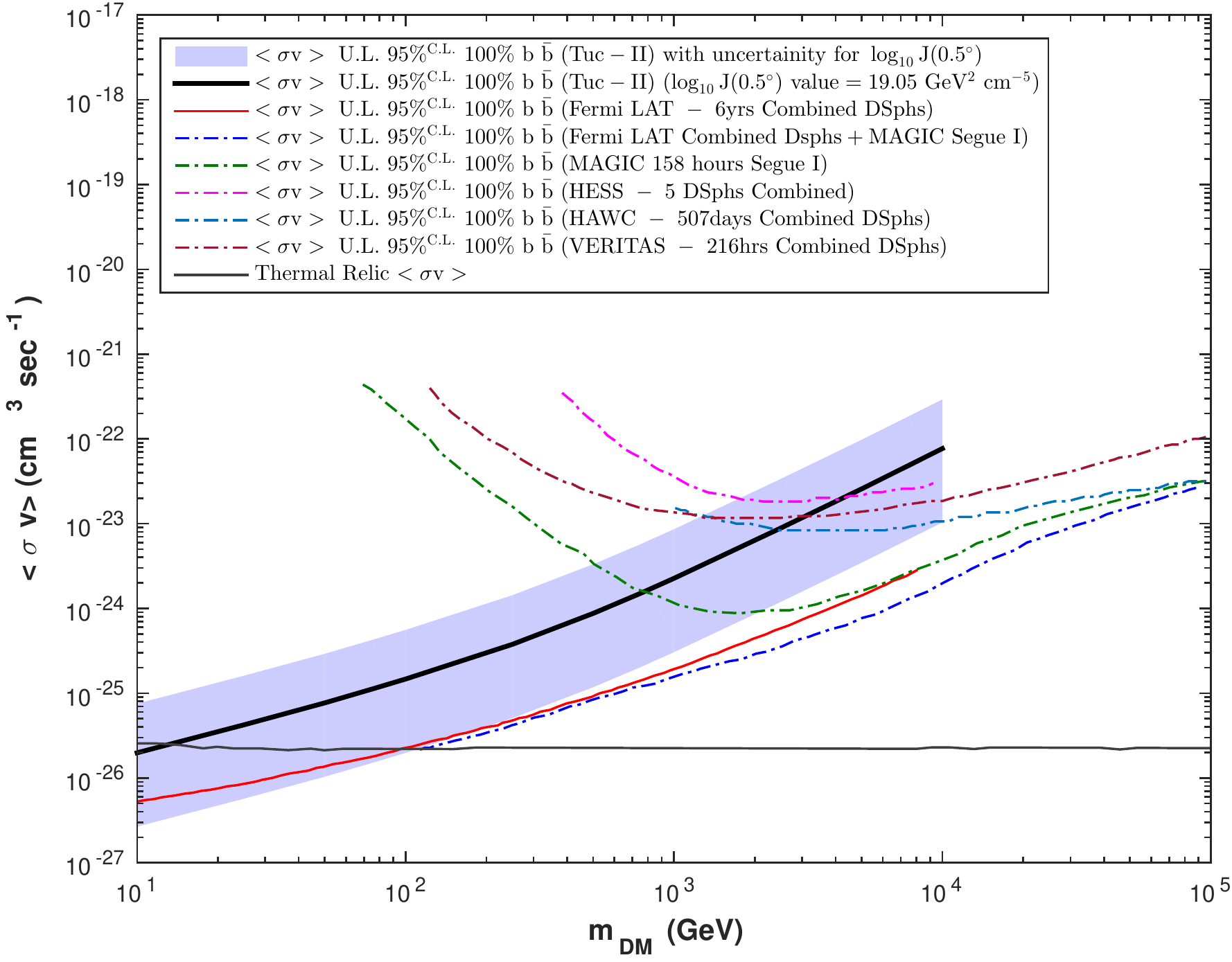}}
\caption{The comparison of the $<\sigma v>$ upper limits for $b\bar{b}$ annihilation final state 
obtained from Tuc-II with the limits obtained from several collaboration work has been shown here. For comparison, we have considered the $<\sigma v>$ upper limits obtained from the single or the combined studies on dSphs by VERITAS, HESS, MAGIC, HAWC, 
\textit{Fermi}-LAT+MAGIC and \textit{Fermi}-LAT, respectively. 
The shaded region denotes the uncertainty associated with the DM profiles for Tuc-II. The relic cross section rate obtained by the Steigman \textit{et al.}, 2012 is represented by the `dashed' sky-blue coloured line.}
\label{fig:fig}
\end{figure}

\section{Conclusions $\&$ Discussions}
\noindent In this work, we have studied nearly nine years of 
\textit{Fermi}-LAT data from the location of Tuc-II to investigate the 
signatures of DM annihilation. 
We have detected a very faint $\gamma$-ray excess from the location of Tuc-II 
for both the power-law spectra and the $\gamma$-ray spectrum from DM 
annihilation. We would also like to report that for $\gamma$-ray spectrum from 
DM annihilation, we 
have shown the variation of the TS values for Tuc-II with DM mass. We 
have also observed that for nine years of \textit{Fermi}-LAT data, TS value of 
Tuc-II peaks at 
~$m_{DM}\sim$~14 GeV for $100\%$ $b\bar{b}$ annihilation channel, while for 
$100\%$ $\tau^{+}\tau^{-}$ TS value peaks at $m_{DM}\sim$~4 GeV. In case of our 
Galactic 
Center, $m_{DM}$ range between 8 GeV to 15 GeV for $\tau^{+}\tau^{-}$ 
annihilation channel 
and the $m_{DM}$ range between 25 GeV to 70 GeV for $b\bar{b}$ annihilation 
channel 
play a crucial role to understand the $\gamma$-ray emission possibly arising 
from 
DM annihilation \cite{Gordon:2013vta, Hooper:2013rwa, Daylan:2014rsa, 
Zhou:2014lva, Calore:2014xka}. 
The mass range for our obtained TS peaks from the analysis for Tuc-II are slightly lower 
than the mass range required to describe the DM interpretation for Galactic 
Center. \\

\noindent From our analysis, we have also confirmed that excess from Tuc-II 
location is increased with increasing the time periods of data and such 
increase 
in TS peak value is approximately
proportional to $\sim\sqrt{t}$ \cite{Charles:2016pgz}; here t is the time 
periods of Fermi-LAT dataset. The most encouraging result of this analysis is 
that such successive increase in TS peak values of Tuc-II with larger time 
periods of the dataset can hint at the existence of any real signal either 
associated with any astrophysical scenario or resulting from DM 
annihilation. In the field of indirect DM detection, such hints of
$\gamma$-ray emission from Tuc-II may open a new path in DM physics. \\

\noindent When we assume the $\gamma$-ray spectra for DM annihilating to 
100$\%$ 
$\tau^{+}\tau^{-}$ channel, we have obtained a p-value of 
$\approx$ 0.003 from Tuc-II location corresponding to the background models provided by Fermi-LAT. It 
can be the result of rare statistical fluctuation in background. The one most 
tantalizing explanations of such excess are the presence of any surrounding 
unresolved bright sources. Among different types of unresolved sources, 
blazars are believed to be the main source of background fluctuation that emits 
$\gamma$-ray 
emission just below the threshold limit for Fermi-LAT. We have searched the 
BZCAT and the CRATES catalog, have found that three nearby radio sources lie
within $1^{\circ}$ ROI of Tuc-II and among all of them, the most nearby source 
i.e., J225455-592606 lies at just $0.55^{\circ}$ away from the location of 
Tuc-II. We have also checked 
the 4FGL catalog of Fermi-LAT (\cite{Fermi-LAT:2019yla}) and have 
noticed that a source, 4FGL 2247.7-5857 lies 0.66 degree away from 
Tuc-II location. Hence, it is very unlikely that the emission detected from 
Tuc-II location would be extremely contaminated by these nearby sources. \\

\noindent We have generated the residual TS maps of Tuc-II for energy $>$ 100 
MeV (Figs.~6.4 and 6.5). 
From these residual TS maps, we have noticed an excess of TS value $\approx$ 
6.5 
that is 
$0.18^{\circ}$ from the location of Tuc-II. We have also shown that whenever we 
have included Tuc-II to our source model, the excess from that location is 
greatly reduced. Thus, there is a very high chance that such emission is 
associated with Tuc-II. We have generated our all residual TS maps for energy 
$>$ 100 MeV. But the PSF of 
Fermi-LAT is comparatively large at lower energies, while at higher energies 
(say for around energy 
$>$ 500 MeV), the 68$\%$ of the photons would be confined within 1 degree of 
the 
location of the source 
\footnote{\tiny{http://www.slac.stanford.edu/exp/glast/groups/canda/lat{\_}Performance.htm}}
Thus, to again check the origin of the excess near Tuc-II location, we have 
produced a residual TS map for energy $>$ 500 MeV. Interestingly, from this new 
TS map 
(Fig 6.11), we could find that after including the Tuc-II to our source model, 
the 
nearby excess region has almost disappeared \cite{Bhattacharjee:2018xem}. This signature would probably hint 
that in Figs.~6.4 and 6.5 
after including Tuc-II to source model, the remaining excess emission is 
associated with weak background modellings. Thus, from our result, we can 
at 
best
conclude that the nearby excess is associated with Tuc-II location and it 
might indicate a DM annihilation signal from our Tuc-II \cite{Bhattacharjee:2018xem}.\\

\noindent Several Fermi collaboration papers observe that in a large region of 
the blank sky, the excess of TS $>$ 8.7 is very common. If we only consider the 
blazars within $1^{\circ}$ from the location of source, they would roughly 
account for 10$\%$ of such excesses. The DM 
subhalos may also be responsible for a $\approx$5$\%$-10$\%$ irreducible 
background. Therefore, we have re-calibrated our obtained significance and it 
decreases the TS peak value of Tuc-II from 8.61 to 4.79, i.e., from p value 0.003 
to 0.029. At present, with nine years of data, the obtained emission from 
Tuc-II is much weaker than Fermi-LAT's threshold detection. But from our work, 
we have also found that the significance of Tuc-II is increased with an increase 
in 
time periods of data and from TS map we have also observed a localized excess 
just beside the Tuc-II. So, in future, with even more time periods of data and 
with better background modelling, we can expect to explain the origin of the $\gamma$-ray excess from the location of Tuc-II.

\begin{figure}
\centering
{\includegraphics[width=.8\linewidth]{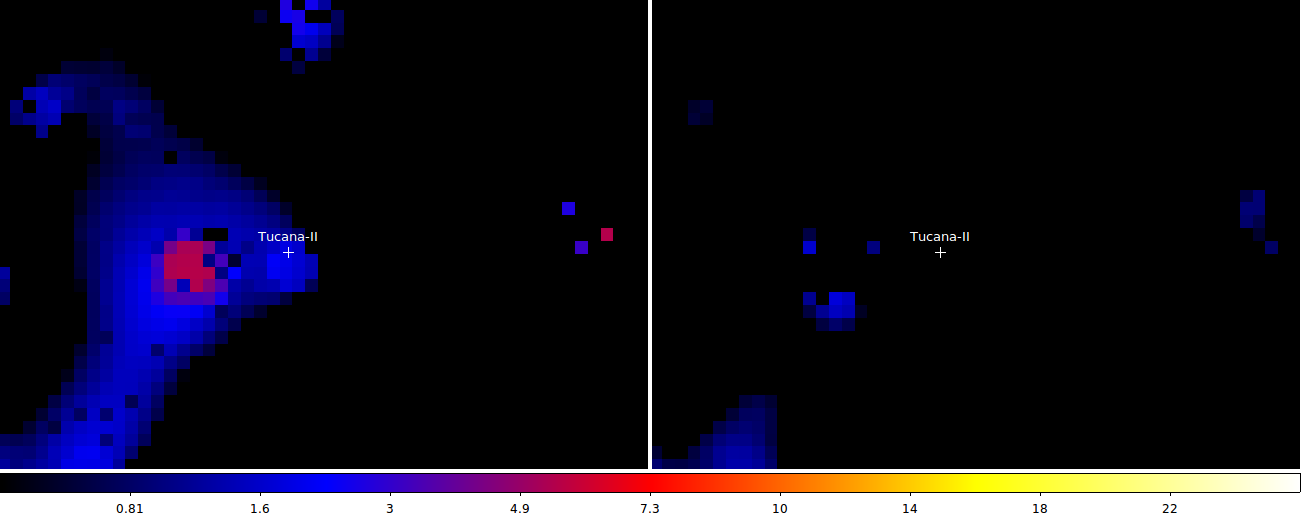}}
\caption{The residual TS maps (between 500 MeV to 300 GeV) for $1^{\circ}~\times~1^{\circ}$ 
ROI centred at Tuc-II extracted from $10^{\circ}$ $\times$ 
$10^{\circ}$ ROI. The image scale for TS map is 0.025 $pixel^{-1}$. In left Fig., Tuc-II is not included in the source model but 4FGL 2247.7-5857 and three CRATES sources are added to our source model; in right Fig., 4FGL 2247.7-5857, Tuc-II and the three CRATES sources are added to our source model.} 
\end{figure}

\noindent As we have already reported, the excess observed from Tuc-II location 
is below the detection threshold for Fermi-LAT. Thus we have derived the 
possible upper-limit of 
pair-annihilation $<\sigma~v>$ of the DM in Tuc-II as a function of 
DM mass and five annihilation channels.  
For our purpose, we have adopted the values J-factor and their uncertainties 
from Evans \textit{et al.}, 2016 \cite{Evans:2016xwx}. \\

\noindent For this, we have analysed the larger periods of compared to 
other previous works performed on Tuc-II and thus from our analysis, we can 
expect to 
provide more stringent limits on the theoretical models. 
We have observed that for median J-factor value, Tuc-II has imposed a strong 
constraint on the blue points in both the mSUGRA and the MSSM model, while the 
uncertainty band of Tuc-II have begun to constrain the red points. Because of 
the large uncertainty band, we may not obtain any impressive limits from Tuc-II 
in parameter space of ($\sigma v$, $m_{DM}$) but our obtained results stress 
that with a more detailed understanding of the internal structure, there is a 
high possibility that in future Tuc-II would provide very strong bounds on 
theoretically favoured DM models. From results show that for $>$100 GeV DM 
mass, 
Tuc-II imposes a stronger bound than the limits obtained from 
Ret-II. Thus, we can expect if we would have larger periods of dataset and more 
detailed information of the internal structure,  we should be able to reduce 
the 
uncertainty band of Tuc-II to a possible narrow band in parameter space of 
($<\sigma~v>$, $m_{DM}$). Then Tuc-II might be considered as one of the most DM 
dominated UFDs.

\chapter{Multiwavelength analysis of low surface brightness galaxies}
\section{Low Surface Brightness Galaxy}
\noindent In this chapter, we have chosen a set of low surface brightness (LSB) 
galaxies which are thought to be an excellent target for indirect DM detection \cite{Bhattacharjee:2019jce}. 
LSB galaxies are the diffuse galaxy whose surface brightness is nearly one 
order 
of magnitude lower than our night sky. Most of the baryonic component in 
LSB galaxies are in form of neutral hydrogen (HI) gas \cite{Burkholder:2001bg, 
ONeil:2004mqj, Du:2015in} 
and that hydrogen disk is extended
up to 2 to 3 times beyond the stellar disks of LSBs\cite{Blok:1996dfr, 
Mishra:2017vgt}. LSB 
galaxies are metal poor and are generally made of dust free stellar disks 
\cite{Impey:1997uc} with a very considerably small amount of molecular hydrogen 
gas 
\cite{Das:2006jp}. Hence, LSB would have a negligible or very little star 
formation rates (SFRs). The $\gamma$-ray emission resulting due to the Star 
formation would not then interfere
much with emission from WIMP annihilation. The 
supermassive black holes or active galactic nuclei (AGN) can also be a source 
of 
$\gamma$-rays emission but AGNs are rarely found in LSB galaxies. Thus, from 
astrophysical perspective, we can consider LSB 
as the clean sources \cite{Bhattacharjee:2019jce}.\\

\noindent The measurements from their HI rotation curves \cite{Das:2019tea} 
indicate
their very high value of mass-to-light ratio \cite{ONeil:2004mqj, 
Honey:2018dfr} i.e., the contribution coming from the stars and the luminous gas 
is very little compare to the total mass in LSB. The studies obtained from the 
observation of rotation curve in LSB galaxies also hint the existence
of massive DM halos \cite{Blok:1997dfr}. Even the centres of LSB galaxies do 
not 
have any large overdensities in stellar components. Therefore, LSB galaxies are 
believed to be DM-dominated even in their centres and that makes them an 
excellent source for the indirect search of the DM signal \cite{Bhattacharjee:2019jce}. 
For indirect detection of DM candidate, LSB galaxies hold two primary criteria 
i.e., (i)~very rich in DM content and 
(ii) do not consist of any strong sources of $\gamma$ radiation, 
for example, AGN and star-forming regions.\\

\noindent The HI rotation curves and gas kinematics of LSB galaxies are also 
used to resolve the 
`cusp-core' problem in the CDM theory of galaxy formation 
\cite{vandenBosch:2000rza, deNaray:2008bs}. The N-body simulation generally 
favours the cuspy profile for 
DM distribution, while for some LSB galaxies the cored profile can provide a 
better fit to their central DM distribution.\\

\noindent Even though the LSB galaxies are very suitable targets for indirect 
DM 
detection, because of their large distances (of the order of Mpc) they are not 
widely studied. There are very few dedicated literatures which have studied the 
possible $\gamma$-ray emission from LSB galaxies \cite{Gammaldi:2017mio, 
Cadena:2017ldx, Hashimoto:2019obg, Bhattacharjee:2019jce}. For our study, we have chosen four LSB 
galaxies that are relatively
close and have applied the multiwavelength approach for investigating the 
possible
DM signal from LSB galaxies at gamma and radio wavelengths \cite{Bhattacharjee:2019jce}.

\section{Sample Selection:}
\noindent In this section, we would give a brief introduction to our selected 
LSB galaxies \cite{Bhattacharjee:2019jce}. They have low B-band luminosity and a large quantity of DM 
contents\cite{vandenBosch:2000rza}. They all are situated within the 15 Mpc 
heliocentric distances. From the optical images, any intense sign of AGN 
activity and new star formation have not been observed. In Table~7.1, we have 
mentioned some observational results for LSB galaxies \cite{Bhattacharjee:2019jce}.\\

1) \textbf{UGC 3371}: UGC 3371, also known as DDO 039, is characterised as the 
irregular dwarf galaxy. Several studies show an impressive agreement between the 
rotational velocities obtained from the H$\alpha$ and the HI, respectively. But 
at the initial, rotation curve from HI disk started to arise more steeply than 
the rotational curve for H$\alpha$. The overcorrection done in the beam smearing 
for HI rotational curve is assumed to the reason for such discrepancies 
\cite{Swaters:2002rx} and thus for UGC 3371, we could not predict the exact 
shape of the rotational curve. Its shape could be either linear or steep. The 
study by ref.~\cite{swaters:PhDthesis} showed that the rotational curve of UGC 
3371 had provided an impressive fit to the DM halo profile with the steep cusp 
at the centre. Thus, for UGC 3371, the DM halo profile is consistent with the 
CDM prediction\cite{vandenBosch:2000rza}.\\

2) \textbf{UGC 11707}: UGC 11707 is characterised as the spiral galaxy which has 
the loosely bound broken arms originating from some individual stellar clusters. 
The observational data points out the very faint bulge (Sd) at the centre of UGC 
11707. But there are not many studies on it and thus because of the insufficient 
data, there are not many sample data to define its rotational curve. But the 
study indicates that between H$\alpha$ and HI rotational curves, the 
inner rise for H$\alpha$ curve is comparatively steeper\cite{swaters:PhDthesis}. 
The rotational curves for UGC 11707, also indicate a discrepancy between the 
approaching and receding values of rotational velocity for radius $\le$ 7 
kpc\cite{swaters:PhDthesis}. The DM halo profile for UGC 11707 is consistent 
with the CDM prediction\cite{vandenBosch:2000rza}.\\

3) \textbf{UGC 12632}: UGC 12632, also known as DD0217, is characterised as the 
weakly barred spiral galaxy (i.e., SABm). Its HI rotational curve follows the 
uniform distribution but a distinguished high-velocity bump has been observed 
from the blue portion of the H$\alpha$ curve. From the velocity map of UGC 
12632, a steep rise in rotational velocity has been observed near its centre and 
that pattern is gradually extended to the outer region. Thus, the observational 
findings, directly indicate that the DM halo profile for UGC 12632 is consistent 
with the CDM prediction\cite{vandenBosch:2000rza}.\\

4) \textbf{UGC 12732}: Like UGC 12632, UGC 12732 is also characterised as the 
weakly barred spiral galaxy (i.e., SABm). The observational data indicates that 
the rotational curve for both the HI and the H$\alpha$ are consistent with each 
other and it is observed that the DM halo profile for UGC 12732 is consistent 
with the CDM prediction\cite{vandenBosch:2000rza, Swaters:2009by}.

\begin{table}
\caption{Properties of LSB galaxies. Column~I: Name of LSB galaxies; Column~II: Galactic longitude and latitude of LSB galaxies; Column~III: The adopted distance of the galaxies, based on a Hubble constant ($H_{\circ}$)= 75 $km~s^{-1}~Mpc^{-1}$. We have obtained the value of distance for each LSB galaxies and their corresponding uncertainties from \textit{NASA/IPAC Extragalactic Database}; Column~IV: Observed rotational velocity at last measured point of rotational curve from van den Bosch \textit{et al.}, 2000; Column~V: Scale length of stellar disk from van den Bosch \textit{et al.}, 2000; Column~VI: B band Luminosity of LSBs from OBrien \textit{et al.}, 2011; Column~VII: Location of the last observed data points of LSB galaxies from Swaters \textit{et al.}, 2009; Column~VIII: Observed HI gas masses of LSB galaxies from Swaters \textit{et al.}, 2002.}
\centering
\begin{minipage}{1.0\textwidth}
\begin{tabular}{|p{1cm}||p{2.6cm}|p{1.5cm}|p{1.5cm}|p{1.5cm}|p{1.5cm}|p{1.5cm}|p{1.5cm}|}
\hline
\hline
Name & (l,b) & D  & $V_{last}$ & $R_{d}$ & $L_{B}$ & $R_{last}$ & $M_{HI}$ \\ 
[0.5ex]
$ $ & [deg],[deg] & (Mpc) & $(km~s^{-1})$ & (Kpc) & $(10^{9}~L_{\odot}^{B})$ & 
(Kpc) & $(10^{8}~M_{\odot})$ \\ [0.5ex]
\hline
UGC 3371 & 138.43,22.81 & $12.73^{+0.90}_{-0.90}$ & ~~~~~86 & 3.09 & 1.54 & 
10.2 & 12.2 \\ [0.5ex]
\hline
UGC 11707 & 74.31,-15.04 & $14.95^{+1.05}_{-1.05}$ & ~~~~~100 & 4.30 & 1.13 & 
15.0 & 37.2 \\ [0.5ex]
\hline
UGC 12632 & 106.77,-19.31 & $8.36^{+0.60}_{-0.60}$ & ~~~~~76 & 2.57 & 0.86 & 
8.53 & 8.7 \\ [0.5ex]
\hline
UGC 12732 & 103.74,-33.98 & $12.38^{+0.87}_{-0.87}$ & ~~~~~98 & 2.21 & 0.71 & 
15.4 & 36.6 \\ [0.5ex]
\hline
\hline
\end{tabular}
\end{minipage}
\end{table}

\pagebreak
\section{\textit{Fermi}-LAT Observation and Data Analysis of LSBs}
\noindent Here we have analysed nearly 9 years of Fermi-LAT data i.e., from 
2008-08-04 to 
2017-10-22 for our each source of targets \cite{Bhattacharjee:2019jce}. For this purpose, we have used the 
Fermi ScienceTools version, 
\textit{v1.2.1}\footnote{\tiny{https://fermi.gsfc.nasa.gov/ssc/data/analysis/software/}} \cite{Bhattacharjee:2019jce}.
Like our other works, for this study we have used the source class 
IRF, $\rm{P8R3\_SOURCE\_V2}$ 
\footnote{\tiny{https://fermi.gsfc.nasa.gov/ssc/data/analysis/documentation/Cicerone/Cicerone{\_}LAT{\_}IRFs/IRF{\_}overview.html}},
\footnote{\tiny{https://fermi.gsfc.nasa.gov/ssc/data/analysis/documentation/Pass8{\_}usage.html}} \cite{Bhattacharjee:2019jce}.
The PSF of 
LAT is yielding to $4^{\circ}$ and $2.5^{\circ}$ for energy around 500 MeV and 1 GeV,  
respectively\footnote{\tiny{https://www.slac.stanford.edu/exp/glast/groups/canda/lat{\_}Performance.html}}. 
Thus in 
order to reduce the possible 
uncertainties at low energies and background contamination at high energies, we 
have used the energy limits or range between 500 MeV to 300 GeV \cite{Bhattacharjee:2019jce}. Here, we have 
extracted the LAT data for a $10^{\circ}$ ROI for each source of interest and 
for generating the source model for likelihood analysis, we have used here 
\textit{Fermi} 4FGL source catalog 
\cite{Fermi-LAT:2019yla} and the most recent version of galactic 
($\rm{gll\_iem\_v07.fits}$) and extragalactic 
($\rm{iso\_P8R3\_SOURCE\_V2\_v1.txt}$) diffuse models \cite{Bhattacharjee:2019jce}.\\

\noindent In sections ~6.2 and 5.2, we have already described the analysis 
methodology for 
Fermi-LAT data.

\subsection{Results from the Power-law Modelling} 
\begin{figure}
\centering
\subfigure[ UGC 3371]
 { \includegraphics[width=0.45\columnwidth]{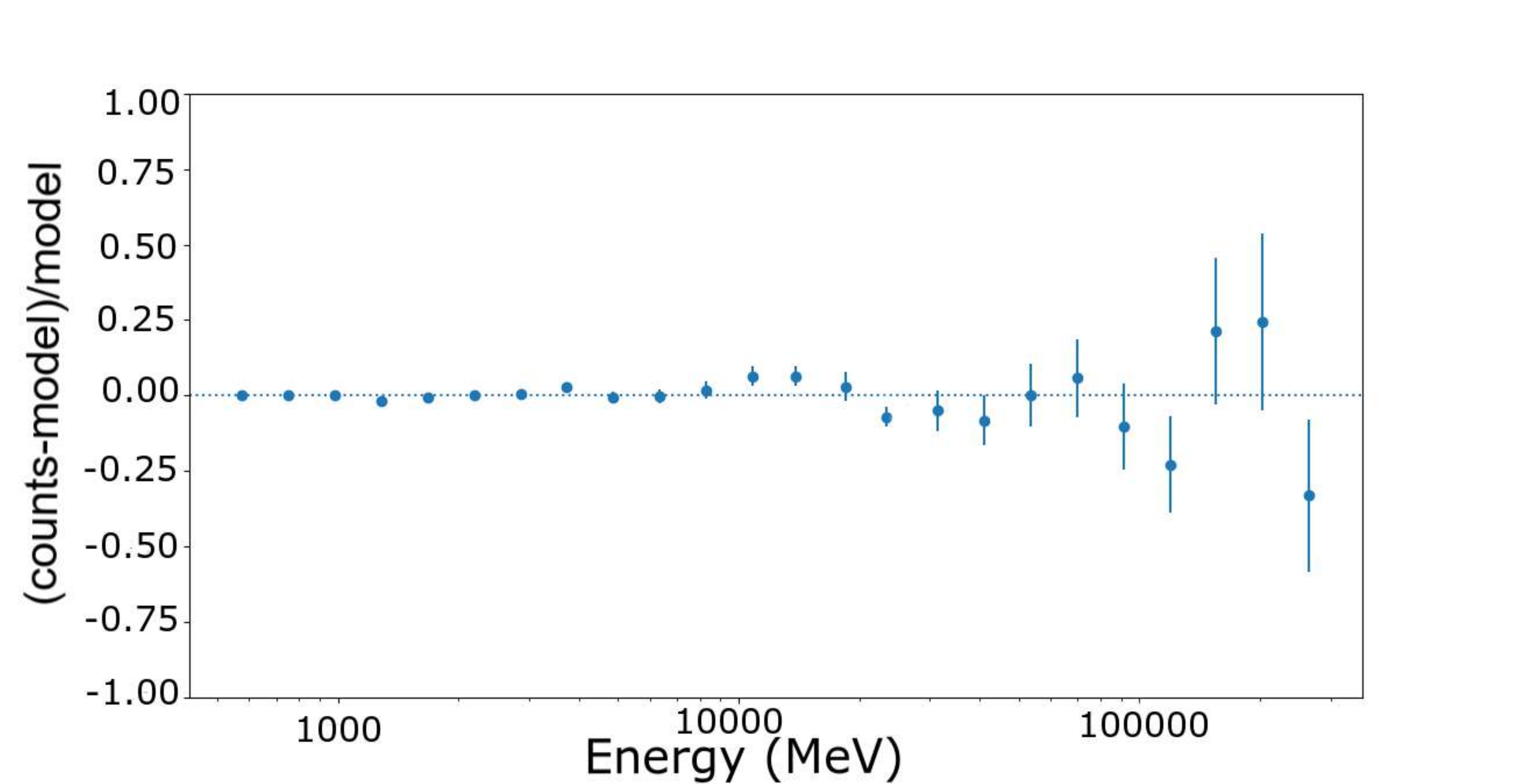}}
\subfigure[ UGC 11707]
 { \includegraphics[width=0.45\columnwidth]{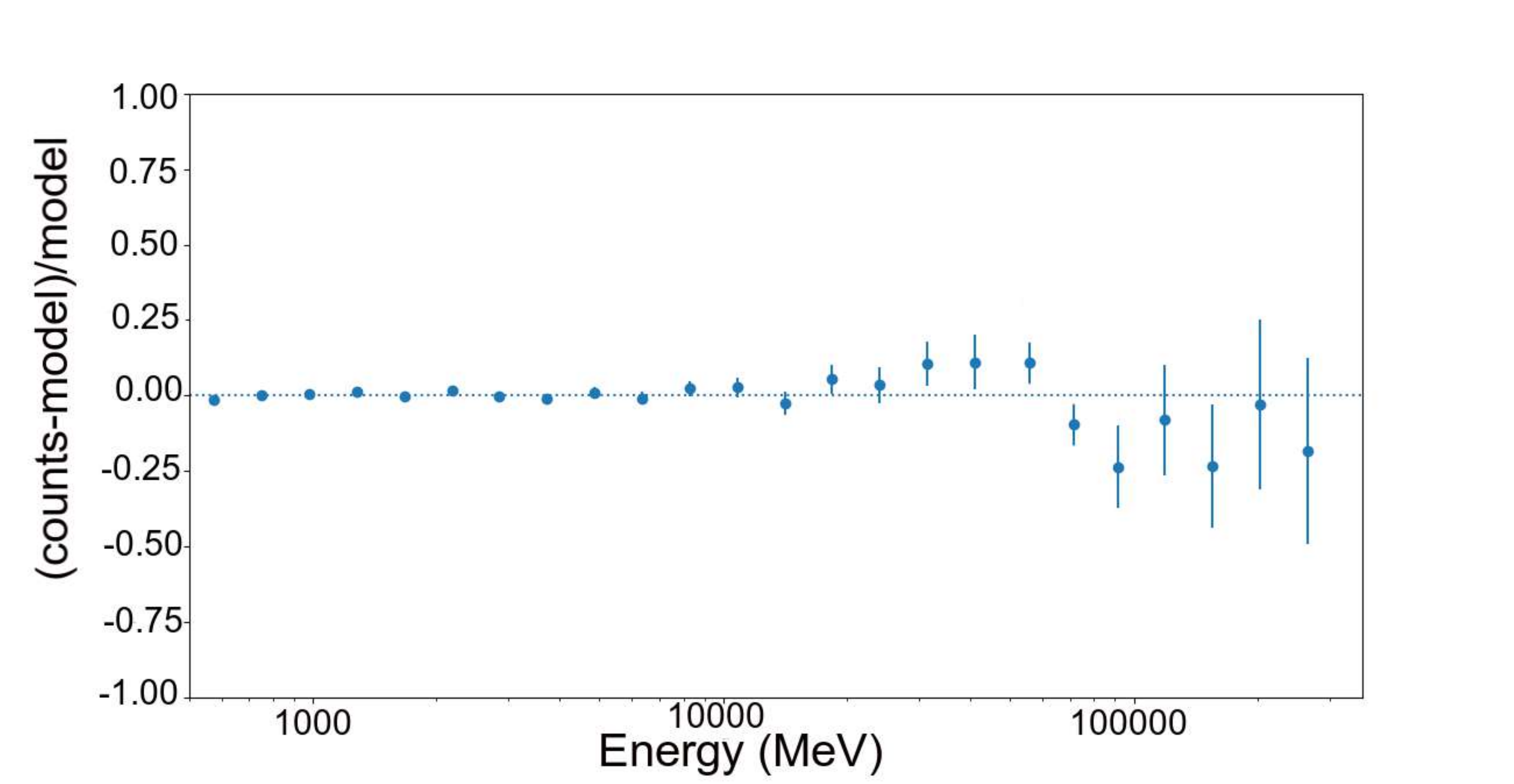}}
\subfigure[ UGC 12632]
 { \includegraphics[width=0.45\columnwidth]{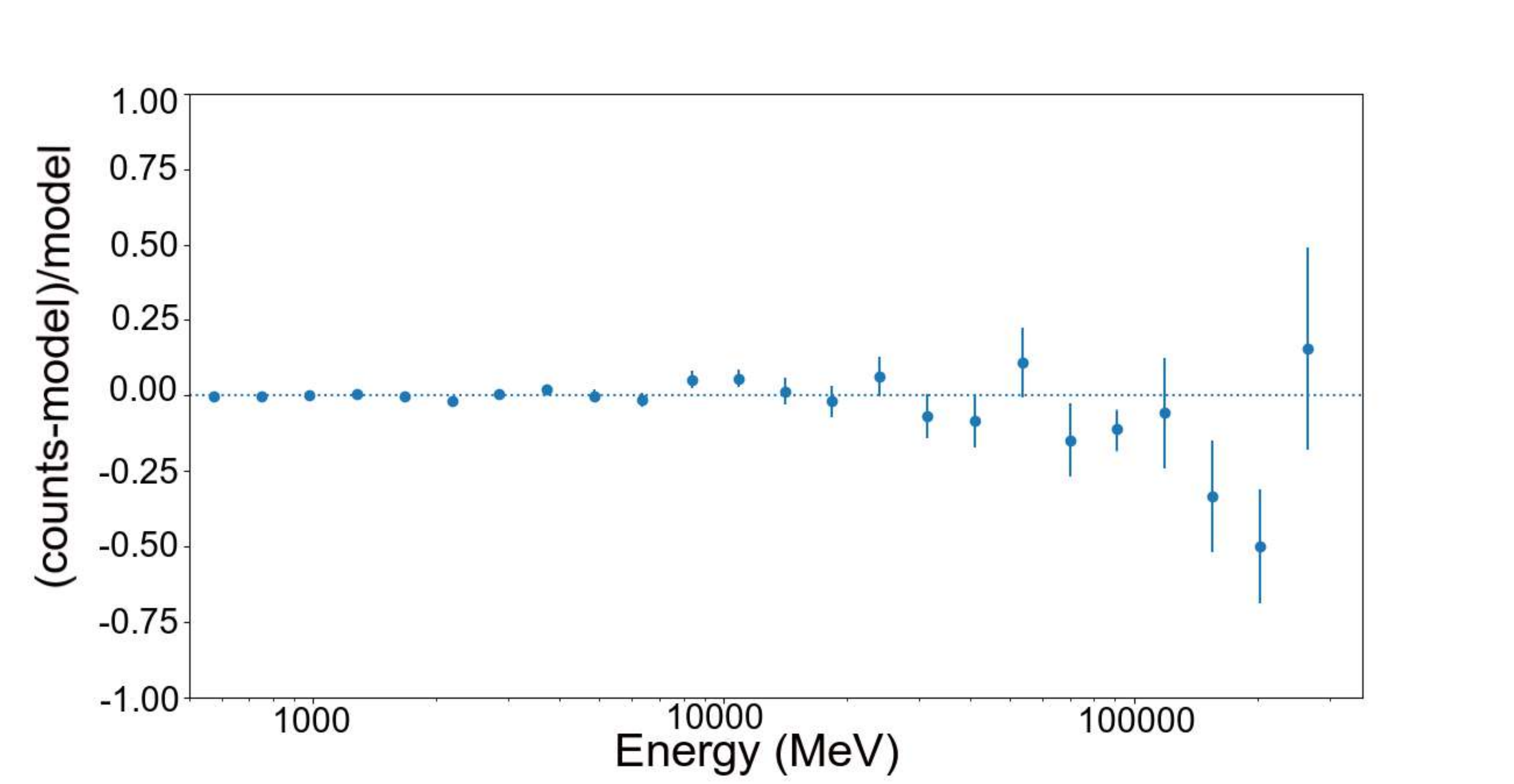}}
\subfigure[ UGC 12732]
 { \includegraphics[width=0.45\columnwidth]{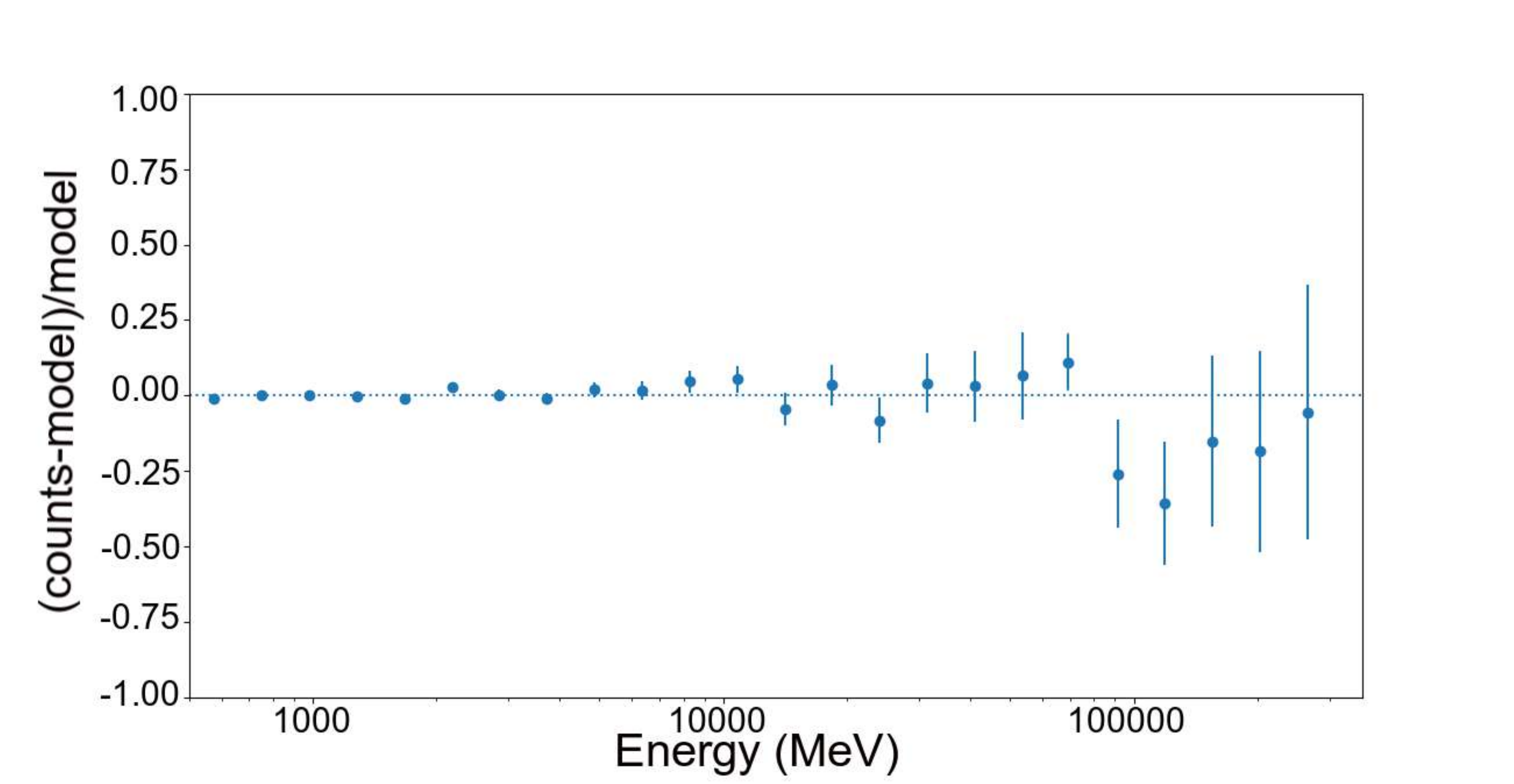}}
\caption{We have shown the residual plots of four LSB galaxies for $10^{\circ} 
\times 10^{\circ}$ ROI. We have modelled them with the power law spectrum for 
$\Gamma = 2$.}
\end{figure}

\noindent In order to check the possible astrophysical constraint from LSB 
galaxies, first we have modelled them
with power-law spectra for spectral index ($\Gamma$) = 2 \cite{Bhattacharjee:2019jce}.\\

\noindent Fig.~7.1(a,b,c,d) shows the residual fit for the four LSB galaxies\cite{Bhattacharjee:2019jce}.
In Table~7.2, we have shown the spectral results obtained from our each LSB 
galaxies \cite{Bhattacharjee:2019jce}. From this table, we can find the best-fitted values for the galactic 
and isotropic components and normalization parameter, $N_{0}$ for each galaxy. 
The best-fitted values for two diffuse models are close to 1 and it strengths 
the reliability of our analysis method \cite{Bhattacharjee:2019jce}. From table 7.2, we can also check that 
$N_{0}$ is always lower than its statistical error by at least an order of 2. 
This signifies that Fermi-LAT has not observed any emission from the location 
of 
four LSB sources \cite{Bhattacharjee:2019jce}.\\

\noindent Next, we have estimated the upper limits of $\gamma$-ray flux by 
profile 
likelihood method \cite{Barbieri:1982eh, Rolke:2004mj, Bhattacharjee:2019jce}. In Table~7.3, we have 
shown the flux upper limits in  95$\%$ C.L.for energy ranges between 500 MeV to 
300 GeV.
  
\begin{table}[!h]
\centering
\caption{The Best-Fit value for the normalization parameter of LSB, diffuse 
galactic and isotropic components.}
\begin{tabular}{|p{2cm}|p{3cm}|p{3cm}|p{4cm}|}
\hline 
\hline
LSB  & Galactic  & Isotropic  & $\rm{N_{0}}\times~10^{-5}$ \\ [0.5ex]
Galaxies &  Component &  Component &  $ $ \\ [0.5ex]
& $cm^{-2} s^{-1} MeV^{-1}$ & $cm^{-2} s^{-1} MeV^{-1}$ & $cm^{-2} s^{-1} MeV^{-1}$ \\ [0.5ex]
\hline 
UGC 3371 & $0.95 \pm 0.011$  & $0.95 \pm 0.035$ & $(6.29\pm 21.55)\times10^{-8}$ \\ [0.5ex]
\hline 
UGC 11707 & $0.92 \pm 0.001$  & $1.06 \pm 0.001$ & $(0.1099\pm6.06)\times10^{-7}$ \\ [0.5ex]
\hline 
UGC 12632 & $0.93 \pm 0.011$  & $1.09 \pm 0.05$ & $(0.334\pm 5.82)\times10^{-6}$ \\ [0.5ex]
\hline 
UGC 12732 & $0.97 \pm 0.001$  & $1.004 \pm 0.017$ & $(0.12\pm2.30)\times10^{-8}$ \\ [0.5ex]
\hline 
\hline
\end{tabular}
\end{table}

\begin{table}
\centering
\caption{The $\gamma$-ray flux upper limits in $95\%$ C.L. of LSB galaxies.}
\begin{tabular}{|p{5cm}|p{5cm}|} 
\hline 
\hline
LSB galaxies & E~$>$~500 MeV \\ [0.5ex]
$ $ & ($cm^{-2} s^{-1}$) \\ [0.5ex] 
\hline 
UGC 3371 & $2.43\times10^{-10}$ \\ [0.5ex] 
\hline 
UGC 11707 & $3.22\times10^{-10}$ \\ [0.5ex] 
\hline
UGC 12732 & $3.54\times10^{-10}$ \\ [0.5ex]
\hline
UGC 12632 & $3.06\times10^{-10}$ \\ [0.5ex]
\hline
\hline
\end{tabular}
\end{table}

\section{A theoretical Framework to Estimate $\gamma$-ray Flux from Pair-annihilation of WIMPs In Case of LSB Galaxies}
\subsection{Modelling with NFW Density Profile}
\noindent We have used the 
NFW density profile \cite{Navarro:1996gj} for modelling the DM distribution in 
LSB galaxies. The rotational curves for our selected LSB galaxies are 
consistent 
with the $\lambda$CDM prediction \cite{vandenBosch:2000rza, vandenBosch:2001bp, 
Swaters:2002rx} and their observational data obtained from the 
ref.~\cite{vandenBosch:2000rza, 
vandenBosch:2001bp, Swaters:2002rx} shows that the cuspy profile can provide a 
good fit to the central region of LSBs. For our J-factor calculation, we have 
taken the necessary parameters from 
ref.~\cite{vandenBosch:2000rza}. The expression of the NFW density 
profile is \cite{Abdo:2010ex, Navarro:1996gj}
\begin{equation}
\rho (r)=\frac{\rho_{s}r_{s}^{3}}{r(r_{s} + r)^{2}}
\end{equation}
\\
\noindent where, $\rho_{s}$ and $r_{s}$ are the characteristic 
density and scale radius, respectively and $r$ is the distance from the center 
of the LSB galaxy. In order to obtain the value of $\rho_{s}$ and $r_{s}$, we 
have used the following relations \cite{Bhattacharjee:2019jce}:\\

\noindent The expression of the $\rho_{s}$ is \cite{Lokas:2000mu, 
Liddle:1998ew}:
\begin{equation}
\rho_{s} = \rho_{c}^{0} \delta_{\rm{char}}
\end{equation}
\\
\noindent where, $\delta_{\rm{char}}$ is the fitting parameter 
and $\rho_{c}^{0}$ is the critical density of Universe. For our calculation, we 
have adopted the Hubble constant of $H_{0}$=$75~\rm{km~s^{-1}Mpc^{-1}}$ = 
$100h~\rm{km~s^{-1}Mpc^{-1}}$ from Ref.~\cite{vandenBosch:2000rza} and thus 
$\rho_{c}^{0}$ can be expressed as $\rho_{c}^{0}$ = $2.78h^{-1}\times10^{11}$ 
$\frac{M_{\odot}}{(h^{-1}Mpc)^{3}}$.\\

\noindent The expression of the $\delta_{\rm{char}}$ is:
\begin{equation}
\delta_{\rm{char}} = \frac{v c^{3}g(c)}{3}
\end{equation}
\noindent where,
\begin{equation}
g(c)=\frac{1}{\ln(1+c)- c/(1+c)}
\end{equation}
\\
\noindent In Eqs.~7.3 $\&$ 7.4, $c$ is the concentration 
parameter that defines the shape of the density profile and the value of the 
virial overdensity, $v$ is assumed to be $\approx$ 178 \cite{vandenBosch:2000rza}.\\

\noindent $R_{\rm{vir}}$ (or we can say $r_{200}$) is the virial 
radius at which mean density is 200 times of present critical density 
($\rho_{c}^{0}$) of our Universe. The circular velocity at $R_{\rm{vir}}$ is 
defined as\cite{Lokas:2000mu, vandenBosch:2000rza}
\begin{equation}
V_{200} = \frac{R_{vir}}{h^{-1}}
\end{equation}
\\
\noindent The expression of scale radius is \cite{vandenBosch:2000rza}:
\begin{equation}
r_{s} = \frac {R_{\rm{vir}}}{c}
\end{equation}
\\
\noindent Thus, using the Eqs.~7.2 to 7.6, we can derive $\rho_{s}$ and 
$r_{s}$. \\

\noindent For our case, we have taken $\theta_{\rm{min}} = 0^{\circ}$ and $\theta_{max} = \rm{sin}^{-1}\Big(\frac{R_{vir}}{d}\Big)$ \cite{Bhattacharjee:2019jce}. The 
J-factor allows us to estimate the
annihilation rate from LSB galaxies for theoretical favored DM models.\\

\noindent In Table~7.4, we have mentioned some necessary 
parameters for estimating the J-factors from Eq.~2.7 \cite{Bhattacharjee:2019jce}. We have adopted the value 
of $c$, $V_{200}$ from 
ref.~\cite{vandenBosch:2000rza}.\\

\noindent In Table~7.4, we have shown the uncertainty associated with the 
J-factors. For deriving the uncertainties in J-factor, we have taken the 
distribution of distance ($d$) and concentration parameter 
($c$) mentioned in Table~7.4 and have developed an algorithm 
to find the limiting values of the J-factor in a 2$\sigma$ limit by a 
Monte Carlo method \cite{Bhattacharjee:2019jce}. As the concentration parameter for LSB galaxies lies within 
asymmetrical limits, we have considered asymmetric normal distribution about 
the 
mean with two different values of the standard deviation on each side of the 
mean\cite{Bhattacharjee:2019jce}. We have first generated random numbers for the user-defined distribution 
and then by performing the 
Smirnov Transform on a set of uniformly distributed random numbers, we have 
generated the uncertainty limits of J-factor for 2$\sigma$ or 95$\%$ C.L. \cite{Bhattacharjee:2019jce}.

\begin{table}
\centering
\caption{The necessary parameter values for calculating the J-factor from 
Eq.~2.7 
($h_{0}=0.75$).}
\begin{tabular}{|p{1.5cm}|p{2cm}|p{1.5cm}|p{2cm}|p{1.5cm}|p{3cm}|p{4cm}|}
\hline \hline
Galaxy & Distance & ~~~~c & ~~$V_{200}$ & $\theta_{max}$ & J~factor\\
name  & ~~Mpc & $ $ & ~~$km~s^{-1}$ & ~~~$^{\circ}$ & $\times10^{16}$~$\frac{GeV^{2}}{cm^{5}}$\\
\hline \hline
UGC 3371  & $12.73^{+0.90}_{-0.90}$ & $14.5^{+14.6}_{-10.2}$ & ~~69.8 & $0.42$ & $0.739^{+2.87}_{-0.63}$  \\
\hline 
UGC 11707 & $14.95^{+1.05}_{-1.05}$ & $14.7^{+14.6}_{-10.3}$ & ~~66.9 & $ 0.34$ & $0.485^{+1.85}_{-0.42}$ \\
\hline 
UGC 12632 & $8.36^{+0.60}_{-0.60}$ & $15.6^{+15.5}_{-10.9}$ & ~~51.4 & $0.47$ & $0.795^{+3.25}_{-0.716}$ \\
\hline 
UGC 12732 & $12.38^{+0.87}_{-0.87}$ & $14.3^{+14.4}_{-10}$ & ~~73.3 & $0.45$ & $0.880^{+3.40}_{-0.75}$ \\
\hline \hline
\end{tabular}
\end{table}

\subsection{J-factor Derived from the Toy Model}
\noindent In this section, we have predicted the J-factor for LSB galaxies by 
using the toy model proposed by 
Charbonnier et al, 2011~\cite{Charbonnier:2011ft}. The sole purpose of using 
the 
toy model is to check the reliability of our derived value for J-factor from 
Eq.~2.7. In Fig.~7.2, we have shown the sketch of the toy model for J-factor 
calculation \cite{Charbonnier:2011ft}. The 
vertical hatched region denotes the contribution from integration, while the 
cross-hatched 
region refers to the toy model.

\begin{figure}
\centering
\includegraphics[width=0.6\columnwidth]{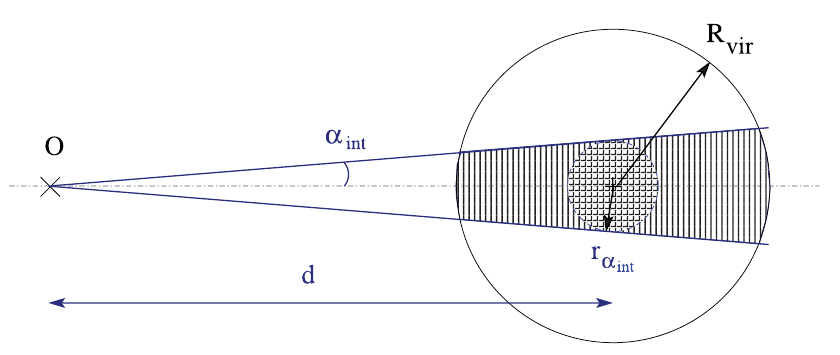}
\caption{The diagram of the toy model for calculating J-factor.}
\end{figure}

\noindent In Fig.~7.2, $d$ is the distance to LSB galaxy from the observer and 
$\alpha_{int}$ defines the angle for integration where, 
$r_{\alpha_{int}} 
= d~\rm{sin}\alpha_{int}$. The toy model assumes that roughly $90\%$ 
of the clump luminosity might contain in scale radius, $r_{s}$ and they do not 
have any direct dependence on the DM density profile. We can rewrite the 
Eq.7.1 as:\\
\begin{equation}
\rho_{approx} = \rho_{s} r_{s}/r~ \rm{for} ~r_{sat}< r \le  r_{s} 
\end{equation}
where, $r_{sat}$ is the saturation distance. The corresponding approximate form 
of J-factor is:
\begin{eqnarray}
J_{approx} &=&\frac{4 \pi} {d^{2}} \int_{0}^{\rm{min}[r_{\alpha_{int}}, r_{s}]} \rho_{approx}^{2} r^{2}~ dr \nonumber\\
              &=&  \frac{4 \pi} {d^{2}} \rho_{s}^{2}r_{s}^{2}(\rm{min}[r_{\alpha_{int}}, r_{s}]).
\end{eqnarray}
\noindent If $r_{\alpha_{int}}\gtrsim r_{s}$, the density profile falls faster 
than $1/r$ for $r \sim r_{s}$. The the toy model advised us stop the 
integration 
at $r_{x}$ where, $\rho_{true} = \frac{\rho_{approx}}{x}$, $x =2$ and $r_{x} = r_{s}[\sqrt2 - 1]$ \cite{Charbonnier:2011ft}.
\begin{equation}
J_{approx}= \frac{4 \pi} {d^{2}} \rho_{s}^{2}r_{s}^{2}(\rm{min}[r_{x}, r_{\alpha_{int}}]).
\end{equation}

\noindent The comparison between the J-values obtained from the toy model and 
integration method has been shown in Table~7.5 \cite{Bhattacharjee:2019jce}. Charbonnier et al, 
2011\cite{Charbonnier:2011ft}, proposed that the difference in the J values 
obtaining from these above-mentioned methods should lie within the factor of 2 
and from Table~7.5, it is clear that our results are consistent \cite{Bhattacharjee:2019jce} with the study 
done by Ref.~\cite{Charbonnier:2011ft}

\begin{table}
\centering
\caption{J-factor obtained from the integration method and the Toy model for 
$h_{0}$=0.75.}
\begin{tabular}{|p{2.5cm}|p{3cm}|p{4.5cm}|}
\hline \hline
Galaxy & Integration method  & Toy model \\
name & ($\rm{GeV^{2}/cm^{5}}$) & ($\rm{GeV^{2}/cm^{5}}$) \\
\hline \hline
UGC 3371  & $0.739^{+2.87}_{-0.63}\times10^{16}$ & $0.918^{+3.47}_{-0.82}\times10^{16}$  \\
\hline
UGC 11707  & $0.485^{+1.85}_{-0.42}\times10^{16}$ & $0.603^{+2.20}_{-0.54}\times10^{16}$  \\
\hline
UGC 12632  & $0.795^{+3.08}_{-0.68}\times10^{16}$ & $0.987^{+3.84}_{-0.88}\times10^{16}$   \\
\hline
UGC 12732 & $0.880^{+3.40}_{-0.75}\times10^{16}$ & $1.09^{+4.37}_{-0.97}\times10^{16}$   \\
\hline
\end{tabular}
\end{table}

\subsection{Constraints on the Annihilation Cross-section}
\begin{figure}
\subfigure[]
 { \includegraphics[width=0.48\linewidth]{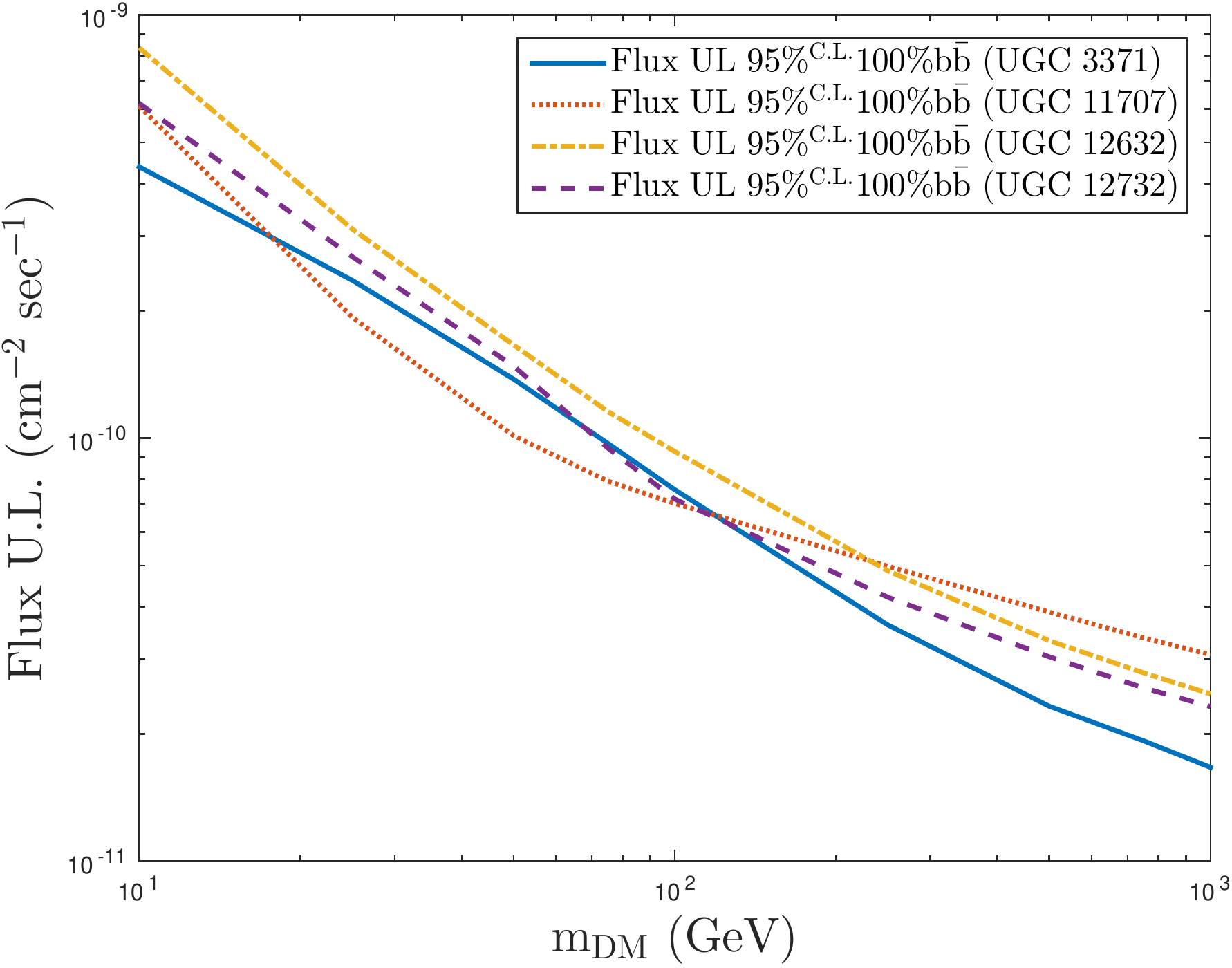}}
\subfigure[]
 { \includegraphics[width=0.48\linewidth]{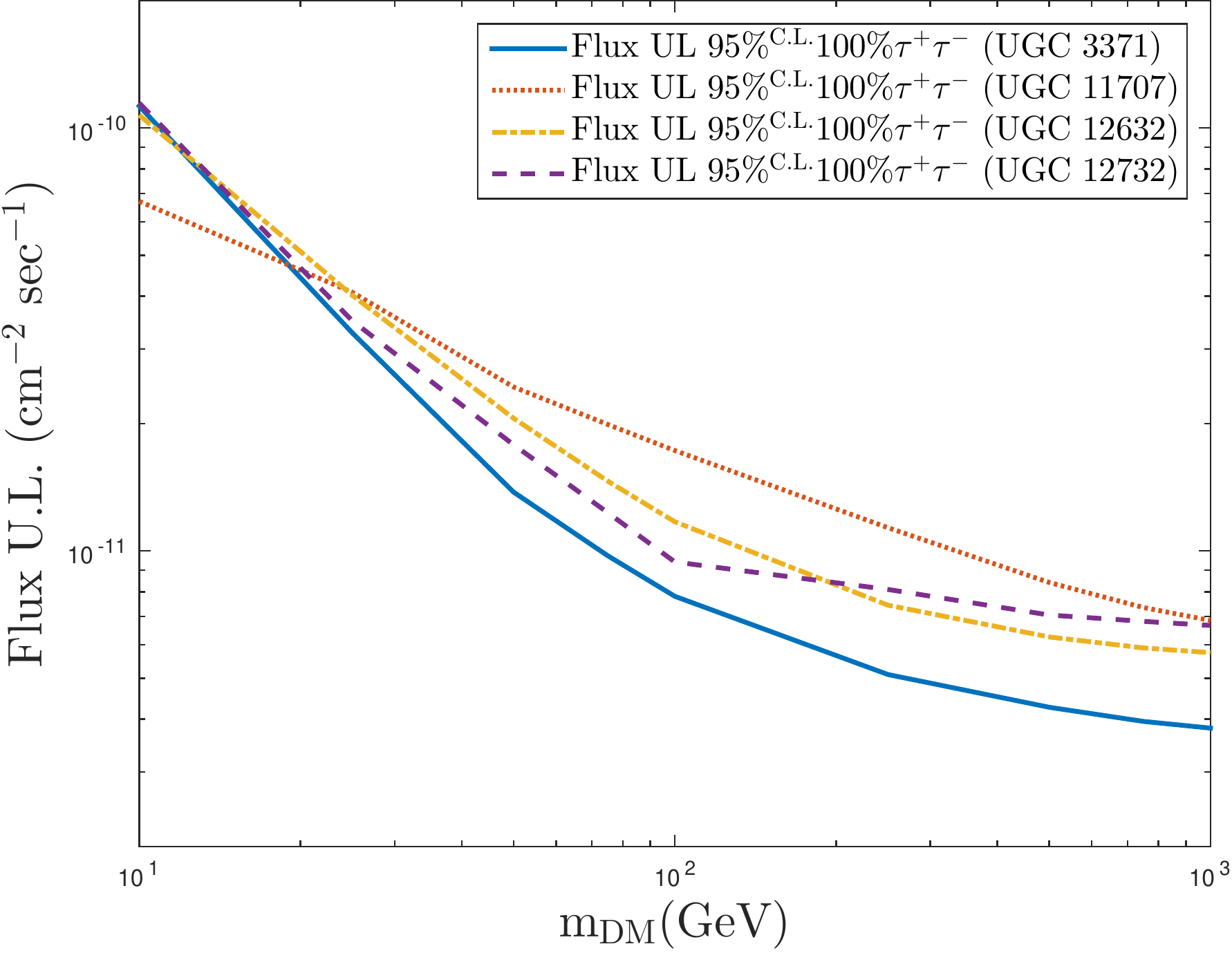}}
 \subfigure[]
 { \includegraphics[width=0.48\linewidth]{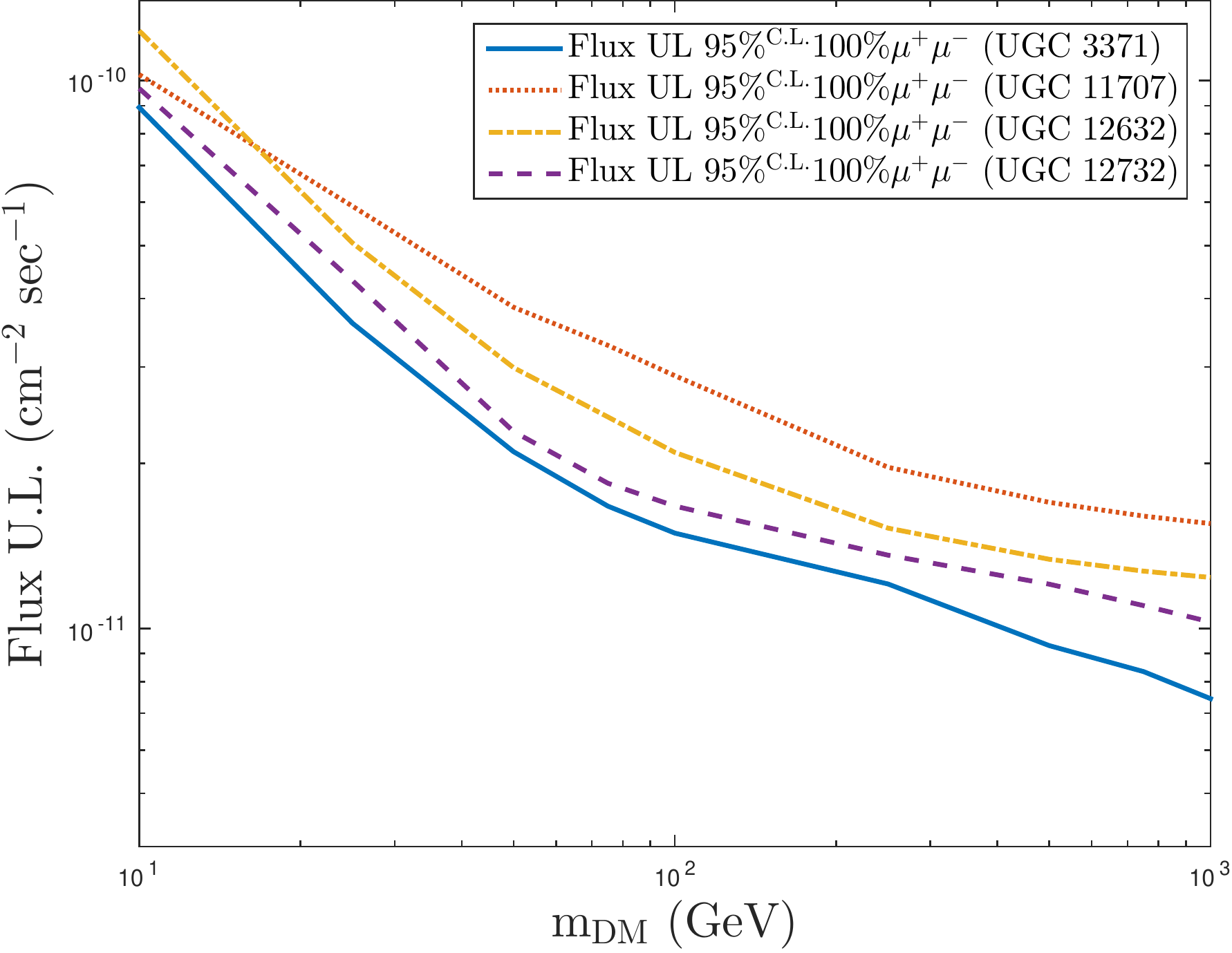}}
\subfigure[]
 { \includegraphics[width=0.48\linewidth]{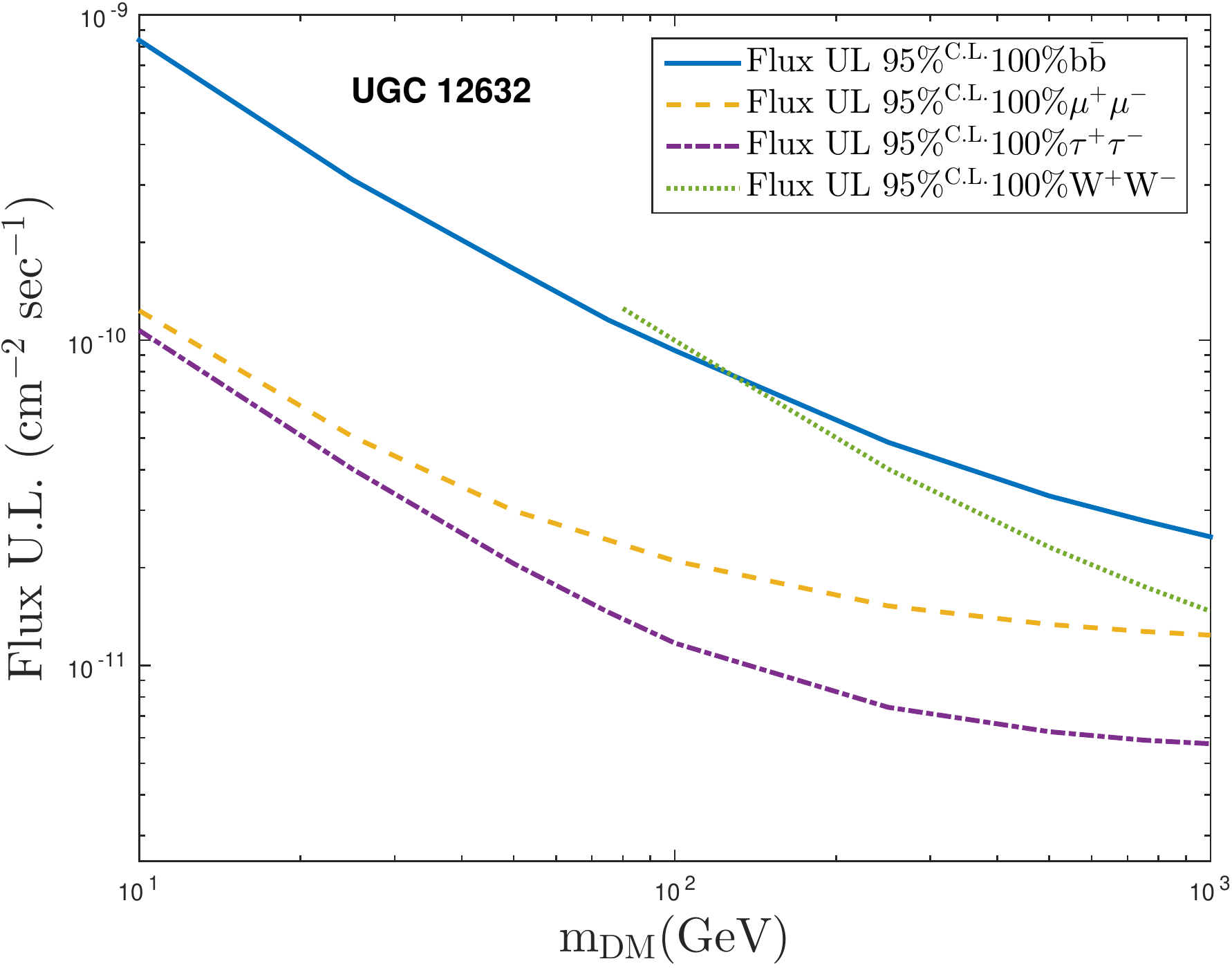}}
\caption{The $\gamma$-ray flux upper limits of all four LSB galaxies for three 
pair 
annihilation channels, such as: (a) the $100\%$ $b\overline{b}$, (b) the 
$100\%$ 
$\rm{\tau^{+} \tau^{-}}$, (c) the $100\%$ $\rm{\mu^{+} \mu^{-}}$. (d) It 
shows the variation of $\gamma$-ray flux upper limits for UGC 12632 with DM 
mass, $m_{DM}$ for
four annihilation channels. We have considered the median J-factor value from 
Table~7.4.}
\end{figure}

\begin{figure}
\subfigure[]
 { \includegraphics[width=0.48\linewidth]{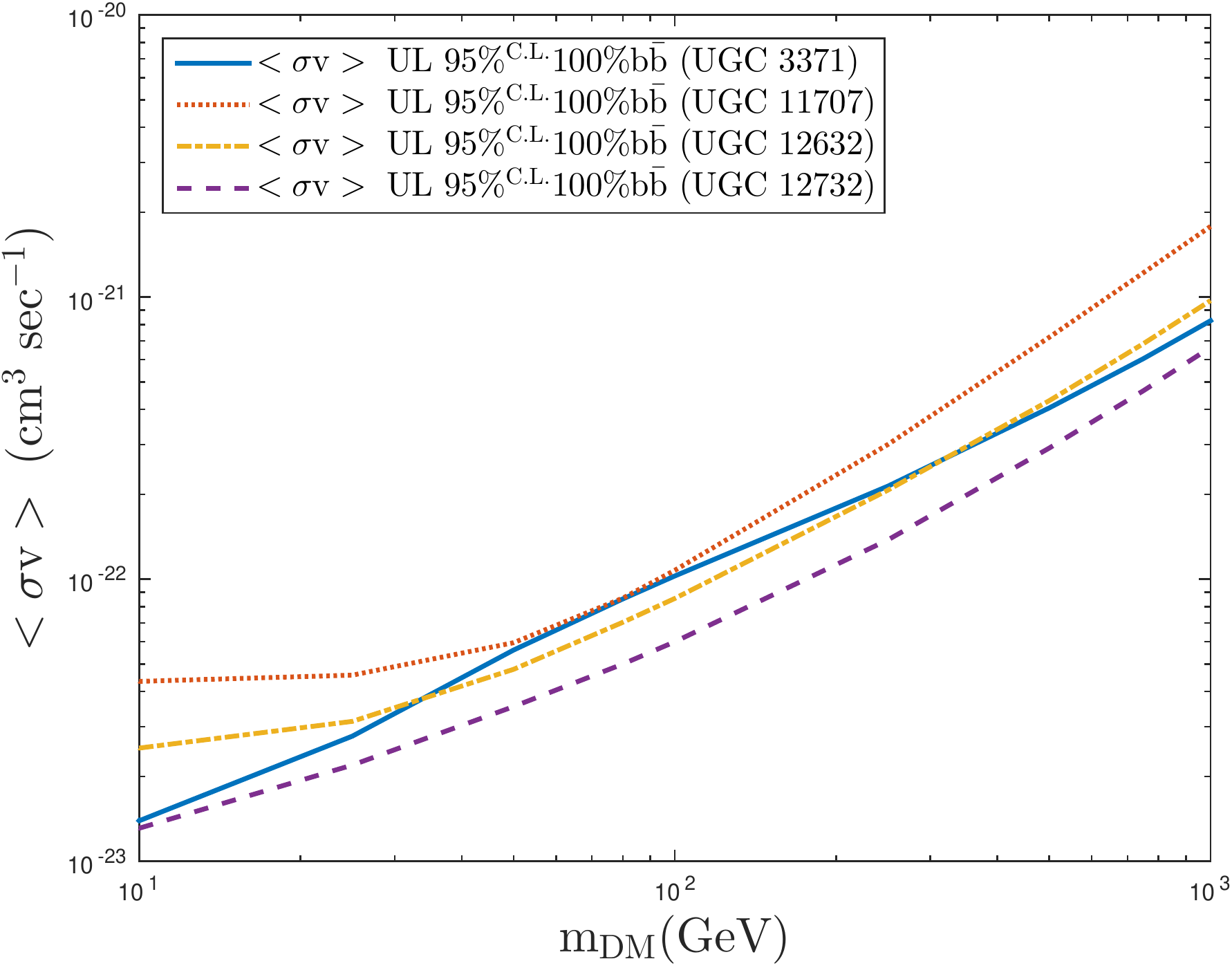}}
\subfigure[]
 { \includegraphics[width=0.48\linewidth]{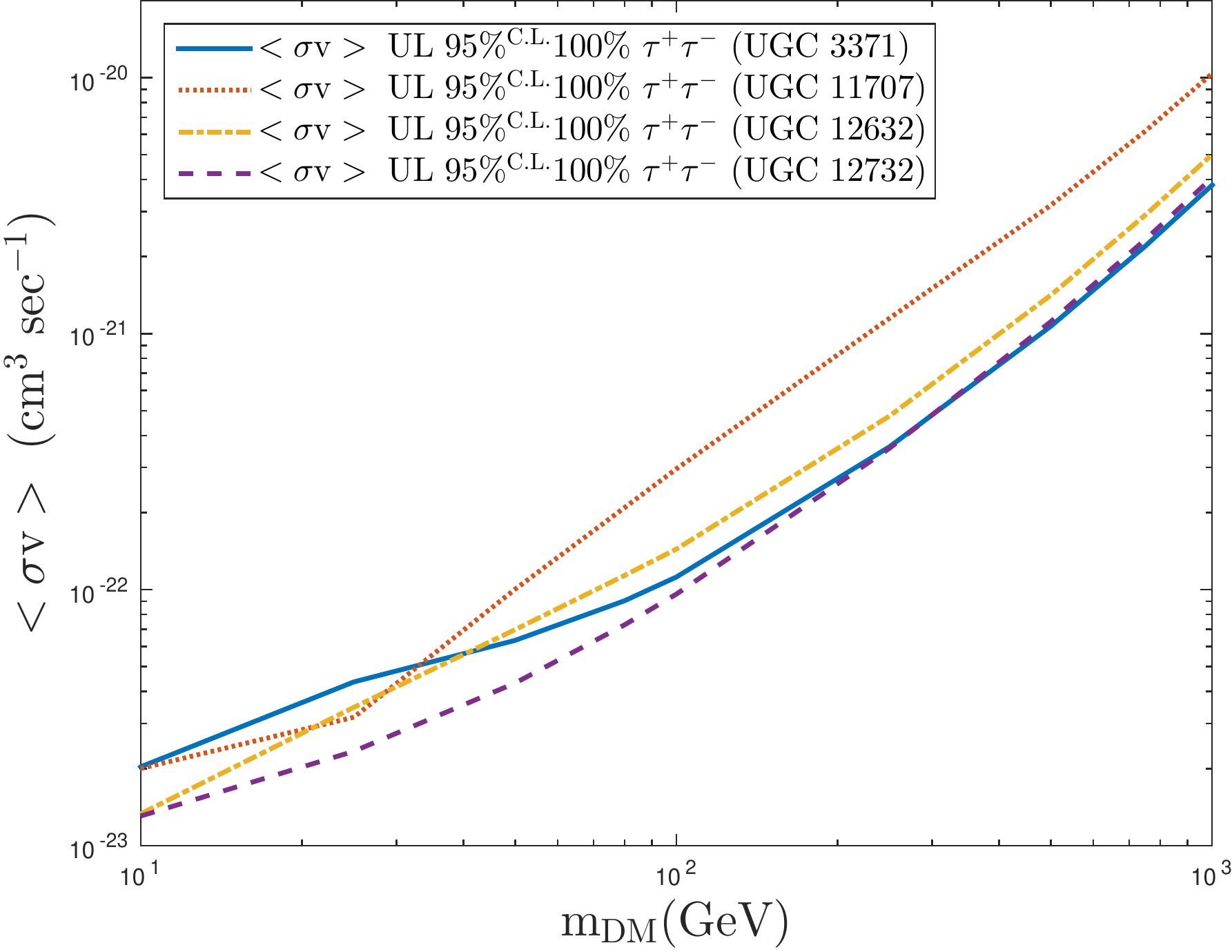}}
 \subfigure[]
 { \includegraphics[width=0.48\linewidth]{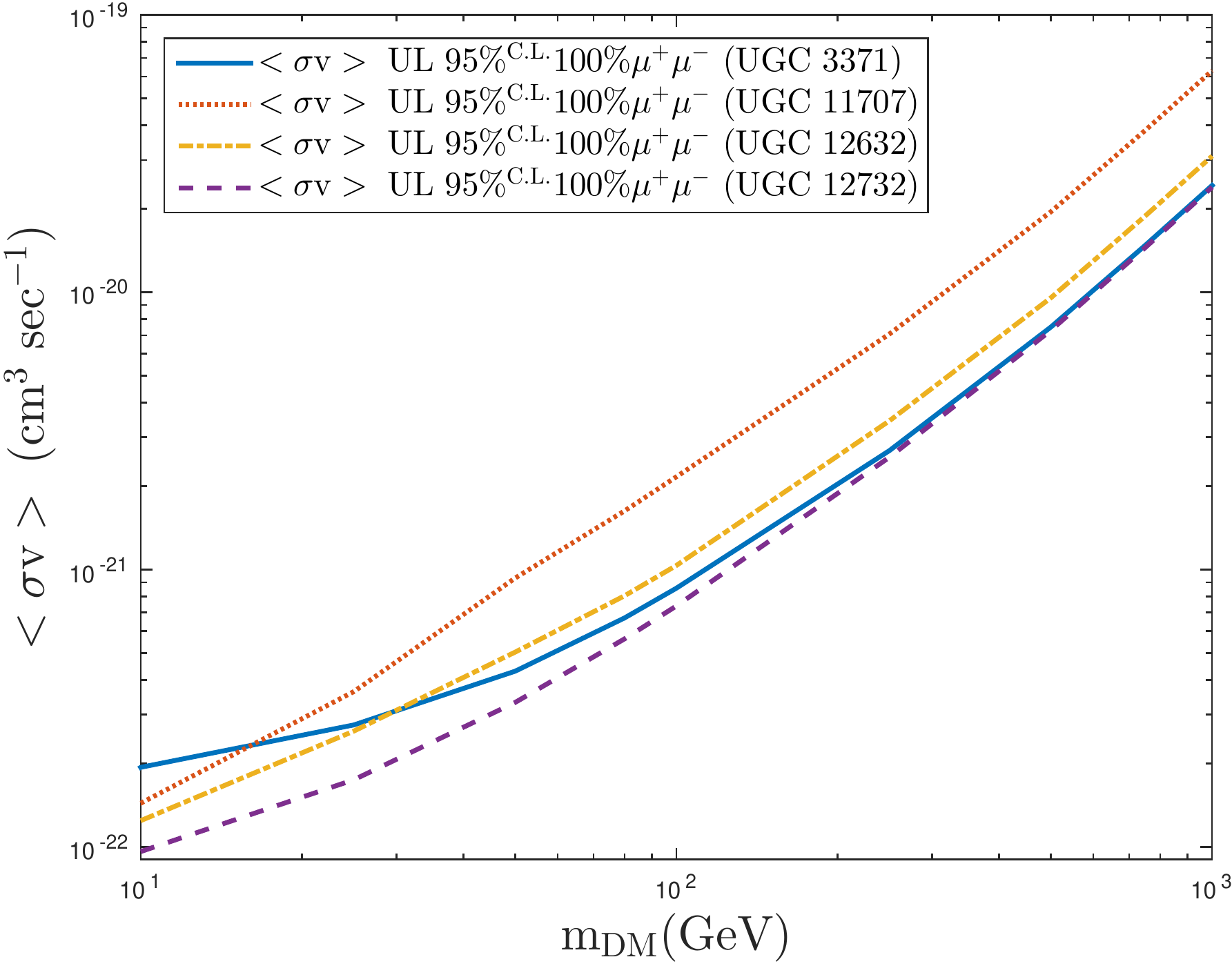}}
\subfigure[]
 { \includegraphics[width=0.48\linewidth]{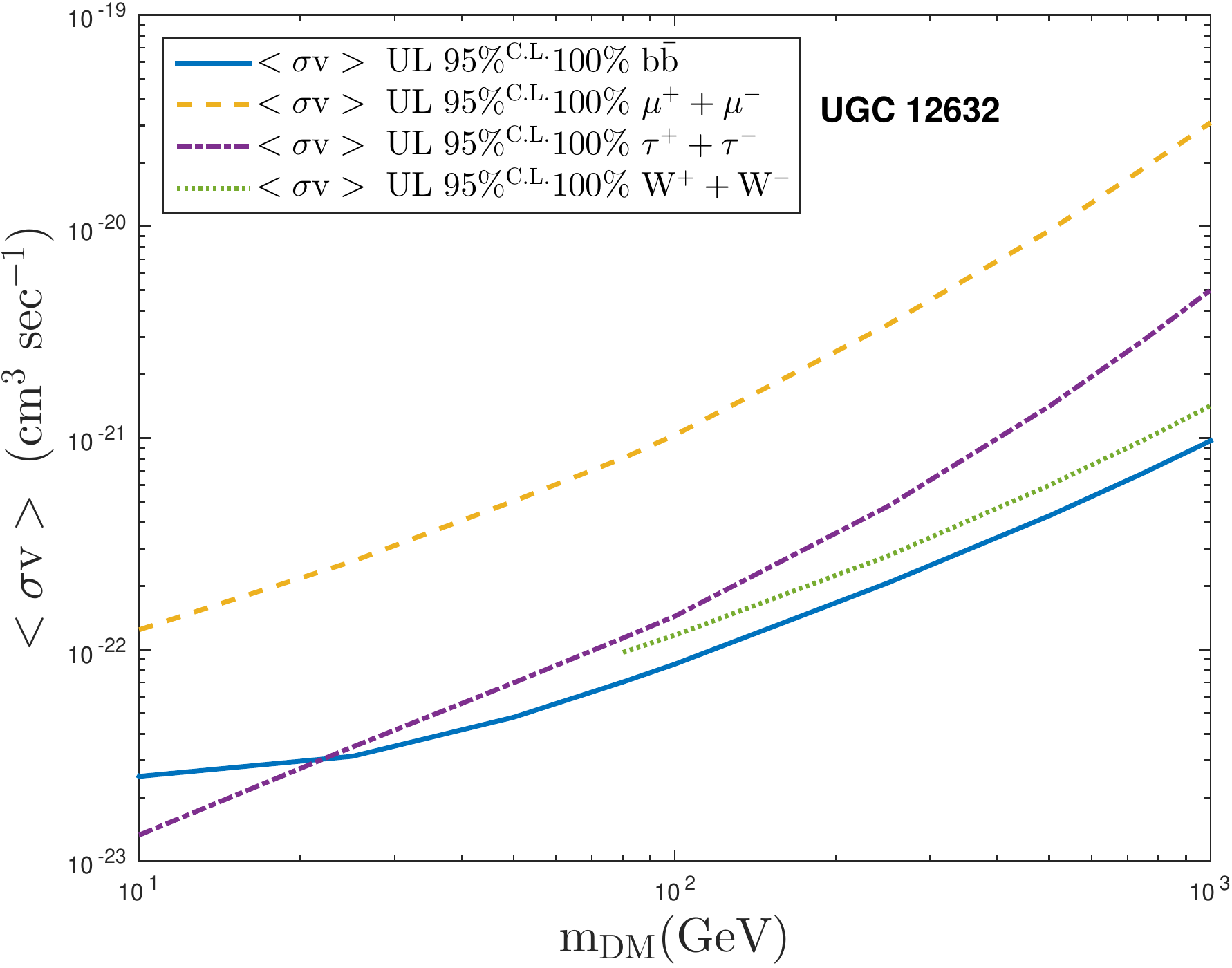}}
\caption{The $<\sigma~v>$ upper limit of all four LSB galaxies for three
annihilation channels, such: (a) the $100\%$ $b\overline{b}$, (b) the $100\%$ 
$\rm{\tau^{+} \tau^{-}}$, (c) the $100\%$ $\rm{\mu^{+} \mu^{-}}$. (d) It 
shows the variation of the upper limits on $<\sigma~v>$ for UGC 12632 with DM 
mas,$m_{DM}$ for 
four annihilation channels. We have considered the median J-factor value from 
Table~7.4.}
\end{figure}

\noindent In this section, we have studied the possible $\gamma$-ray flux 
upper limits in 95$\%$ C.L. resulting from WIMP annihilation and its relative 
thermally averaged 
pair-annihilation cross-section $<\sigma v>$ as a function of DM mass 
($\rm{m_{DM}}$) and WIMP annihilation final states (f) \cite{Bhattacharjee:2019jce} for each LSB galaxy 
using 
the 
DMFit tool \cite{Jeltema:2008hf, 
Gondolo:2004sc} and for that purpose, we have chosen four WIMP pair 
annihilation 
final states 
(f), such as, $100\%$ $b\overline{b}$, $100\%$ $\tau^{+}\tau^{-}$, $100\%$ 
$\rm{\mu^{+} \mu^{-}}$ and $100\%$ $W^{+}W^{-}$, 
respectively \cite{Jungman:1995df}. \\

\noindent In Figs.~7.3 (a,b,c) and 7.4 (a,b,c), we have displayed the 
$\gamma$-ray flux and the $<\sigma~v>$ upper limits as a function of DM mass, 
$m_{DM}$ for three pair annihilation channels, respectively, while in Figs.~7.3 
(d) and 7.4 (d), we have presented the variation for UGC 12632 \cite{Bhattacharjee:2019jce}. From 
Fig.~7.3(d), we find that at high energies where, the diffuse background is 
comparatively less, $\rm{\mu^{+} \mu^{-}}$ and $\rm{\tau^{+} \tau^{-}}$ channels 
provide the best $\gamma$-ray flux limits \cite{Bhattacharjee:2019jce}. From fig.~7.3(d), we can also notice that at 
around 1~TeV DM mass, the gamma-ray flux upper limits for four annihilation 
channels varies within a factor of $2$, whereas for low DM mass, this variation 
is increased to a factor of 4 \cite{Bhattacharjee:2019jce}. All our sources show the same nature, thus in 
figs.~7.3(d) and 7.4(d), we have only shown the result for UGC 12632. For 
obtaining the Figs~7.3 and 7.4, we have used the median J values (see 
Table~7.4) \cite{Bhattacharjee:2019jce}.

\begin{figure}
\begin{center}
\includegraphics[width=0.5\linewidth]{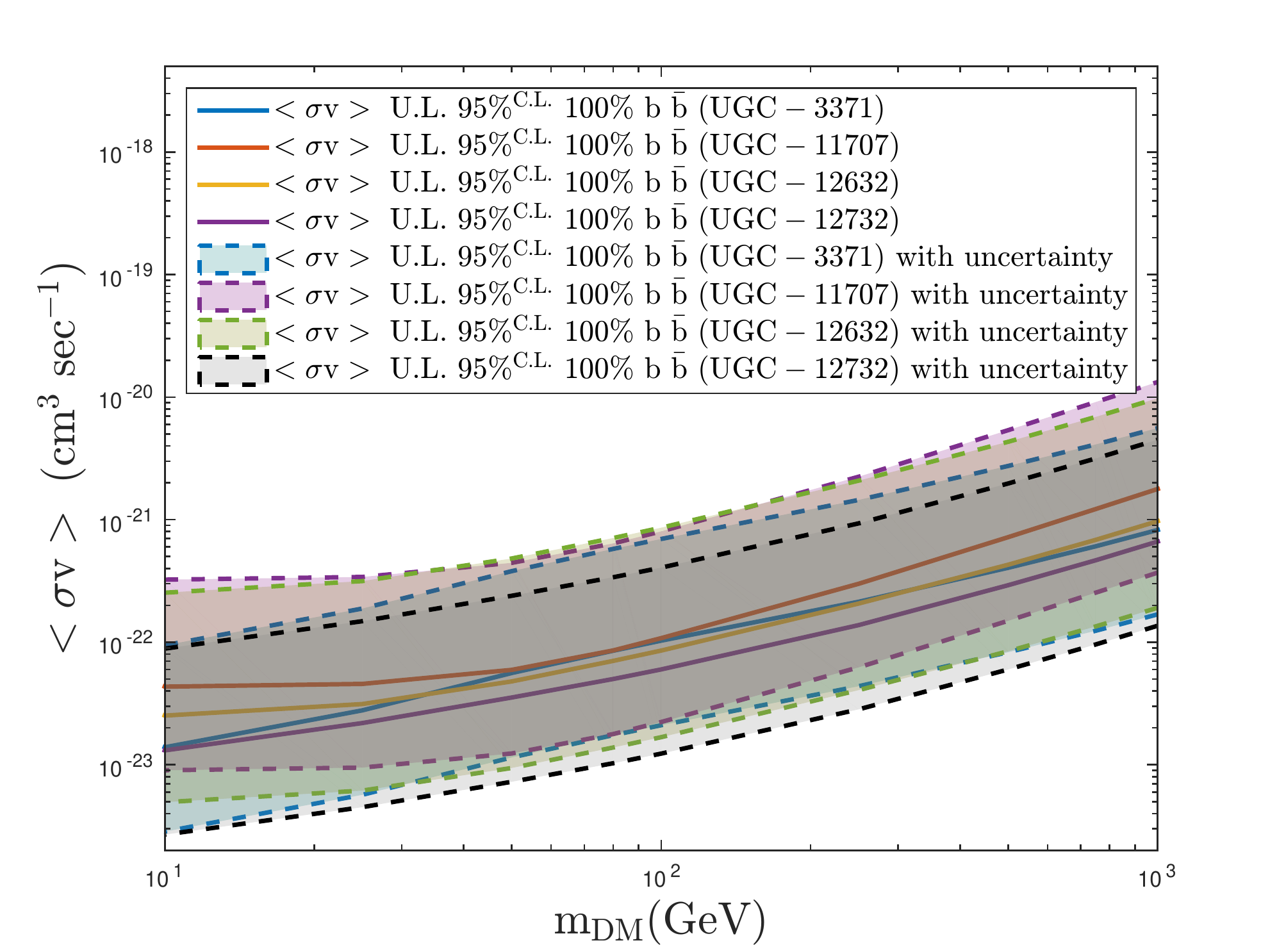}
\caption{The variation of $<\sigma v>$ upper limits in 95$\%$ C.L. with $m_{DM}$ 
for $b\overline{b}$ annihilation channels of four LSB galaxies is shown in the 
plane of ($m_{DM}$, $<\sigma v>$) for the median value of J-factor along with 
the
uncertainties. The shaded region refers to the uncertainty of the DM 
profiles for our LSB galaxies.}
\end{center}
\end{figure}

\begin{figure}
\begin{center}
\includegraphics[width=0.5\linewidth]{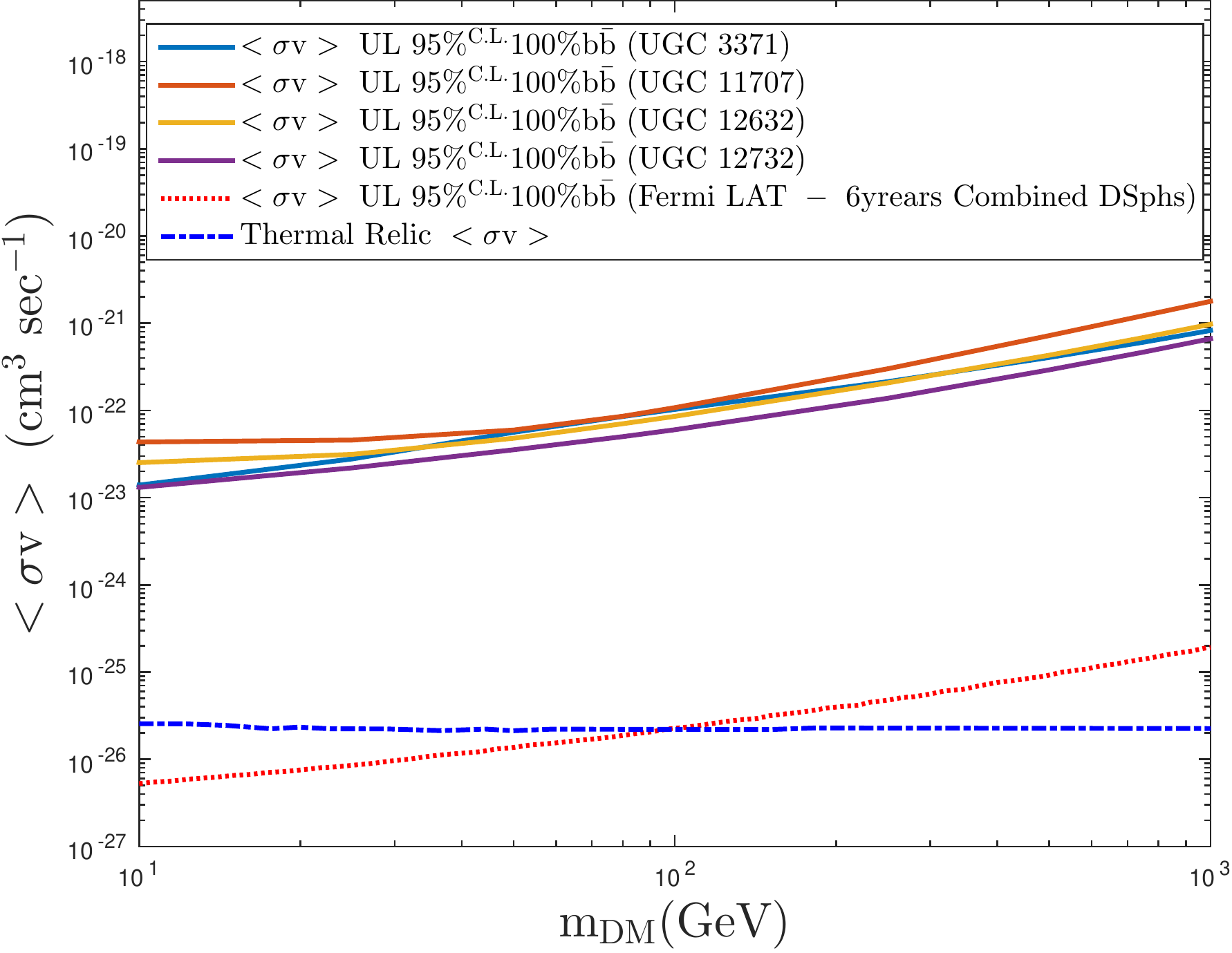}
\caption{The variation of $<\sigma v>$ upper limits in 95$\%$ C.L. with $m_{DM}$ 
for $b\overline{b}$ annihilation channels of four LSB galaxies is shown in the 
plane of ($m_{DM}$, $<\sigma v>$) for the median value of J-factor. The relic 
abundance 
cross-section rate i.e., $2.2~\times~10^{-26}~cm^{3}~s^{-1}$ derived by Steigman \textit{et al.}, 2012 and the combined 
$<\sigma v>$ upper limits obtained from the Fermi-LAT analysis of 15 dSphs by Ackermann \textit{et al.}, 2015 are overplotted here.}
\end{center}
\end{figure}

\noindent From Fig.~7.5, we have shown the variation of
$<\sigma v>$ in 95$\%$ C.L. with ${m_{DM}}$ for median value of
J-factor and its 2$\sigma$ C.L. uncertainties \cite{Bhattacharjee:2019jce}. We have only considered the
100$\%$ $b\overline{b}$ annihilation channel because for gamma-ray analysis, 
they put the most stringent limits on the parameter space of 
the (<$\sigma$~v>, $m_{DM}$). From Fig.~7.5, it is evident that LSB galaxies 
impose large uncertainty on the 
parameter space of the (<$\sigma$~v>, $m_{DM}$) and the uncertainty bands for 
all LSB galaxies are 
overlapping with each other \cite{Bhattacharjee:2019jce}. Thus, from this plot,
we won't be able to favour one LSB galaxies between all four \cite{Bhattacharjee:2019jce}. 
The very low rate of star formation and poor nuclear activity are considered as 
the primary 
reasons for the large uncertainties associated with the DM distribution in LSB 
galaxies. \\

\noindent Next, we have performed a comparative study between the $<\sigma v>$ 
limits obtained 
from the LSB galaxies with the limits derived by the 
Ackermann et al.\cite{Ackermann:2015zua} and the same is 
shown in Fig.~7.6 \cite{Bhattacharjee:2019jce}. The limits obtained from Ackermann et al.\cite{Ackermann:2015zua} 
performed the analysis on 15 dSphs with six years of LAT data. In Fig.~7.6, we 
have also 
compared the limits from LSB galaxies with the thermal relic cross-section rate 
estimated by~Steigman et al.\cite{Steigman:2012nb}. \\

\noindent In Fig.~7.6, the thermal cross-section rate 
obtained by the study of~Steigman et al.\cite{Steigman:2012nb} is denoted by the blue 
``dot-dashed ''line while the $<\sigma v>$ limits 
derived by Ackermann et al.\cite{Ackermann:2015zua} is represented by the red ``dotted''
line. From Fig.~7.6, it is clear that the $<\sigma v>$ limits obtained from our 
four LSB galaxies are roughly 3 orders of magnitude weaker than the limits 
achieved by the 
Ackermann et al.\cite{Ackermann:2015zua} and the Steigman et al.\cite{Steigman:2012nb}. In our next 
section, we would estimate the stacking limits for LSB galaxies \cite{Bhattacharjee:2019jce}.

\subsection{Stacking Analysis}
\noindent In Section 7.4.3, we have estimated the $<\sigma v>$ upper limits for 
individual LSB galaxies and from Fig. 7.6,
we have checked that the individual $<\sigma v>$ limits are around 3 orders of 
magnitude weaker \cite{Bhattacharjee:2019jce} than the limits estimated by the combined analysis of 
Ackermann et al.\cite{Ackermann:2015zua} and the annihilation rate for relic abundances 
derived by Steigman et al.\cite{Steigman:2012nb}. In this section, in order to increase the sensitivity of the limits, we have preferred to derive 
the stacking limits on the individual $<\sigma v>$ limits obtained from each LSB galaxies \cite{Bhattacharjee:2019jce}. In Chapter 4, we have already 
discussed the formalism for stacking likelihood function. For this work, to 
estimate the stacking limits, we have used the Eq.~4.11. \\

\noindent The J-factor provides a rough estimation on WIMP signal coming from 
the DM rich sources, thus the stacking analysis would be able to generate a more 
stringent result than the limits obtained from any individual LSB galaxy \cite{Bhattacharjee:2019jce}. Even 
for the combined analysis,
we have not observed any gamma-ray emission from the location of LSB. Thus we 
have computed the $<\sigma v>$ upper limits in 95$\%$ C.L. by the 
delta-likelihood method \cite{Bhattacharjee:2019jce}. In Fig.~7.7(a), we have shown the $<\sigma v>$ upper 
limits as a function of $m_{DM}$ obtained from the stacking analysis and have 
compared it with the individual limits of LSB galaxies for $100\%$ 
$b\overline{b}$ final state. From Fig.~7.7(b), we can find the comparison 
between the stacking $<\sigma v>$ limits for LSBs and the $<\sigma v>$ limits 
taken from Ackermann et al.\cite{Ackermann:2015zua} with the thermal annihilation rate from 
Steigman et al.\cite{Steigman:2012nb}. For Fig.~7.7(b), the 2$\sigma$ uncertainty band 
associated with stacking limits of $<\sigma v>$ has been displayed \cite{Bhattacharjee:2019jce}.

\begin{figure}
\subfigure[]
 { \includegraphics[width=0.48\linewidth]{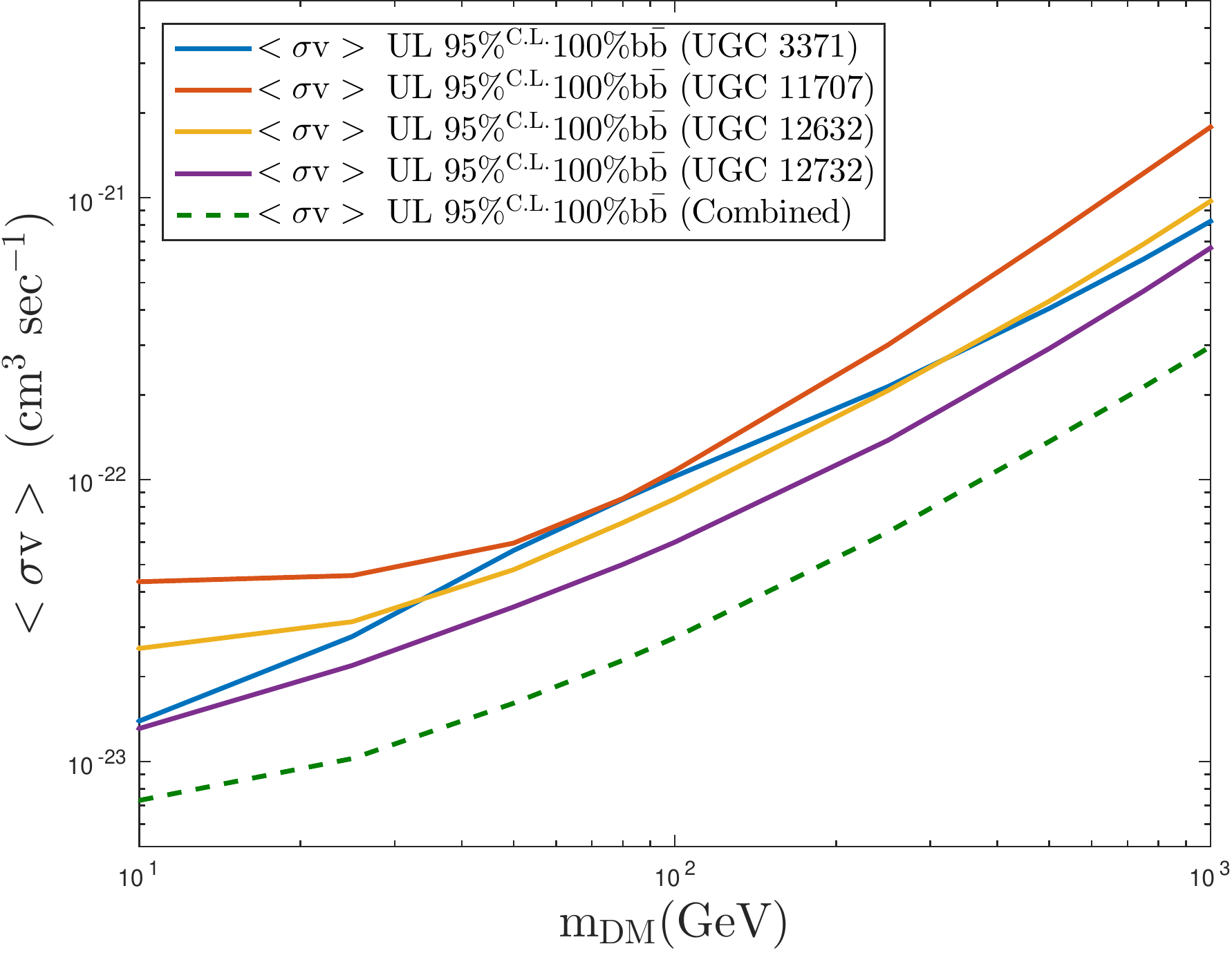}}
\subfigure[]
 { \includegraphics[width=0.48\linewidth]{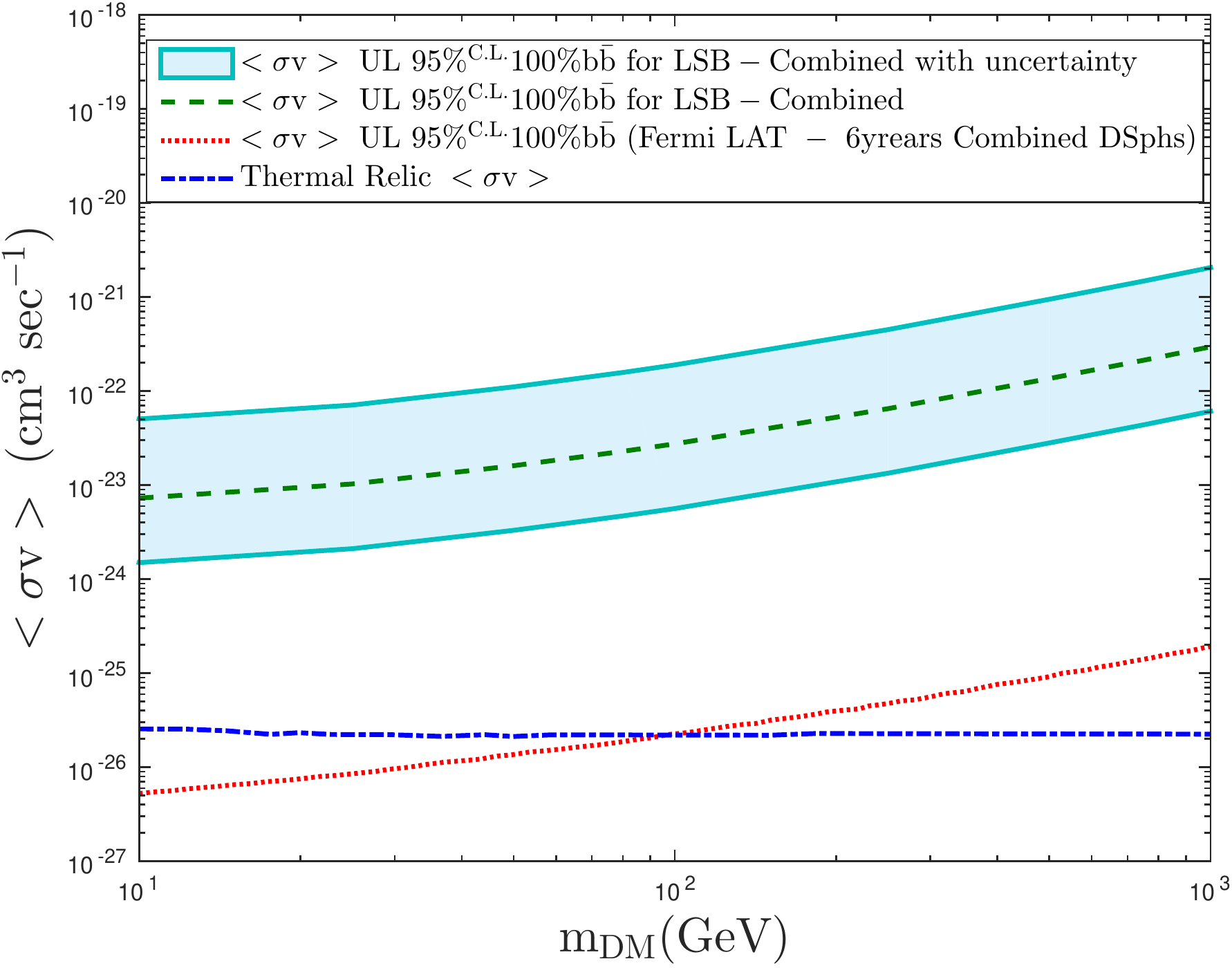}}
\caption{(a) The comparison between the upper limits on $<\sigma v>$ obtained 
from LSB for
the stacking analysis and the $<\sigma v>$ limits obtained from individual LSB 
for the $100\%$ $b\overline{b}$
annihilation channel. 
b) The comparison between the upper limits on $<\sigma v>$ obtained from LSB for
the stacking analysis and the relic abundance 
cross-section rate i.e., $2.2~\times~10^{-26}~cm^{3}~s^{-1}$ derived by Steigman \textit{et al.}, 2012 and the combined 
$<\sigma v>$ upper limits obtained from the Fermi-LAT analysis of 15 dSphs by Ackermann \textit{et al.}, 2015. The shaded region refers to the uncertainty associated 
with the stacking limits.}
\end{figure}

\noindent From Fig.~7.7, it is evident that the stacking limit of $<\sigma~v>$ 
has been improved by a factor of $\approx$ 4 from the individual limit obtained 
from LSB galaxies, but it is still nearly two orders of magnitude weaker \cite{Bhattacharjee:2019jce} than 
the limits obtained from
Ackermann et al. \cite{Ackermann:2015zua} and Steigman et al. \cite{Steigman:2012nb}. \\

\noindent We may then conclude that at present due to low 
J-values (roughly 2-3 orders weaker than the standard values for dSphs/UFDs), the 
$\gamma$-ray $<\sigma v>$ limits obtained for the LSB galaxies, are unable to 
produce any stringent limits on the theoretical WIMP models. But in the future, the next generation optical surveys such as Large Synoptic Survey Telescope (LSST) are designed to
discover many new LSB galaxies. Thus, the constraint limits on theoretical DM models obtained from LSB galaxies might
improve significantly. 

\subsection{Possible Radio Constraint Obtained from LSB Galaxies}

\begin{figure}
\subfigure[UGC 3371]
 { \includegraphics[width=0.48\linewidth]{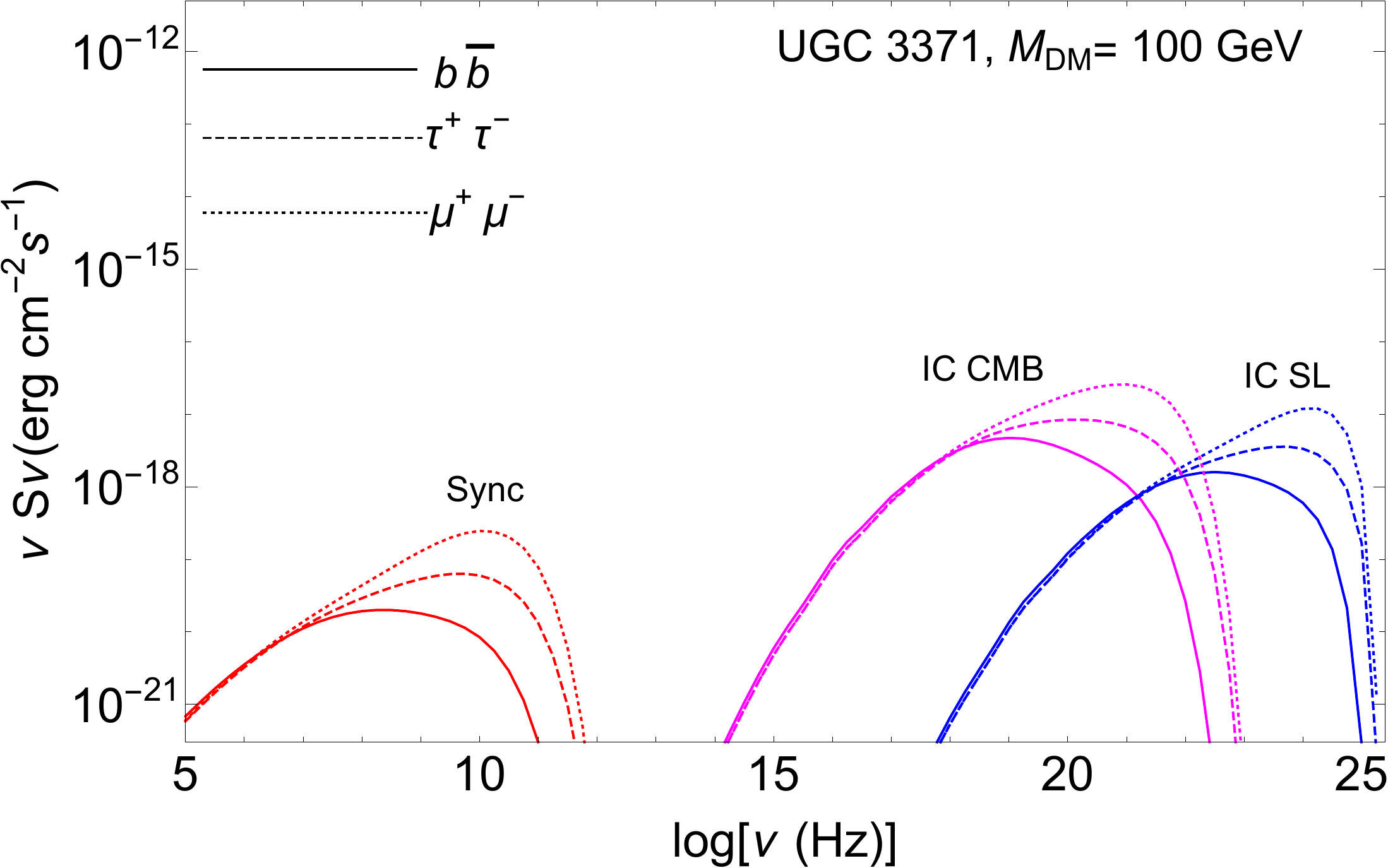}}
\subfigure[ UGC 11707]
 { \includegraphics[width=0.48\linewidth]{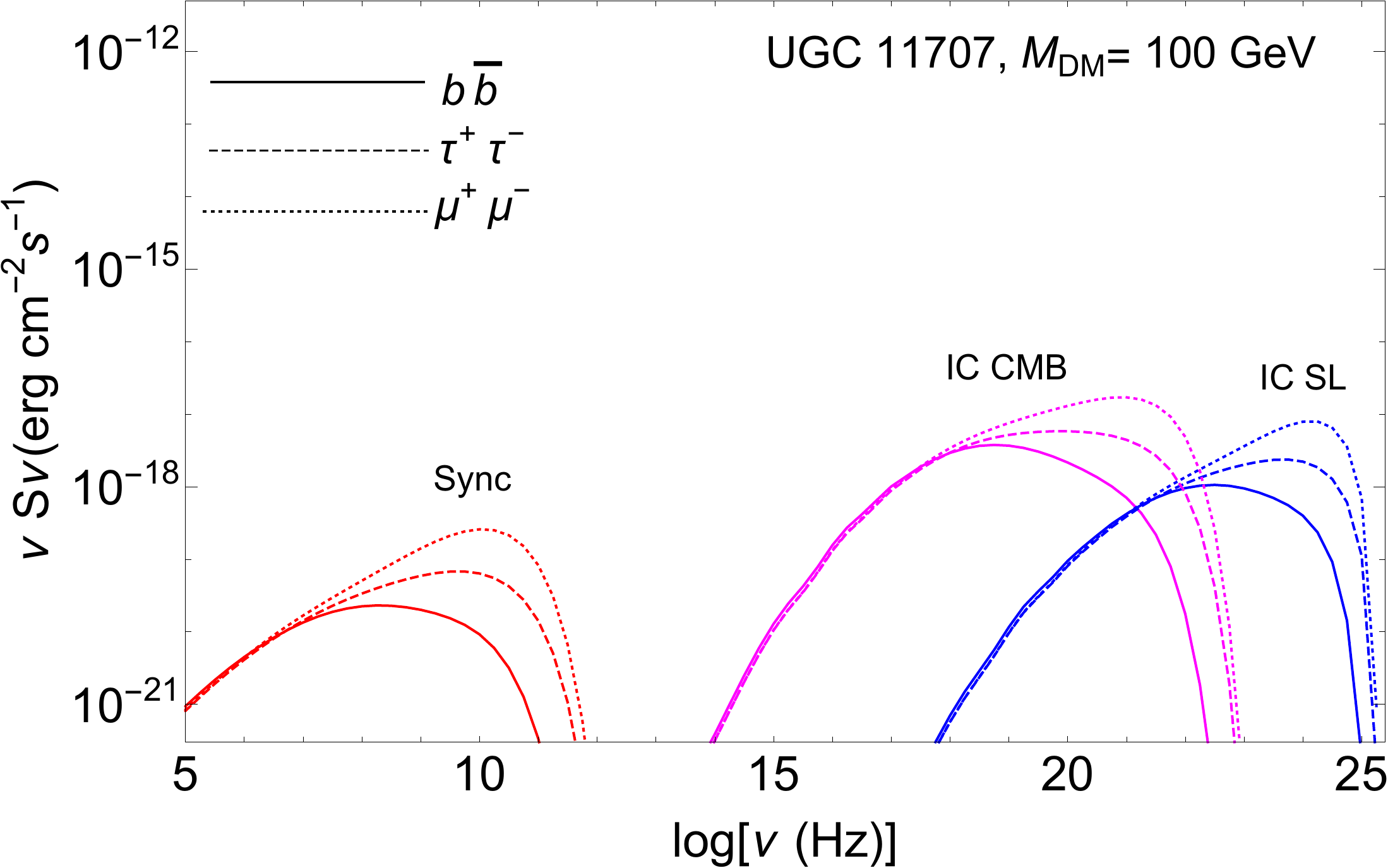}}
\subfigure[ UGC 12632]
 { \includegraphics[width=0.48\linewidth]{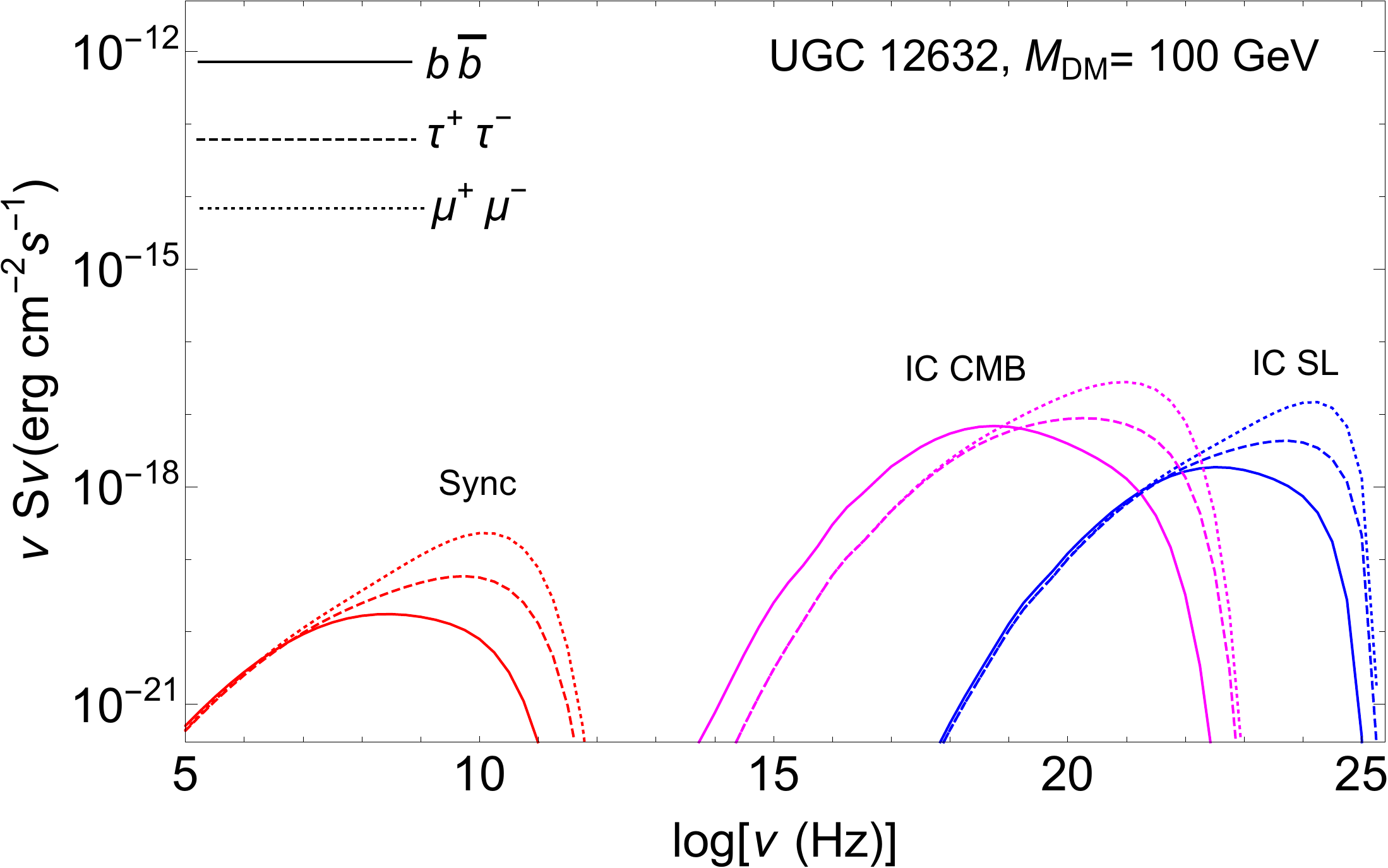}}
\subfigure[ UGC 12732]
 { \includegraphics[width=0.48\linewidth]{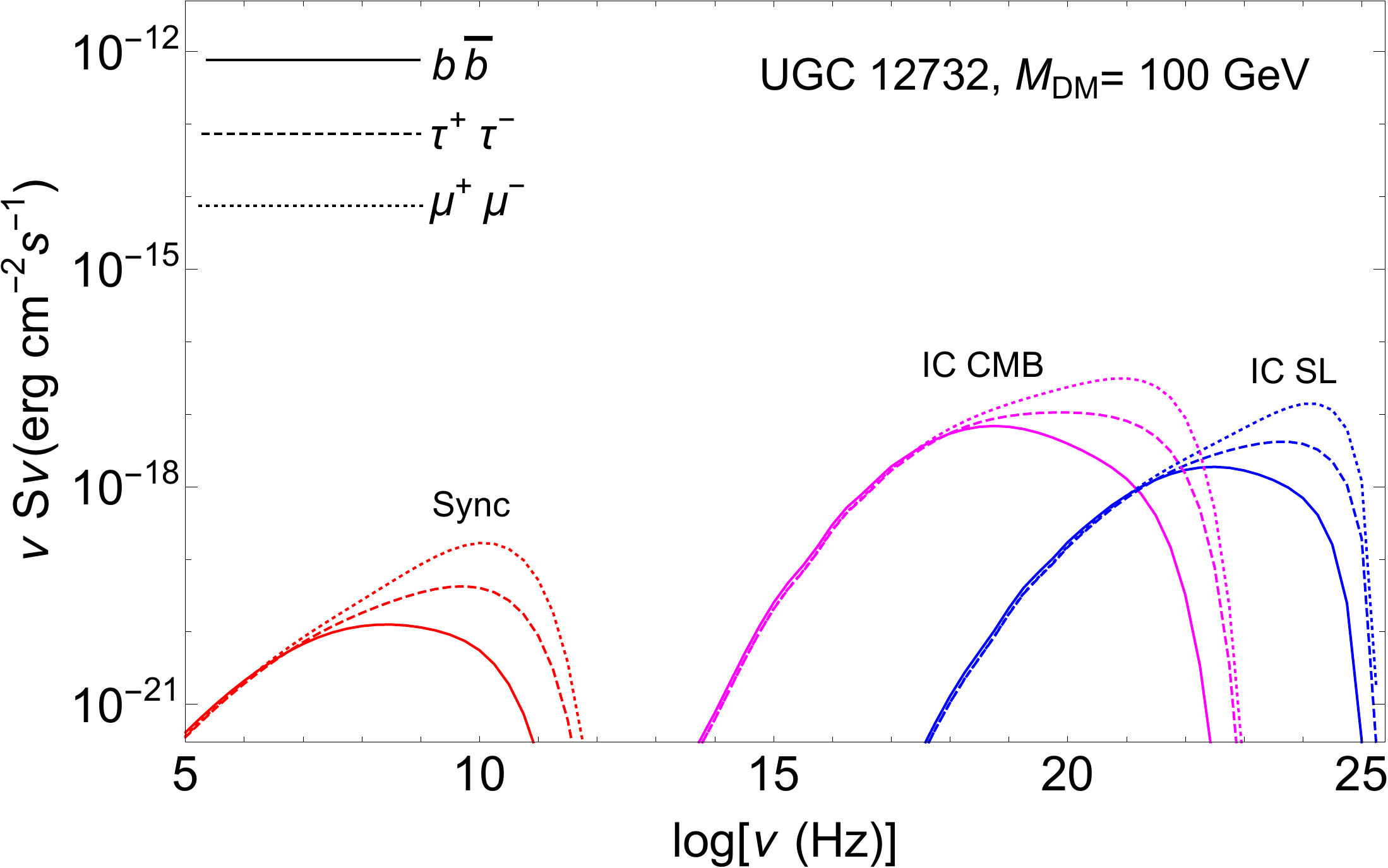}}
\caption{The multiwavelength SED of four LSB galaxies for three DM 
annihilation final states, such as the $b\overline{b}$ (solid), the 
$\tau^{+}\tau^{-}$ (dashed) and the $\mu^{+}\mu^{-}$ (dotted).
We have considered $m_{DM}$=100 GeV , $B_{0}$ = 1$\mu$G and 
$D_{0}$=$3\times10^{28}~(cm^{2}s^{-1})$.}
\end{figure}

\begin{figure}
\centering
\subfigure[Variation with $B_{0}$]
 { \includegraphics[width=0.49\linewidth]{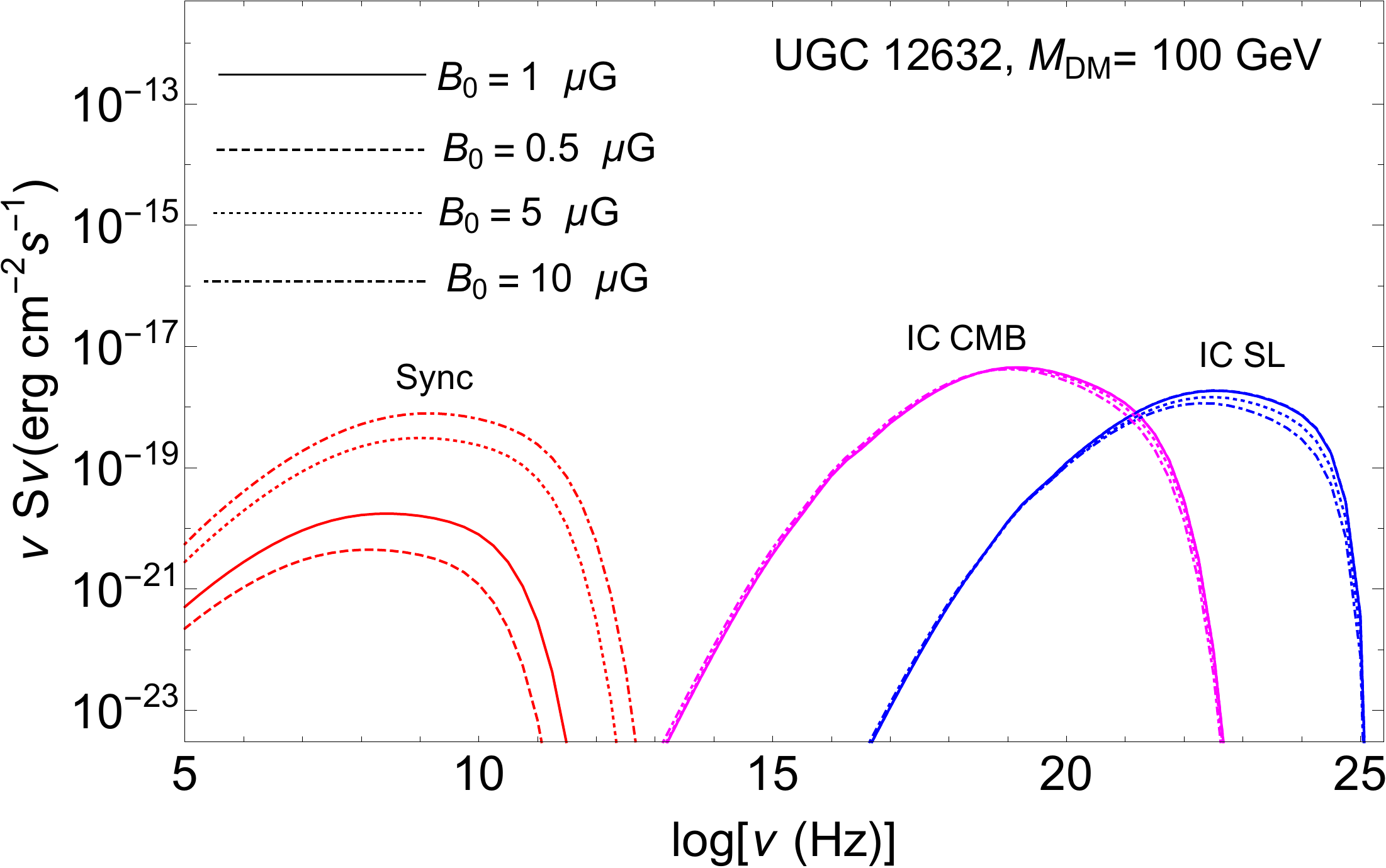}}
\subfigure[Variation with $D_{0}$]
 { \includegraphics[width=0.49\linewidth]{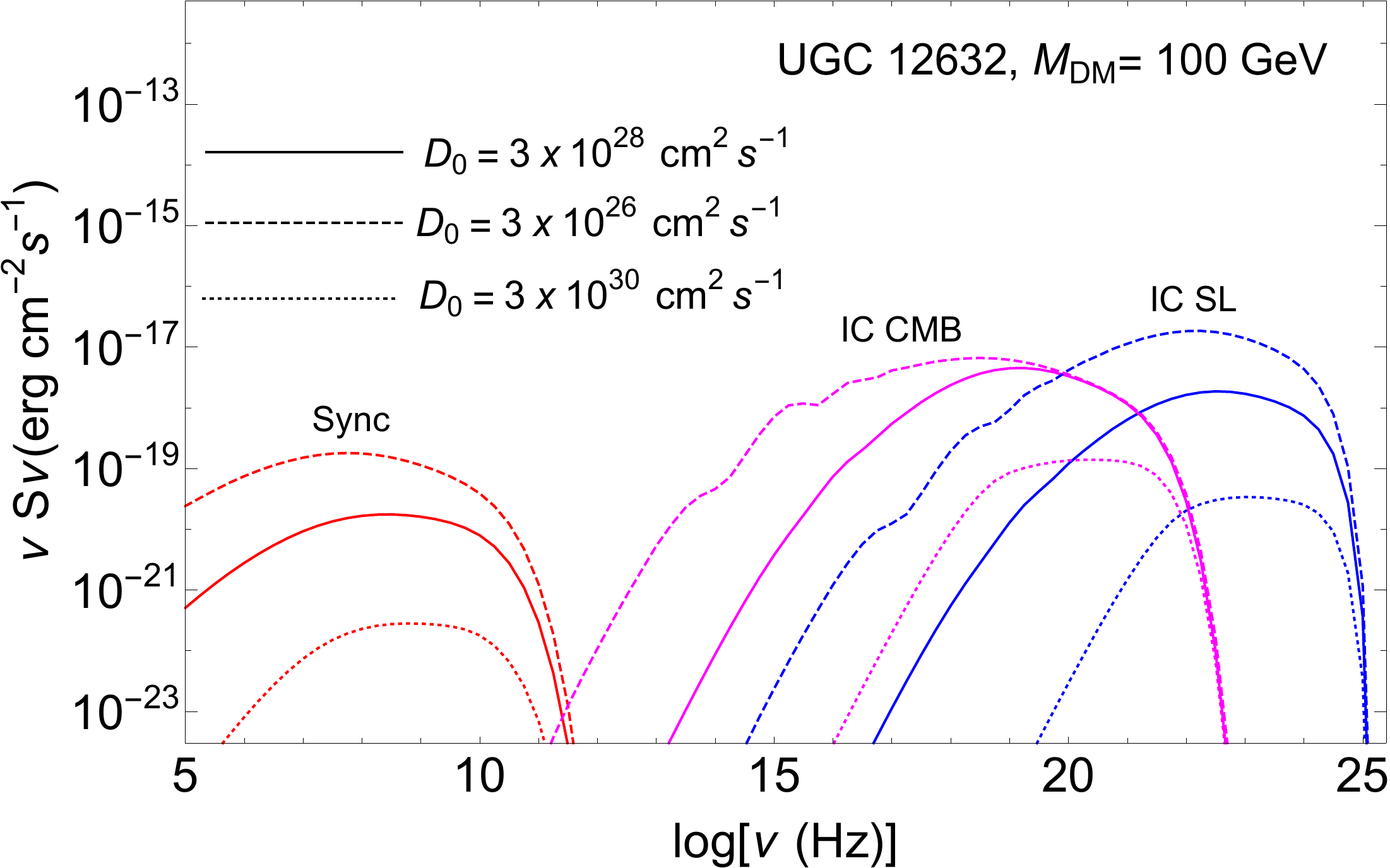}}
\subfigure[Variation with $\gamma_{D}$]
 { \includegraphics[width=0.5\linewidth]{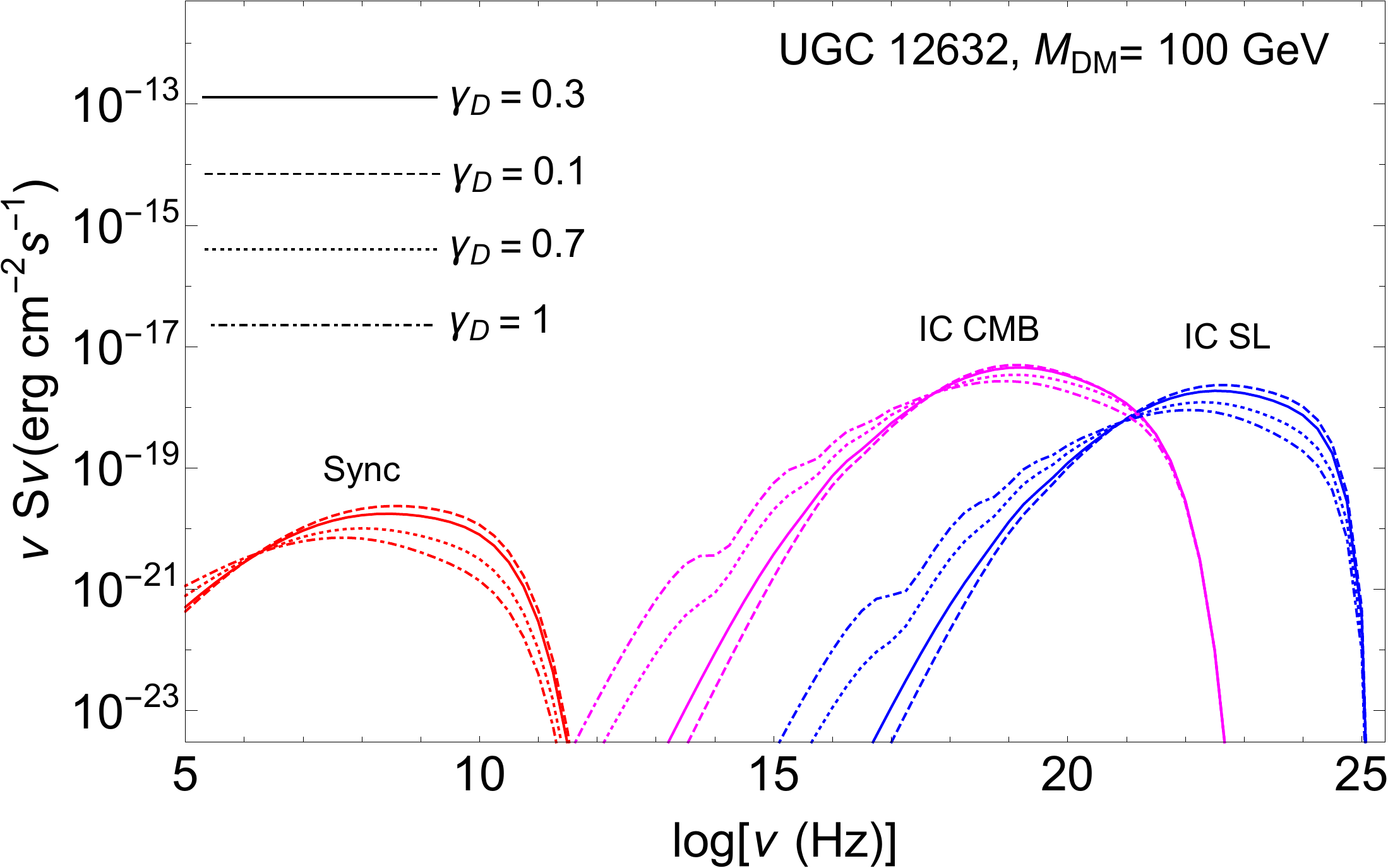}}
\caption{The variation of the multiwavelength SED of UGC 12632 for (a) four 
values of 
$B_{0}$, (b) three values of $D_{0}$ and (c) four values of 
$\gamma_{D}$. We have considered $m_{DM}$=100 GeV, $B_{0}$ = 1$\mu$G, 
$D_{0}$=$3\times10^{28}~(cm^{2}s^{-1})$ and have fixed the thermal 
averaged $<\sigma v>$) to $3 
\times 10^{26}~cm^{3}s^{-1}$.}
\end{figure}

\noindent In the earlier section, we find that with $\gamma$-ray data, LSB 
galaxies could not impose strong limits on DM models. Thus, in this section, we 
have tried to investigate the radio emission that might come from the WIMP 
annihilation \cite{Bhattacharjee:2019jce}.\\

\noindent In order to 
estimate the radio and the X-ray emission resulting from the DM annihilation, 
we 
have solved 
the diffusion equation for the secondary electron spectrum (Eq.~2.11). In 
chapter~2, we 
have already defined the formulation for the radio and the X-ray emission 
through DM annihilation \cite{Colafrancesco:2005ji, 
Colafrancesco:2006he}.\\

\noindent Here, we have used a publicly accessible 
code, 
RX-DMFIT \cite{McDaniel:2017ppt}. 
This code is an extension of the DMFit tool \cite{Jeltema:2008hf, Gondolo:2004sc} that we have earlier used to investigate the DM signal
from $\gamma$-ray data. With the RX-DMFIT code, it might be 
possible to predict the flux limits from the radio and the X-ray emission 
resulting 
from the secondary charged particles which are assumed to be generated from the 
DM annihilation. For radio analysis, we have modelled the DM 
density distribution of LSB galaxies with the NFW profile \cite{Bhattacharjee:2019jce}. In order to calculate 
the source term for DM signal i.e., $Q_{e}$ (check Eq.~2.11), RX-DMFIT tools uses 
the set of Fortran packages from the DarkSUSY 
v5.1.2 which is designed to estimate the $e^{+}/e^{-}$ injection 
spectrum per DM 
annihilation event (i.e., $\sum_{f} \frac{dN^{e}_{f}}{dE}B_{f}$) for any approved 
range of DM masses and DM annihilation final states \cite{Bhattacharjee:2019jce}. \\

\noindent RX-DMFIT gives us access to customize the wide range of parameter sets 
for the astrophysical and the particle components \cite{McDaniel:2017ppt}. With this code, we can check 
how does the diffusion mechanism, magnetic field, DM distribution etc. can 
possibly affect the radio and the X-ray emission from LSB galaxies.\\

\noindent As we already mentioned in sections 7.1 and 7.2, there are not many 
observational studies on the LSB galaxies 
and thus it is difficult to preciously have any information on 
their magnetic fields and diffusion mechanism. But fortunately, the systematics 
of the dSphs are not very different from the LSB galaxies, so for our 
calculation, we have used the values of diffusion constant ($D_{0}$) and 
magnetic field (B) that are generally favoured for the dSphs \cite{Bhattacharjee:2019jce}. We have defined 
the diffusion coefficients of the LSB galaxies by the Kolmogorov form (i.e., 
$\rm{D(E) = D_{0} 
E^{\gamma}}$) where, the diffusion zone i.e., $r_{h}$ is assumed to be equal to 
the $2~\times~R_{last}$ (see Table~7.1). We have also fixed the values of $D_{0}$ and  $\gamma_{D}$ at 
$3~\times~10^{-28}~cm^{-2}~s^{-1}$ and 0.3, respectively \cite{Bhattacharjee:2019jce}. For LSB galaxies, 
there is no such detailed study on the distribution of the magnetic field and 
thus we do not have any knowledge on the spatial extension of their magnetic fields \cite{Bhattacharjee:2019jce}. 
Thus we have used the exponential form to define the magnetic field of the LSBs. 
The expression for magnetic field is, $\rm{B(r) = B_{0}~e^{\frac{-r}{r_{c}}}}$, 
where, we have fixed the $B_{0}$ at 1$\mu G$\cite{Fitt:1993dfr} and $r_{c}$ 
defines the core radius of the LSB which is equal to the $r_{d}$ (see 
Table~7.1) \cite{Bhattacharjee:2019jce}. Here, we have also fixed the $<\sigma v>$ at the 
$3\times10^{-26} cm^{3} s^{-1}$. In Table~7.6, we have shown all the parameter 
values that we have used for our radio analysis \cite{Bhattacharjee:2019jce}.\\

\noindent Using the parameter set mentioned in Table~7.6, we have tried to 
predict the spectral energy distribution i.e., SED in multiwavelength range for 
all four LSB galaxies at 100 GeV of DM mass \cite{Bhattacharjee:2019jce}. From Fig. 7.8, we can find the SED 
plots for three DM annihilation channels, where, the synchrotron emission is 
defined by `Sync' and the IC emission due to starlight and CMB photons are 
defined by `IC SL' and `IC CMB', respectively \cite{Bhattacharjee:2019jce}. The SED plots that we have shown 
in Fig.7.8 are dependent on our choice of parameter sets. So, next, we would try 
to find how the SED plot would be affected by changing the astrophysical 
parameters \cite{Bhattacharjee:2019jce}. In Fig.~7.9, we have shown the variation of SED plots with $B_{0}$, 
$D_{0}$ and $\gamma_{D}$ and for our purpose, we have only chosen UGC 12632 and 
$b\overline{b}$ final state \cite{Bhattacharjee:2019jce}. From Fig.~7.9(a), it is evident that the magnetic 
field has the direct impact on the synchrotron emission and high B field would 
increase the emission, while the IC emission is not much affected by the 
variation of B field \cite{Bhattacharjee:2019jce}. Next, from Fig.~7.9(b), we can find that both synchrotron 
and IC emission are strongly dependent on the $D_{0}$ \cite{Bhattacharjee:2019jce}. Last, from Fig.~7.9(c), 
we can check how SED would vary with changing the $\gamma_{D}$. Here we would like 
to mention that $\gamma_{D}$ is associated with the Kolmogorov form of 
diffusion coefficient \cite{Bhattacharjee:2019jce}.

\begin{table}
\begin{center}
\caption{The parameter set used as the input of RX-DMFIT tool.}
\begin{tabular}{|p{1.5cm}|p{1cm}|p{1cm}|p{2cm}|p{1cm}|p{1cm}|p{1cm}|p{1.5cm}|p{1cm}|}
\hline 
\hline
Galaxy & d & $r_{h}$ & $D_{0}$ & $\gamma_{D}$ & $B_{0}$ & $r_{c}$ & $\rho_{s}$ & $r_{s}$ \\
$ $ & Mpc & Kpc & $cm^{2}s^{-1}$ & & $\mu G$ & Kpc & $\frac{GeV}{cm^{3}}$ & kpc\\
\hline
UGC 3371 & 13.1 & 20.4 & $3\times10^{28}$ & 0.3 & 1 & 3.09 & 0.5725 & 6.5151 \\
\hline
UGC 11707 & 15.4 & 30.0 & $3\times10^{28}$ & 0.3 & 1 & 4.30 & 0.5875 & 6.2529 \\
\hline
UGC 12632 & 8.59 & 17.06 & $3\times10^{28}$ & 0.3 & 1 & 2.57 & 0.6825 & 4.5223 \\
\hline
UGC 12732 & 12.72 & 30.8 & $3\times10^{28}$ & 0.3 & 1 & 2.21 & 0.5676 & 6.5556 \\
\hline
\hline
\end{tabular}
\end{center}
\end{table}

\noindent With  RX-DMFIT tool, from observed X-ray or radio flux density, it 
would be possible to estimate the corresponding $<\sigma v>$ as a function of 
$m_{DM}$ and WIMP annihilation channels \cite{Bhattacharjee:2019jce}. 
The star formation rate in LSB galaxies are extremely low and that makes them an 
ideal room for examining the radio emission which might dominantly come from the 
DM annihilation/decay \cite{Bhattacharjee:2019jce}.
For our purpose, we have taken the observed value of radio flux density for all 
LSB galaxies from the NVSS survey 
\cite{Condon:1998iy}. NVSS is the `NRAO VLA Sky Survey' which has performed a 
sky survey at the frequency($\nu$)= 1.4 GHz. The Very Large Array (VLA) is 
located in southwestern New Mexico. It has the 27 elements of the 
interferometric array which generates the radio images of the sky for a very 
broad range of resolutions and frequencies. 
The spatial 
size of the NVSS images\footnote{\tiny{https://www.cv.nrao.edu/nvss/}} around 
UGC 3371, UGC 11707, UGC 12632 and UGC 12732 
are 185.40", 300.70", 66.00" and 307.70", respectively. But except UGC 11707, 
other three LSB galaxies only provide the upper limits of the radio flux 
density. The flux limits from the NVSS survey are shown in Table~7.7 \cite{Bhattacharjee:2019jce}.

\begin{table}
\begin{center}
\caption{The radio flux density limit obtained from the NVSS at frequency 1.4 GHz.}
\begin{tabular}{|p{5cm}|p{5cm}|}
\hline 
\hline
Galaxy & Observed Flux density in mJy \\
\hline
UGC~3371 & $<$ 0.45 mJy \\
\hline
UGC~11707 & $1.17$ mJy \\
\hline
UGC~12632 & $<$ 0.45 mJy \\
\hline
UGC~12732 & $<$ 0.45 mJy \\
\hline
\hline
\end{tabular}
\end{center}
\end{table}

\begin{figure}
\subfigure[]
 { \includegraphics[width=0.48\linewidth]{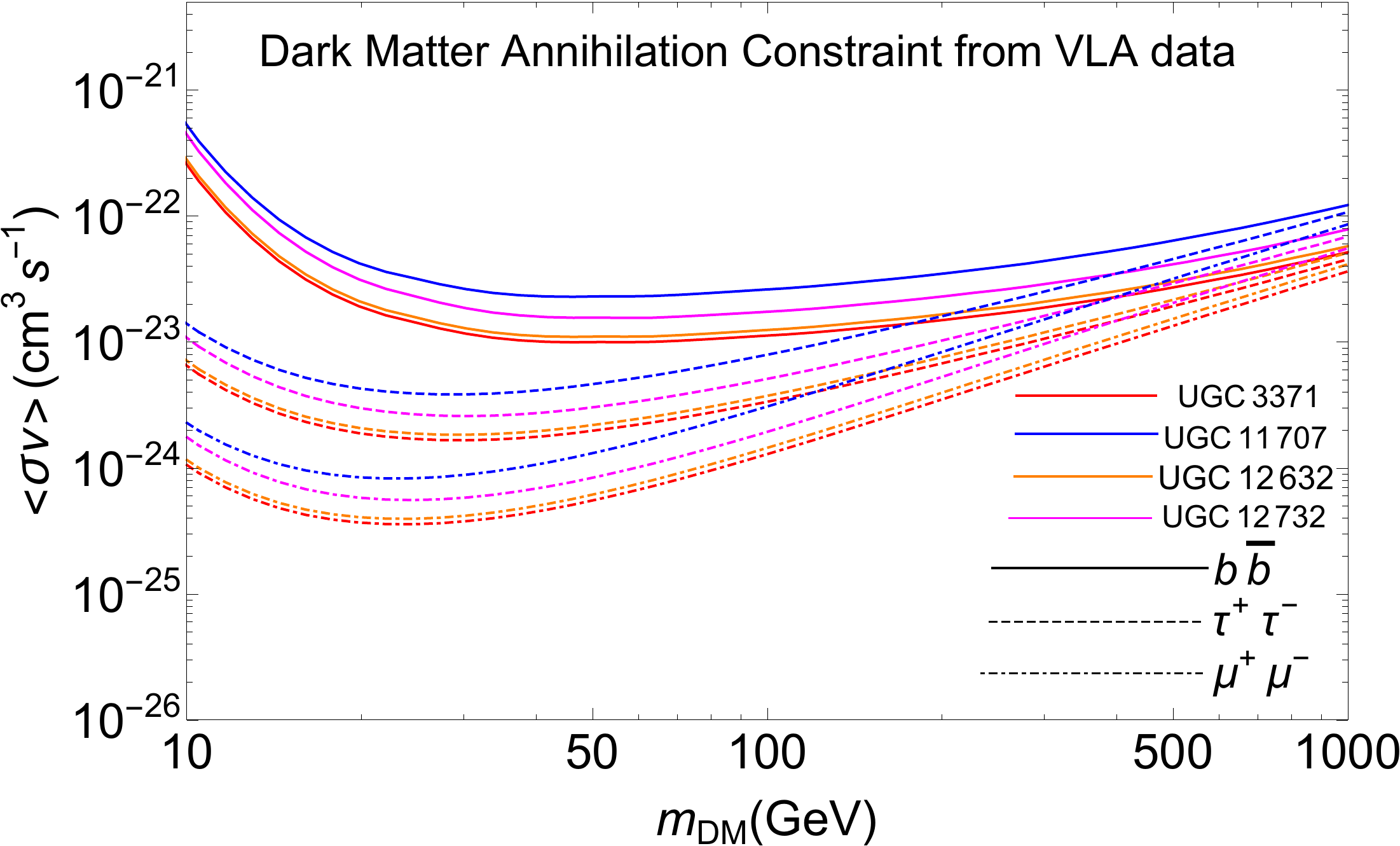}}
\subfigure[]
 { \includegraphics[width=0.48\linewidth]{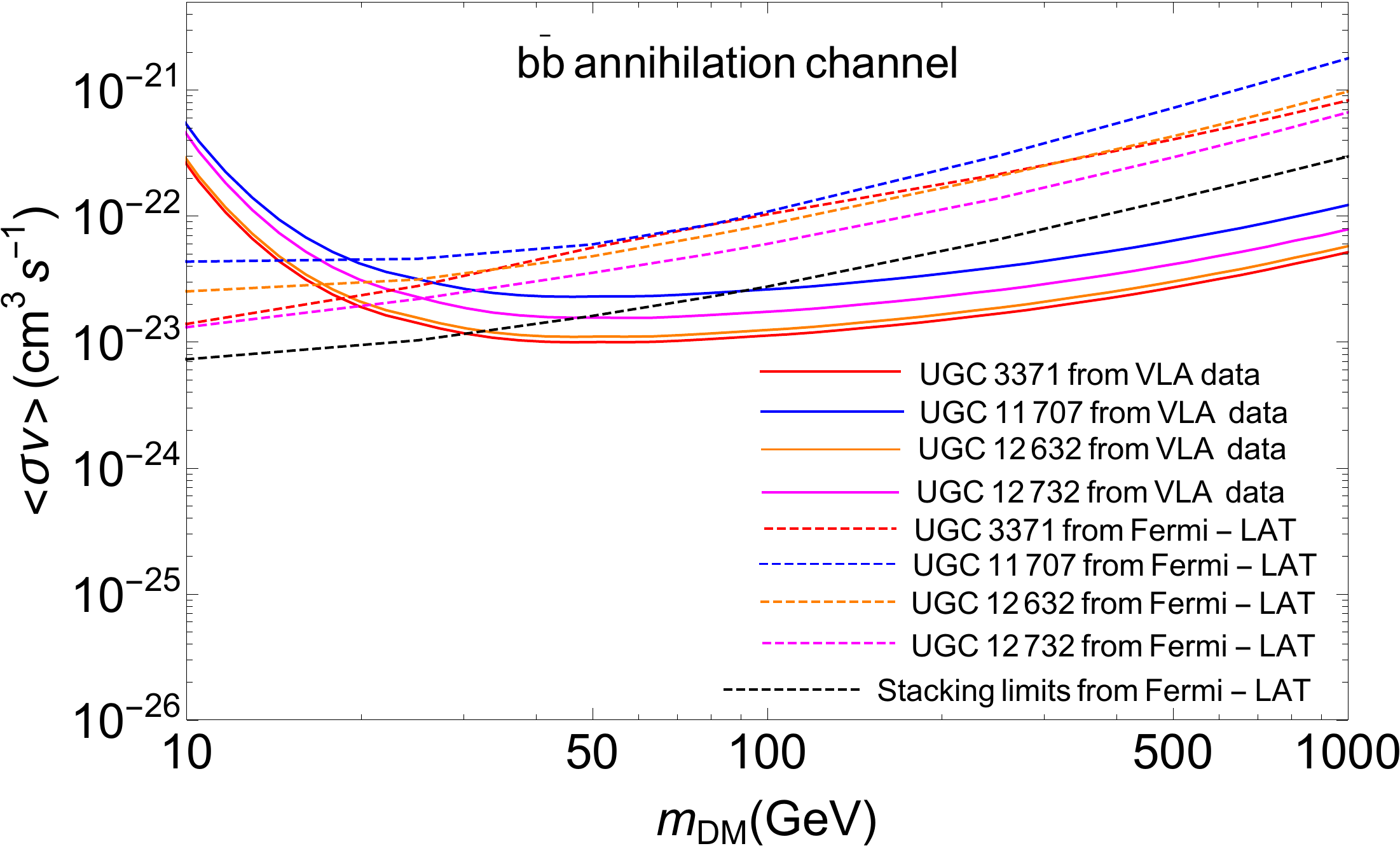}}
\subfigure[]
 { \includegraphics[width=0.48\linewidth]{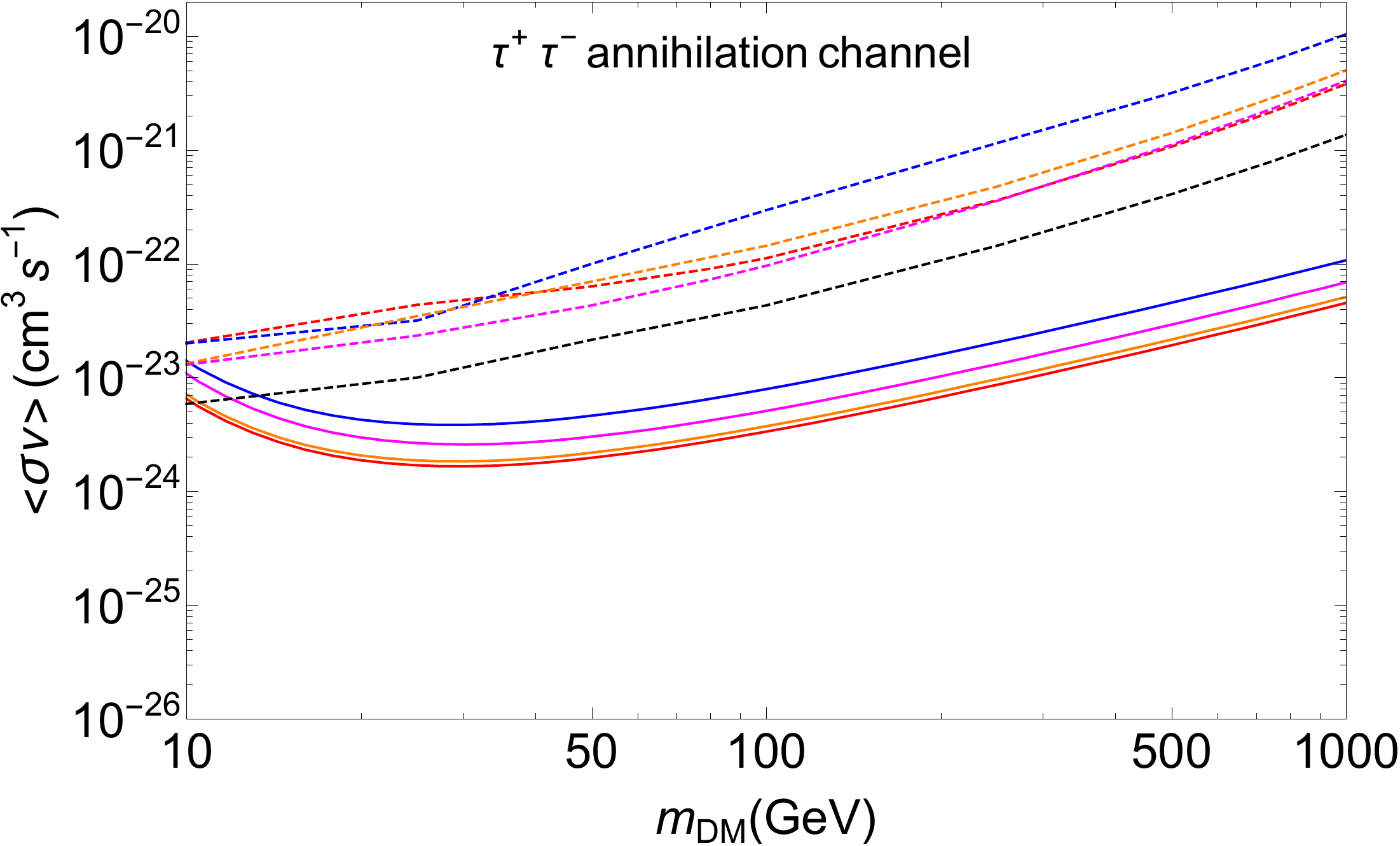}}
\subfigure[]
 { \includegraphics[width=0.48\linewidth]{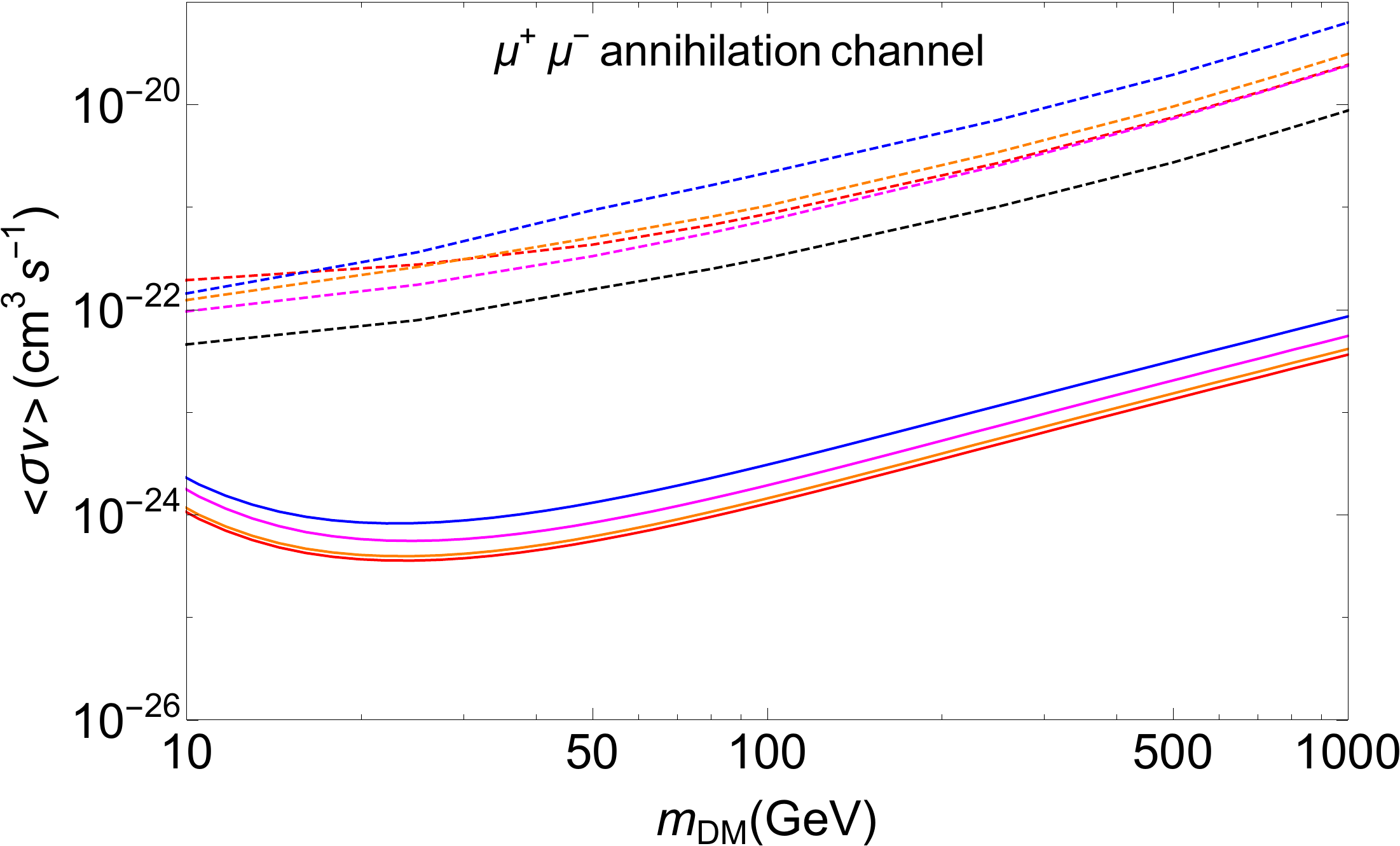}}
\caption{(a) The limits on $<\sigma v>$ by using radio flux density obtained 
from the NVSS images and for three annihilation channels are shown here. The (a) 
solid, 
dashed and dot-dashed linestyle denote the $b\overline{b}$, 
the $\tau^{+}\tau^{-}$ and the$\mu^{+}\mu^{-}$ channels, respectively. 
Comparison of the radio $<\sigma v>$ limits obtained from NVSS data with 
$\gamma$-ray $<\sigma v>$ limits obtained from the individual and the stacked 
analysis for (b) the $b\overline{b}$, (c) the 
$\tau^{+}\tau^{-}$ and (d) the $\mu^{+}\mu^{-}$ annihilation channels. we have 
chosen the
NFW profile.We have considered $m_{DM}$=100 GeV, $B_{0}$ = 1$\mu$G, 
$D_{0}$=$3\times10^{28}~(cm^{2}s^{-1})$ and have fixed the thermal averaged 
$<\sigma v>$ to $3 \times 10^{26}~cm^{3}s^{-1}$. Like (a), the same linestyles 
have been used for (b), (c) and (d).}
\end{figure}

\begin{figure}
\begin{center}
\subfigure[$b\overline{b}$]
 { \includegraphics[width=0.9\linewidth]{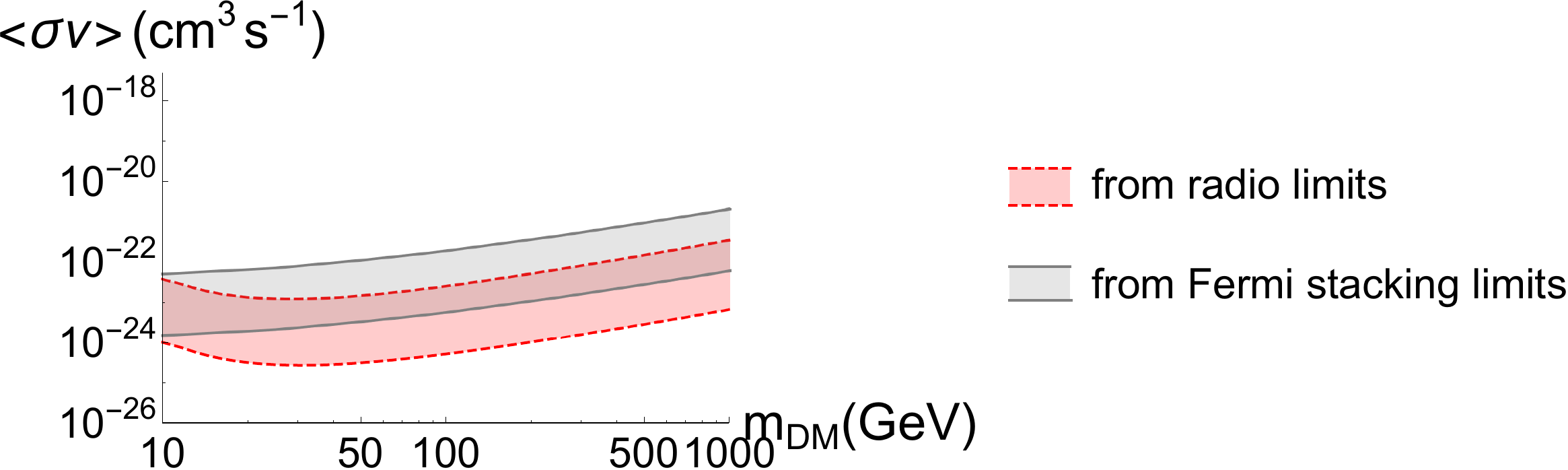}}
\subfigure[$\tau^{+}\tau^{-}$]
 { \includegraphics[width=0.9\linewidth]{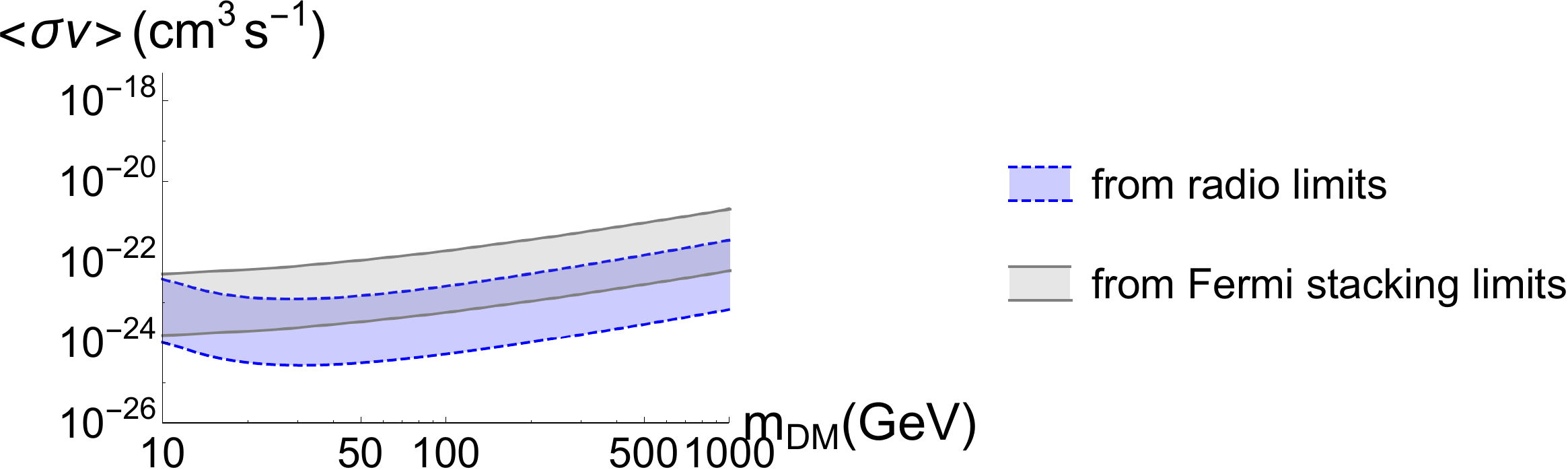}}
\subfigure[$\mu^{+}\mu^{-}$]
 { \includegraphics[width=0.9\linewidth]{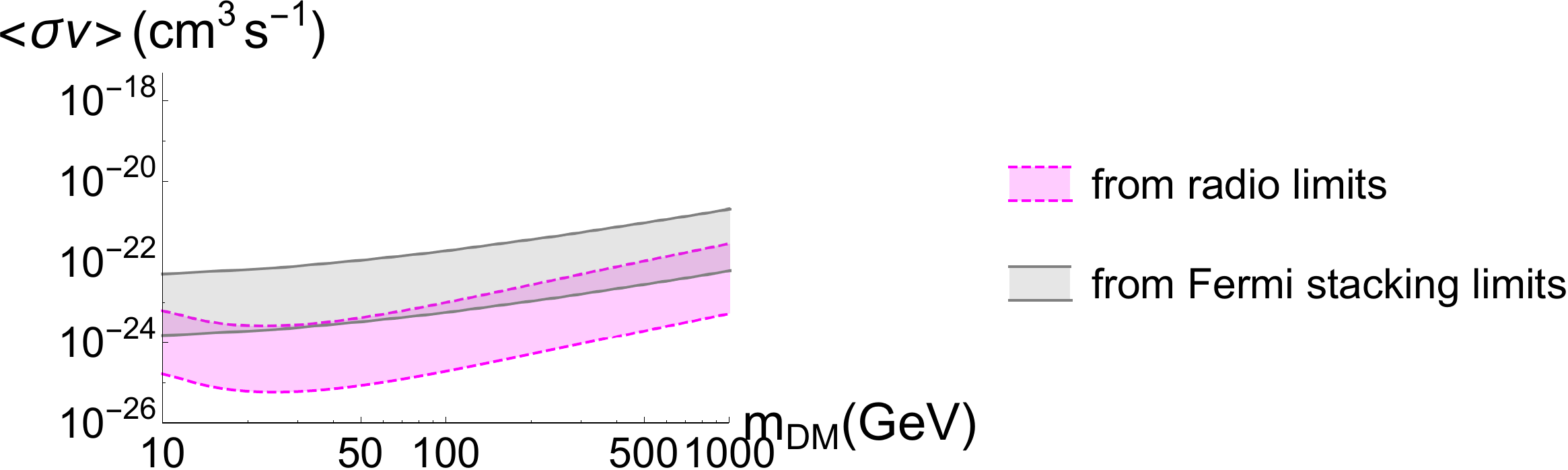}}
\caption{The uncertainties associated with the $<\sigma v>$ limits obtained 
from 
NVSS images for (a) the $b\overline{b}$, (b) the $\tau^{+}\tau^{-}$ and (c) 
the $\mu^{+}\mu^{-}$ final states are shown here. The radio limits for each 
annihilation channels are compared with the uncertainty band associated with 
$\gamma$-ray stacking limits for $b\overline{b}$. The shaded region between 
dashed lines displays the uncertainty band for radio limits, while the shaded 
region between solid lines shows the uncertainty band for $\gamma$-ray stacking 
limits.}
\end{center}
\end{figure}

\begin{figure}
\subfigure[UGC 3371]
 { \includegraphics[width=0.48\linewidth]{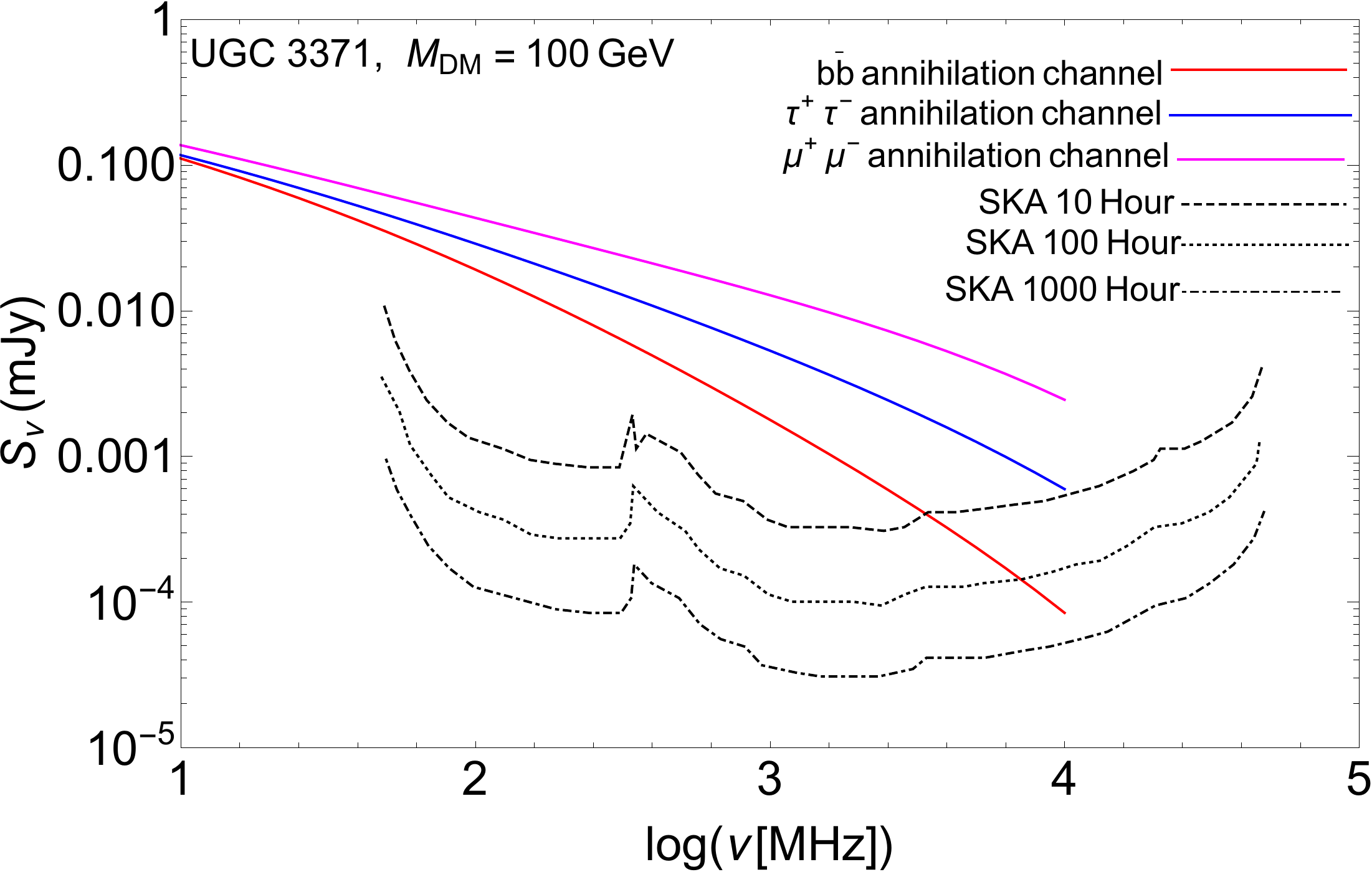}}
\subfigure[ UGC 11707]
 { \includegraphics[width=0.48\linewidth]{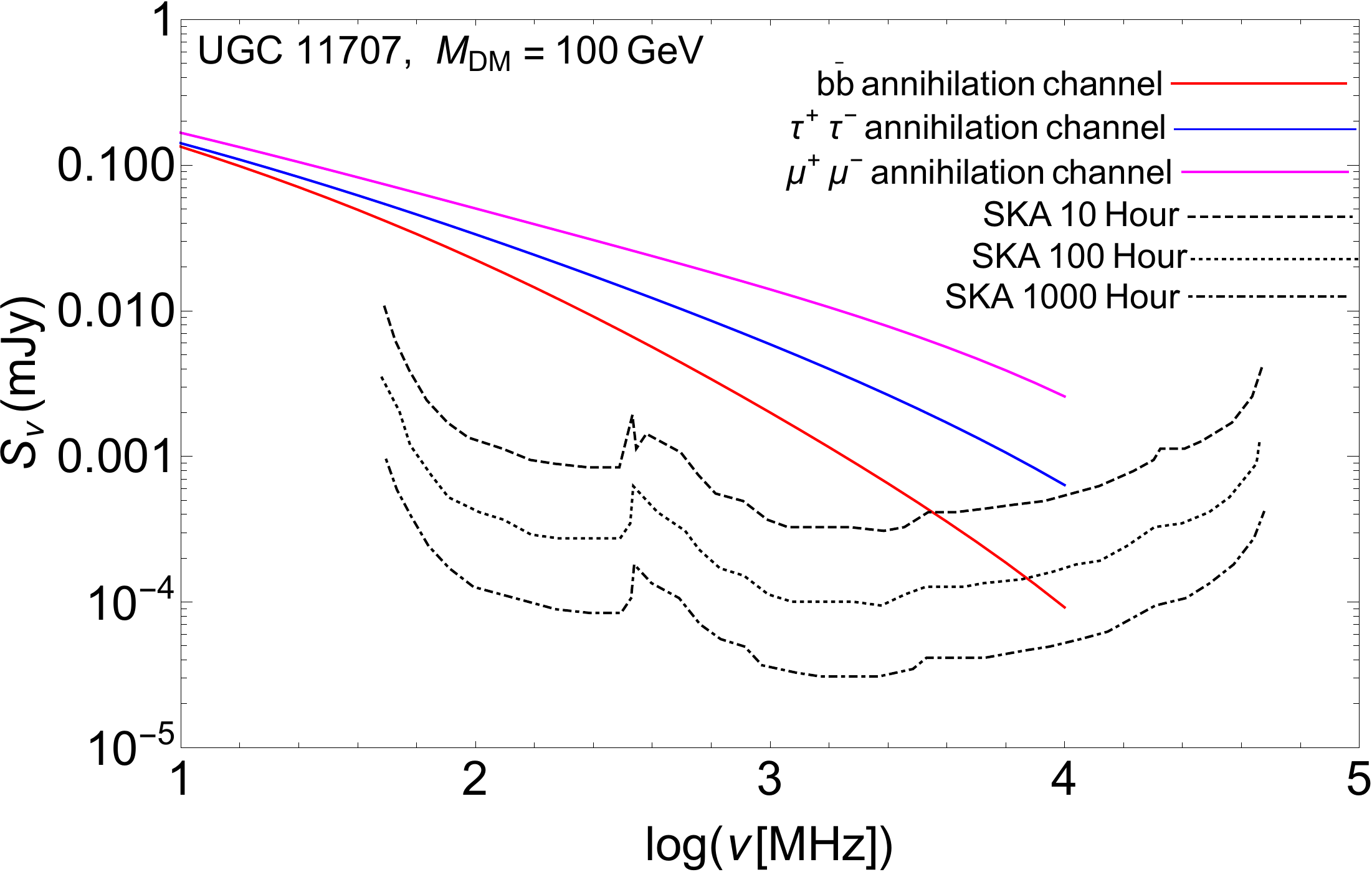}}
\subfigure[ UGC 12632]
 { \includegraphics[width=0.48\linewidth]{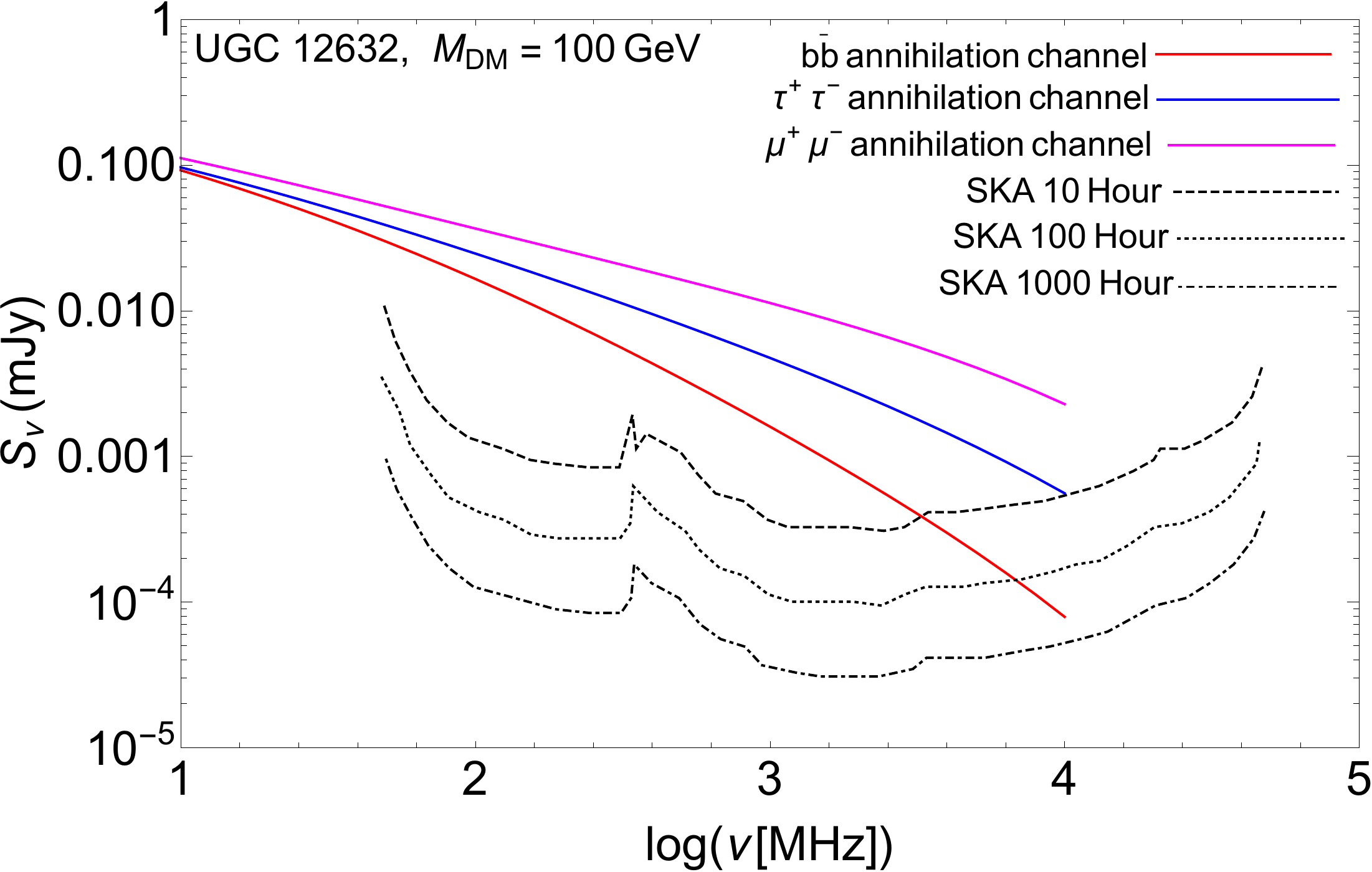}}
\subfigure[ UGC 12732]
 { \includegraphics[width=0.48\linewidth]{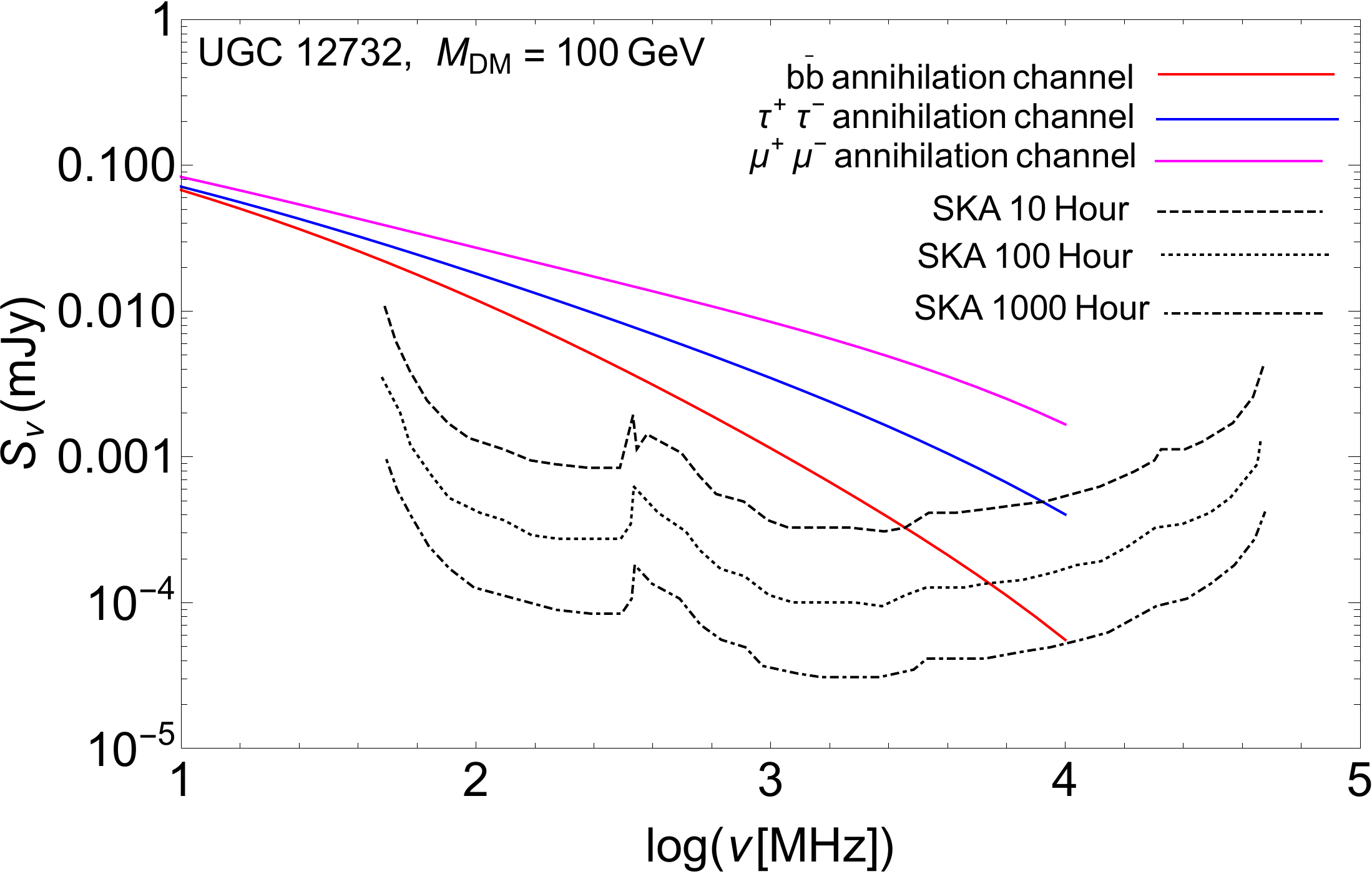}}
\caption{The flux density predicted for our LSB galaxies that annihilates into 
the
$b\overline{b}$, the $\mu^{+}\mu^{-}$ and the $\tau^{+}\tau^{-}$ channels. we 
have chosen the
NFW profile.We have considered $m_{DM}$=100 GeV, $B_{0}$ = 1$\mu$G, 
$D_{0}$=$3\times10^{28}~(cm^{2}s^{-1})$ and have fixed the thermal averaged 
$<\sigma v>$ to $3 \times 10^{26}~cm^{3}s^{-1}$. We have overplotted the SKA 
sensitivity curve for its 10, 
100 and 1000 hours of observation time with the dashed, the dotted and the 
dot-dashed black 
curves, respectively.}
\end{figure}

\noindent Before proceeding to our next analysis, we would like to report that the signal observed from the location of
UGC 11707 is roughly less than 3$\sigma$ and for data analysis, such faint emission
is assumed to be mostly originated from the fluctuation in some unknown astrophysical sources. Thus a more sensitive survey at $\nu$=1.4 GHz
is needed to examine the real nature of signal coming from the location of UGC 11707.
Thus, for our analysis, even though UGC 11707 produces the physical flux limits, we have performed the same method for all our targets.\\

\noindent By using the VLA data (mentioned in Table~7.7) we have estimated the 
$<\sigma v>$ limits as the function of $m_{DM}$ for three annihilation channels 
and the relevant plot is shown in Fig.~7.10 \cite{Bhattacharjee:2019jce}. From Fig.~7.10(a), we can find that 
for radio data $\tau^{+}\tau^{-}$ and $\mu^{+}\mu^{-}$ final states provide the 
more stringent limits than $b\overline{b}$ \cite{Bhattacharjee:2019jce}. But from $\gamma$-ray data (see 
Fig.~7.4(d)), $b\overline{b}$ final state put the most stringent limits. 
Theoretically most of the $b\overline{b}$ final state first annihilates to the 
$\pi^{\circ}$ and they decays to $\gamma$-ray photons, while $\tau^{+}\tau^{-}$ 
and $\mu^{+}\mu^{-}$ final states (i.e., leptonic channel) mostly decay to the 
$e^{+}$/$e^{-}$ \cite{Bhattacharjee:2019jce}. Hence, for gamma-ray analysis, $b\overline{b}$ annihilation 
channel is expected to produce stronger limits than leptonic channels but for 
radio analysis, we would get the reverse result \cite{Bhattacharjee:2019jce}. For Figs.~7.10 (b,c,d), The 
comparison between the radio $<\sigma v>$ limits with the limits obtained from 
the $\gamma$-ray analysis (from sections~7.4.2 and 7.4.3) for three annihilation 
final states has been shown in Figs.~7.10 (b,c,d) \cite{Bhattacharjee:2019jce}. For Fig.~7.10, we have used 
the other necessary parameter values from Table 7.6. For Fig.~7.10, we have not 
considered the uncertainty associated with radio and gamma-ray limits and for 
now, we can observe that for 100 GeV DM mass, the radio data might provide the 
stronger limits than gamma-ray \cite{Bhattacharjee:2019jce}. For the $\mu^{+}\mu^{-}$ channel, the radio 
limits even provide nearly 2 orders of the more stringent limits than the 
stacking $<\sigma v>$ limits for Fermi-LAT data \cite{Bhattacharjee:2019jce}. \\

\noindent Next we would try to check the uncertainty associated with the radio 
$<\sigma v>$ limits for LSB galaxies \cite{Bhattacharjee:2019jce}. As we already mentioned that there is not 
much detailed study for our selected LSB galaxies and thus with the inadequate 
kinematics data of LSB galaxies, they can produce the large uncertainty bands \cite{Bhattacharjee:2019jce}. 
For radio data, we have estimated the uncertainty band in 2$\sigma$ C.L. and 
then compared the radio limits with the stacked limits obtained from the 
gamma-ray data for $b\overline{b}$ (Fig.~7.11) \cite{Bhattacharjee:2019jce}. For gamma-ray data, we have 
chosen the $b\overline{b}$ final states as this channel produces the strongest 
limits, while for radio data we have shown the uncertainty band associated with 
UGC 12632 for $b\overline{b}$ (Fig.~7.11 (a)), $\tau^{+}\tau^{-}$ (Fig.~7.11 
(b)) and $\mu^{+}\mu^{-}$ (Fig.~7.11 (c)) final states \cite{Bhattacharjee:2019jce}. 
Now, from Fig.~7.11, we can observe that for LSB galaxies the magnitude of the 
uncertainty band associated to radio $<\sigma v>$ is at the order of 2 and for 
each annihilation channels uncertainty band corresponding to both radio and 
gamma-ray data is overlapping with each other \cite{Bhattacharjee:2019jce}. Unlike the Fig.~7.10, whenever we 
would consider the uncertainty band, it would not be possible for us to strongly 
favour the radio analysis over gamma-ray. Our result at best shows that radio 
and gamma-ray limits are competitive with each other \cite{Bhattacharjee:2019jce}.\\

\noindent We next have tried to explore whether, with Square Kilometre Array 
(i.e., SKA), it would be possible to detect any radio emission from LSB galaxies \cite{Bhattacharjee:2019jce}. 
SKA is the next generation radio telescope and because of its wide F.O.V and 
resolution \cite{Proceedings:2015yra}, we expect that from next decade SKA would 
be able to explore many resolved problems in cosmology. Searching for the DM 
signal would be one of the most intriguing parts of it\cite{braun:2015sd}.\\

\noindent We have predicted the possible flux density $S(\nu)$ for each LSB 
galaxies in a form of synchrotron emission with RX-DMFIT tool \cite{Bhattacharjee:2019jce}. In Fig.~7.12, we 
have shown the variation of $S(\nu)$ with frequency ($\nu$) for three WIMP 
annihilation channels and have compared it with the sensitivity curve of SKA for 
its 10, 100 and 1000 hours of observations \cite{Bhattacharjee:2019jce}. 
From Fig.~7.12 (a,b,c,d), we find that it might be possible for SKA to detect 
the radio emission from the LSB galaxies, especially with its 1000 hours of 
sensitivity curve \cite{Bhattacharjee:2019jce}. In Fig.~7.9, we have presented how does the `Sync' SED depend 
on the astrophysical parameters, especially on the B and the D \cite{Bhattacharjee:2019jce}. Hence, accurate 
knowledge on B, D and DM density distribution is very necessary. Otherwise, we 
could not strongly state whether SKA would be able to detect any positive signal 
from LSB. Thus our study, at best, hints that SKA would definitely play a very 
major part to investigate the radio emission (most possibly from DM 
annihilation) \cite{Bhattacharjee:2019jce}.

\subsection{Comparison between the NFW, Burkert and Pseudo Isothermal Density Profiles}
\begin{table}
\centering
\caption{J-factors derived for three DM density profiles at $h_{0}=0.75$.}
\begin{tabular}{|p{2cm}|p{3cm}|p{5cm}|}
\hline \hline
Galaxy name &  Density Profile & J-factor ($\rm{GeV^{2}/cm^{5}}$)\\
\hline \hline
UGC & NFW & $0.739^{+2.87}_{-0.63}\times10^{16}$ \\
3371 & ISO & $0.188^{+0.775}_{-0.169}\times10^{16}$  \\
& BURKERT & $0.385^{+1.594}_{-0.346}\times10^{16}$ \\
\hline \hline
UGC & NFW & $0.485^{+1.85}_{-0.42}\times10^{16}$  \\
11707 & ISO & $0.123^{+0.501}_{-0.110}\times10^{16}$  \\
& BURKERT & $0.253^{+1.03}_{-0.227}\times10^{16}$  \\
\hline \hline
UGC & NFW & $0.795^{+3.08}_{-0.68}\times10^{16}$  \\
12632 & IS0 & $0.202^{+0.835}_{-0.182}\times10^{16}$  \\
& BURKERT & $0.414^{+1.717}_{-0.373}\times10^{16}$  \\
\hline \hline
UGC & NFW & $0.880^{+3.40}_{-0.75}\times10^{16}$  \\
12732 & ISO & $0.223^{+0.919}_{-0.1997}\times10^{16}$  \\
& BURKERT & $0.459^{+1.888}_{-0.411}\times10^{16}$ \\
\hline \hline
\end{tabular}
\end{table}

 \begin{figure}
\subfigure[]
 { \includegraphics[width=0.48\linewidth]{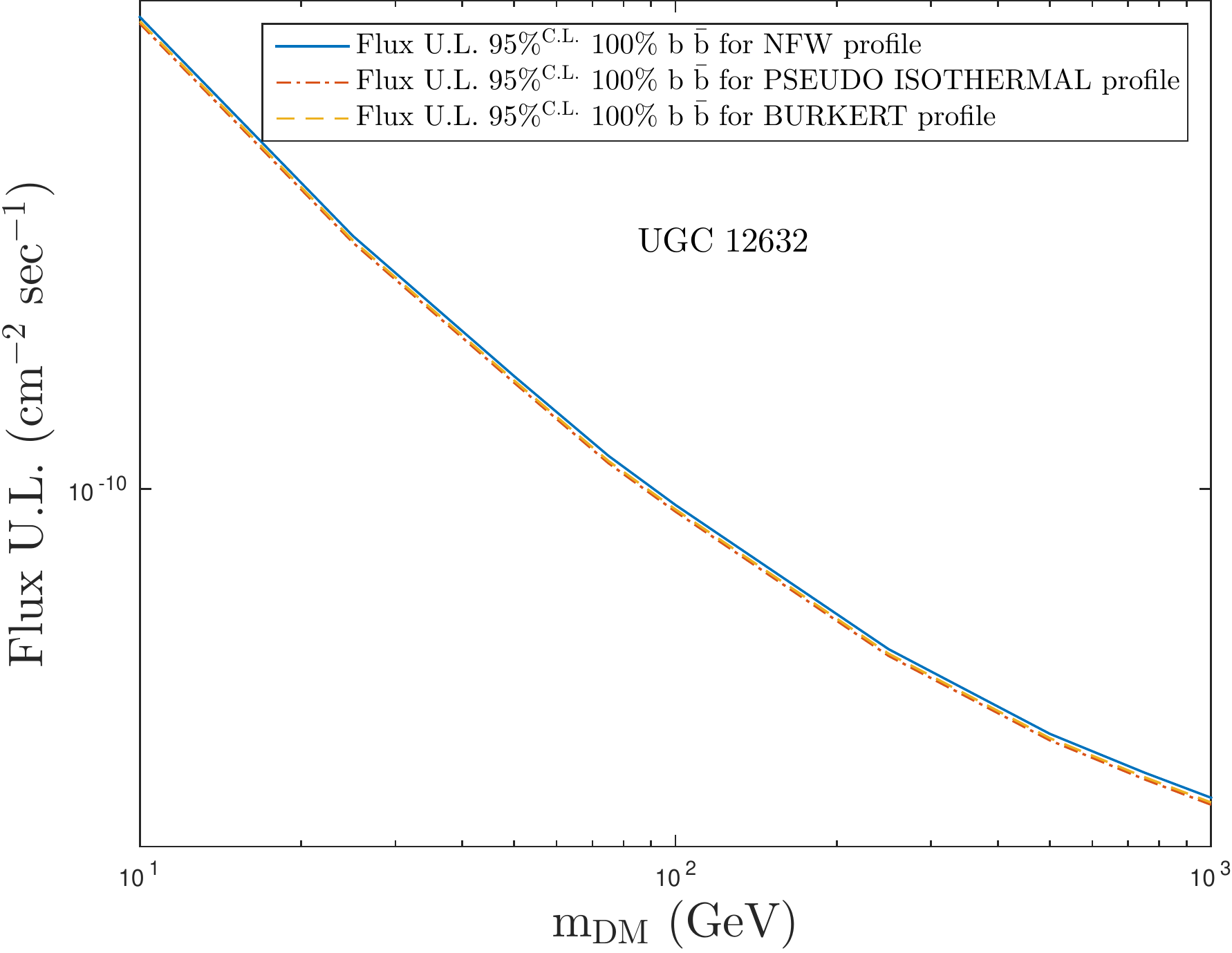}}
\subfigure[]
 { \includegraphics[width=0.53\linewidth]{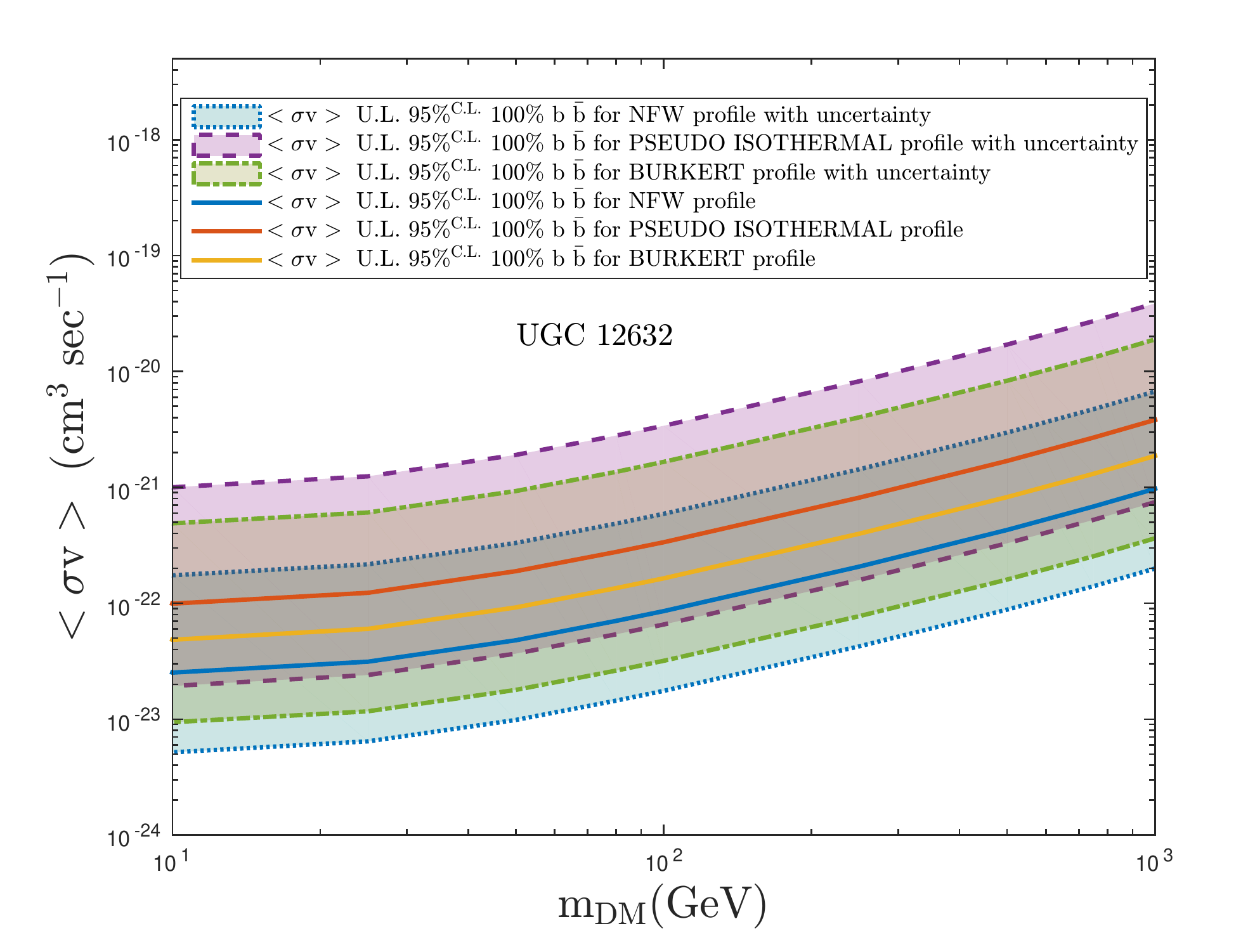}}
\caption{(a) The upper limit on the $\gamma$-ray flux for three different 
density profiles. (b) The comparison between the upper limits on the $<\sigma 
v>$ for three density profiles estimated for the median value of J-factor along 
with the uncertainty. The shaded region refers to the uncertainty in the DM 
density for LSB galaxies. In both figures, we have chosen UGC 12632 that 
annihilates into the $b\overline{b}$ channel.}
\end{figure}

\noindent In this section, we have done a comparative study between three 
popular density profiles, i.e., between NFW\cite{Navarro:1996gj}, Pseudo 
Isothermal (ISO)\cite{Gunn:1972sv} and Burkert (BURK) 
profile\cite{Burkert:1995yz, Salucci:2011ee}. For examining the distribution of 
the DM, two types of profiles are widely used in the literatures. Those are 
cuspy (e.g. NFW) and cored-like (e.g. BURK, ISO) profiles. N-body simulation 
results strongly support the cuspy-like distribution of DM distribution, while 
the observational study i.e., the rotational curves for several irregulars and 
dwarf galaxy favour the cored profile \cite{Bhattacharjee:2019jce}. This problem is known as the 
``cuspy-core'' problem. Before presenting the comparison between three density 
profiles, we would like to mention that for our sources we have preferred to use 
the NFW profile \cite{Bhattacharjee:2019jce}. Because the available rotational curves for our LSB galaxies 
showed that the NFW profile produced an acceptable fit to the rotational curves 
\cite{vandenBosch:2000rza, vandenBosch:2001bp, Swaters:2002rx} and their study 
was not able to differentiate between the $1/r$ cusps and constant cores.\\

\noindent The mathematical form of these three DM density profiles are described 
in Chapter~2. Using the Eqs.~2.1, 2.2, 2.3 and 2.7, we have calculated the 
J-factors of UGC 12632 for all three profiles. The values are mentioned in Table 
7.8. We can notice that, among three density profiles, NFW produces the largest 
J-factor \cite{Bhattacharjee:2019jce}.\\

\noindent Next, we have estimated the $\gamma$-ray flux upper limits and the 
corresponding $<\sigma v>$ limits for three density profiles. The J-factors of 
UGC 12632 has been taken from Table~7.8 and for our purpose, in Fig.~7.13 we 
have shown the result for $b\overline{b}$ final state \cite{Bhattacharjee:2019jce}. In Fig~7.13 (a), we have 
shown the gamma-ray flux upper limits for three density profiles. The flux upper 
limits have no direct dependence on the J-factor, so from for all three 
profiles, we have obtained the same order of flux limits \cite{Bhattacharjee:2019jce}. In Fig.~7.13 (b), we 
have displayed the $<\sigma v>$ upper limits along with its 2$\sigma$ 
uncertainty band for three DM density profiles \cite{Bhattacharjee:2019jce}. From this figure, we could 
notice that the uncertainty band for each profile are overlapping and thus 
without reducing the uncertainty band, from this Fig.~7.13 (b) we could not 
comment which density profile can produce the most stringent limit in the space 
of ($m_{DM}$, $<\sigma v>$) \cite{Bhattacharjee:2019jce}.

\section{The Future of LSB Galaxies for Dark Matter Searches and the Impact of the CTA}
\noindent It is much expected that from next decade, the Cherenkov Telescope 
Array (in short CTA) would come as the most advanced and sensitive $\gamma$-ray 
telescope for high-energies. CTA would study the $\gamma$ rays between 20 GeV to 
300 TeV energy range and because of its large angular resolution (say around 2 
arc minutes) and improved energy resolution (much below than $\sim$10$\%$), it 
might be possible for CTA to detect the $\gamma$ rays even from a very weak and 
distant target.
CTA has very wide F.O.V (for small and medium sized telescopes it is around 
$\sim$ 8 degree) and the encouraging part of this instrument is that its 
effective area would increase with energies. Thus all of those quantities make 
CTA the foremost sensitive comparing to all or any currently working space-based 
and ground-based telescopes and that also gives us a hope that in future it 
might be possible for CTA to identify the DM signals. \\

\noindent For our work, we would like to check whether in the future CTA can 
detect any emission from the LSB galaxies and for that purpose we have compared 
the differential flux of LSB galaxies obtained from the Fermi-LAT with the 
sensitivity curve of CTA \cite{Bhattacharjee:2019jce}. 
Our adopted CTA sensitivity curve (\cite{Maier:2017cjy}) is derived from the 
point-like sources where, they are modelled with the power-law function and have 
the 5$\sigma$ significance for 50 hours of CTA
observation. For Fermi-LAT we have used the sensitivity curve 
for 10 years of LAT observation for the high-Galactic-latitude sources 
\footnote{\tiny{http://www.slac.stanford.edu/exp/glast/groups/canda/lat{\_}Performance.html}}.
The sensitivity curve for the Fermi-LAT is also estimated for the point-like 
sources that are modelled with the power law and have the 5$\sigma$ detection 
significance\footnote{\tiny{http://www.slac.stanford.edu/exp/glast/groups/canda/lat{\_}Performance.html}}.\\

\noindent The comparison between the differential flux for all LSBs with the 
sensitivity curves for CTA and Fermi-LAT instruments are shown in Fig.~7.14 \cite{Bhattacharjee:2019jce}. In 
order to estimate the differential fluxes for LSB galaxies, they were modelled 
with the power-law spectrum for $\Gamma$=2 (see section~7.3.1) \cite{Bhattacharjee:2019jce}. From Fig.~7.14, 
it is quite evident that between the energy range of 100 GeV to 1 TeV, CTA might 
be able to observe the emission from LSB galaxies. This is the really very 
intriguing part of this study but we should also keep in mind that from 
Fig.~7.14 we can only hint that above 100 GeV with the 50 hours of observation 
there are chances that CTA would detect the emission from LSB galaxies \cite{Bhattacharjee:2019jce}. But that 
emission can either come from any astrophysical sources or from the DM 
annihilation.
A detailed simulation study is needed to check whether such emission is 
resulting from the DM annihilation but that part is currently beyond the scope 
of the analysis \cite{Bhattacharjee:2019jce}. 
Hence, from our study, we can only at best comment that in the next decade CTA 
would be very important tools for the gamma-ray analysis and would be especially 
ideal for the indirect DM searching \cite{Bhattacharjee:2019jce}.

\begin{figure}
\begin{center}
 { \includegraphics[width=0.5\linewidth]{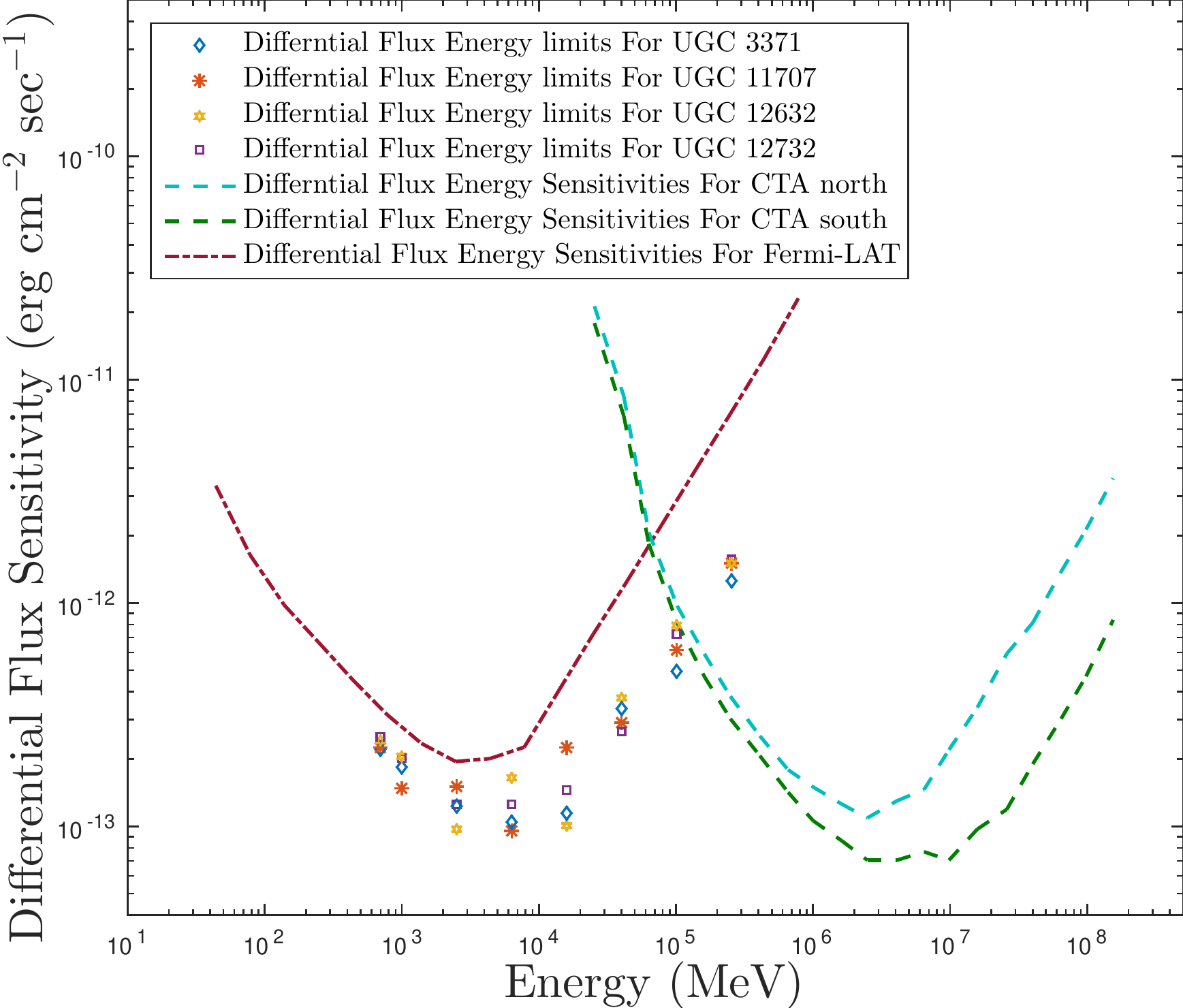}}
\caption{The comparison of the differential $\gamma$-ray flux obtained from our 
LSB galaxies with the detection-sensitivity curves for the Fermi-LAT and CTA.}
\end{center}
\end{figure}

\section{Conclusions \& Discussions}
\noindent For this work, we have studied for nearly nine years of LAT data but 
have not detected any emission from the location of LSB. With DMFit tools, we 
estimated the $\gamma$-ray and $<\sigma v>$ upper limits for four annihilation 
states. But because of their low J-factors, individual limits obtained from the 
LSB galaxies have not put any stringent limits on the DM theoretical models.  
With the hope of increasing the LAT sensitivity, we have then performed the 
joined likelihood on the set of four LSB galaxies. As expected, the stacking 
method has improved the $<\sigma v>$ by the factor of 4 than the individual 
limits obtained from LSB galaxies. But, the combined $<\sigma v>$ were still 
around two orders of magnitude weaker than the $<\sigma v>$ limits obtained from 
refs.~Ackermann et al.\cite{Ackermann:2015zua} and Steigman et al.\cite{Steigman:2012nb}. \\

\noindent The observation data for our chosen LSB galaxies could not 
particularly favour cored profile over the cuspy profile. The rotational curves 
for LSBs are in an agreement with the prediction from $\lambda$CDM and some 
study also indicated that the cuspy profile could also provide a reasonable fit 
to the DM distribution at the internal core. Thus, motivated by all the 
observation indications, we have modelled the DM distribution of LSB galaxies 
with the NFW profile. We have also performed a comparative study between NFW, 
ISO and BURK DM density profiles (check Fig.~7.13) and find that the $<\sigma v>$ 
limits for each density profiles are overlapping with other. Thus, from our 
study, we could not favour one profile between all three but for the median 
value of J-factor, the most stringent limits would come from the NFW profile.\\

\noindent For this study, we have used the multiwavelength approach which is 
considered as the complementary of the $\gamma$-ray detection method and is very 
popular in nowadays for the indirect searching of the DM signal. For our 
analysis, we have preferred to focus on the radio signal and for that purpose, 
we have followed the code RX-DMFIT. RX-DMFIT is the extension of DMFIt package 
and is specially designed to investigate the possible radio and X-ray emission 
from DM annihilation. LSB galaxies have very low nuclear activity and poor star 
formation rates and that makes them suitable targets for examining the diffuse 
radio emission most possibly coming from the DM annihilation/decay. We have 
estimated the multiwavelength SED plots for LSB galaxies and have also checked 
how the nature of SED varies with varying the parameter sets (check Figs.~7.8 
$\&$ 7.9). We have searched for the radio flux limits for all LSB galaxies from 
the NVSS sky survey data but only the location of UGC 11707 gives detected flux 
density values and other thee LSBs only provide the upper limits to the flux 
density. With the VLA flux density, we have tried to predict the radio $<\sigma 
v>$ limits in parameter space of ($<\sigma v>$, $m_{DM}$) (check Fig.~7.10). If 
we consider the 2$\sigma$ uncertainty band associated with the radio limits, we 
have noticed that the radio limits are overlapping with the limits obtained from 
stacking analysis for LAT data (check Fig.~7.11) and all three annihilation 
channels have shown the same nature. Hence, from our analysis, we could, at 
best, comment that the radio data is competitive with the gamma-ray data. With 
more detailed observational data and precise analysis, in future, it might be 
possible for LSB galaxies to impose strong limits on DM models.\\

\noindent We have checked whether with the next generation radio (SKA) and 
gamma-ray (CTA) telescopes it would be possible to detect any emission from the 
location of LSB galaxies. We have noticed (check 7.12) that SKA might be able to 
detect the emission from the location of LSB galaxies and its 1000 hours of 
observation would have the highest possibility to detect the emission from LSBs. 
But we would also like mention that in order to claim that SKA would detect the 
emission from DM annihilation, we first need to perform a simulation study. 
Besides, the estimated radio emission is also dependent on the various 
astrophysical scenario. We need to have a well-defined knowledge on the 
distribution of diffusion zone, magnetic fields, DM density profile, etc.. 
Hence, from our analysis, we could, at best, hint the possibility of observing 
the radio signal from LSB galaxies by SKA. We have also found (Fig.~7.14) that 
for energy ranges between 100 GeV to 1 TeV, it might be possible for CTA to 
observe the $\gamma$-ray emission with the 50 hours of sensitivity curve. But 
like SKA, the same conclusion also holds for CTA. A simulation study is needed 
to examine whether it would be possible for CTA to detect the emission resulting 
from the DM annihilation/decay.\\

\noindent Hence, from our work, we can conclude that the $\gamma$-ray data 
obtained from the Fermi-LAT could not impose the strong $<\sigma v>$ limits on 
the WIMP models. We find that the radio signal possibly originated from WIMP 
annihilation is quite competitive with the $\gamma$-ray emission observed by the 
Fermi-LAT. Our analysis, at best, indicates that to study $\gamma$-ray and radio 
signal from the LSB galaxies, SKA and CTA would play a very significant role in 
future. 

\chapter{Synchrotron and gamma-ray radiation from few ultra faint dwarf galaxies}
\section{Source Details}\label{section:source_details}

\noindent In this chapter, we have investigated the gamma-ray and radio emission 
possibly resulting from DM annihilation \cite{Bhattacharjee:2020phk}. For this purpose, we have chosen 
several UFDs based on their very high mass to light ratio, large velocity 
dispersion of their stars, etc. and thus they are very likely to be rich in DM 
\cite{Baumgardt:2008zt}. The observed spectroscopic and photometric properties 
of our selected UFDs are described in Table~8.1 \cite{Bhattacharjee:2020phk}, where, $M/L$, $\sigma$, $d$, 
$r_{1/2}$ and $\theta_{max}^o$ refers to mass-to-light
ratio, velocity dispersion, heliocentric distance, half light radius and maximum 
galactocentric distance of each UFDs, respectively\cite{Pace:2018tin}.

\begin{table}[!h]
\begin{center}
\begin{tabular}{||p{2.3cm}|p{1.8cm}|p{1.9cm}|p{1.5cm}|p{2cm}|p{2cm}||}
\hline 
\hline
Galaxy & M/L $(M_{\odot}/L_{\odot})$ & d (Kpc) & $r_{1/2}~(pc)$ & $\sigma~(km~s^{-1})$ & $\theta_{max}^o$ \\
\hline
Aquarius~II & $1330^{+3242}_{-227}$ & $107.9^{+3.3}_{-3.3}$ & $123^{+22}_{-21}$ & $6.2^{+2.6}_{-1.7}$ & 0.11134 \\
\hline
Carina~II & $369^{+309}_{-161}$ & $37.4^{+0.4}_{-0.4}$ & $77^{+8}_{-8}$ & $3.4^{+1.2}_{-0.8}$ & 0.23\\
\hline
Draco~II & $501^{+1083}_{-421}$ & $20.0^{+3.0}_{-3.0}$ & $12^{+5}_{-5}$ & $3.4^{+2.5}_{-1.9}$ & 0.1\\
\hline
Eridanus~II & $420^{+210}_{-140}$ & $366.0^{+17.0}_{-17.0}$ & $176^{+14}_{-14}$ & $7.1^{+1.2}_{-0.9}$ & 0.062 \\
\hline
Grus~I & $<~2645$ & $120.2^{+11.1}_{-11.0}$ & $52^{+26}_{-26}$ & $4.5^{+5.0}_{-2.8}$ & 0.093\\
\hline
Horologium~I & $570^{+1154}_{-112}$ & $79.0^{+7.0}_{-7.0}$ & $32^{+5}_{-5}$ & $5.9^{+3.3}_{-1.8}$ & 0.0619 \\
\hline
Hydra~II & $<~315$ & $151.0^{+8.0}_{-8.0}$ & $71^{+11}_{-11}$ & $<6.82$ & 0.08509 \\
\hline
Leo~V & $264^{+326}_{-264}$ & $173.0^{+5.0}_{-5.0}$ & $30^{+17}_{-17}$ & $4.9^{+3.0}_{-1.9}$ & 0.077 \\
\hline
Pegasus~III & $1470^{+5660}_{-1240}$ & $215.0^{+12}_{-12}$ & $37^{+14}_{-14}$ & $7.9^{+4.4}_{-3.1}$ & 0.03049\\
\hline
Pisces~II & $370^{+310}_{-240}$ & $183.0^{+15}_{-15}$ & $48^{+10}_{-10}$ & $4.8^{+3.3}_{-2.0}$ & 0.06861\\
\hline
Reticulum~II & $467^{+286}_{-168}$ & $30^{+2}_{-2}$ & $32^{+3}_{-3}$ & $3.4^{+0.7}_{-0.6}$ & 0.24\\
\hline
Tucana~II & $1913^{+2234}_{-950}$ & $57.5^{+5.3}_{-5.3}$ & $115^{+32}_{-32}$ & $7.3^{+2.6}_{-1.7}$ & 0.225\\
\hline
Tucana~III & $<~240$ & $25.0^{+2}_{-2}$ & $43^{+6}_{-6}$ & $<2.18$ & 0.2\\
\hline
Triangulum~II & $<~2510$ & $30^{+2}_{-2}$ & $28^{+8}_{-8}$ & $<6.36$ & 0.15\\
\hline
\hline
\end{tabular}
\end{center}
\caption{Properties of the UFDs.}
\label{table:astro_fundamental_param_dwarfs}
\end{table}

\subsection{Dependence of $J$ on the Density Profiles}
\label{sec:DM_profile}

\noindent As we have already discussed, NFW density profile is the benchmark 
choice for the DM distribution which is mainly favoured by the $N$-body 
simulations~\cite{Navarro:2008kc, Wagner:2020opz}, while some observational 
studies~\cite{de_Blok_2001} prefer the cored profile. Thus for this work, we 
have performed a comparative study between the NFW~\cite{Navarro:1996gj}, 
Burkert (BURK)~\cite{Burkert:1995yz, Salucci:2011ee} and Pseudo-Isothermal 
(ISO)~\cite{Gunn:1972sv} profiles \cite{Bhattacharjee:2020phk}. We have estimated the J-factor of each UFDs 
for three density profiles \cite{Bhattacharjee:2020phk}. From Table 8.2, we could find that Burkert provides 
stronger limits than NFW, while ISO imposes the weakest limits \cite{Bhattacharjee:2020phk}. In Table 8.2, we 
have also compared our estimated J values for NFW profile with the J values 
derived by the Pace {\em et al}\cite{Pace:2018tin}. 

\begin{table}[!t]
\centering
\begin{tabular}{|p{2.5cm}|c|c|c|c|}
\hline \hline
Galaxy & \multicolumn{4}{c|}{$\log_{10}(J(0.5^{\circ})/{\rm GeV}^2\, {\rm cm}^{-5})$}\\
\cline{2-5}
& Pace {\em et al}\cite{Pace:2018tin} & \multicolumn{3}{c|}{Direct Integration}\\
\cline{3-5}
& (NFW) & NFW & Burkert & ISO \\
\hline \hline
Aquarius II & $18.27^{+0.65}_{-0.59}$ & $18.11^{+0.68}_{-0.63}$   & $18.53^{+0.72}_{-0.66}$ & $18.01^{+0.73}_{-0.66}$ \\     
\hline \hline                                                                    
Carina II & $18.24^{+0.53}_{-0.53}$ & $18.16^{+0.55}_{-0.53}$     & $18.45^{+0.60}_{-0.56}$ & $18.05^{+0.58}_{-0.54}$ \\     
\hline \hline                                                                    
Draco II & $18.97^{+1.29}_{-1.69}$ & $19.07^{+1.33}_{-1.69}$      & $19.54^{+1.35}_{-1.70}$ & $18.90^{+1.34}_{-1.70}$ \\     
\hline \hline                                                                    
Eridanus II & $17.29^{+0.35}_{-0.26}$ & $17.14^{+0.35}_{-0.30}$   & $17.68^{+0.35}_{-0.31}$ & $17.06^{+0.35}_{-0.31}$ \\     
\hline \hline                                                                    
Grus-I & $16.87^{+1.52}_{-1.68}$ & $16.94^{+1.57}_{-1.74}$        & $17.48^{+1.60}_{-1.75}$ & $16.76^{+1.54}_{-1.67}$ \\     
\hline \hline                                                                    
Horologium I & $19.25^{+0.79}_{-0.70}$ & $19.01^{+0.83}_{-0.73}$  & $19.37^{+0.85}_{-0.75}$ & $18.73^{+0.85}_{-0.75}$ \\     
\hline \hline                                                                    
Hydra II & $<~17.71$ & $<~17.92$  &  $<~18.46$   &  $<~17.84$ \\                                                
\hline \hline                                                                    
Leo V & $17.69^{+0.93}_{-0.99}$ & $17.91^{+1.03}_{-1.06}$   & $18.51^{+1.02}_{-1.08}$ & $17.84^{+1.01}_{-1.07}$ \\     
\hline \hline                                                                    
Pegasus III & $18.41^{+0.89}_{-1.07}$ & $18.46^{+0.94}_{-1.05}$   & $19.06^{+1.02}_{-1.07}$ & $18.39^{+1.03}_{-1.05}$ \\     
\hline \hline                                                                    
Pisces II & $17.31^{+0.97}_{-0.107}$ & $17.53^{+1.02}_{-1.09}$    & $18.10^{+1.04}_{-1.09}$ & $17.45^{+1.03}_{-1.09}$ \\     
\hline \hline                                                                    
Reticulum II & $18.95^{+0.57}_{-0.52}$ & $18.76^{+0.53}_{-0.48}$ & $19.21^{+0.53}_{-0.54}$ & $18.66^{+0.53}_{-0.53}$ \\     
\hline \hline                                                                    
Triangulum II & $<~19.72$ & $<~19.74$  &$<~20.18$ & $<~19.64$ \\                                                
\hline \hline                                                                    
Tucana II & $19.02^{+0.57}_{-0.52}$ & $18.93^{+0.62}_{-0.58}$   & $19.22^{+0.64}_{-0.61}$ & $18.83^{+0.66}_{-0.62}$ \\     
\hline \hline                                                                    
Tucana III & $<~17.68$ & $<~17.87$                        & $<~18.20$   & $<~17.76$ \\                                                
\hline \hline                                                                    
Draco & $18.83^{+0.10}_{-0.10}$ & $18.85^{+0.12}_{-0.12}$ & $19.08^{+0.13}_{-0.13}$ & $18.75^{+0.13}_{-0.13}$ \\     
\hline \hline
\end{tabular}
\caption{The astrophysical factors (J-factor) of our selected UFDs deriving from 
the Eq.~2.7 for NFW, Burket and ISO DM density profiles at 
$\theta_{max}=0.5^{\circ}$. Also mentioned J-factors of NFW profile estimated by 
the scaling relation from Pace \textit{et al.}, 2019.} 
\label{table:table-1}
\end{table}

\section{Analysis of $\gamma$-ray Fluxes from UFDs}
\label{sec:analysis}

\noindent Since the last decade, several dSphs/UFDs have been studied in order 
to investigate the DM signal but no strong emission has been detected from their 
location. But even the null detection can provide an intriguing knowledge on the 
DM signature \cite {Ackermann:2011wa,GeringerSameth:2011iw,Ackermann:2013yva, 
Ackermann:2015zua, Fermi-LAT:2016uux}. With all these keeping in mind, we have 
chosen a recently discovered 14 UFDs and have analyzed nearly eleven years 
(2008-09-01 to 2019-02-04) of Fermi-LAT data \cite{Bhattacharjee:2020phk}. For our analysis, we have used the 
Fermi ScienceTools version, v1.2.1 and have accessed the source class IRF, 
$\rm{P8R3\_SOURCE\_V2}$ processed data \cite{Bhattacharjee:2020phk}. We have considered the energy range $E$, 
{\em viz.} $E\in [0.1, 300]$~GeV and have extracted data within the $15^{\circ}$ 
ROI around the location of each UFDs \cite{Bhattacharjee:2020phk}. We have then generated the source model 
file where, we have included our `source of interest' along with all the sources 
within $20^{\circ}$ ROI from the 4FGL catalog \cite{Fermi-LAT:2019yla}. In 
addition, we have also added the galactic ($\rm{gll\_iem\_v07.fits}$) and 
isotropic ($\rm{iso\_P8R3\_SOURCE\_V2\_v1.txt}$) diffuse models to our source 
model \cite{Bhattacharjee:2020phk}. Next, we have performed the binned likelihood analysis \cite{Cash:1979vz, 
Mattox:1996zz} on our extracted dataset and during the process, the spectral 
parameters of all the sources within $15^{\circ}~\times~15^{\circ}$ ROI and the 
normalization parameters for two diffuse backgrounds models have been left free. 
The necessary information for Fermi-LAT analysis is mentioned in TABLE~8.3 \cite{Bhattacharjee:2020phk}.

\begin{table}
    \caption{Parameters used for the analysis of \textit{Fermi}-LAT data.}
    \begin{tabular}{||p{7 cm}p{8 cm}||}
        \hline \hline
        {\bf Parameter for data extraction}  &\\ 
                \hline\hline
        Parameter & Value\\
        \hline \hline
        Radius of interest (ROI) &  $15^{\circ}$\\
        TSTART (MET) & 241976960 (2008-09-01 15:49:19.000 UTC)\\
        TSTOP (MET) & 570987500 (2019-02-04 15:38:15.000 UTC)\\
        Energy Range & 100 MeV - 300 GeV\\
        \textit{Fermitools} version & \texttt{1.2.1}\\ 
        \hline \hline
        \texttt{gtselect} for event selection &\\ 
                \hline \hline
        Event class & Source type (128)\\ 
                Event type & Front+Back (3)\\ 
        Maximum zenith angle cut & $90^{\circ}$\\
        \hline \hline
        \texttt{gtmktime} for time selection &\\ 
                \hline \hline
        Filter applied & $\textit{(DATA\_QUAL>0)\&\&(LAT\_CONFIG==1)}$\\ 
                ROI-based zenith angle cut & No\\ 
        \hline \hline
        \texttt{gtltcube} for livetime cube &\\ 
                \hline \hline
        Maximum zenith angle cut ($z_{cut}$) & $90^{\circ}$\\ 
                Step size in $cos(\theta)$ & 0.025\\
        Pixel size (degrees) & 1\\
        \hline \hline
        \texttt{gtbin} for 3-D counts map &\\ 
                \hline \hline
        Size of the X $\&$ Y axis (pixels) & 140\\
        Image scale (degrees/pixel) & 0.1\\
        Coordinate system & Celestial (CEL)\\
        Projection method & AIT\\
        Number of logarithmically uniform energy bins & 24\\ 
        \hline \hline
        \texttt{gtexpcube2} for exposure map &\\ 
                \hline \hline
        Instrument Response Function (IRF) & $\rm{P8R3\_SOURCE\_V2}$\\ 
                Size of the X and Y axis (pixels) & 400\\
        Image scale (degrees/pixel) & 0.1 \\
        Coordinate system & Celestial (CEL)\\
        Projection method & AIT\\
        Number of logarithmically uniform energy bins & 24\\ 
        \hline \hline
        diffuse models and Source model XML file &\\ 
                \hline \hline
        Galactic diffuse emission model & $\rm{gll\_iem\_v07.fits}$\\ 
                Extragalactic isotropic diffuse emission model & $\rm{iso\_P8R3\_SOURCE\_V2\_v1.txt}$\\
        Source catalog & 4FGL\\
        Extra radius of interest &  $5^{\circ}$\\
        Spectral model &  DMFit Function\cite{Jeltema:2008hf}\\ 
        \hline \hline
    \end{tabular}
    \label{table:fermi_lat_parameters}
\end{table} 

\subsection{Constraints on DM Annihilation with Eleven Years of Fermi-LAT Data}\label{section:gamaray_sigmav_constraint}

\noindent In order to investigate the $\gamma$-ray signal from the location of 
our `source of interest', we have modelled our targets with the 
power-law spectrum (i.e., $dN/dE \propto E^{-\Gamma}$) for spectral index 
$\Gamma$ = 2 \cite{Ackermann:2013yva, Ackermann:2015zua, Fermi-LAT:2016uux, 
Bhattacharjee:2018xem}. Unfortunately, we have not observed any strong emission 
from the location of UFDs.

\begin{figure}[h!]
\begin{minipage}[c]{0.5\textwidth}
{\small
\begin{center}
    \begin{tabular}{|l|r|r|}
          \hline
  UFD & $TS_{\rm peak}(b)$ & $TS_{\rm peak}(\tau)$ \\[-0.3ex]
  \hline
  Aquarius II  & 2.88 & 2.94 \\[-0.3ex]
  Carina II    & 1.24 & 1.81 \\[-0.3ex]
  Draco II     & 1.37 & 1.88 \\[-0.3ex]
  Eridanus II  & 0.81 & 1.23 \\[-0.3ex] 
  Grus I       & 1.59 & 1.65 \\[-0.3ex]
  Horologium I & 4.21 & 4.71 \\[-0.3ex]
  Hydra II     & 2.21 & 2.31 \\[-0.3ex]
  Leo V        & 0.88 & 0.92 \\[-0.3ex]
  Pegasus III  & 1.91 & 2.13 \\[-0.3ex]
  Pisces II    & 1.22 & 1.96\\[-0.3ex]
  Reticulum II & 4.85 & 4.95 \\[-0.3ex]
  Tucana II    & 11.87 & 12.47 \\[-0.3ex]
  Tucana III    &  4.36 & 4.53 \\[-0.3ex]  
  Triangulum II & 1.19 & 1.25 \\[-0.3ex] 
  \hline
  \end{tabular}
\end{center} 
}

\end{minipage}
\hfill
\begin{minipage}{0.5\textwidth}
\includegraphics[width=\textwidth,clip,angle=0]{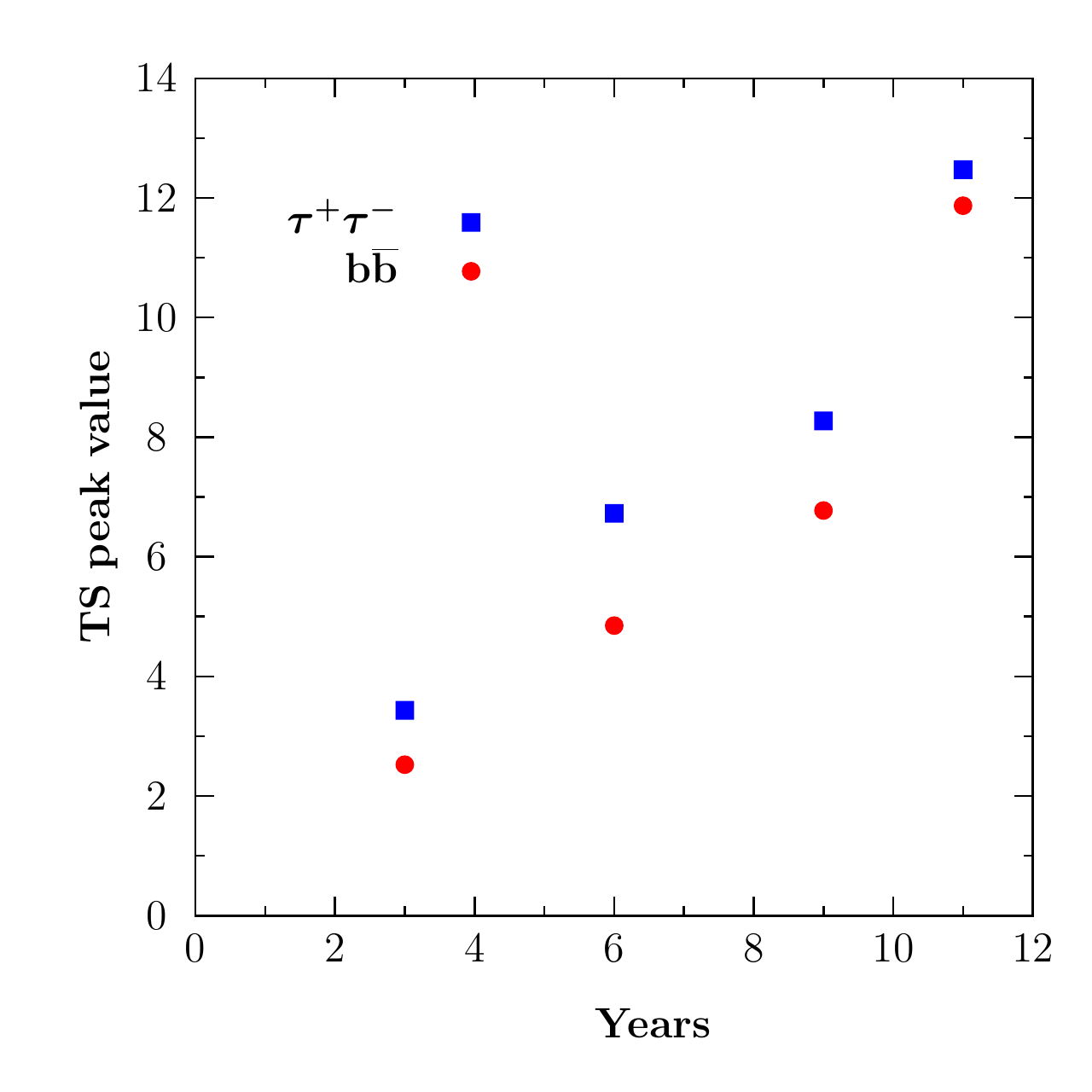}
\end{minipage}
\caption{The maximum TS values (or peak value) detected from location of our 
selected UFDs for $b\bar{b}$ and $\tau^{+}\tau^{-}$ final states with eleven 
years of LAT data (left). The TS peak value observed from the location of Tucana 
II for three, six, nine and eleven years of LAT data (right).}
\label{figure:ts_tucana}
\end{figure}

\noindent We would like to mention out that except for Tucana II, we have not 
observed any faint emission from the location of other UFDs (i.e., for them TS 
$\le$ 5). In Fig.~8.1(a), we have listed the TS peak value of UFDs for 
$b\bar{b}$ and $\tau^{+}\tau^{-}$ annihilation channels. An intriguing hint of a 
faint emission had been reported from the direction of Tucana-II in a recent 
publication (ref.~\cite{Bhattacharjee:2018xem}). The significance of this faint 
emission was shown to increase with time. In Fig.~8.1(b), we have shown the peak 
TS value as a function of time for Tucana-II. As was seen in 
ref.~Bhattacharjee et al. \cite{Bhattacharjee:2018xem}, the significance seems to grow even with a
11 years of LAT data. But the observed significance with eleven years of 
Fermi-LAT data is still faint (i.e., TS $<$ 25) enough to state any strong claim of the 
existence of a signal.\\

\noindent As we have not detected any strong emission from the direction of 
UFDs, we have then derived the 95$\%$ C.L. gamma-ray flux upper limits from the 
region of these objects \cite{Bhattacharjee:2020phk}. For this purpose, we have used the Bayesian approach 
(\cite{Helene:1990yi}), which is sensitive~\cite{Rolke:2004mj, Barbieri:1982eh} 
for the low statistics analysis. The approach was developed by Helene 
\cite{Helene:1990yi} and is implemented in the Fermi-\texttt{ScienceTools}.\\

\begin{figure}[h!]
\centering
 \includegraphics[width=.49\linewidth]{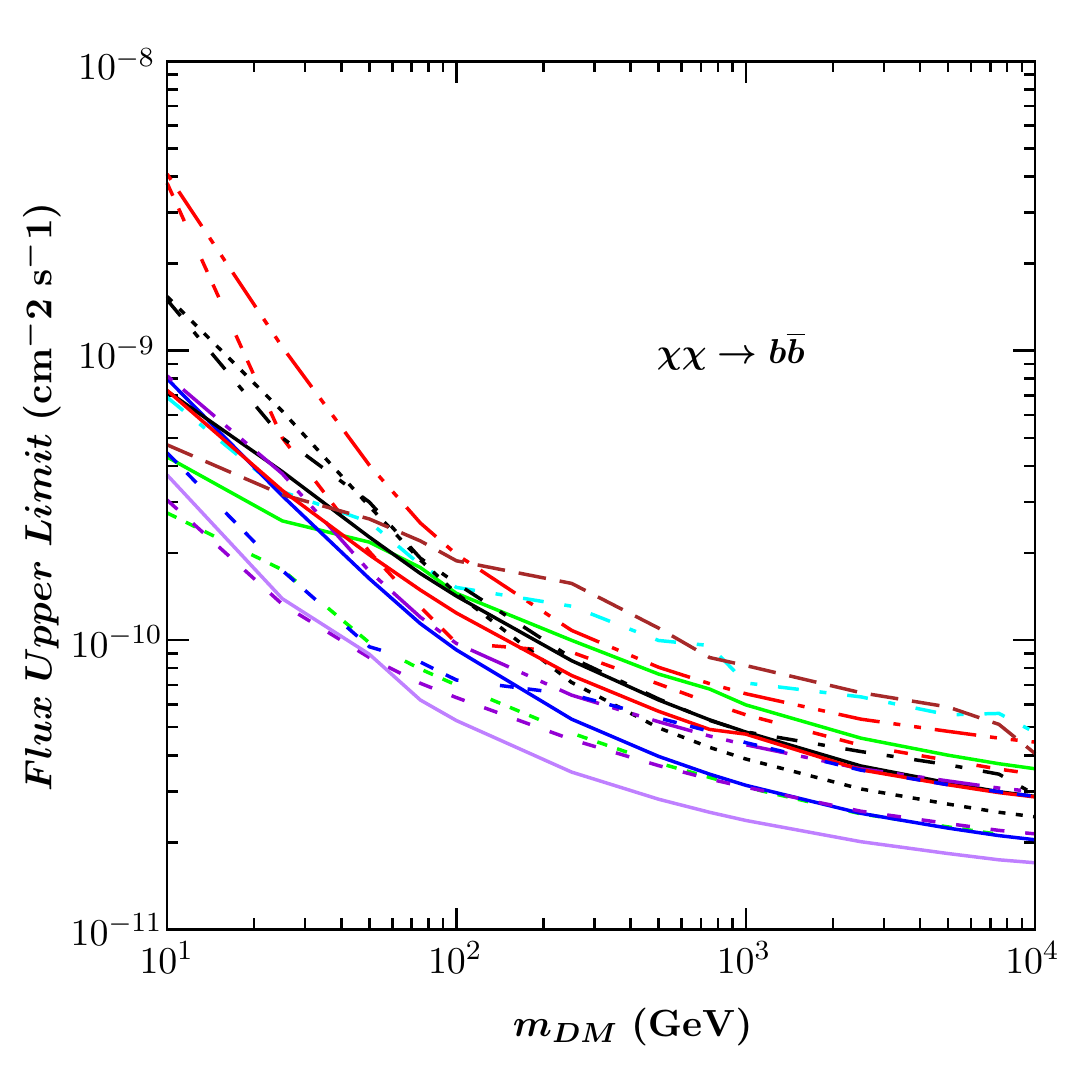}
 \includegraphics[width=.49\linewidth]{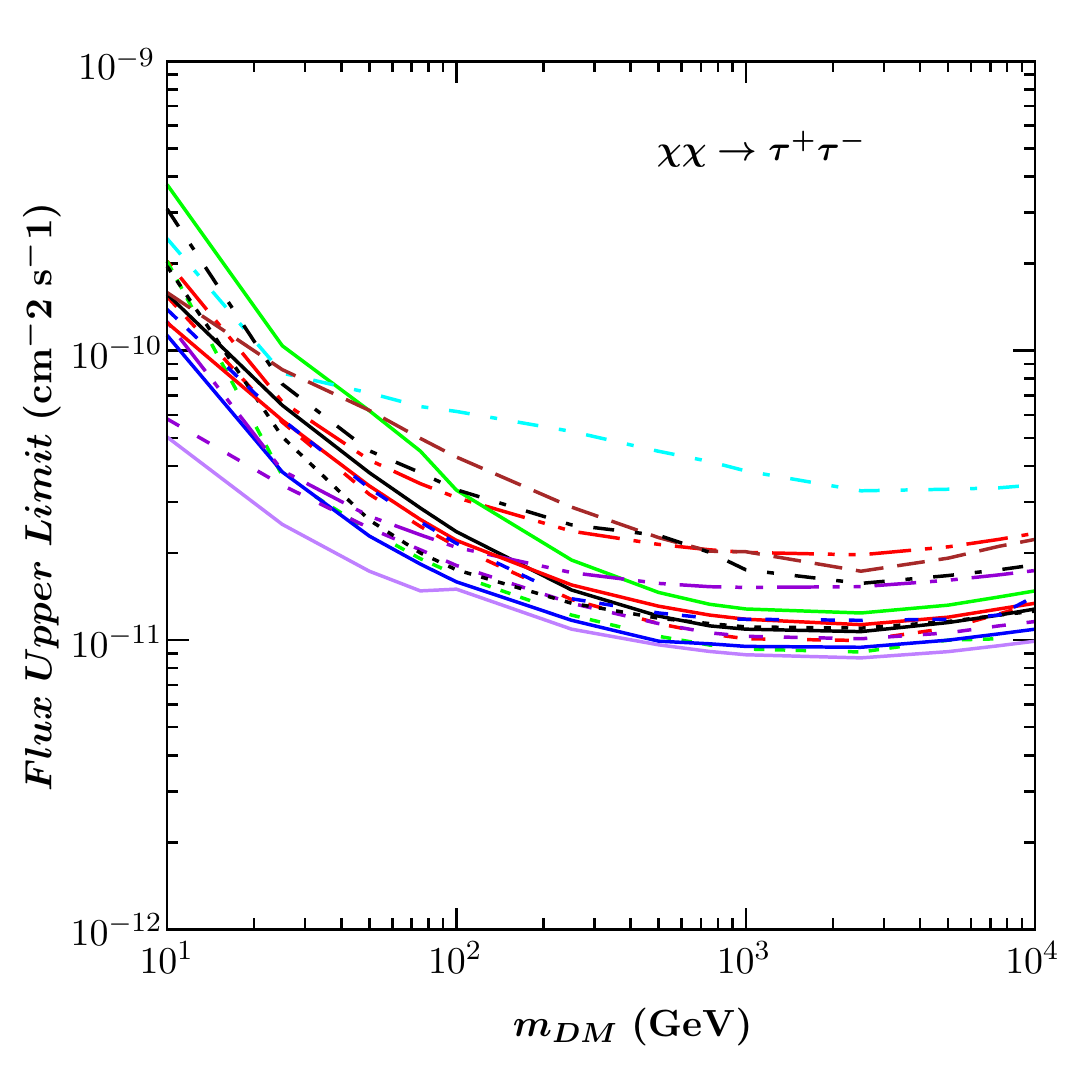}
 \includegraphics[width=.5\linewidth]{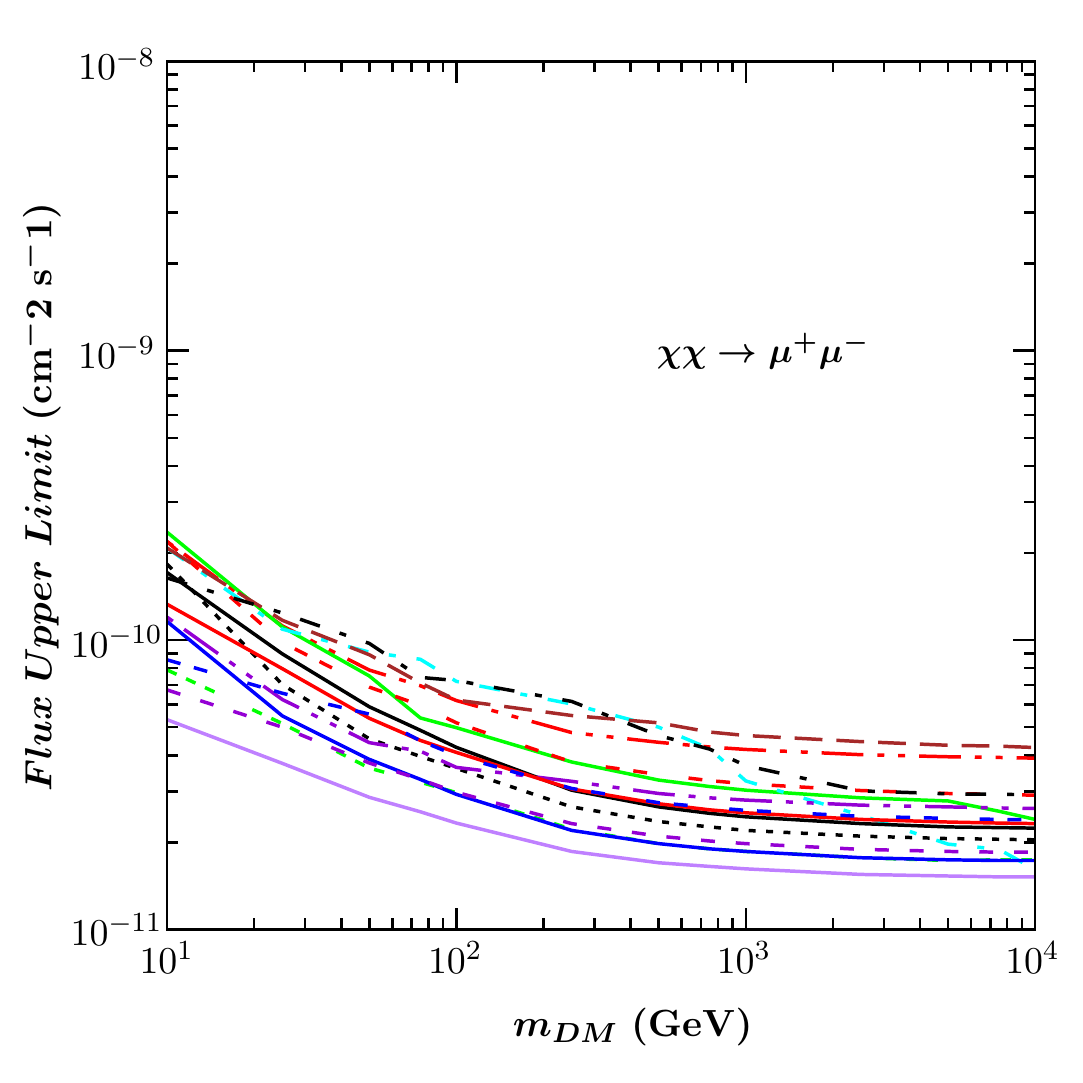}
 \hskip 10pt
 \includegraphics[width=1.0\linewidth]{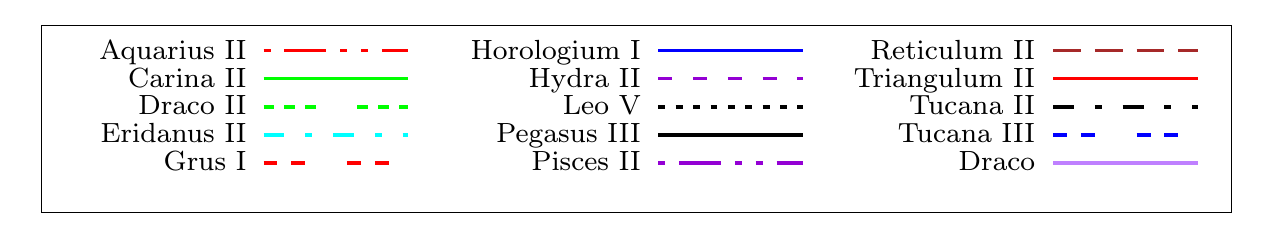}
  \caption{$95\%$ C.L. $\gamma$-ray flux upper limits of our selected UFDs for 
$b\bar{b}$, $\tau^+\tau^-$ and $\mu^+\mu^-$ pair-annihilation channels.} 
   \label{figure:fermi_flux}
\end{figure}

\begin{figure}[h!]
  \centering
  \includegraphics[width=0.49\linewidth]{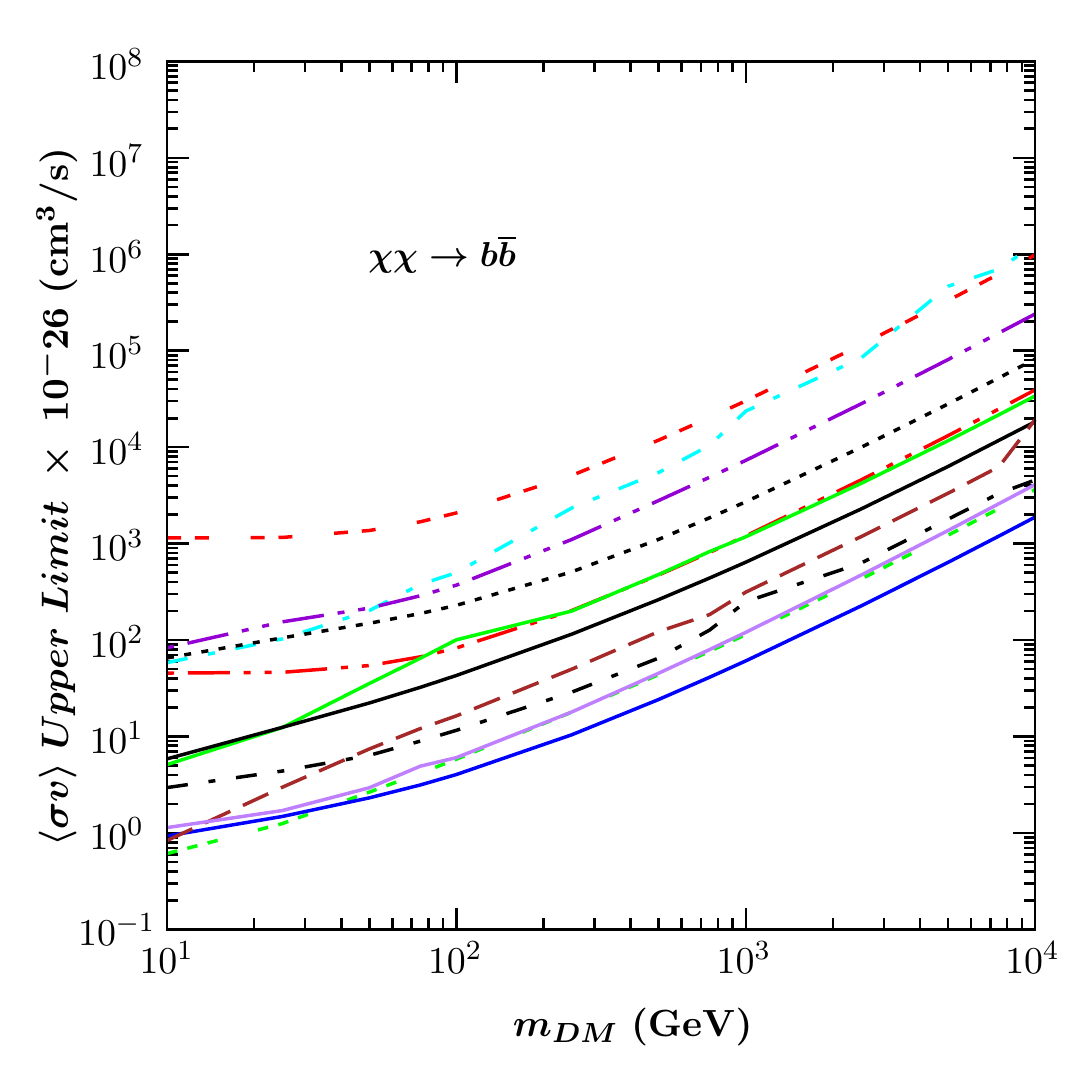}
  \includegraphics[width=0.49\linewidth]{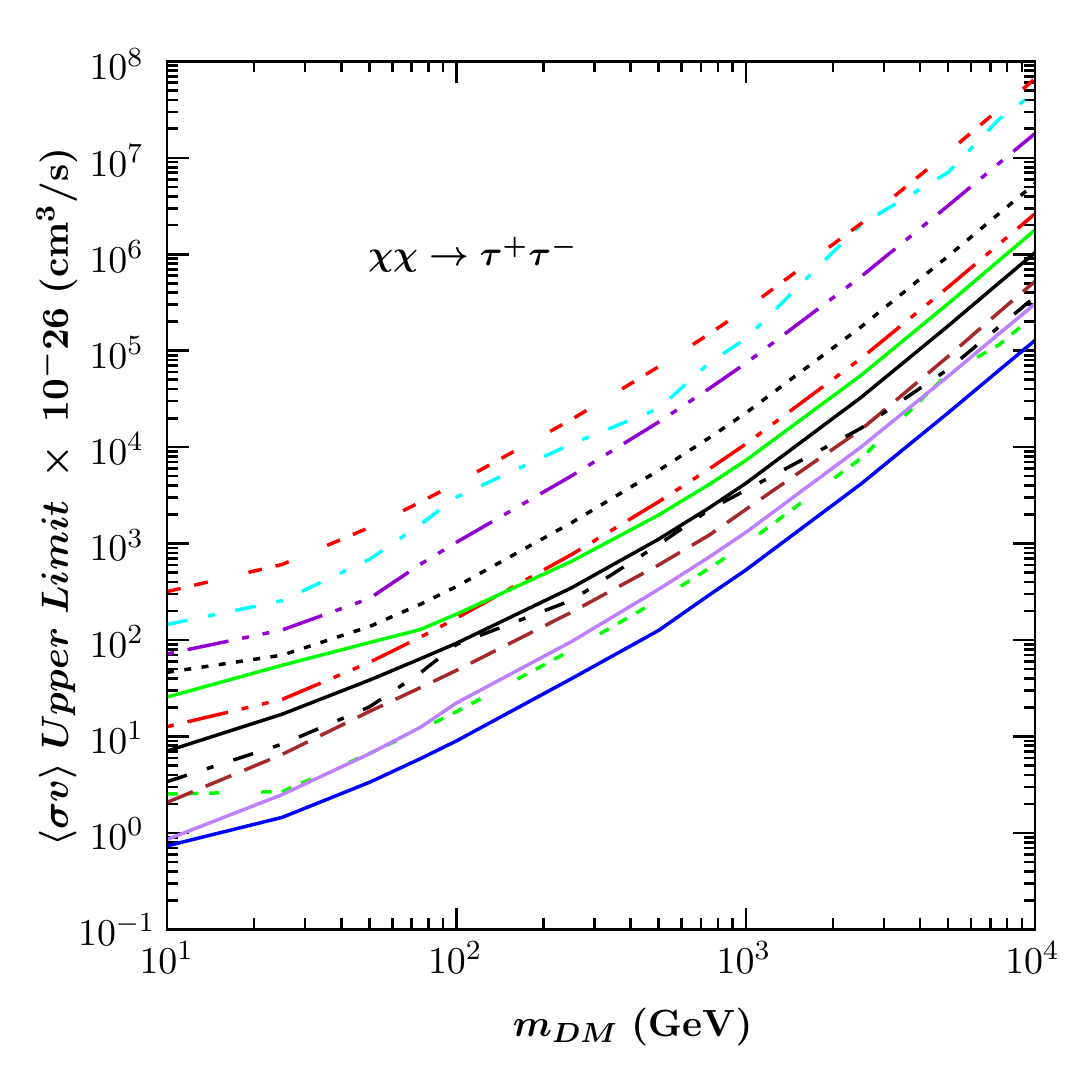} 
  \includegraphics[width=0.5\linewidth]{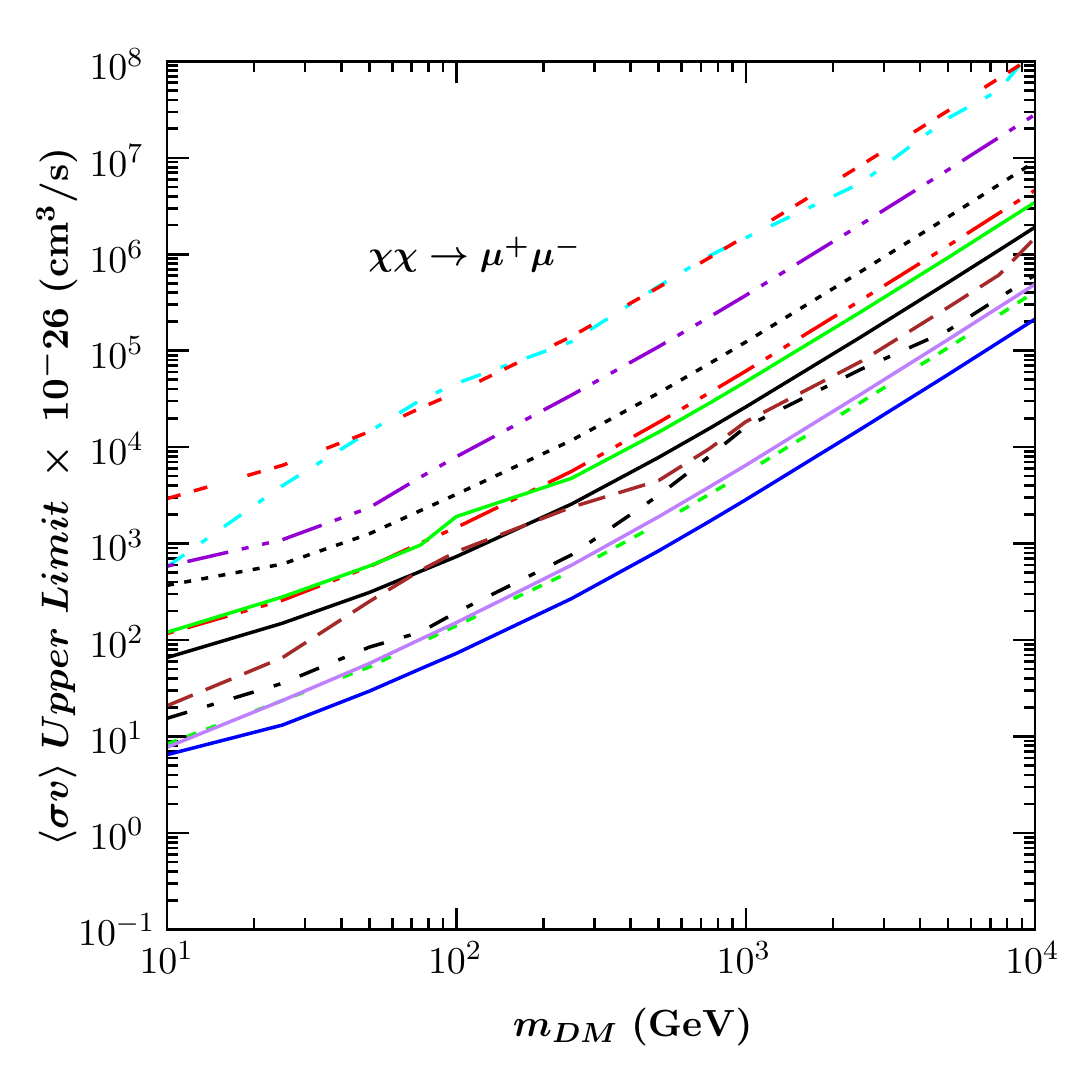}
  \hskip 10pt
  \includegraphics[width=1.0\linewidth]{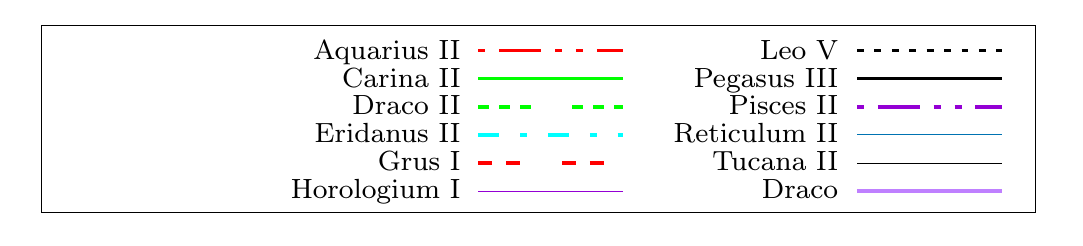}
  \label{fig:cross_legends}
 \caption{$95\%$ C.L. $\langle \sigma v \rangle$ upper limit of our selected 
UFDs for $b\bar{b}$, $\tau^+\tau^-$ and $\mu^+\mu^-$ pair-annihilation channels. 
We have not included the limits from Triangulum II, Hydra II and Tucana III
 as they only have the upper limits of $J$-factor.}
   \label{figure:fermi_cross}
\end{figure}

\noindent The aforesaid $\gamma$-ray flux upper limits obtained from the location of our targets can be translated 
to the WIMP pair-annihilation cross-section, $\langle \sigma v\rangle$ as a 
function of DM mass and WIMP annihilation channels \cite{Bhattacharjee:2020phk}. We have adopted three pair 
annihilation final states; such as $b \bar b$, $\tau^+\tau^-$ and $\mu^+\mu^-$. 
For estimating the 95$\%$ C.L. limits on $\langle \sigma v \rangle $, we have 
modelled the 
$\gamma$-ray flux upper limits with the DMFitFunction 
\cite{{Jeltema:2008hf}}\footnote{\tiny{https://fermi.gsfc.nasa.gov/ssc/data/analysis/scitools/source{\_}models.html}}\\

\noindent The consequent upper limits on $\gamma$-ray flux and $\langle \sigma v 
\rangle$ limits for all three annihilation channels are shown in Fig.~8.2 and 8.3 \cite{Bhattacharjee:2020phk}. 
From Fig~8.2, we can observe that for most of DM mass range, Draco provides the 
strongest limits for all three channels. In Fig.~8.3, we have shown the LAT 
sensitivity in  ${\rm m_{DM}}- \langle \sigma v \rangle$ plane for all 15 UFDs \cite{Bhattacharjee:2020phk}. 
The obtained limits in Fig.~8.3 depend on the $J$-factor and the DM density 
profiles. Among all our considered UFDs, Horologium I, due to its largest 
$J$-factor, has imposed the most stringent limit on ${\rm m_{DM}}- \langle 
\sigma v \rangle$ plane for
all three annihilation final states \cite{Bhattacharjee:2020phk}. But, we also should not ignore the large 
uncertainties associated with the J-factor of Horologium I. Thus, the limit 
obtained from Horologium I might not be as robust as we can expect from Draco. 
In Fig.~8.3, we have not showed the $\langle \sigma v \rangle $ limits for Triangulum II, Hydra II and 
Tucana III, because they can only produce the limiting values for $\langle 
\sigma v \rangle$ due to their upper 
limits of J-factor \cite{Bhattacharjee:2020phk}.

\section{Synchrotron Radiation from UFDs}
\label{sec:synchr}

\noindent As we have seen above, the limits obtained from the $\gamma$-ray data 
are directly dependent on the J-factor but this is not the case for synchrotron 
emission. The radio emission generating from synchrotron emission strongly 
depends on the diffusion coefficient ($D_{0}$), magnetic field ($B$) and energy 
loss mechanism, etc. The magnetic field of dSphs are not well-studied but 
several studies suggest to consider the $B$ $\approx$ 1 $\mu$G for dSphs 
\cite{Colafrancesco:2006he, McDaniel:2017ppt, Spekkens:2013ik}. For our 
analysis, we have also assumed the same \cite{Colafrancesco:2006he, 
Jeltema:2008hf}. For diffusion coefficient, we have considered the simplified 
form of it, i.e., $D(E) = D_{0} \left(\frac{E}{1 \GeV}\right)^{\gamma_D}$, where, 
$D_{0}$ is the diffusion constant. For galaxy clusters, $D_0$ lies between the 
range of $10^{28}$--$10^{30}\, {\rm cm}^2/{\rm s}$~\cite{Natarajan:2015hma, 
Jeltema:2008ax}, while for Milky Way it stands between $10^{27}$--$10^{29}\, 
{\rm cm}^2/{\rm s}$~\cite{Webber:1992bn, Baltz:1998xv, Maurin:2001sj}. 
Similarly, $\gamma_D$ is expected to lie between $0\leq \gamma_D \leq 1$ 
\cite{Jeltema:2008ax}. For our analysis, we have fixed $D_0$ and $\gamma_{D}$ at 
$3 \times 10^{28}\, {\rm cm}^2/{\rm s}$ and $0.3$\cite{McDaniel:2017ppt}, 
respectively. \\

\noindent For a specific DM mass, the synchrotron emission would also depend on 
the WIMP pair-annihilation channels and their relative cascades. Just like our 
$\gamma$-ray analysis, here we have again considered three annihilation final 
states; such as $b\bar{b}$, $\tau^+ \tau^-$ and $\mu^+ \mu^-$. Next, in order to 
predict the possible synchrotron emission resulting from the DM annihilation, we 
have used a publicly accessible code, RX-DMFIT \cite{McDaniel:2017ppt} which is 
an extension of the DMFit tool \cite{Jeltema:2008hf, Gondolo:2004sc}. As a 
default we have used the NFW density profile and have fixed the pair 
annihilation cross-section, $\langle \sigma v \rangle$ at $10^{-26} \, {\rm 
cm}^3/{\rm s}$ \cite{Bhattacharjee:2020phk}. In addition, we have used the thermal electron density
$n_{e} \approx 10^{-6}$ cm$^{-3}$ \cite{Colafrancesco:2006he,McDaniel:2017ppt} 
for all our selected UFDs.\\

\noindent Using the parameters, $d$, $r_{1/2}$, and $\sigma$ listed in Table 
8.1, we have calculated characteristic density ($\rho_{s}$), scale radius 
($r_{s}$) and diffusion zone ($r_{h}$). The parameter values mentioned in Table 
8.4 are derived from the `central values' of $d$, $r_{1/2}$, and $\sigma$ \cite{Bhattacharjee:2020phk}.

\begin{table}[!ht]
\begin{center}
 \begin{tabular}{c c c c c}
  dSphs        & d(Kpc) & $r_h(Kpc)$ & $\rho_s (GeV/cm^3)$ & $r_s (Kpc)$ \\ 
\hline
  Aquarius II  & 107.9& 0.42  & 2.27  & 0.615  \\
  Carina II    & 37.4 & 0.3 & 1.78 & 0.38 \\
  Draco II     & 20 & 0.07 & 71.73  & 0.06 \\
  Eridanus II  & 366 & 0.792 & 1.454 & 0.88 \\ 
  Grus I       & 120.2 & 0.39 & 6.7 & 0.26 \\
  Horologium I & 79 & 0.188 & 30.55 & 0.16 \\
  Hydra II      & 151 & 0.448 & < 8.24 & 0.335 \\
  Leo V        & 173 & 0.465 & 23.83 & 0.15 \\
  Pegasus III  & 215 & 0.228 & 40.73  & 0.185  \\
  Pisces II    & 183 & 0.438 & 8.93 & 0.24 \\
  Reticulum II & 30 & 0.251 & 10.08  & 0.16 \\
  Tucana II    & 57.5 & 0.452 & 3.6  & 0.575 \\
  Tucana III    &  25 & 0.174 & < 2.29 &  0.215 \\  
  Triangulum II & 30 & 0.157 & < 46.1 &  0.14  \\ 
  \hline \hline
  \textbf{Draco} & 76 & 2.5   & 1.4    & 1.0 \\  
  \hline
 \end{tabular}
 \caption{The astrophysical parameters for our selected UFDs along with the 
classical dSphs Draco. The values for $r_h$, $\rho_s$ and 
$r_s$ has been derived from the `central values' of the astrophysical parameters 
listed in Table 8.1.}
\label{table:astro_param_dwarfs}
\end{center} 
\end{table}

\begin{figure}[!h]
\centering
  \subfigure[]
{\includegraphics[width=0.49\linewidth]{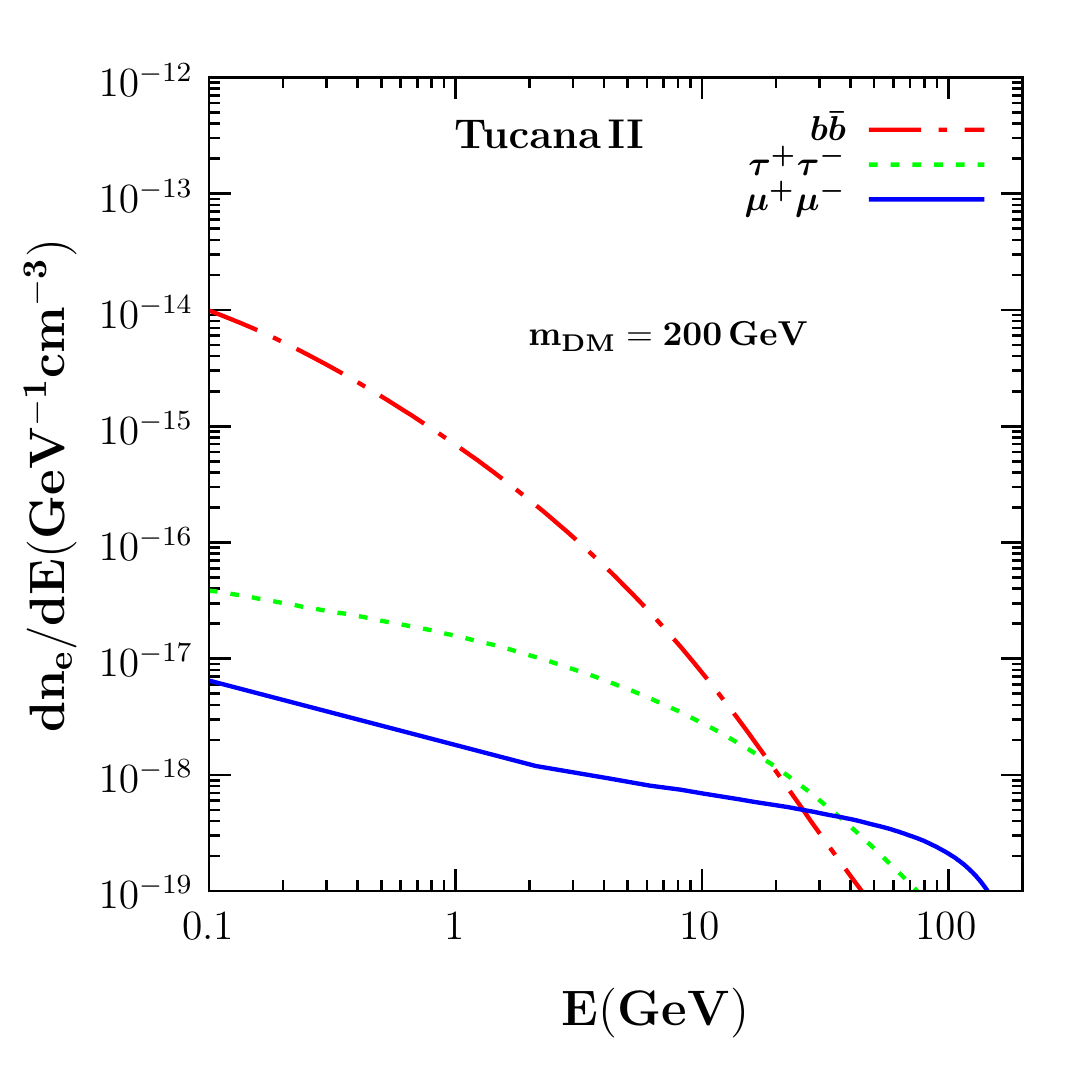}}
    \label{fig:dnde_200GeV}
  \subfigure[]
{\includegraphics[width=0.49\linewidth]{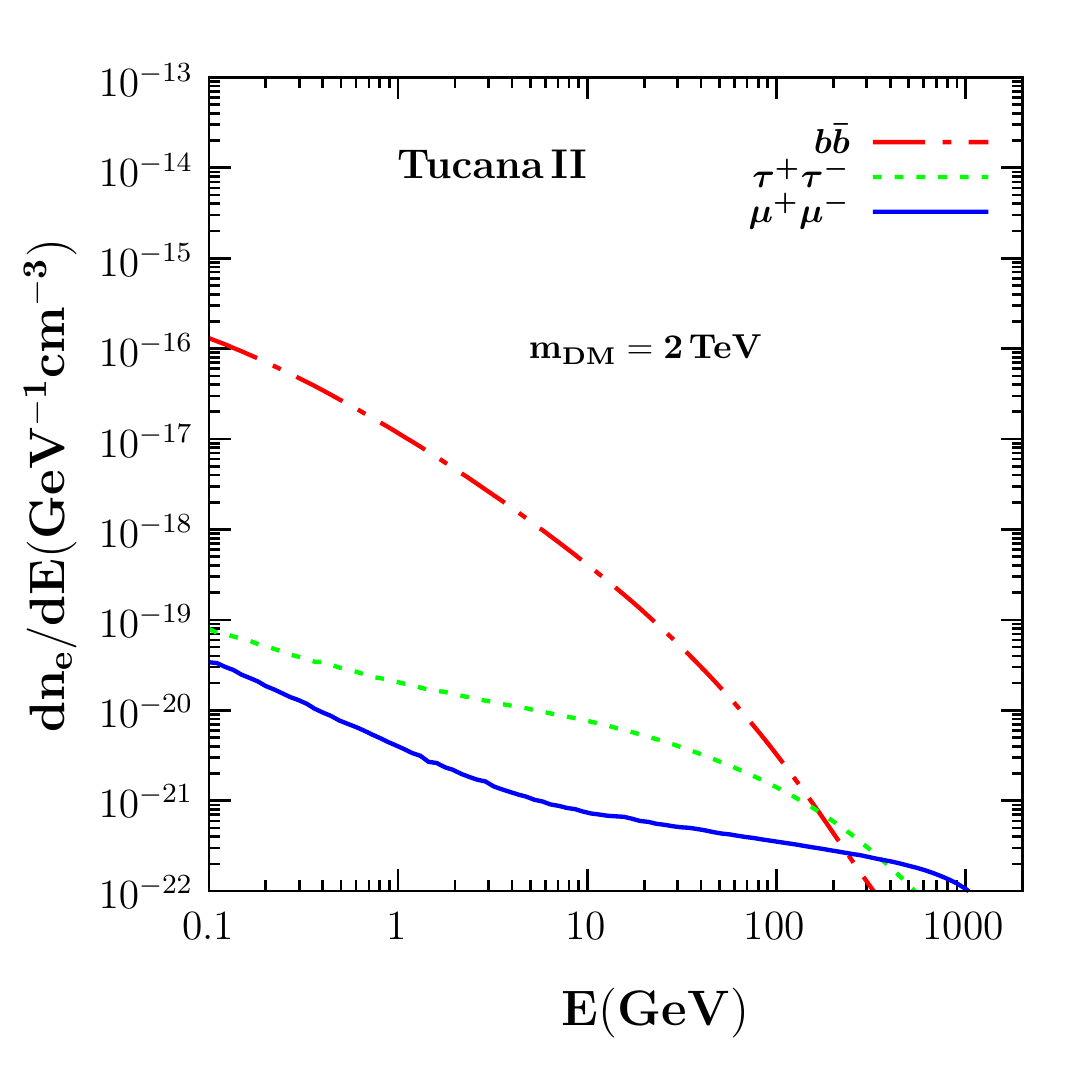}}
    \label{fig:dnde_2TeV}
 \caption{The $e^{\pm}$ distribution spectrum of Tucana II at equilibrium for 
radial distance $r=0.1$~kpc and three pair-annihilation channels, such as: 
$b\bar{b}$ (red), $\mu^+ \mu^-$ (blue) and $\tau^+ \tau^-$ (green). We have 
considered NFW profile and fixed the parameters at $\langle \sigma v\rangle \, = 
10^{-26}$ cm$^3$/s, $B \, = \, 1\, \mu$G, $D_0 = 3 \times 10^{28}$ cm$^2$/s, 
$\gamma_D = 0.3$. The spectrum for DM masses 200 GeV and 2 TeV have been shown 
in left and right panels, respectively.} 
  \label{figure:dnde}
 \end{figure}

\noindent In Fig.~8.4, we have shown the $e^{\pm}$ distribution spectrum of Tucana II
at a radial distance 0.1 kpc and for DM masses, 2 TeV and 200 GeV \cite{Bhattacharjee:2020phk}. The cascade 
channels resulting from the $b\bar{b}$ annihilation could produce a large amount 
of $e^\pm$ that we can expect from the $\tau^{+}\tau^{-}$ or the 
$\mu^{+}\mu^{-}$ annihilation channel. Thus, the integrated spectrum obtained 
from the $b\bar{b}$ channel would be larger than the $\tau^{+}\tau^{-}$ and the 
$\mu^{+}\mu^{-}$ \cite{Bhattacharjee:2020phk}. From Fig.~8.4, we can also explain the relative softness 
between three annihilation channels.

 \begin{figure}[!h]
\centering
  \subfigure[]
   {\includegraphics[width=0.50\linewidth]{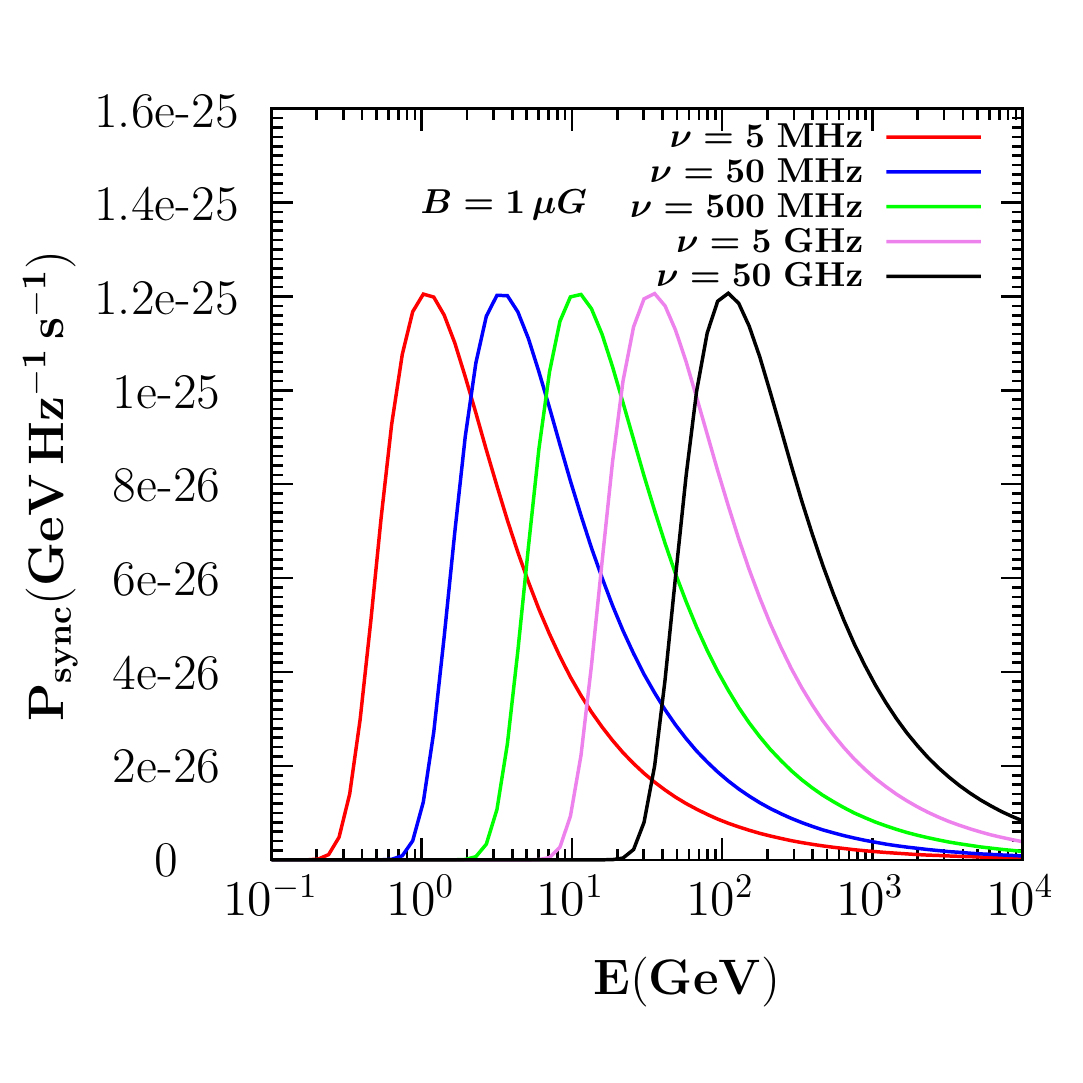}}
    \label{fig:synpower}
  \subfigure[]
 {\includegraphics[width=0.49\linewidth]{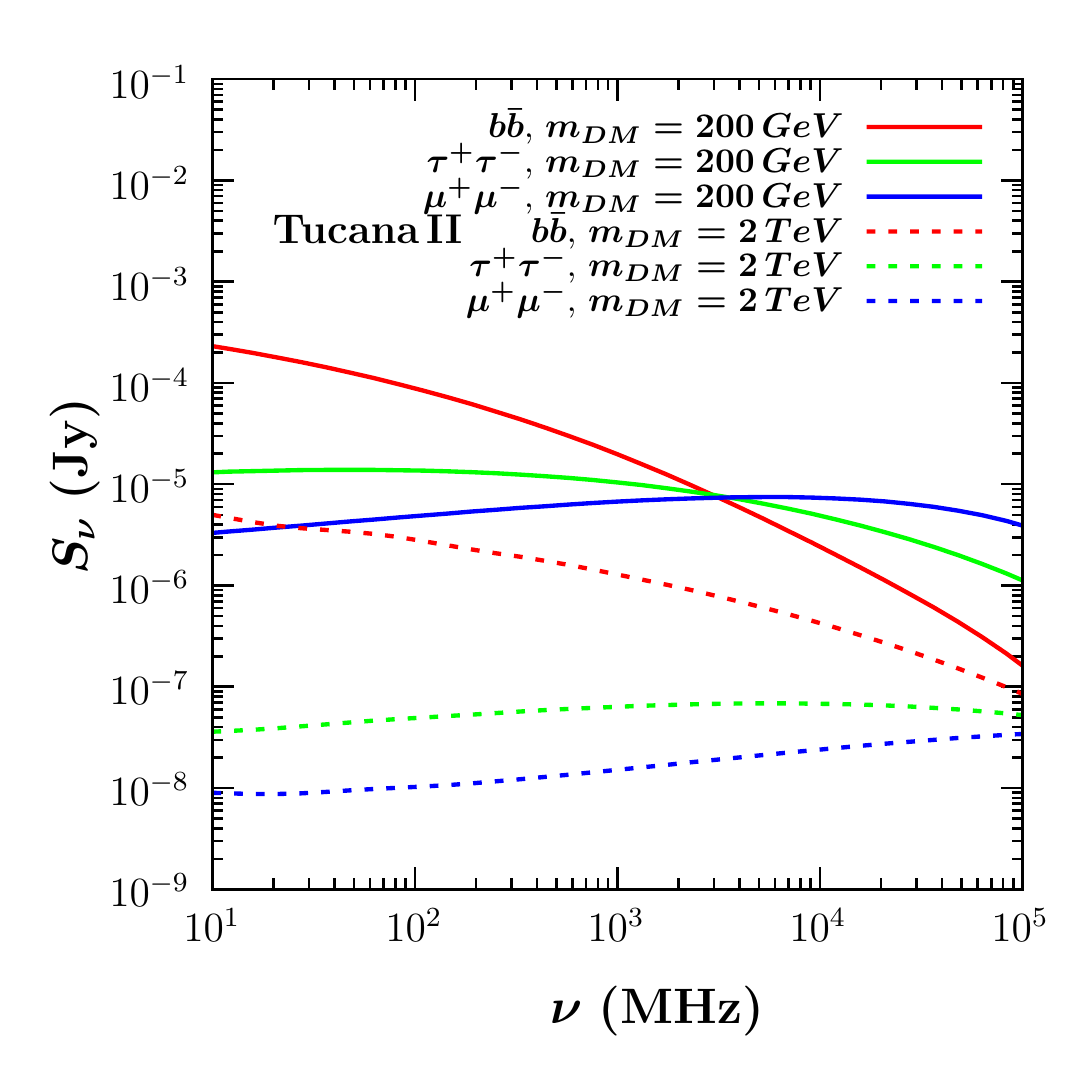}}
    \label{fig:bbbartautau_flux_comp}
 \caption{(a) The power-spectrum at five different frequency values for $B \, = 
\, 1\, \mu$G. (b) The synchrotron flux densities for $b\bar{b}$ (red), $\mu^+ 
\mu^-$ (blue) and $\tau^+ \tau^-$ (green) annihilation channels. The fluxes for 
DM masses 200 GeV and 2 TeV have been denoted with solid and dashed lines, 
respectively. We have considered NFW profile and fixed the parameters at 
$\langle \sigma v\rangle \, = 10^{-26}$ cm$^3$/s, $B \, = \, 1\, \mu$G, $D_0 = 3 
\times 10^{28}$ cm$^2$/s, $\gamma_D = 0.3$.} 
 \end{figure}

\noindent In Fig.~8.5(a), we have shown the power-spectrum ($P_{\rm 
synch}(\nu,E,B)$) of Tucana II at $B= 1$ $\mu$G for frequency range between 5 
MHz to 50 GHz. We find that $P_{\rm synch}$ for higher frequency values peaks at 
comparatively higher energies \cite{Bhattacharjee:2020phk}.
For a specific frequency value, the annihilation channel which produces a large 
number of $e^\pm$ with increasing energies would generate a larger amount of 
synchrotron flux. Thus, for a higher value of frequencies, the leptonic 
annihilation channel would dominate over the hadronic final states. We can 
observe this feature from Fig.~8.5(b) \cite{Bhattacharjee:2020phk}. In that figure, for $=$ 200 GeV mass
and high frequency value, $\tau^+ \tau^-$ annihilation channel dominates over $b 
\bar{b}$ final state, while for the low frequency value, $b \bar{b}$ 
annihilation channel dominates over $\tau^+ \tau^-$ final state. The $e^\pm$ 
resulting as the end product of WIMP annihilation final states could
possess the maximum energy, $\sim$ $M_{DM}$ and thus for higher DM mass values, we would obtain 
the harder the $e^\pm$ spectrum (have already shown in Fig.~8.4). Therefore, 
from Fig.~8.5(b), we can observe the crossover between $b\bar b$ dominance and 
$\tau^+ \tau^-$ dominance with changing the frequencies \cite{Bhattacharjee:2020phk}.

\subsection{Results Pertaining to the UFDs}
\label{sec:synchro_ufd}

\noindent In this section, we have considered the radio data observed by two 
popular radio telescopes; such as: 
\begin{itemize}
\item 1) The sky-survey data observed by the Giant Metrewave Radio Telescope 
(GMRT) \cite{Intema:2016jhx}. It covers the sky between  $-53^{\circ}$ to 
$+90^{\circ}$ declination at $\nu =0.1475~{\rm GHz}$
\item 2) The NVSS survey data the Very Large Array 
(VLA) telescope \cite{condon1998}. It covers the sky between $-40^{\circ}$ to 
$+90^{\circ}$ declination at $\nu = 1.4~{\rm GHz}$.
\end{itemize}

\noindent Unfortunately, no excess emission has been detected from the location 
of UFDs by both the telescopes \cite{Bhattacharjee:2020phk}. Thus the noise obtained from the direction of 
UFDs is translated to upper limits on flux density for 95$\%$ C.L. same as 
listed in the Table~8.5 \cite{Bhattacharjee:2020phk}. Here we would like to mention that the radio images are 
generally prepared in per unit beam-size, where, the beam-size is convolved with 
the PSF of the respective telescope. Thus, for the final processed radio images, 
the unit for flux density is in Jy. As both of our considered telescopes do not 
cover the full sky, we do not have the information for some UFDs, e.g. 
Tucana II \cite{Bhattacharjee:2020phk}. The observed upper limits on flux density are then translated to the 
$\langle
\sigma v \rangle$ upper limits for three annihilation final states. In Fig.~8.6, 
we have shown our results \cite{Bhattacharjee:2020phk}.

\begin{table}[!h]
\centering
\begin{tabular}{|p{2.5cm}|p{4cm}|p{4cm}|}
\hline \hline
Galaxy &  GMRT ($\nu = 147.5$ MHz) & VLA ($\nu = 1.4$ GHz)  \\
\hline \hline
Aquarius II & $6.8 $ & $0.86$ \\
\hline \hline
Draco II & $9 $ & $1.1$ \\
\hline \hline
Eridanus II & $7.8 $ & No Data \\
\hline \hline
Grus I & $4.1$ & No Data \\
\hline \hline
Hydra II & $8.8 $ & $1.1 $ \\
\hline \hline
Leo V & $6 $ & $0.98$ \\
\hline \hline
Pegasus III & $10$ & $0.96$ \\
\hline \hline
Pisces II & $3.5$ & $0.88$ \\
\hline \hline
Triangulum II & $6 $ & $1$ \\
\hline \hline
Draco & $7.2$ & $9.2$ \\
\hline \hline
\end{tabular}
\caption{2$\sigma$ upper limits on radio flux densities detected by the 
sky-survey performed by GMRT and VLA. 
The location of Carina II, Reticulum II, Horologium II and Tucana II\&III are 
not covered by both the surveys.}
\label{table:radio_flux_upper_limits}
\end{table}

\begin{figure}[h!]
\centering
\includegraphics[width=.49\linewidth]{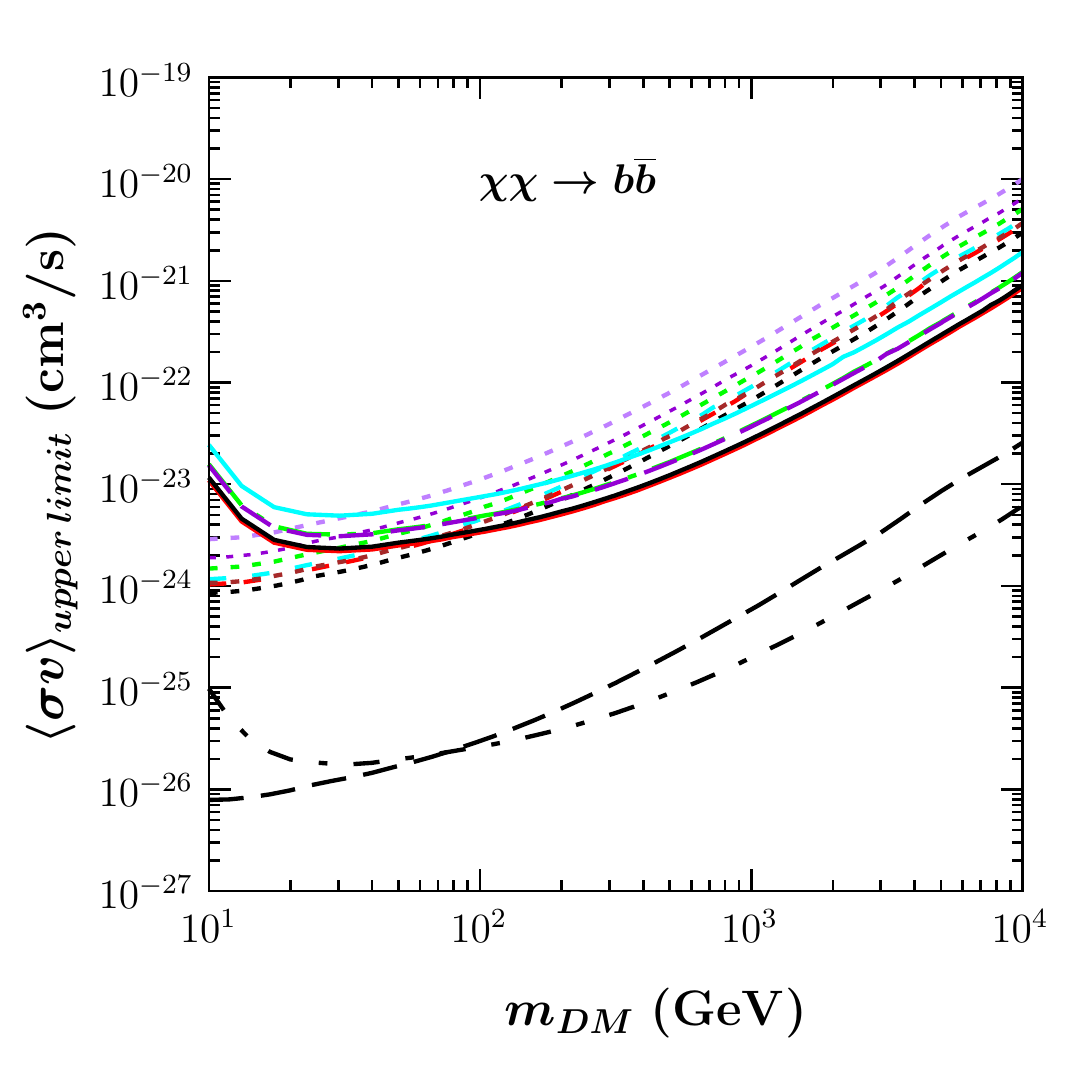}
\includegraphics[width=.49\linewidth]{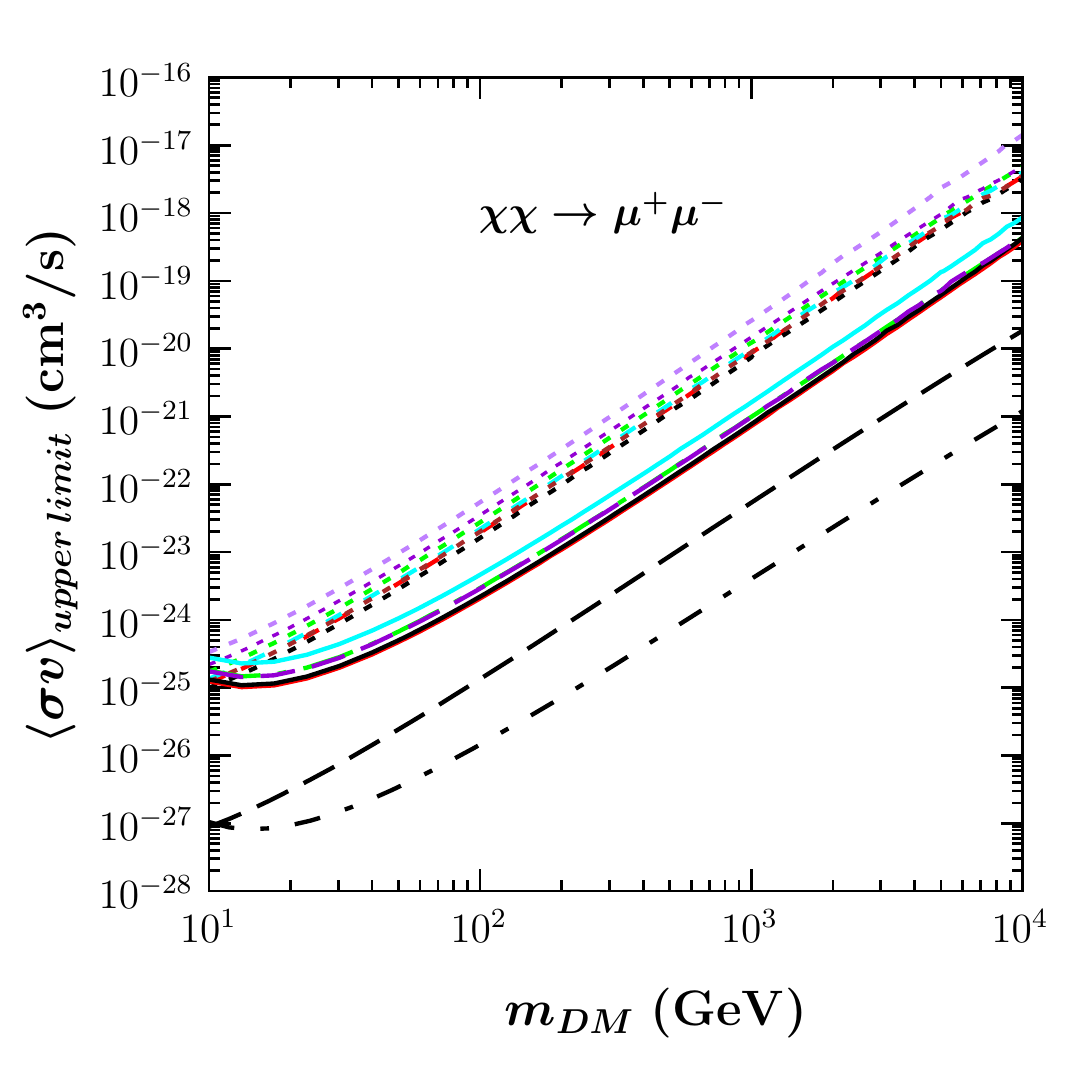}
\includegraphics[width=.5\linewidth]{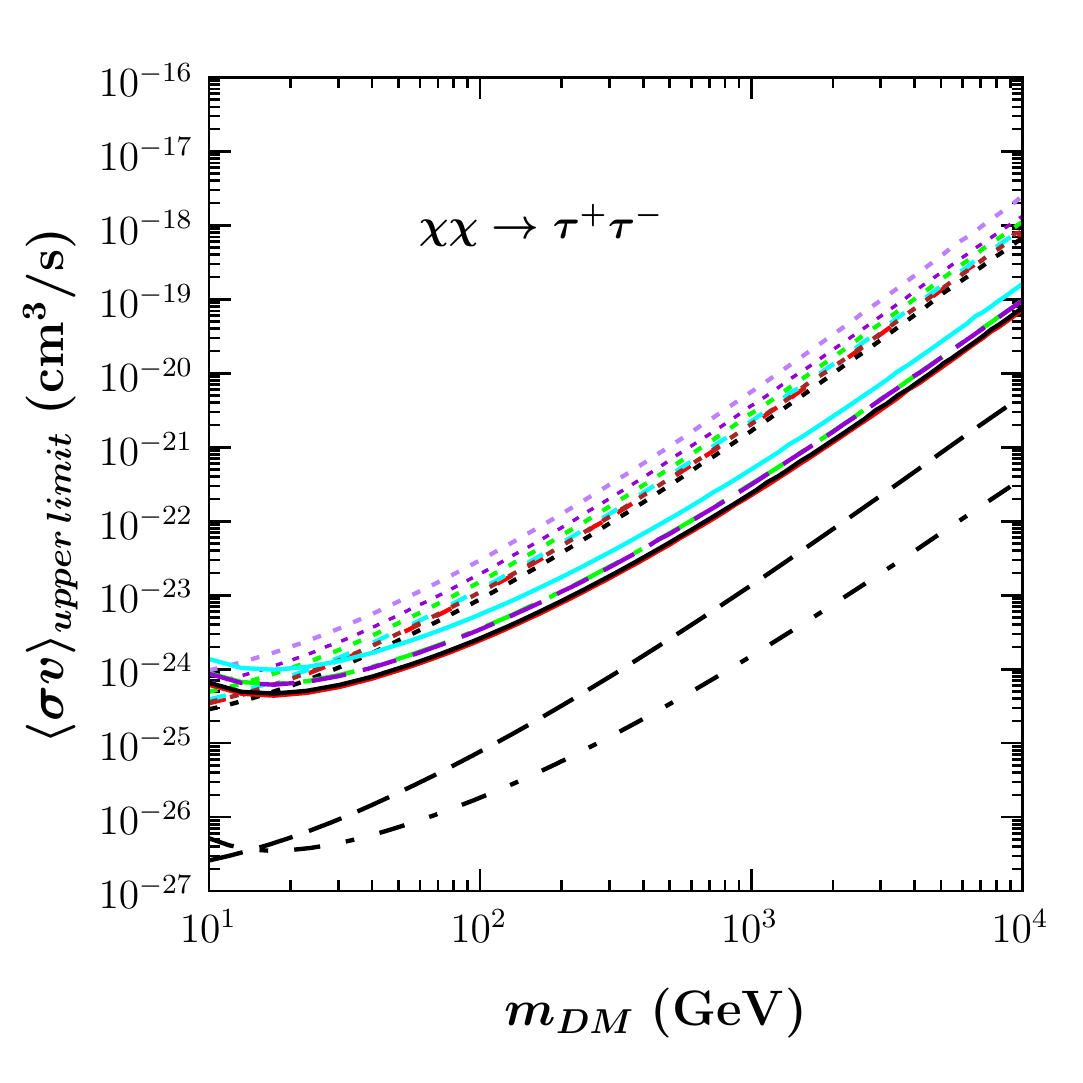}
 \hskip 10pt
\includegraphics[width=1.0\linewidth]{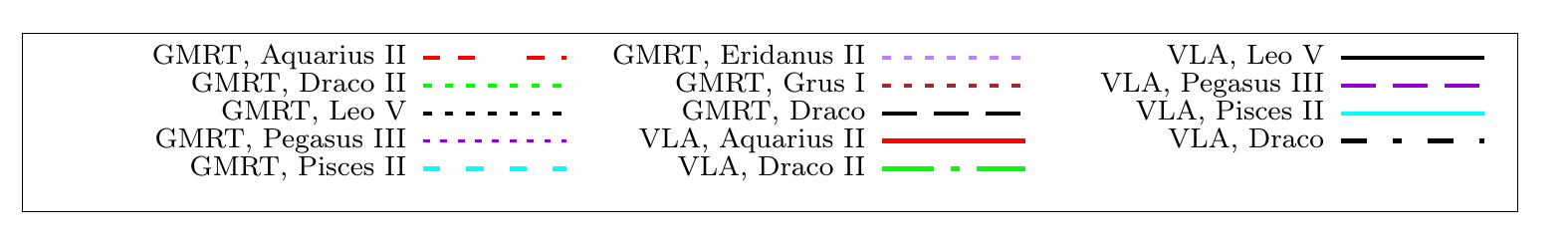}
\caption{95 $\%$ C.L. $\langle \sigma v \rangle$ limit of UFDs using upper 
limits on flux densities observed by GMRT and VLA for  $b\bar{b}$, 
$\tau^+\tau^-$ and $\mu^+\mu^-$ pair-annihilation channels. We 
have considered NFW profile and fixed the parameters at $\langle \sigma v\rangle 
\, = 10^{-26}$ cm$^3$/s, $B \, = \, 1\, \mu$G, $D_0 = 3 \times 10^{28}$ 
cm$^2$/s, $\gamma_D = 0.3$.}
\label{figure:sigmav_Exclusion_curve}
\end{figure}

\noindent Compared to GMRT, the VLA telescope has a wider effective area and 
operates in one order of magnitude higher frequency range which reduces the 
contribution from the galactic background. From Fig.~8.6, we find that for large 
DM mass, the $\langle\sigma v\rangle$ limits obtained from the NVSS images are 
stronger than the limits obtained from the GMRT data, while for low DM mass, 
GMRT data imposes the strongest limits \cite{Bhattacharjee:2020phk}. This result is the outcome of the 
comparative efficiencies between two telescopes and the dependence of the 
$e^\pm$ spectrum on DM mass.

\subsection{Future Projections}

\noindent SKA is expected to operate for a wide range of radio frequency i.e., 
between 50 MHz - 50 GHz. This enables SKA to observe the synchrotron emission 
from DM annihilation in dSphs/UFDs \cite{Bull:2018lat, Colafrancesco:2015ola}. 
We have calculated the 
synchrotron flux from our considered UFDs and have examined the possibility of 
observing these signals by SKA \cite{Bhattacharjee:2020phk}. Fig. 8.7, shows the estimated synchrotron 
fluxes, for the UFDs listed in Table 8.4, for $b\bar{b}$, $\tau^+ \tau^-$ and 
$\mu^+ \mu^-$ annihilation channels \cite{Bhattacharjee:2020phk}. In Fig. 8.7, we have also shown the 
sensitivity of SKA~\cite{braun2017ska,braun2019anticipated}
for its 10, 100 and 1000 hours of observation time. Here we would like to 
mention that SKA would process a very wide effective area and thus we can expect 
it to cover all our selected UFDs \cite{Bhattacharjee:2020phk}.

\begin{figure}[h!]
\centering
\includegraphics[width=.3\linewidth]{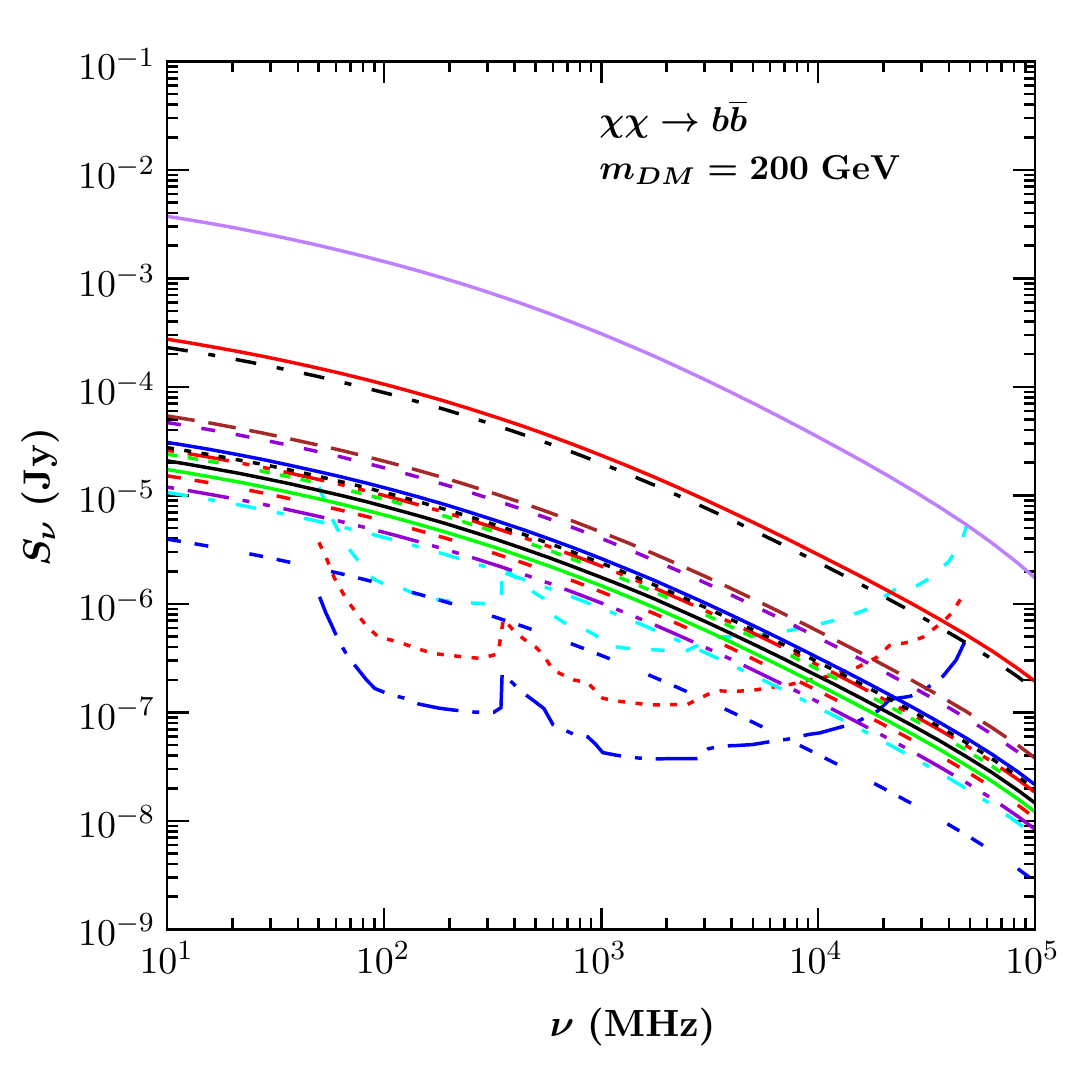}
\hskip 10pt
\includegraphics[width=.3\linewidth]{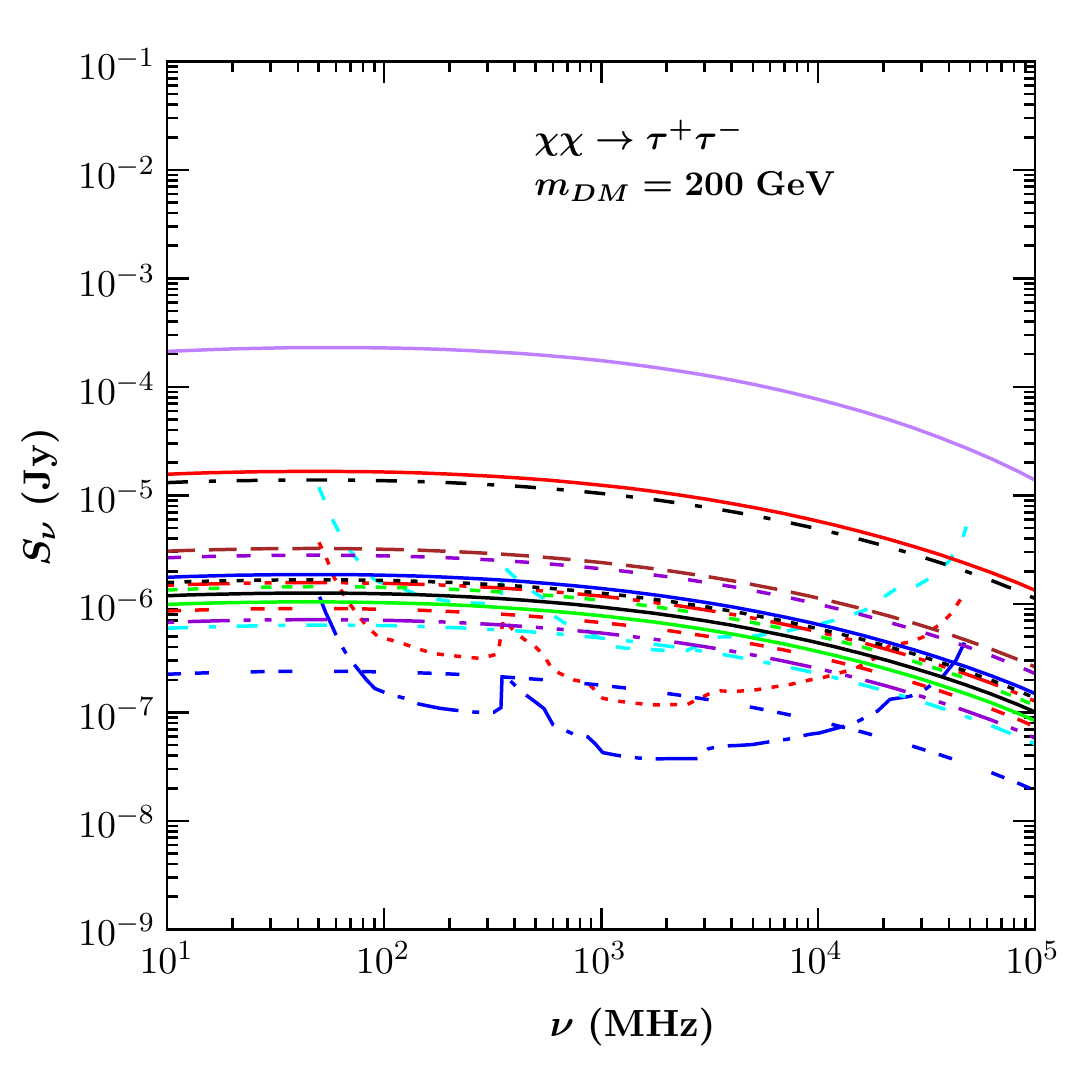}
\hskip 10pt
\includegraphics[width=.3\linewidth]{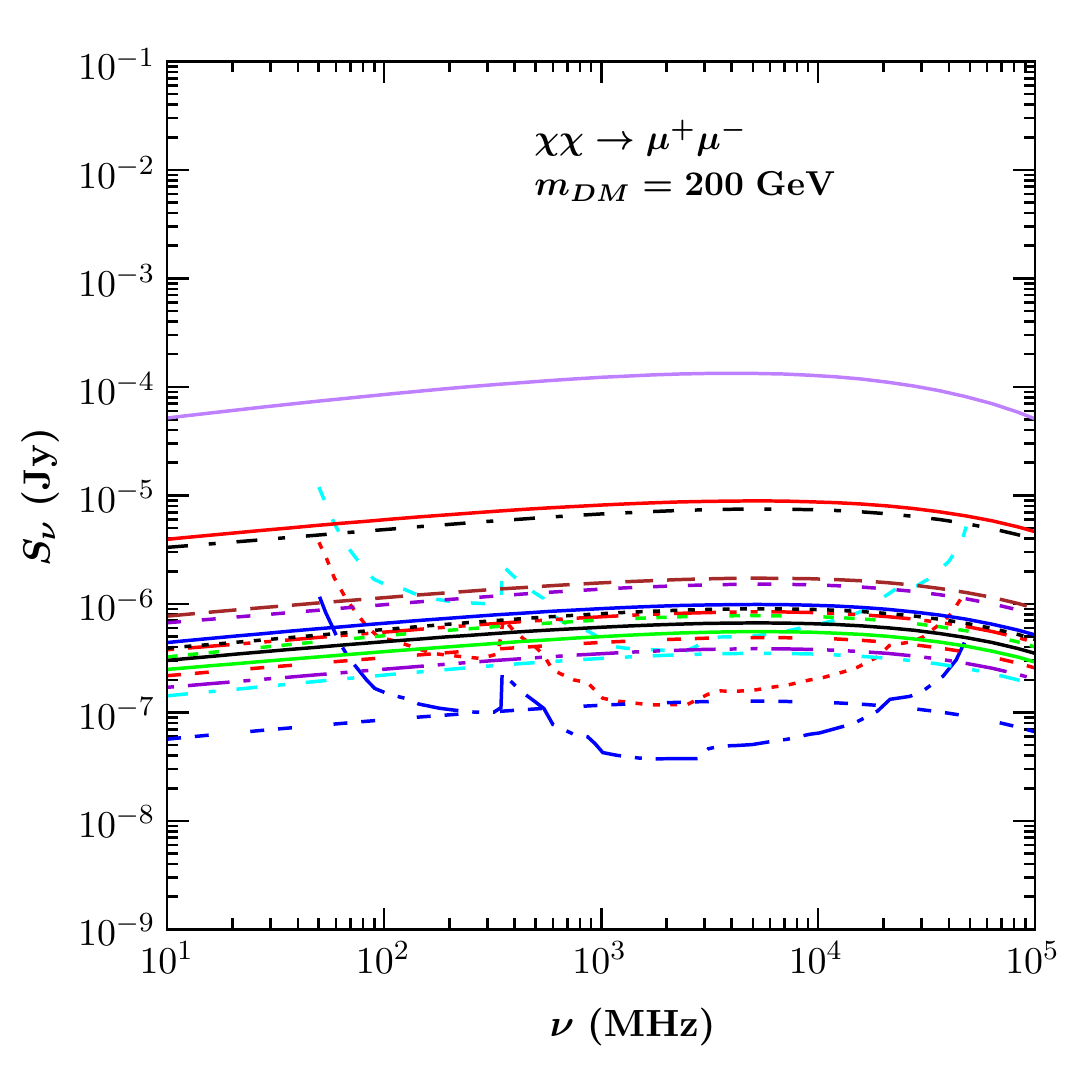}
\includegraphics[width=.3\linewidth]{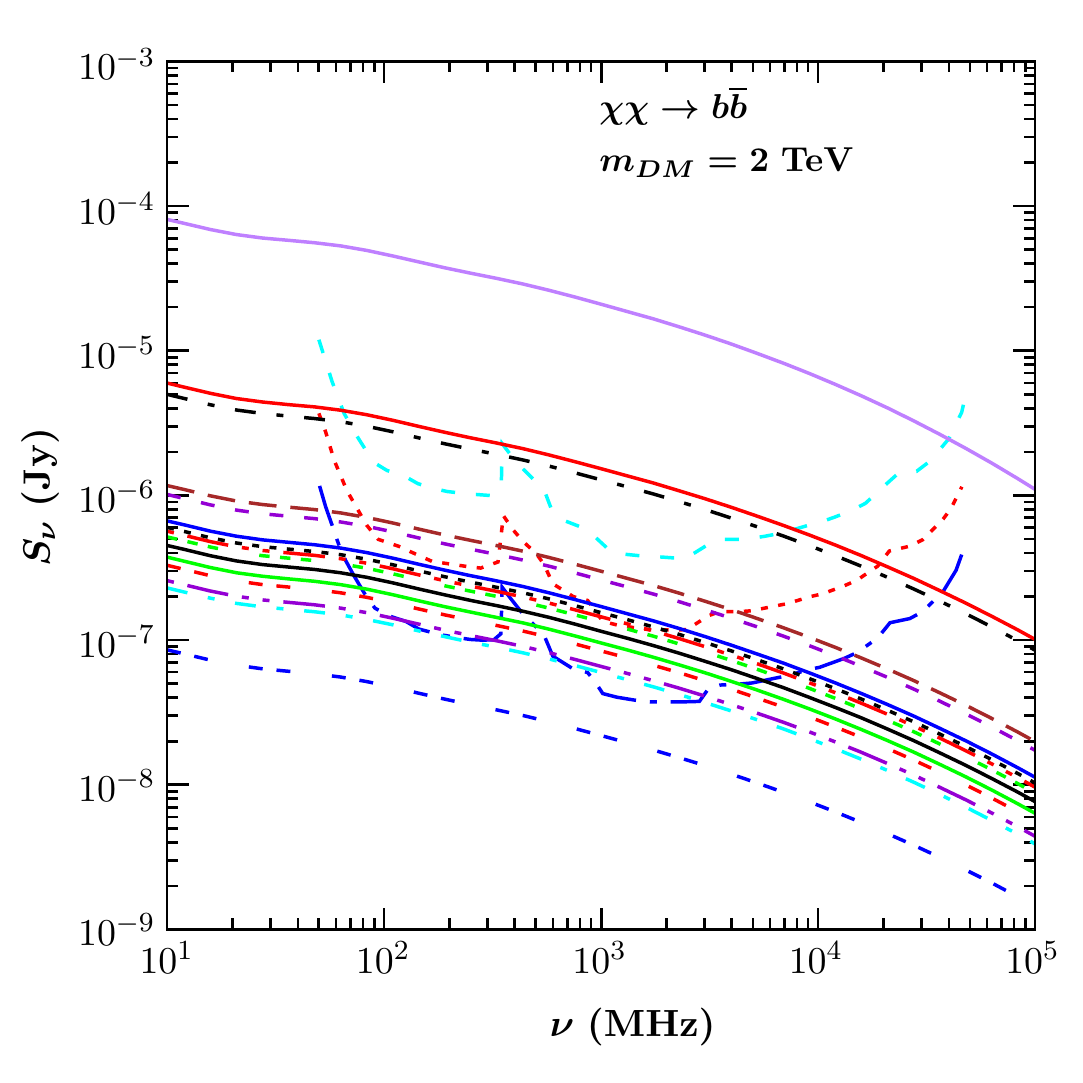}
\hskip 10pt
\includegraphics[width=.3\linewidth]{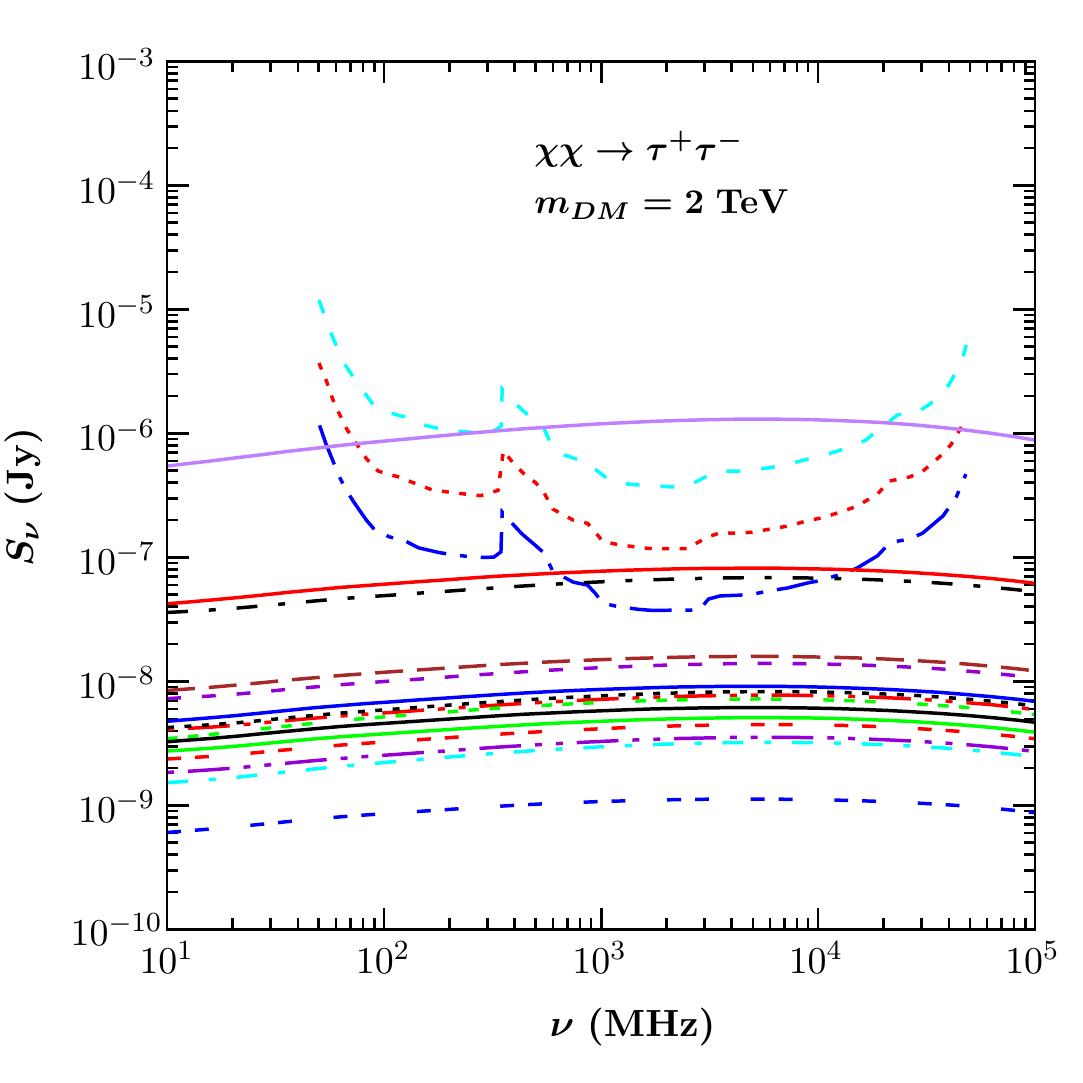}
\hskip 10pt
\includegraphics[width=.3\linewidth]{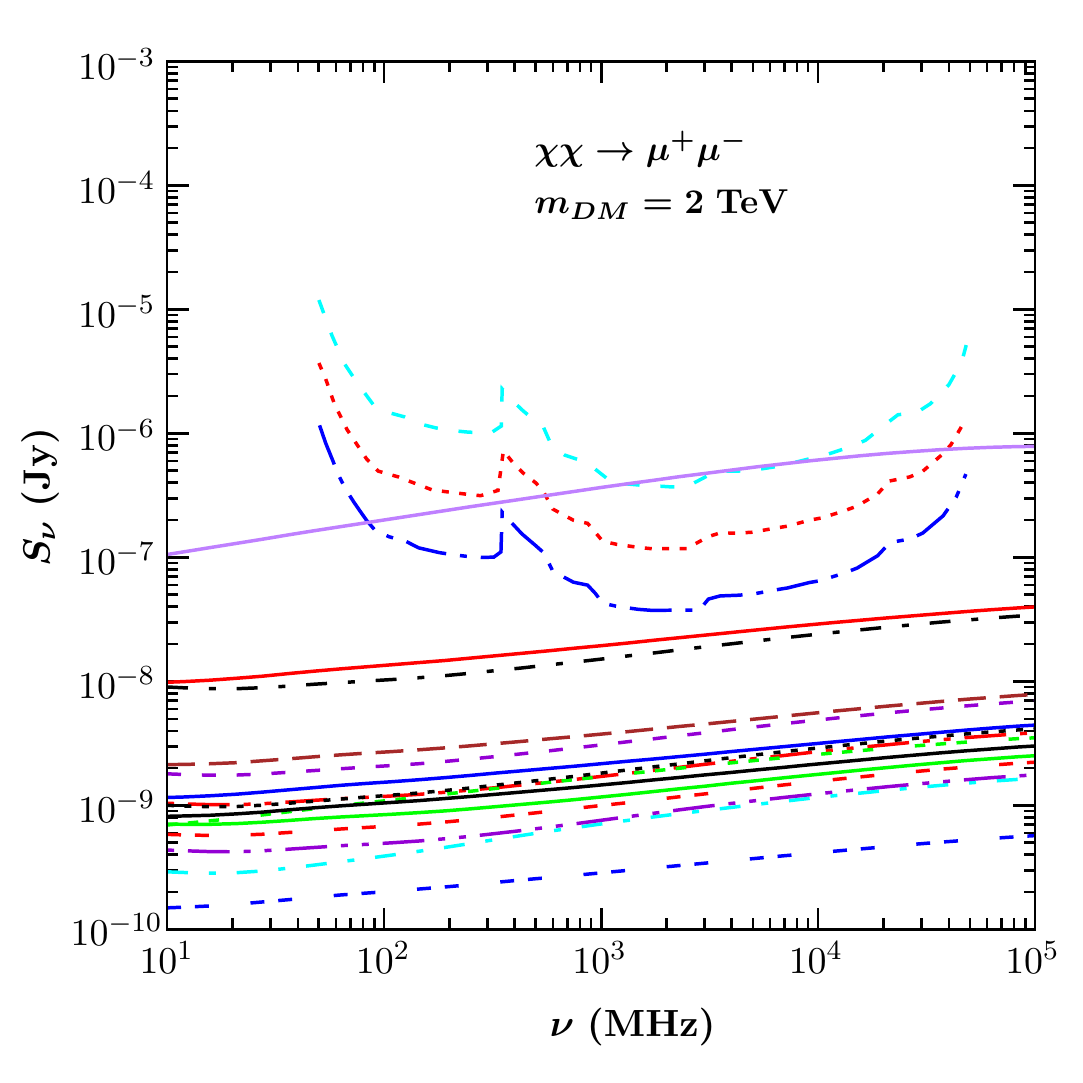}
\hskip 10pt
\includegraphics[width=0.9\linewidth]{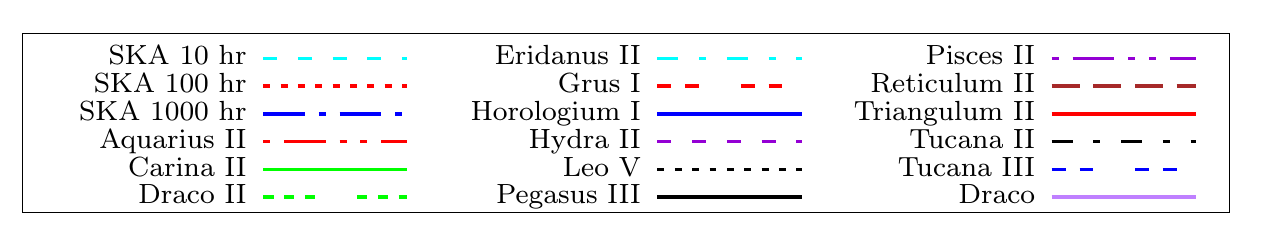}
\caption{The synchrotron flux densities of our considered UFDs and classical 
dSphs, Draco has been determined for three annihilation channels, such as: 
$b\bar{b}$ (left), $\tau^+\tau^-$ (center) and $\mu^+\mu^-$ (right) and for two 
particular DM masses, such as: 200 GeV (top) and 2 TeV (bottom). For each 
figure, we have considered NFW profile and fixed the parameters at $\langle 
\sigma v\rangle \, = 10^{-26}$ cm$^3$/s , $B \, = \, 1\, \mu$G, $D_0 = 3 \times 
10^{28}$ cm$^2$/s, $\gamma_D = 0.3$. The values of $\rho_s$, $r_s$, $d$ and 
$r_h$  have been taken from Table 8.4. For Hydra II, Triangulum II and Tucana 
III, we only have the upper limits on $\rho_s$ (Table 8.4), thus they can only 
provide the upper limits on synchrotron flux densities.}
\label{figure:synflux_newgalaxies}
\end{figure}

\noindent For high DM mass, the $e^\pm$ produces the hard spectrum and the 
resulting synchrotron flux can go beyond the detection range of SKA. Thus, that 
would also consequently reduce the detection feasibility at SKA. From Fig.~8.7, 
we can observe the same \cite{Bhattacharjee:2020phk}. In Fig.~8.7, we find that for 200 GeV DM mass, the 
radio emission of 12 UFDs originating from three annihilation channels can be 
detected by the 100 hours of SKA sensitivity, while for 2~TeV DM mass, the 
synchrotron emission only for the $b\bar{b}$ annihilation channel can be 
observed by the 1000 hours of sensitivity curve \cite{Bhattacharjee:2020phk}. Interestingly, from Fig.~8.7, 
we can also notice that for both 200 GeV and 2 TeV DM masses, only Draco can be 
detected by the SKA (too with the $\sim 10$ hours of sensitivity curve) for all 
three annihilation channels \cite{Bhattacharjee:2020phk}.
 
\section{Astrophysical Uncertainties and the Constraints}
\label{sec:uncertainty}

\noindent The limits that we have derived in the earlier sections are based on 
the central values of the parameters listed in Tables~8.1 and 8.2. But there is 
not much detailed spectroscopic study for the newly discovered UFDs and due to 
the inadequate observation, the astrophysical parameters associated with them 
might process very large uncertainty. Thus, in order to draw any strong conclusion 
from the analysis of UFDs, we need to address the possible uncertainty related 
to the constraints that we obtained from gamma-ray and radio data.

\subsection{Uncertainties in the $\gamma$-ray Bounds}
\label{section:uncertainties_horo_tuc}

\noindent For gamma-ray analysis, our insufficient knowledge of the shape of DM 
distribution is the main source for large uncertainties. Especially for the 
newly discovered UFDs, a few member stars have been detected and that creates 
the prime obstacle to assume the DM distribution in them\cite{Funk:2013gxa}. As 
we have already mentioned, the $N$-body
simulation results favour the cuspy-NFW profile but the observation data from 
some particular galaxies favour the cored profile for DM distribution (e.g. 
Pseudo isothermal and Burkert\cite{de_Blok_2001}). Hence, we would like to 
investigate the role of the DM profiles for UFDs and for that purpose we have 
chosen Horologium I as it provides the strongest $\gamma$-ray limits for 
$b\bar{b}$ annihilation channel \cite{Bhattacharjee:2020phk}. 

\noindent We have used the median value of $J$-factor from Table~8.2 and have 
derived the $\langle \sigma v \rangle$ upper limits for three DM density 
profiles (Fig.~8.8). From Fig.~8.8, we observe that Burkert profile imposes the 
strongest limits, while Pseudo isothermal profile provides the weakest 
constraints \cite{Bhattacharjee:2020phk}. Though we have only shown the result for Horologium I, all our 
selected UFDs would show the same nature.

\begin{figure}[h!]
\begin{center}
\includegraphics[width=0.5\textwidth,clip,angle=0]{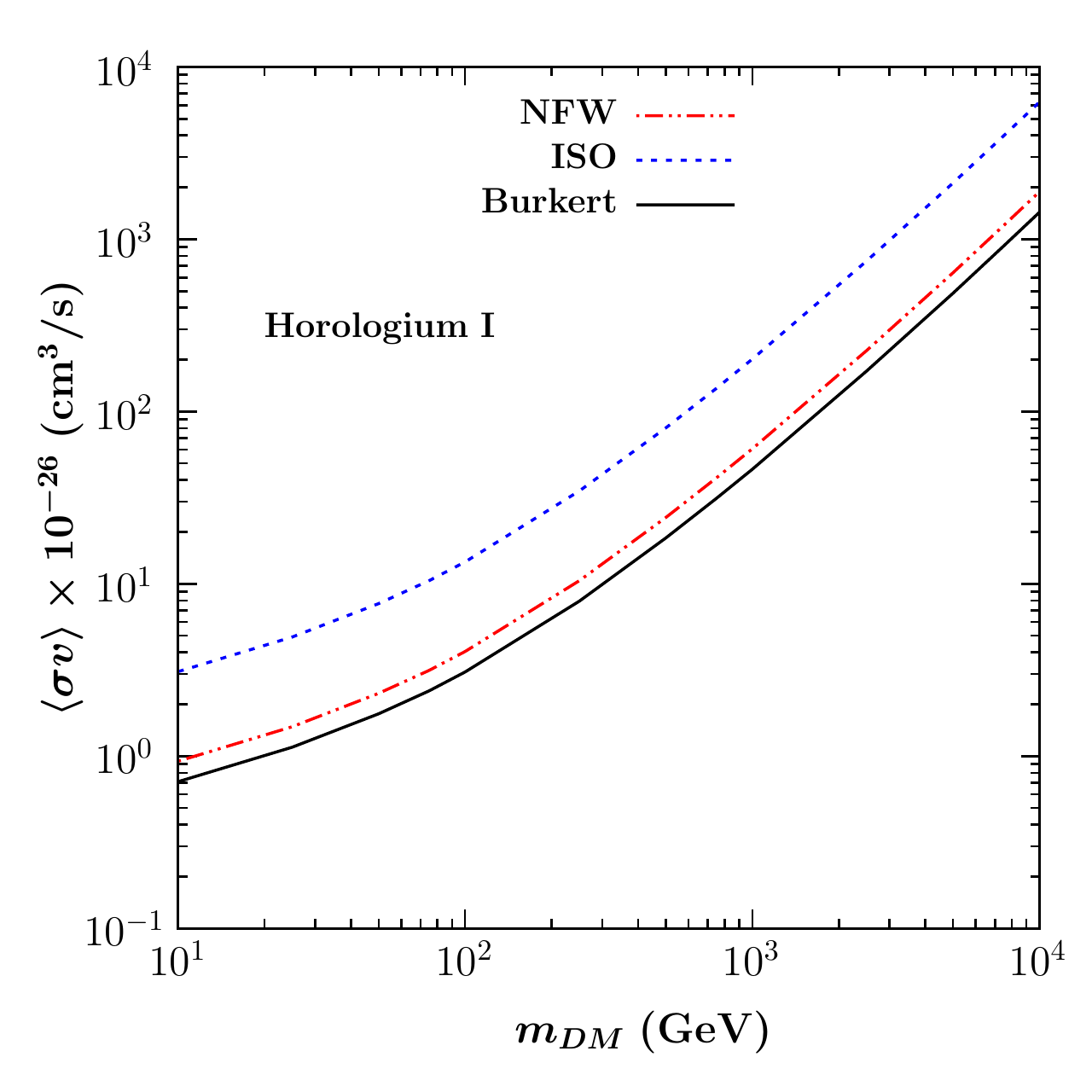}
\end{center}
\caption{Comparison between the upper limits on $\langle \sigma v \rangle $ 
obtained from the Fermi-LAT data for three density profiles for $b\overline{b}$ 
final state.}
\label{figure:profile_comparison}
\end{figure}

\noindent Next, we have checked how the uncertainties associated with the 
J-factor for NFW profile can influence our results. We have considered the 
$1\sigma$ uncertainty band associated with NFW profile (Table~8.2), and in 
Fig.~8.9 we have shown their corresponding limits. Here, we again consider only 
the  $b\bar{b}$ annihilation channel as for $\gamma$-ray data this channel 
provides the most stringent limits \cite{Bhattacharjee:2020phk}. From Fig.~8.9, we could find that UFDs have 
possessed a large uncertainty band in the parameter space of ($m_{DM}$, $\langle 
\sigma v \rangle$) \cite{Bhattacharjee:2020phk}. If we check the Eq.~2.5, we find that $\gamma$-ray flux 
resulting from the WIMP annihilation is proportional to J-factor and thus the 
large uncertainty in J-factor would always translate to the large uncertainties 
in the $\langle \sigma v\rangle$ upper limits. In the future, with more detailed 
spectroscopic studies, it might be possible to reduce the uncertainty band for 
newly discovered UFDs. 

\begin{figure}[!h]
\begin{center}
\includegraphics[width=.49\linewidth]{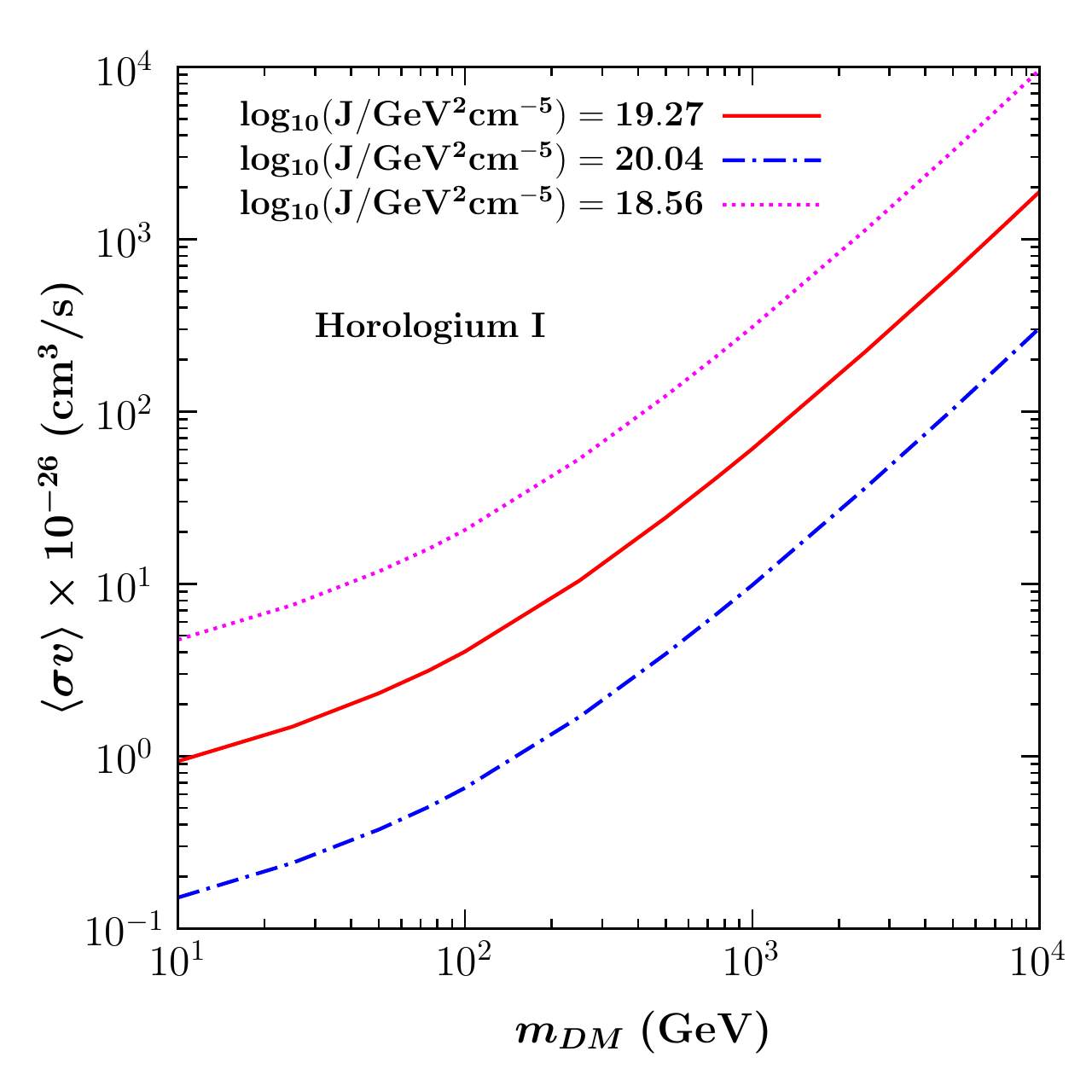}
\includegraphics[width=.49\linewidth]{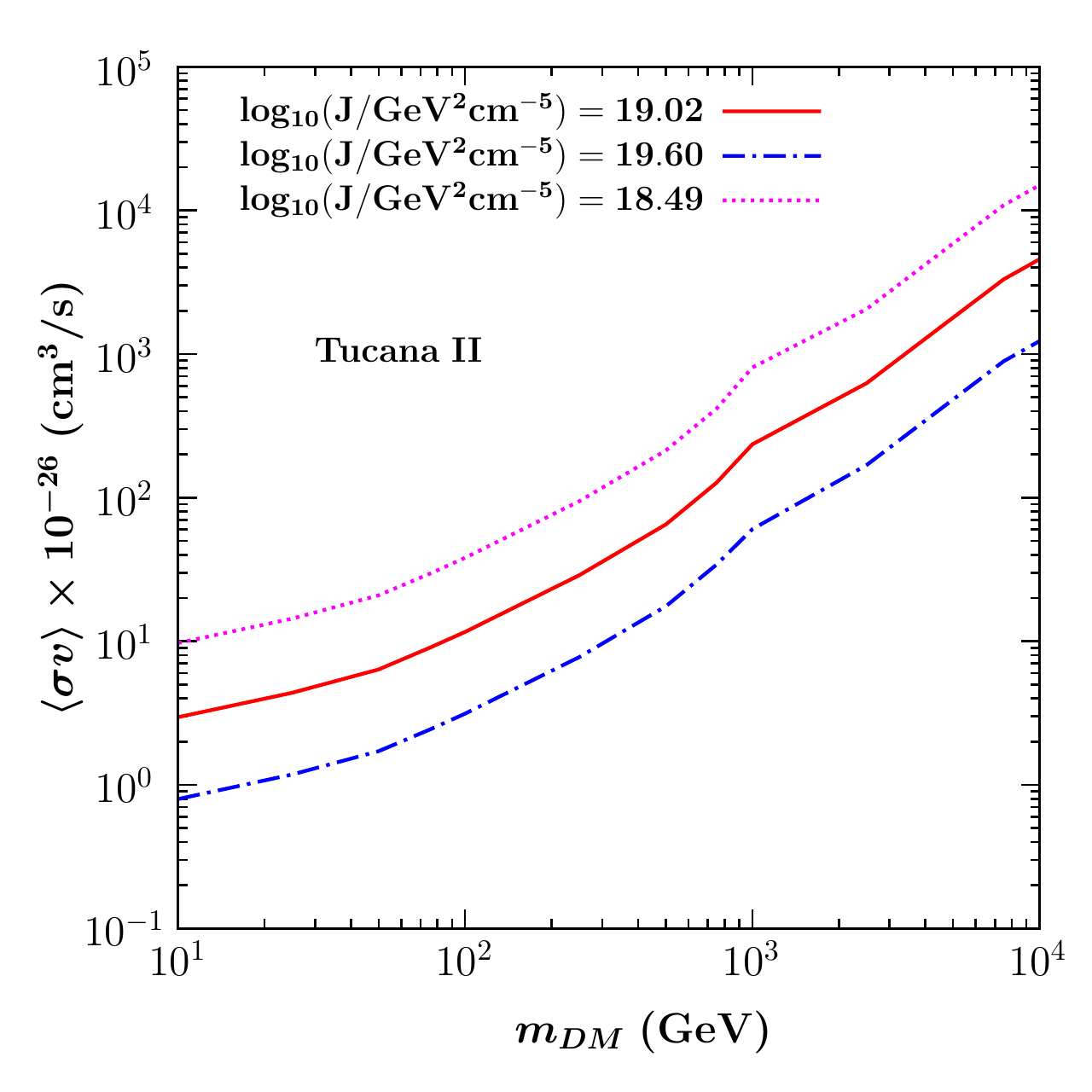}
\end{center}
\caption{95$\%$ C.L. upper limits of $\langle \sigma v \rangle $ as a function 
of DM mass, $m_DM$ for the `central value' of J-factor derived by Pace \textit{et al.}, 
2019 and its relative uncertainties (Table~8.2). The 
$\langle \sigma v \rangle $ limits for Horologium I and Tucana II have been 
shown in the left and the right panels, respectively.}
\label{figure:crossuncertainty}
\end{figure}


\subsection{Uncertainties in the Synchrotron Fluxes}
\label{sec:synchrotron_uncertainty}

\noindent Like the gamma-ray fluxes, the uncertainties in the astrophysical 
parameters (e.g. $d$, $r_{1/2}$ and $\sigma$) can also affect the synchrotron 
fluxes. Thus, in this subsection, we would again check the possible 
uncertainties associated with radio limits and for that purpose, we have used 
the $1\sigma$ uncertainties of the parameters listed in Table 8.1. From 
Fig.~8.10, we have shown the uncertainties related to the synchrotron flux in 
Tucana II for DM mass 200 GeV and $b\bar{b}$ annihilation channel. Here we have 
particularly chosen Tucana II as it shows the highest synchrotron emission \cite{Bhattacharjee:2020phk}.

\begin{figure}[!h]
\centering
\subfigure[]
{\includegraphics[width=0.49\linewidth]{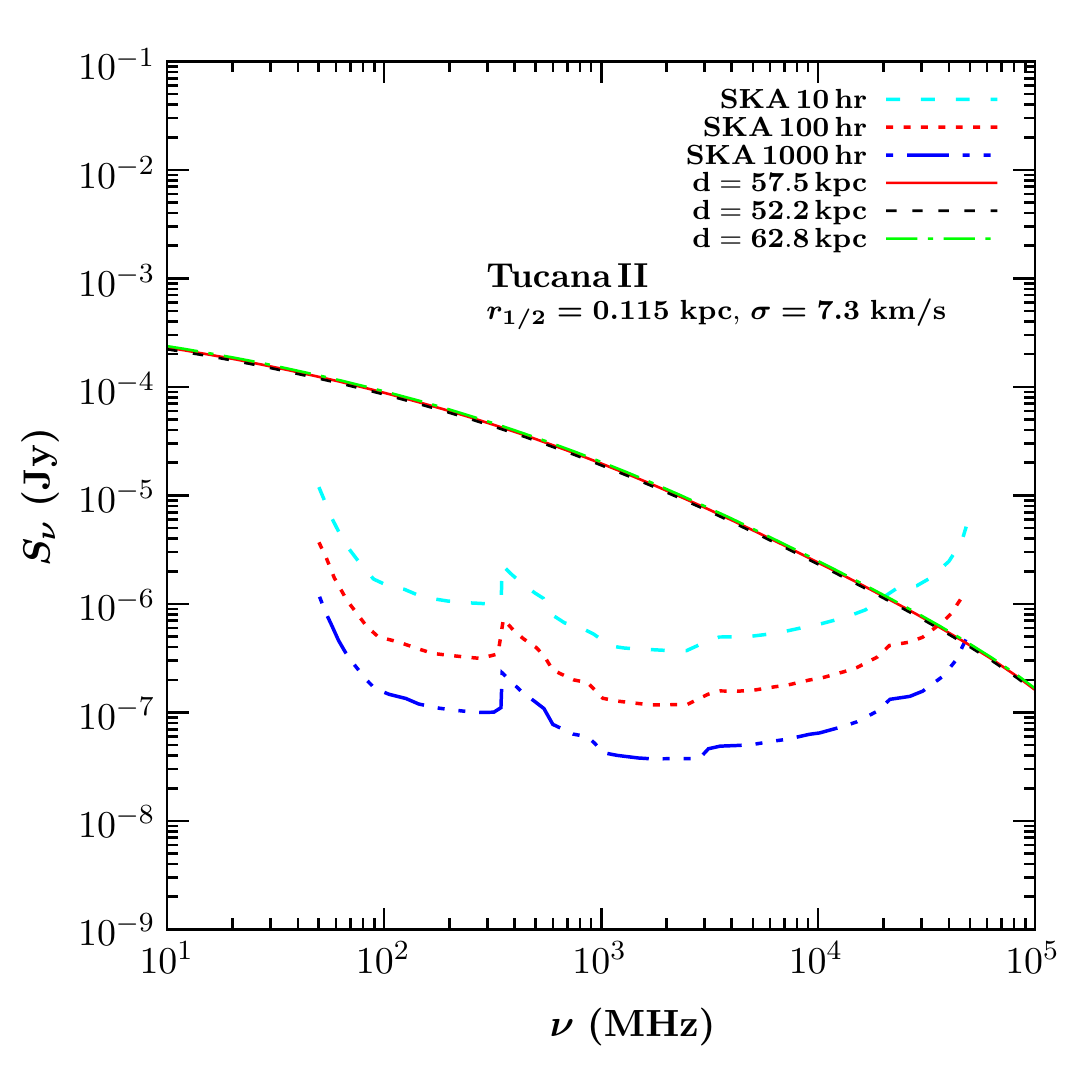}}
\label{fig:d_uncertainty}
\subfigure[]
{\includegraphics[width=0.49\linewidth]{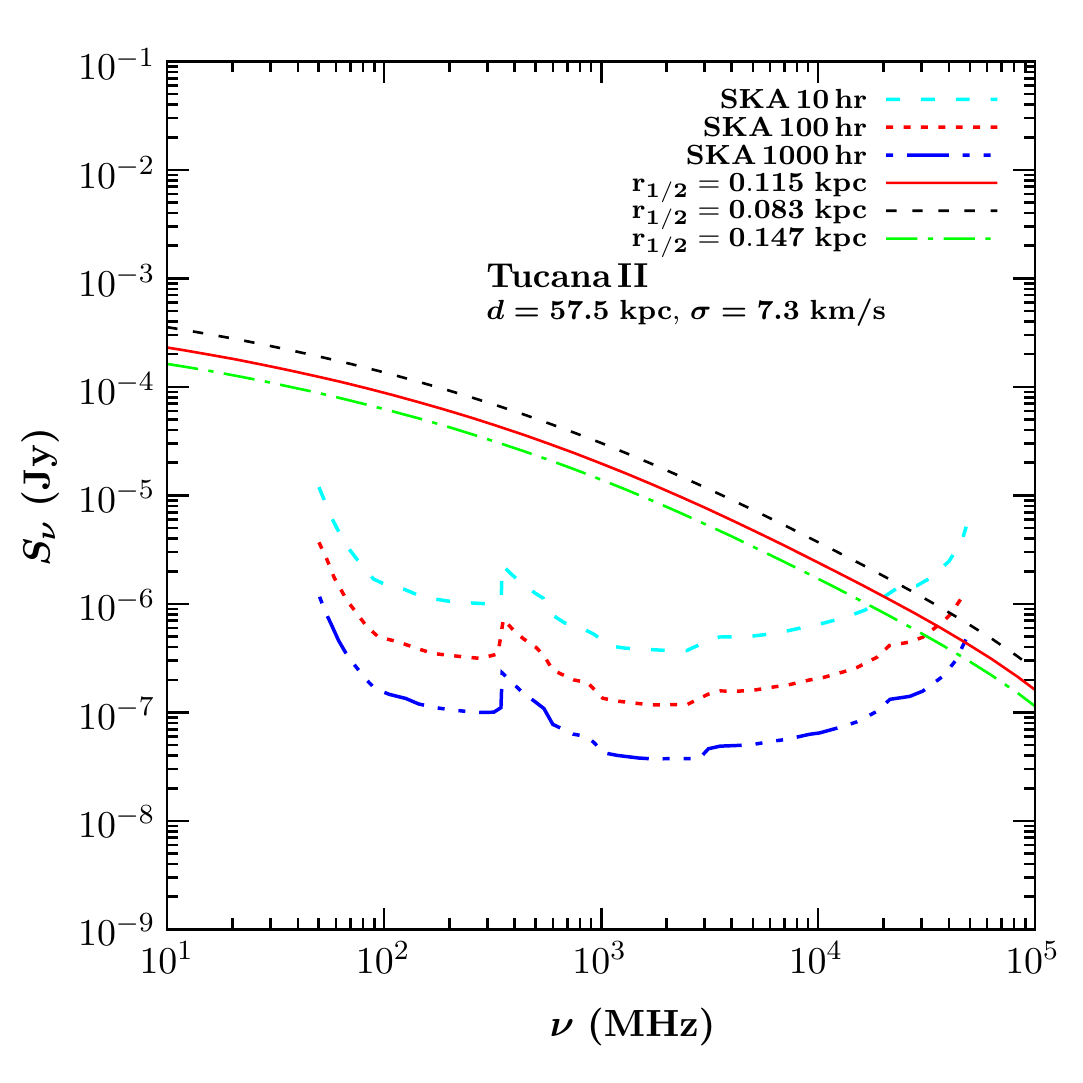}}
\label{fig:rhalf_uncertainty}
\subfigure[]   
{\includegraphics[width=0.49\linewidth]{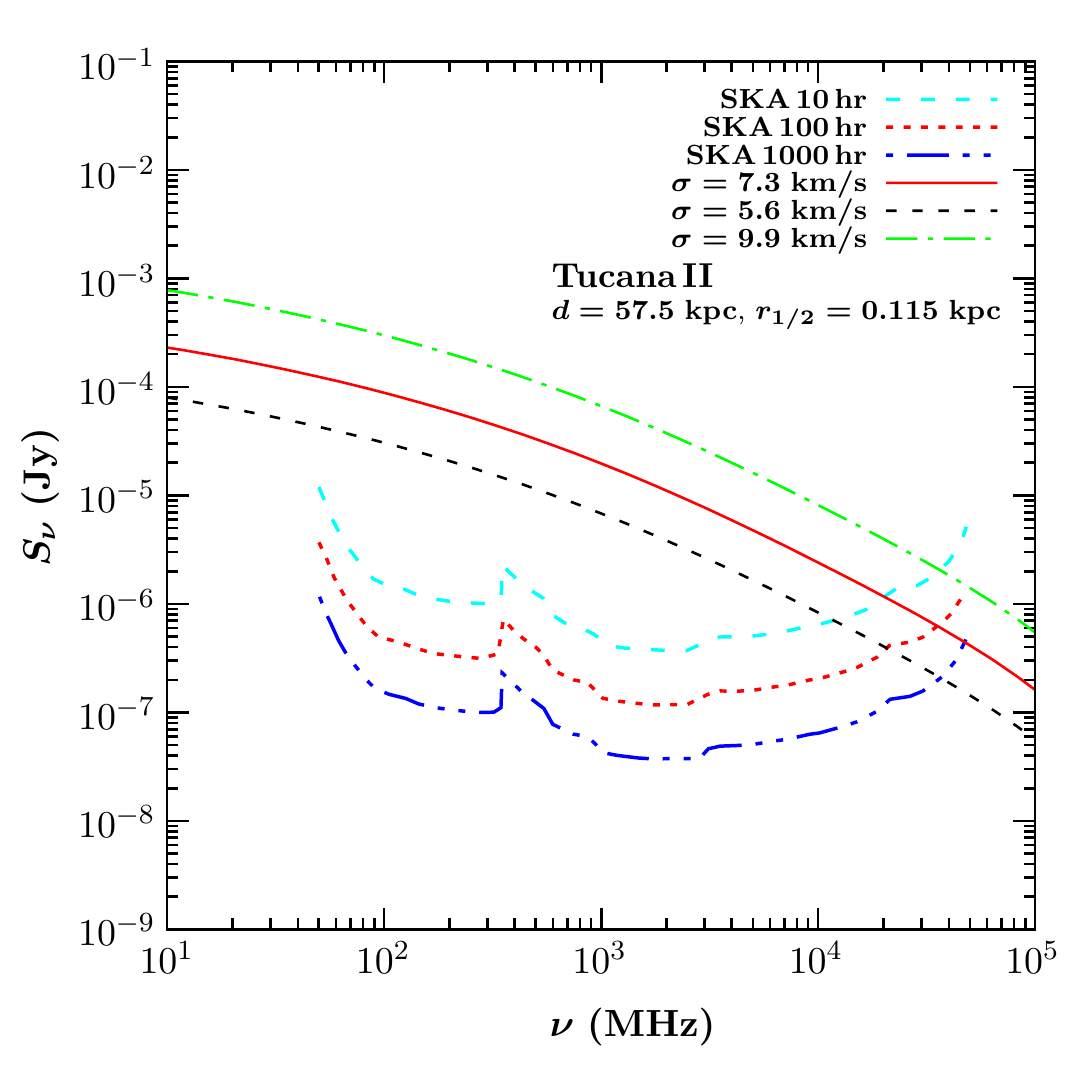}}
\label{fig:sigma_uncertainty}
\caption{The variation of synchrotron flux densities in Tucana II, for 200 GeV 
$m_{DM}$ and $b\bar b$ final state, with 1$\sigma$ uncertainties in (a) $d$, (b) 
$r_{\frac{1}{2}}$, (c) $\sigma$. We have considered NFW profile and fixed the 
parameters at $\langle \sigma v\rangle \, = 10^{-26}$ cm$^3$/s , $B \, = \, 1\, 
\mu$G, $D_0 = 3 \times 10^{28}$ cm$^2$/s, $\gamma_D = 0.3$.}
\label{figure:uncertainty}
\end{figure}

\noindent The range of uncertainties that we have shown in Fig.~8.10 is the 
combination of the errors associated with $d$, $r_{1/2}$ and $\sigma$ \cite{Bhattacharjee:2020phk}. From 
Table~8.1, we can notice that compared to $r_{1/2}$ and $\sigma$, the error in 
$d$ is relatively small and, thus $d$ would not impose large uncertainty in 
synchrotron flux. But both  $r_{1/2}$ and $\sigma$ contribute a significant 
amount to the uncertainties in flux \cite{Bhattacharjee:2020phk}. Hence, in order to reduce the uncertainties 
level, the accuracy in the measurements of these two parameters would play a 
very crucial role. From Fig.~8.10, we can check the same \cite{Bhattacharjee:2020phk}.\\

\noindent  A 
further source of uncertainty is due to the density distribution for the DM. 
Until now, we have only used the NFW density profile in order to predict the 
synchrotron flux, while in Fig.~8.11, we have shown the synchrotron flux from 
Tucana II predicted for NFW, ISO and Burkert DM profiles. Here, we would like to 
mention that unlike the $\gamma$-ray limits, for synchrotron flux NFW profile 
provides the highest flux, while Burkert imposes the lowest \cite{Bhattacharjee:2020phk}. \\

\begin{figure}[h!]
\begin{center}
\includegraphics[width=.5\linewidth,height=3in]{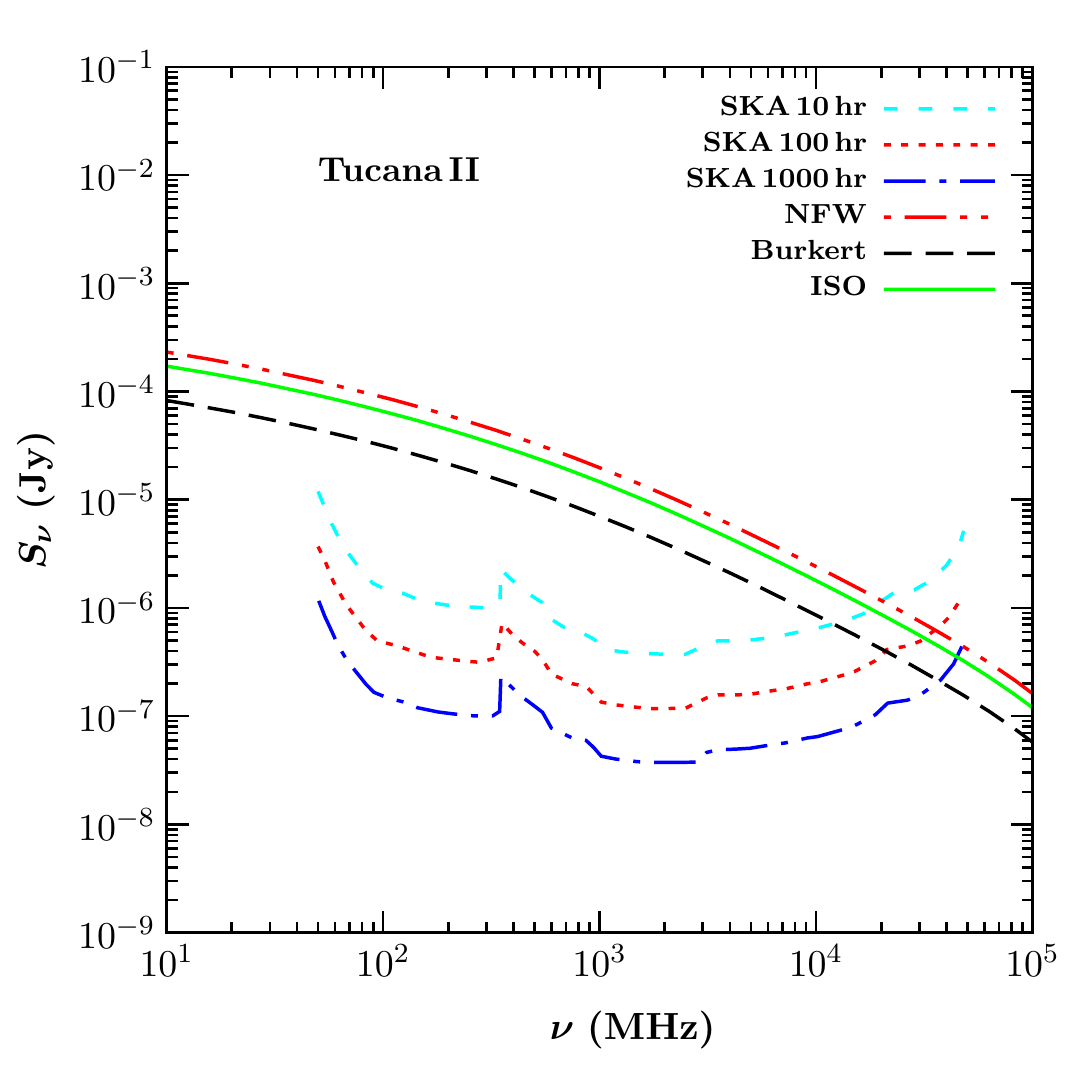}
\end{center}
\caption{The synchrotron flux densities in Tucana II predicted from NFW, 
Burkert and ISO density profiles. We have considered 200 GeV $m_{DM}$ and $b\bar 
b$ final state. Besides, we have fixed the parameters at $\langle \sigma 
v\rangle \, = 10^{-26}$ cm$^3$/s , $B \, = \, 1\, \mu$G, $D_0 = 3 \times 
10^{28}$ cm$^2$/s, $\gamma_D = 0.3$.}
\label{figure:sync_profile_dependence}
\end{figure}

\noindent Besides, the synchrotron fluxes also strongly depend on magnetic field 
($B$), diffusion constant $D_{0}$ and its exponent ($\gamma_D$). Unfortunately, 
for UFDs, we do not have any precise knowledge of them.
In section 8.3, we discussed the possible values for $B$, $D_{0}$ and 
$\gamma_{D}$ but to predict the possible synchrotron flux limits, we had only 
used the central values of them. Hence, To check the effect of these parameters 
on the prediction of the amount of synchrotron flux, in Fig. 8.12 we show the 
synchrotron flux for different values of $B$, $D_0$ and $\gamma_D$ within their 
plausible ranges \cite{Bhattacharjee:2020phk}. We have taken the values of $B$ in the range of $0.5$-$10$ 
$\mu$G, $D_0$ in the range of $3\times 10^{26}$-$10^{30}$ $cm^2/s$ and $\gamma_D$ 
in the range of $0.1$-$1$ \cite{Bhattacharjee:2020phk}. Since the magnetic field is the cause of synchrotron 
radiation, the flux increases as we go higher in the magnetic field (Fig.~8.12 
(a)), while the diffusion constant shows the reverse effect (Fig.~8.12 (b)). For 
a large value of $D_0$, the synchrotron would decrease as most of the 
relativistic charged particles then leave the diffusion region without radiating 
their complete energy.
For $\gamma_D$, its large value would suppress or enhance the $D(E)$, depending on the energy value. Since for synchrotron emission, the high energy $e^\pm$ 
accelerates in the presence of the magnetic field (check Fig.~8.5 (a)), the large value of $\gamma_D$ 
would strongly suppress flux at high frequency, while for frequency below $\sim 
1$~MHz, the synchrotron flux would be enhanced \cite{Bhattacharjee:2020phk}. Between 1 MHz to 5 MHz, the flux 
would rise to its peak value at $E(e^\pm) = 1$~GeV (Fig.~8.5 (a)). But the 
effect of $\gamma_D$ relatively less crucial than $B$ and $D_{0}$ (Fig.~8.12 
(c)) \cite{Bhattacharjee:2020phk}.

\begin{figure}[!h]
\centering
\subfigure[]
{\includegraphics[width=0.49\linewidth]{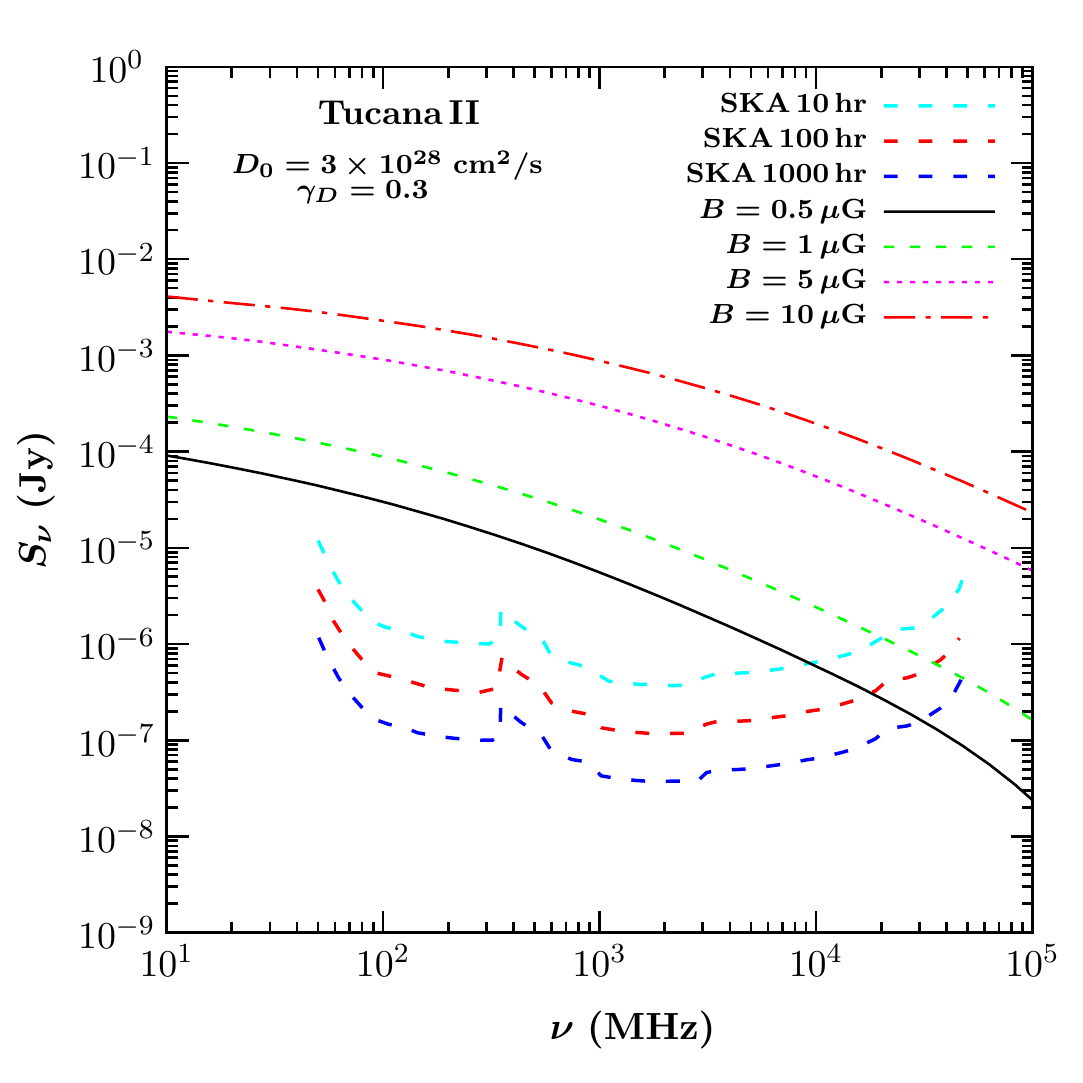}}
\label{fig:flux_different_B}
\subfigure[]
{\includegraphics[width=0.49\linewidth]{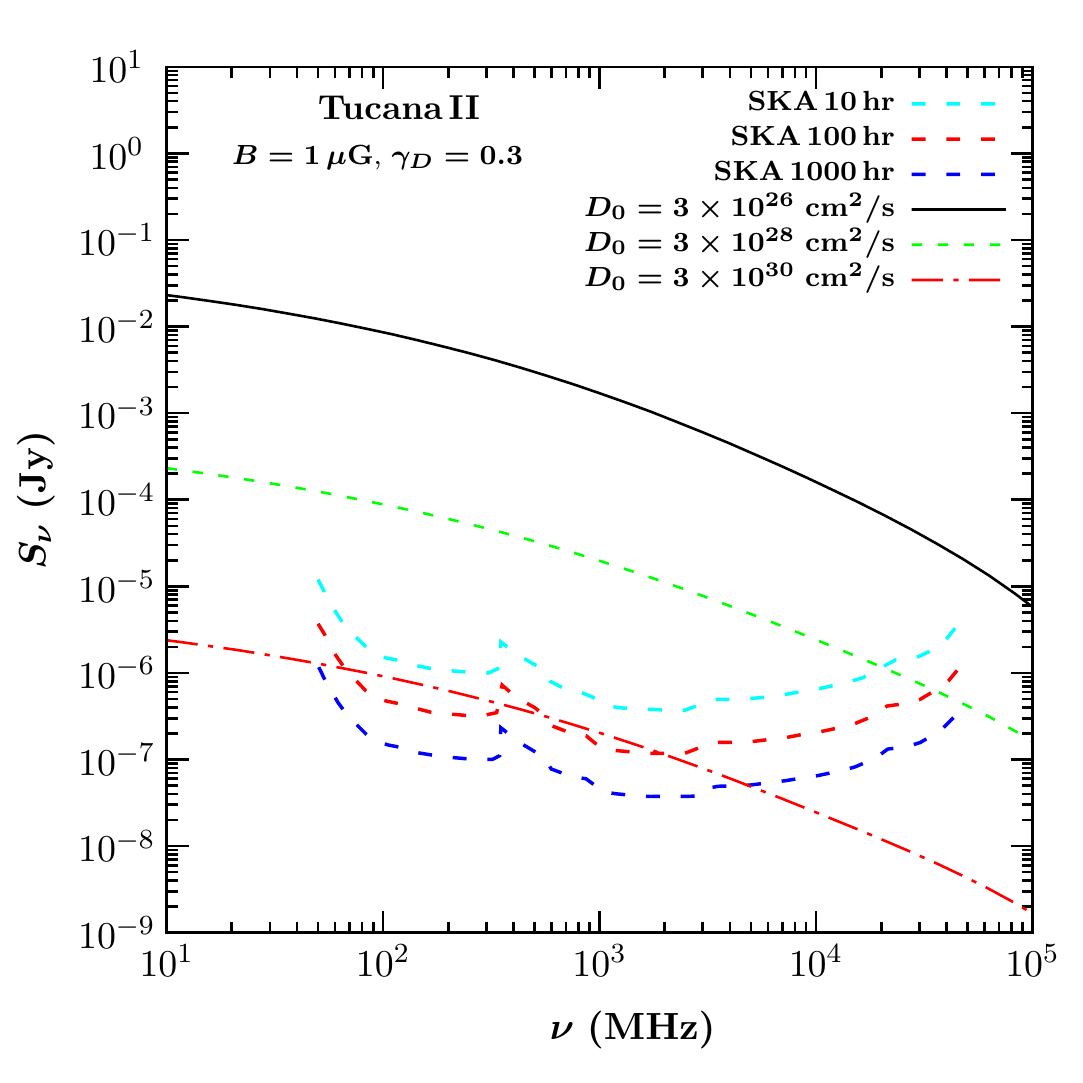}}
\label{fig:flux_different_D0}
\subfigure[]
{\includegraphics[width=0.49\linewidth]{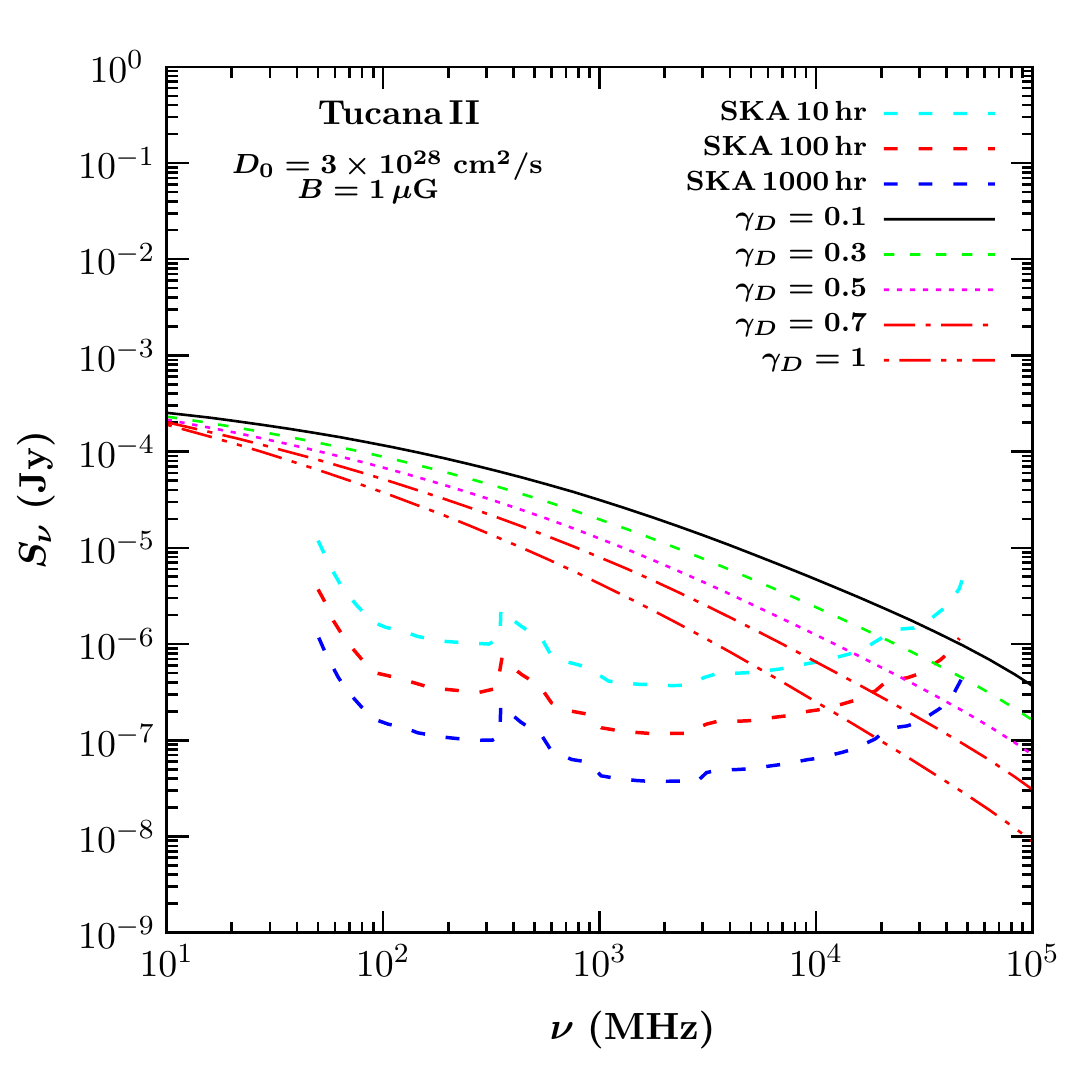}}
\label{fig:flux_different_gamma}
\caption{Variation of synchrotron flux densities in Tucana II with (a) $B$, 
(b)  $D_0$ and (c) $\gamma_D$. We have considered 200 GeV $m_{DM}$, $b\bar b$ 
final state and NFW density profile. Besides, we have fixed the parameters at 
$\langle \sigma v\rangle \, = 10^{-26}$ cm$^3$/s , $B \, = \, 1\, \mu$G, $D_0 = 
3 \times 10^{28}$ cm$^2$/s, $\gamma_D = 0.3$.}
\label{fig:flux_diff_BD0gamma}
\end{figure}

\section{Conclusions $\&$ Discussions}
\label{section:conclusion}
\noindent The UFDs, dominated by the DM content, can also possess the moderately 
large value of the magnetic field and that makes them an ideal target for the 
indirect detection of DM signals through the multiwavelength approach. In recent 
times, several literatures have tried to derive the strong limits on 
annihilation $\langle \sigma v \rangle$ from gamma-ray and radio data 
\cite{Hoof:2018hyn, DiMauro:2019frs, Fermi-LAT:2016uux, Beck:2019ukt, 
Regis:2017oet, Regis:2014tga}. For our work, we have considered the newly 
discovered UFDs detected by the observations performed by Pan-STARRS, DES and 
some other spectral survey \cite{Bhattacharjee:2020phk}.
Using both the gamma-ray (detected by Fermi-LAT) and the radio data (detected by 
VLA and GMRT), we have searched for the WIMP annihilation signal in 15 UFDs. We 
have also predicted the possible spectra associated with the radio emission and 
have checked whether it would be possible for SKA to detect any emission from 
them \cite{Bhattacharjee:2020phk}.

\noindent With eleven years of Fermi-LAT data, we have not detected any 
significant emission from the location of UFDs. Thus, we have then derived the 
upper limits on $\langle\sigma v\rangle$ as a function of DM mass for our chosen 
DM annihilation channels. We have estimated the limits for 12 UFDs. Because, for 
Triangulum II, Hydra II and Tucana III, we only have the upper limits on 
J-factor, so they could not provide any robust limits on the parameter space of 
($m_{DM}$, $\langle\sigma v\rangle$) \cite{Bhattacharjee:2020phk}. For gamma-ray data, Holorologium I 
provided the most stringent constraints but our obtained limits strongly depend 
on the distribution of DM. Using the NFW profile, we have derived most of the 
results. Besides, we have also performed a comparative study between NFW, 
Burkert and ISO profiles. In view of gamma-ray analysis, the Burkert profile 
imposed the strongest limits on $\langle \sigma v\rangle$, while the ISO imposed 
the weakest limits \cite{Bhattacharjee:2020phk}. 

\noindent In view of synchrotron emission, we have considered the radio-flux 
limits observed by GMRT and VLA and have predicted the respective $\langle\sigma 
v\rangle$ upper limits for $b\bar{b}$, $\tau^+ \tau^-$ and $\mu^+ \mu^-$ final 
states. We have compared our obtained radio limits with the limits obtained from 
gamma-ray data and found that the VLA telescope has the potential to impose more 
stringent limits than Fermi-LAT. 

\noindent We have derived the possible the synchrotron fluxes in UFDs for a wide 
range of frequencies, i.e., between 10 MHz to 100 GHz and compared these with the 
sensitivity curves of SKA. We find that for 200 GeV DM mass and $b \bar b$ final 
state, it might be possible for SKA to detect the radio emission from our 
considered UFDs, even with its 10 hours of sensitivity curve. For $\tau^+\tau^-$ 
and $\mu^+ \mu^-$ final states, the emission could be detected with the 100 
hours of exposure curve of SKA. On the other side, for comparatively heavy DM 
masses, (say $\sim$ 2 TeV), the synchrotron spectrum would become harder, and 
thus a longer observation time would be necessary to detect the radio signal. 

\noindent We also need to remember that the synchrotron fluxes have strong 
dependences on several astrophysical components, such as magnetic field, 
diffusion coefficient, distance, etc. But, due to insufficient observation, the 
values are not very precise. Thus, in order to predict the synchrotron fluxes in 
UFDs, we must have the most accurate information of the astrophysical 
parameters, especially the magnetic field and the diffusion coefficient. We have 
checked how the synchrotron flux in Tucana II varies with $B$, $D_0$ and 
$\gamma_D$ for DM mass 200 GeV and $b\bar{b}$ annihilation channel. We have 
noticed that synchrotron emission strongly depends on these. Besides, the 
emission is also controlled by the choice of DM density distribution in UFDs. We 
have found that for Tucana II, NFW density profile could produce the maximum 
amount of radio flux between all three density profiles. Our considered UFDs 
process a large uncertainties in $r_{1/2}$, $d$ and $\sigma$. The uncertainties 
in these astrophysical parameters can also affect the synchrotron emission 
arising from UFDs. We have performed the respective checks and have found that 
the largest contribution is coming from the uncertainties in $\sigma$. 

\noindent Despite the dependence on these uncertainties, we can safely conclude 
that a very intriguing aspect of indirect searching for DM signal from UFDs has 
been discussed in our study. In Fig.~8.13, we have compared the most stringent 
obtained from the VLA sky-survey with the best limits obtained from the 
Fermi-LAT data for three final states. From Fig.~8.13, we could notice that for 
$\mu^+ \mu^-$ and $\tau^+ \tau^-$ final states, VLA imposes the better limits 
that Fermi-LAT, while for $b\bar{b}$ final state Fermi-LAT provides the stronger 
limits that VLA \cite{Bhattacharjee:2020phk}.

\begin{figure}[h!]
\centering
\includegraphics[width=.49\linewidth]{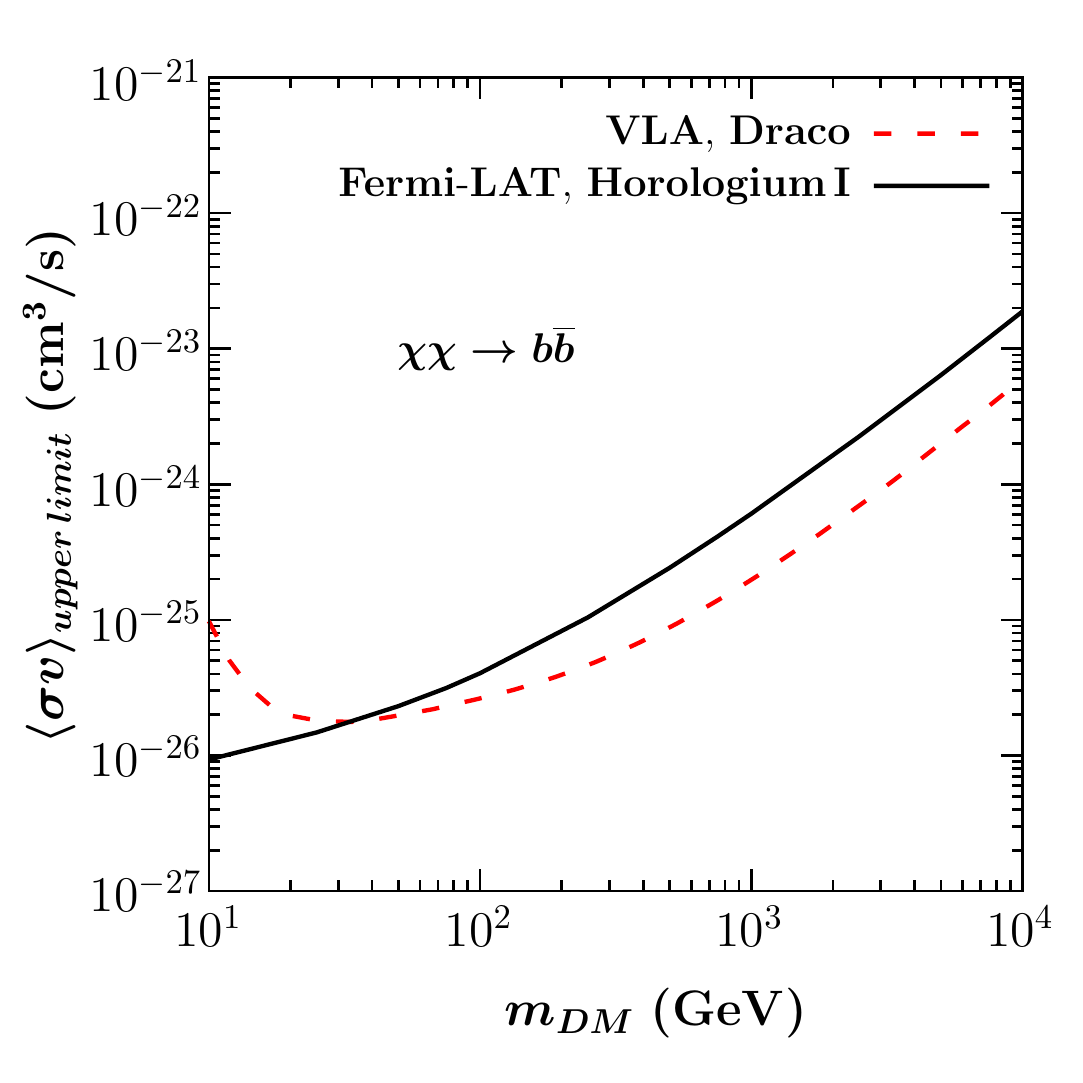}
\includegraphics[width=.49\linewidth]{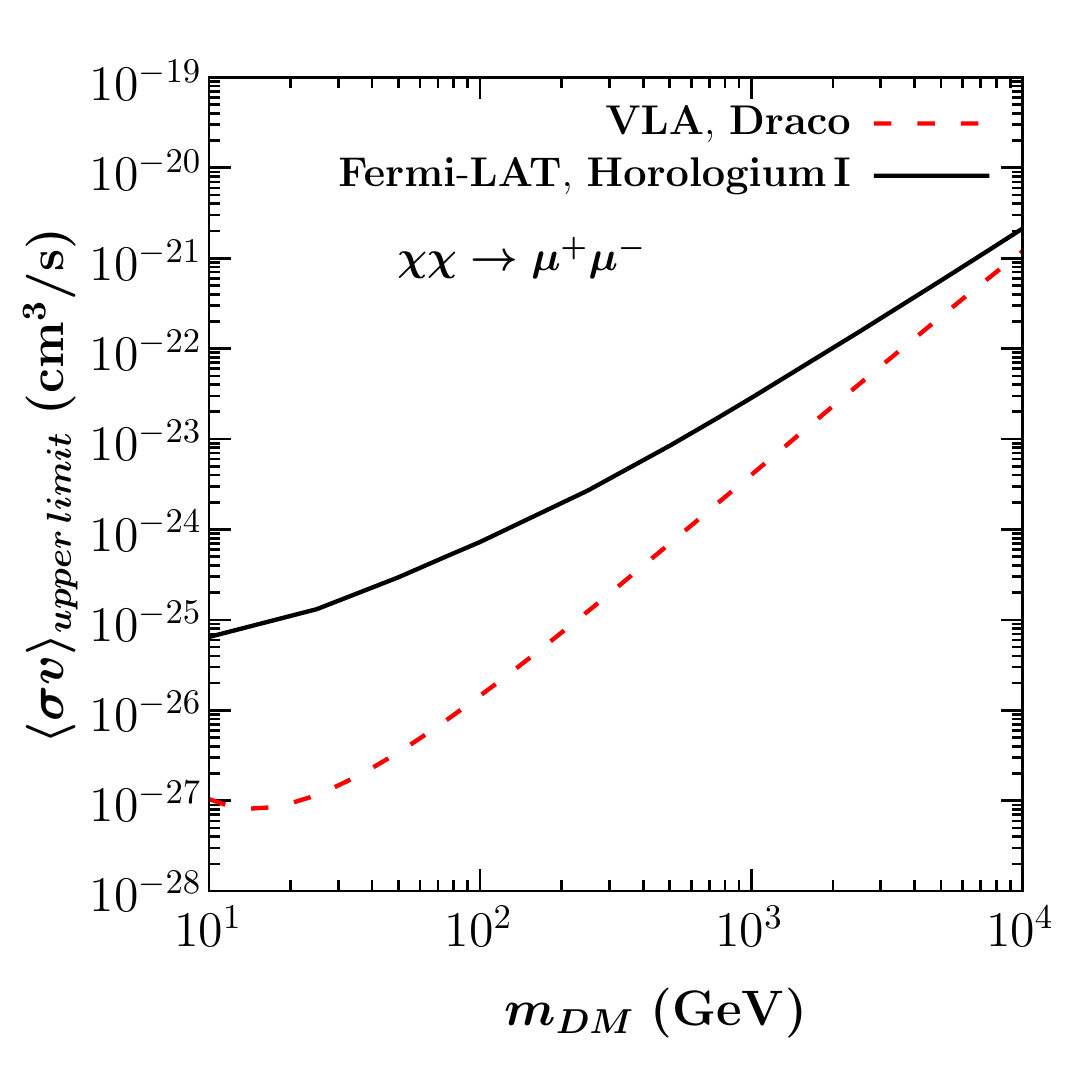}
\includegraphics[width=.49\linewidth]{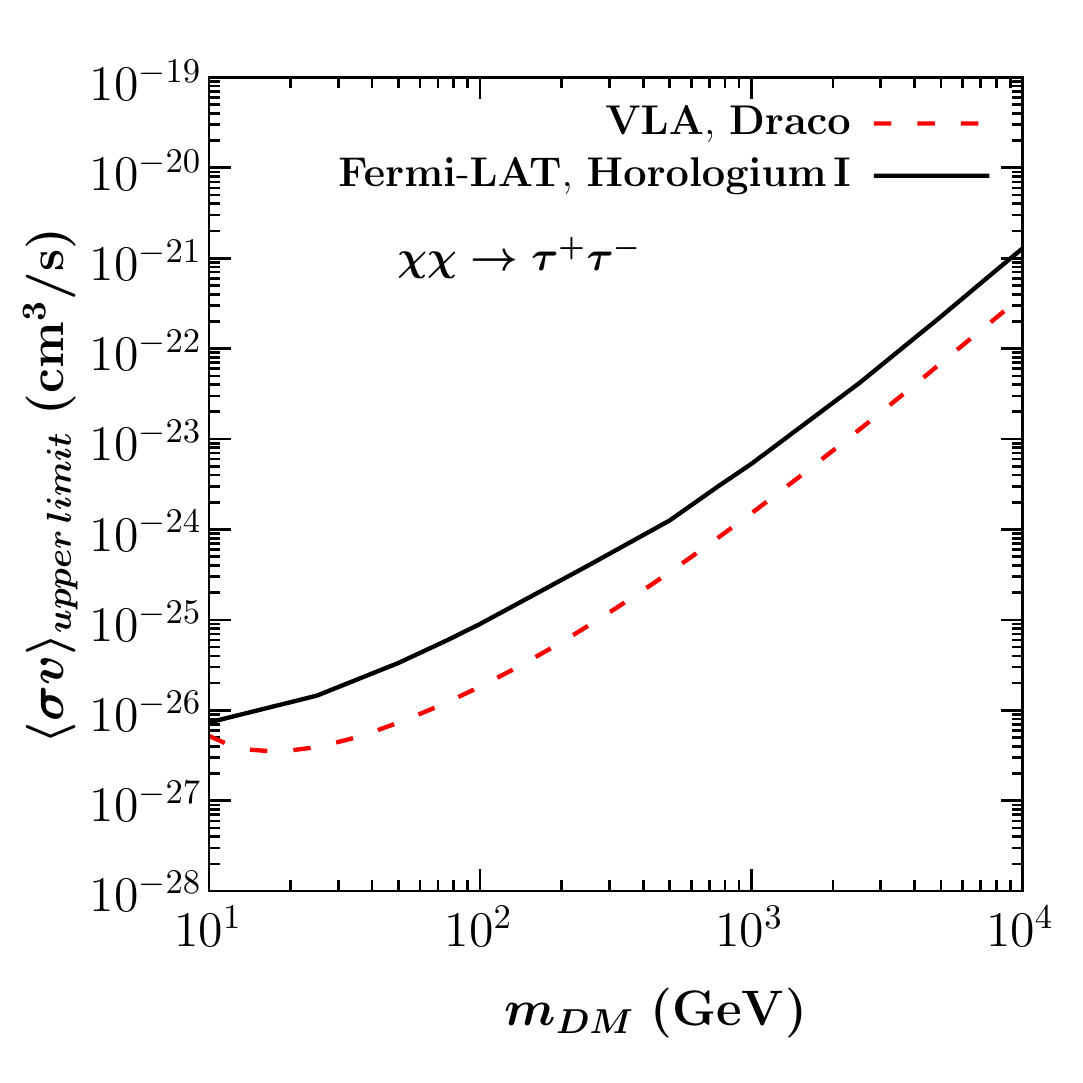}
\caption{Comparison between the 95 $\%$ C.L. $\langle \sigma v \rangle$ limits 
obtained from the VLA and the Fermi-LAT data for three annihilation channels, 
such as: $b\bar{b}$, $\tau^+\tau^-$ and $\mu^+\mu^-$. 
For comparison, we have considered the strongest radio and gamma-ray limits from 
obtained from Draco and Horologium I, respectively. We have considered NFW profile and fixed the parameters at 
$\langle \sigma v\rangle \, = 10^{-26}$ cm$^3$/s , $B \, = \, 1\, \mu$G, $D_0 = 
3 \times 10^{28}$ cm$^2$/s, $\gamma_D = 0.3$.}  
\label{figure:Exclusion_curve_VLA_Fermilat_comparison}
\end{figure}

\noindent In view of indirect DM search, we expect that the next-generation 
$\gamma$-ray telescope, CTA would play a very crucial role. CTA would have the 
deepest sensitivity for a very wide range of 
energies\cite{CTAConsortium:2018tzg} and would be able to investigate the 
thermal $\langle \sigma v \rangle$ rate from several of DM rich targets. Along 
with the CTA, in radio sky, SKA is expected to become the most sensitive radio 
telescopes in the future. Besides, Low-Frequency Array (LOFAR) such as MeerKAT 
and ASKAP would also be complementary to the CTA and SKA. We can, at best, 
expect that all of these next-generation telescopes would be able to solve 
several crucial aspects of dark matter physics.

\chapter{Discussion and Concluding Remarks}
\section{Discussion and Concluding Remarks}

\noindent Finally, in this concluding section, we want to wrap up the thesis. We have performed a detailed study on the indirect detection for DM signature that the aims to investigate the signal originating from the self-annihilation of DM candidates. The methods for targeting the DM signal is two-fold, on one hand, we explore the gamma rays resulting from DM particles. On the other hand, we focus on complementary radio properties. In the earlier chapters, we already summarized the outcomes of each work at the end of the chapter. In this following section, we briefly cover that again for the sake of completeness. Let us take a quick look back at what has been accomplished so far and what may be possible going forward.

\noindent In our work for Triangulum-II (Tri-II), we analysed nearly seven years of \texttt{Fermi-LAT} data but could not observe any $\gamma$-ray excess from its location. We then derived the upper limit of $\gamma$-ray flux for two possible scenarios, such as for $\rm{\sigma_{v}= 4.2~km~s^{-1}}$  and $\rm{\sigma_{v} = 3.4~km~s^{-1}}$. For $\rm{\sigma_{v}= 4.2~km~s^{-1}}$, Tri-II constrain the mSUGRA and the MSSM models with low thermal relic densities, whereas the limits constrain the Kaluza-Klein DM in UED and the AMSB models for masses $\lesssim 230$~GeV and $\lesssim 375$~GeV, respectively. Even for $\rm{\sigma_{v}~=~3.4~km~s^{-1}}$, Tri-II can constrain the MSSM model with low thermal relic densities and the AMSB model for masses $\lesssim 300$~GeV. Besides, from our work, we found that $\gamma$-ray data from Tri-II can even put stronger limits on the theoretical DM models than UMi. We would also like to point out that our results are entirely based on the standard NFW profile and we do not consider any effects of boost factor related to substructures in Tri-II or the Sommerfeld effect in accord to the annihilation cross-section. Finally, from our work, we can state that with more precise observations of Tri-II, in future we can establish Tri-II as a very strong DM candidate for indirect DM searching.\\

\noindent In case of Tucana-II (Tuc-II), we studied for a longer period of (nearly nine years) \textit{Fermi}-LAT data to investigate the signatures of DM annihilation. Unlike the Tri-II, we detected a very faint $\gamma$-ray excess from the location of Tuc-II for both the power-law spectra and the $\gamma$-ray spectrum from DM annihilation. We checked the variation of the gamma-ray excess with DM mass and observed that for nine years of data, TS value of Tuc-II peaks at $m_{DM}\sim$~14 GeV for $100\%$ $b\bar{b}$ annihilation channel, while for $100\%$ $\tau^{+}\tau^{-}$, it peaks at $m_{DM}\sim$~4 GeV. Apart from that, our study also confirmed the successive increase in TS peak values with increasing the time periods of data. This hints its association with any real signal either astrophysical or resulting from DM annihilation. We also produced a residual TS map for energy $>$ 500 MeV (Fig 6.11). From the residual map, we can at best conclude that the nearby excess is associated with Tuc-II location and it indicates its connection with DM annihilation signal from our Tuc-II \cite{Bhattacharjee:2018xem}. In the field of indirect DM detection, such hints of $\gamma$-ray emission from Tuc-II may open a new path in DM physics. \\

\noindent For Low Surface Brightness (LSB) galaxies, we studied for nearly nine years of LAT data but did not detect any emission from the location of LSB.  With DMFit tools, we estimated the $\gamma$-ray and $<\sigma v>$ upper limits for four annihilation channels. But because of their low J-factors, individual limits obtained from the LSB galaxies could not provide any stringent limits on the DM theoretical models. With the hope of increasing the LAT sensitivity, we then performed the joined likelihood on the set of four LSB galaxies. As expected, the stacking method improved the $<\sigma v>$ bounds by the factor of 4 than the individual limits obtained from LSB galaxies. But, the combined $<\sigma v>$ were still around two orders of magnitude weaker than the $<\sigma v>$ limits obtained from refs.~Ackermann et al.\cite{Ackermann:2015zua} and Steigman et al.\cite{Steigman:2012nb}. With the gamma-ray data for our chosen LSB galaxies, we could not particularly favour cored profile over the cuspy profile. The rotational curves for LSBs were in an agreement with the prediction from $\lambda$CDM and some study also indicated that the cuspy profile could also provide a reasonable fit to the DM distribution at the internal core. Thus, motivated by all the observational evidences, we modelled the DM distribution of LSB galaxies with the NFW profile. We also performed a comparative study between NFW, ISO and BURK DM density profiles (check Fig.~7.13) and find that the $<\sigma v>$ limits for each density profiles were overlapping with other. Thus, from our study, we could not favour one profile between all three but for the median value of J-factor, the most stringent limits would come from the NFW profile.\\

\noindent For this study, we also studied the complementary radio flux upper limits for the indirect search of the DM signal. For radio analysis, we used the RX-DMFIT tool which is an extension of DMFIt package. We estimated the multiwavelength SED plots for LSB galaxies and observed how their SED varies with the parameter sets (check Figs.~7.8 $\&$ 7.9). We surveyed the NVSS all-sky data and searched for the radio flux density for all LSB galaxies. But only the location of UGC 11707 appeared as an excess and other thee LSBs provide the upper limits to the flux density. With the VLA flux density, we have tried to predict the radio $<\sigma v>$ limits in parameter space of ($<\sigma v>$, $m_{DM}$) (check Fig.~7.10). When we considered the 2$\sigma$ uncertainty band associated with the radio limits, we noticed that the radio limits were overlapping with the limits obtained from stacking analysis for LAT data (check Fig.~7.11). Hence, from our analysis, we could, at best, comment that the radio data is competitive with the gamma-ray data. With more detailed observational data and precise analysis, in future, it might be possible for LSB galaxies to impose strong limits on DM models. We also checked whether with the next generation radio (SKA) and gamma-ray (CTA) telescopes, it would be possible to detect any emission from the location of LSB galaxies. We noticed (check 7.12) that SKA might be able to detect the emission from the location of LSB galaxies and its 1000 hours of observation would have the highest possibility to detect the emission from LSBs. But we would also like mention that in order to claim that SKA would detect the emission from DM annihilation, we first need to perform a simulation study. Besides, the estimated radio emission is also dependent on the various astrophysical scenario. We need to have a well-defined knowledge on the distribution of diffusion zone, magnetic fields, DM density profile, etc.. Hence, from our analysis, we could, at best, hint the possibility of observing the radio signal from LSB galaxies by SKA. We also found (Fig.~7.14) that for energy ranges between 100 GeV to 1 TeV, it might be possible for CTA to observe the $\gamma$-ray emission with the 50 hours of sensitivity curve. But like SKA, the same conclusion also holds for CTA. A simulation study is needed to examine whether it would be possible for CTA to detect the emission resulting from the DM annihilation/decay. From our work, we can ultimately conclude that the $\gamma$-ray data obtained from the Fermi-LAT could not impose the strong $<\sigma v>$ limits on the WIMP models. We found that the radio signal possibly originated from WIMP annihilation is quite competitive with the $\gamma$-ray emission observed by the Fermi-LAT. Our analysis, at best, indicates that to study $\gamma$-ray and radio signal from the LSB galaxies, SKA and CTA would play a very significant role in future. \\

\noindent In our last chapter, we considered the newly discovered UFDs detected by the observations performed by Pan-STARRS, DES and some other spectral survey \cite{Bhattacharjee:2020phk}. In recent times, several literatures tried to derive the strong limits on annihilation $\langle \sigma v \rangle$ from gamma-ray and radio data \cite{Hoof:2018hyn, DiMauro:2019frs, Fermi-LAT:2016uux, Beck:2019ukt, Regis:2017oet, Regis:2014tga}. Using both the gamma-ray (detected by Fermi-LAT) and the radio data (detected by VLA and GMRT), we searched for the WIMP annihilation signal in 15 UFDs. We also predicted the possible spectra associated with the radio emission and checked whether it would be possible for SKA to detect any emission from them \cite{Bhattacharjee:2020phk}. With eleven years of Fermi-LAT data, we did not detect any significant emission from the location of UFDs. Thus, we then derived the upper limits on $\langle\sigma v\rangle$ as a function of DM mass for our chosen DM annihilation channels. We estimated the limits for 12 UFDs. Because, for Triangulum-II, Hydra-II and Tucana-III, we only have the upper limits on J-factor, so they could not provide any robust limits on the parameter space of ($m_{DM}$, $\langle\sigma v\rangle$) \cite{Bhattacharjee:2020phk}. For gamma-ray data, Holorologium I provided the most stringent constraints but our obtained limits strongly depend on the distribution of DM. Using the NFW profile, we derived most of the results. Besides, we also performed a comparative study between NFW, Burkert and ISO profiles. In view of gamma-ray analysis, the Burkert profile imposed the strongest limits on $\langle \sigma v\rangle$, while the ISO imposed the weakest limits \cite{Bhattacharjee:2020phk}. \\

\noindent In view of synchrotron emission, we considered the radio-flux limits observed by GMRT and VLA and predicted the respective $\langle\sigma v\rangle$ upper limits for $b\bar{b}$, $\tau^+ \tau^-$ and $\mu^+ \mu^-$ final states. We compared our obtained radio limits with the limits obtained from gamma-ray data and found that the VLA telescope has the potential to impose more stringent limits than Fermi-LAT. We have derived the possible the synchrotron fluxes in UFDs for a wide range of frequencies, i.e., between 10 MHz to 100 GHz and compared these with the sensitivity curves of SKA. We found that for 200 GeV DM mass and $b \bar b$ final state, it might be possible for SKA to detect the radio emission from our considered UFDs, even with its 10 hours of sensitivity curve. For $\tau^+\tau^-$ and $\mu^+ \mu^-$ final states, the emission could be detected with the 100 hours of exposure curve of SKA. On the other side, for comparatively heavy DM masses, (say $\sim$ 2 TeV), the synchrotron spectrum would become harder, and thus a longer observation time would be necessary to detect the radio signal. We also need to remember that the synchrotron fluxes have strong dependences on several astrophysical components, such as magnetic field, diffusion coefficient, distance, etc. But, due to insufficient observation, the values are not very precise. Thus, in order to predict the synchrotron fluxes in UFDs, we must have the most accurate information of the astrophysical parameters, especially the magnetic field and the diffusion coefficient. We checked how the synchrotron flux in Tucana-II varies with $B$, $D_0$ and $\gamma_D$ for DM mass 200 GeV and $b\bar{b}$ annihilation channel. We noticed that synchrotron emission strongly depends on these. Besides, the emission is also controlled by the choice of DM density distribution in UFDs. We found that for Tucana II, NFW density profile could produce the maximum amount of radio flux between all three density profiles. Our considered UFDs process a large uncertainties in $r_{1/2}$, $d$ and $\sigma$. The uncertainties in these astrophysical parameters can also affect the synchrotron emission arising from UFDs. We performed the respective checks and found that the largest contribution was coming from the uncertainties in $\sigma$. \\

\noindent Despite the dependence on these uncertainties, we can safely conclude that a very intriguing aspect of indirect searching for DM signal from UFDs has been discussed in our study. In Fig.~8.13, we compared the most stringent obtained from the VLA sky-survey with the best limits obtained from the Fermi-LAT data for three final states. From Fig.~8.13, we could notice that for $\mu^+ \mu^-$ and $\tau^+ \tau^-$ final states, VLA imposed the better limits that Fermi-LAT, while for $b\bar{b}$ final state Fermi-LAT provided the stronger limits that VLA \cite{Bhattacharjee:2020phk}. In view of indirect DM search, we expect that the next-generation $\gamma$-ray telescope, CTA would play a very crucial role. CTA would have the deepest sensitivity for a very wide range of energies\cite{CTAConsortium:2018tzg} and would be able to investigate the thermal $\langle \sigma v \rangle$ rate from several of DM rich targets. Along with the CTA, in radio sky, SKA is expected to become the most sensitive radio telescopes in the future. Besides, Low-Frequency Array (LOFAR) such as MeerKAT and ASKAP would also be complementary to the CTA and SKA. We can, at best, expect that all of these next-generation telescopes would be able to solve several crucial aspects of dark matter physics.

\chapter{Appendix}
\section*{A: T-TEST for Unequal Variance}
                                                 
\noindent T-TEST is the statistical hypothesis test that is generally applied to 
check whether any significant deviation lies between two populations 
\cite{student:1908df, Miodrag:2011gh, Kim:2015gh}. When we specially deal with 
the small set of data, e.g. $n_{1}$ or/and $n_{2}$ $<$ 30, T-test is 
favoured\cite{student:1908df, Miodrag:2011gh, Kim:2015gh}. The shape of the 
T-TEST distribution is very similar to the Gaussian distribution and for 
implying the T-TEST, the variables of each sample must be drawn from the 
Gaussian distribution.
 
\noindent To obtain the output of T-TEST statistics, such as t-value and degree 
of freedom (d.o.f), we need to provide the values of mean, standard deviation 
and number of counts of each sample as inputs. This t-value is also defined as 
the test statistics value which is derived from the two-sample dataset at the 
time of performing the hypothesis test for T-TEST.  \\
 
\noindent Depending on the standard deviation values from two samples, there are 
generally three types of T-TEST hypothesis check. For our analysis \cite{Bhattacharjee:2018xem}, 
(ref.~Chapter~6), we have performed the T-TEST on two independent samples and 
they have the unequal variance. 
Thus, we have considered the T-TEST for the unequal variance which is also known 
as the Welch's T-TEST\cite{Welch:1947df, Ruxton:2006dh}. The Welch's T-TEST is 
generally favoured when two samples have different values of variance and their 
sample size could be same or not. In that case, the formula for evaluating 
t-value and d.o.f are:\\

\begin{equation}
\rm{t-value} = 
\frac{mean_{1}-mean_{2}}{\sqrt{\frac{(var_{1})^{2}}{n_{1}}+\frac{(var_{2})^{2}}{
n_{2}}}}
\end{equation}

\begin{equation}
\rm{d.o.f.} = 
\frac{\Big(\frac{(var_{1})^{2}}{n_{1}}+\frac{(var_{2})^{2}}{n_{2}}\Big)^{2}}{
\frac{(\frac{var_{1}^{2}}{n_{1}})^{2}}{n_{1}-1}+\frac{(\frac{var_{2}^{2}}{n_{2}}
)^{2}}{n_{2}-1}} ,
\end{equation}

\noindent where,\\
$\rm{mean_{1}}$ and $\rm{mean_{2}}$ are the mean values of $\rm{sample_{1}}$ and 
$\rm{sample_{2}}$, respectively. \\
$\rm{var_{1}}$ and $\rm{var_{2}}$ are the variance of $\rm{sample_{1}}$ and 
$\rm{sample_{2}}$, respectively. \\
$\rm{n_{1}}$ and $\rm{n_{2}}$ are the number of counts in $\rm{sample_{1}}$ and 
$\rm{sample_{2}}$, respectively.  \\

\noindent Once we obtain the t-value, we can then derive the probability i.e., 
the p-value by following the two-tailed t-distribution curve. We also have to 
assign the significance level i.e., $\alpha$ and for our purpose, we have used 
the $\alpha$ = 5$\%$. Now, from the p-value, we can determine whether our two 
samples agree with the null-hypothesis. The p-value lies between 0 and 1. The 
small p-value i.e., p$\le$ 0.05 indicates that we might reject the 
null-hypothesis, while the large p-value i.e., p$>$ 
0.05 hints that it might not be possible to reject the null-hypothesis.\\

\noindent For Fig.~6.3(b), we find that for both the full energy residual 
spectrum and the positive bump for energy above 500 MeV, we obtain the p-values 
$>$ 0.05. This indicates that for both cases, we are not able to reject the 
null-hypothesis \cite{Bhattacharjee:2018xem}. 
Thus, from our analysis, we can, at best, conclude that the DM annihilation 
spectra can provide an acceptable fit to the residual energy-spectrum of Tuc-II \cite{Bhattacharjee:2018xem}.
                                                                               
\section*{B: Normality Test of Dataset}
 
\noindent As we already have mentioned in the above section, we can only perform 
the T-TEST hypothesis test, if both the sample set follow the Gaussian 
distribution \cite{student:1908df, Miodrag:2011gh, Kim:2015gh}. Thus, we have 
tried to check whether our sample data from Fig~6.3(b) follow the normal distribution \cite{Bhattacharjee:2018xem}. To check 
the normality of the sample dataset, there are various statistical tests such as 
Kolmogorov–Smirnov test \cite{Kolmogorov:1933gh, Smirnov:1948ff}, Shapiro–Wilk 
\cite{Shapiro:1695hd}, normal quantile plot\cite{loy:2015fh, wilk:1968fb},etc. 
For our analysis, we have generated the quantile-quantile (Q-Q) plot. The Q-Q 
plot is the graphical representation that helps to check whether the dataset 
from two samples originate from the same population which follow a common 
distribution \cite{loy:2015fh, wilk:1968fb}.  \\

\noindent In order to check the normality of the sample, this Q-Q plot shows the 
quantiles of our dataset versus the quantiles values of an ideal Gaussian 
distribution \cite{loy:2015fh, wilk:1968fb}. The quantile values obtained the 
theoretical Gaussian distribution are plotted on the horizontal axis, while the 
quantile values obtained from our samples are plotted on the y-axis. If our 
sample dataset follows the Gaussian distribution, from Q-Q plot we should obtain 
a straight line which indicates the correlation between our sample data and the 
theoretical data from Gaussian distribution \cite{loy:2015fh, wilk:1968fb}. In 
order to find the exact correlation between the paired dataset used for Q-Q 
plot, they are fitted with the regression equation (y=ax) and that fitting would 
return to the value of the coefficient of determination ($R^{2}$). Once we 
obtain the value of $R^{2}$, we can then calculate the Pearson correlation 
coefficient (r), i.e., r=$(R^{2})^{1/2}$=R \cite{asuero:2006cju, schober:2018fg}. 
Ideally, the value of r should lie between 0.9 to 1 which indicates the high 
correlation between our sample set and the Gaussian distribution 
\cite{asuero:2006cju, schober:2018fg}. The r-value close to the 1 indicates that 
the sample set has very less deviation from the normality, while r-value close 
to the 0 denotes the large deviation from normality \cite{asuero:2006cju, 
schober:2018fg}.\\

\noindent It is true that no experimental dataset would have the r-value = 1 but 
for statistical hypothesis check, the dataset should roughly follow the Gaussian 
distributed i.e., r-value should be $>$ 0.9. For our study (Chapter~6), we have 
produced the Q-Q plot and have evaluated the respective r-value for the dataset 
that we have considered for the T-TEST \cite{Bhattacharjee:2018xem}. We find that the residual energy 
spectrum for both i) full energy range and ii) energy range $>$ 500 MeV, 
produce a straight line in the Q-Q plots and their corresponding r-value lies 
$>$ 0.94 \cite{Bhattacharjee:2018xem}. Thus, from this test, we can find that our sample has some deviation 
from the normality but the r-value $>$ 0.94 indicates that we can safely use our 
sample set for checking the T-TEST goodness of fitting \cite{Bhattacharjee:2018xem}.

\bibliographystyle{unsrt}
\bibliography{ref}


\end{document}